%% file: thesis.tex
\documentclass[12pt,preprint]{aastex}
\usepackage{natbib}
\usepackage{epsfig}

\usepackage{times}

\catcode`\@=11
\newcommand{\gapprox}{\mathrel{\mathpalette\@versim>}}
\newcommand{\lapprox}{\mathrel{\mathpalette\@versim<}}
\newcommand{\propapprox}{\mathrel{\mathpalette\@versim\propto}}
\newcommand{\@versim}[2]
  {\lower3.1truept\vbox{\baselineskip0pt\lineskip0.5truept
\ialign{$\m@th#1\hfil##\hfil$\crcr#2\crcr\sim\crcr}}}

\def\spitzer{\:{\it Spitzer}}
\def\hubble{\:{\it Hubble}}
\def\chandra{\:{\it Chandra}}

\def\gsim{\mathrel{\rlap{\lower4pt\hbox{\hskip1pt$\sim$}}
    \raise1pt\hbox{$>$}}}                

\newcommand{\msun}{M_\odot}
\newcommand{\rsun}{R_\odot}
\newcommand{\lsun}{L_\odot}
\newcommand{\kms}{km s$^{-1}$}

\begin{document}

\title{Supernova Remnants as a Probe of Dust Grains in the Interstellar Medium}

\author{Brian J. Williams, \altaffilmark{1}
}

\altaffiltext{1}{Physics Dept., North Carolina State U.,
    Raleigh, NC 27695-8202; bjwilli2@ncsu.edu}

\begin{abstract}

The Ph.D. Thesis of Brian J. Williams, submitted to and
accepted by the Graduate School of North Carolina State
University. Under the direction of Stephen P. Reynolds and Kazimierz
J. Borkowski.\\

Interstellar dust grains play a crucial role in the evolution of the
galactic interstellar medium (ISM). Despite its importance, however,
dust remains poorly understood in terms of its origin, composition,
and abundance throughout the universe. Supernova remnants (SNRs)
provide a laboratory for studying the evolution of dust grains, as
they are one of the only environments in the universe where it is
possible to observe grains being both created and destroyed. SNRs
exhibit collisionally heated dust, allowing dust to serve as a
diagnostic both for grain physics and for the plasma conditions in the
SNR. I present theoretical models of collisionally heated dust which
calculate grain emission as well as destruction rates. In these
models, I incorporate physics such as nonthermal sputtering caused by
grain motions through the gas, a more realistic approach to sputtering
for small grains, and arbitrary grain compositions porous and
composite grains. I apply these models to infrared and X-ray
observations of Kepler's supernova and the Cygnus Loop in the galaxy,
and SNRs 0509-67.5, 0519-69.0, and 0540-69.3 in the LMC. X-ray
observations characterize the hot plasma while IR observations
constrain grain properties and destruction rates. Such a
multi-wavelength approach is crucial for a complete understanding of
gas and dust interaction and evolution. Modeling of both X-ray and IR
spectra allows disentangling of parameters such as pre and postshock
gas density, as well as swept-up masses of gas and dust, and can
provide constraints on the shock compression ratio. Observations also
show that the dust-to-gas mass ratio in the ISM is lower by a factor
of several than what is inferred by extinction studies of
starlight. Future observatories, such as the James Webb Space
Telescope and the International X-ray Observatory, will allow testing
of models far beyond what is possible now.

\end{abstract}

\section{Introduction to Supernovae and Supernova Remnants}

Supernova explosions are among the most energetic events in the
universe since the Big Bang, releasing more energy ($\sim
10^{51}-10^{53}$ ergs) than the Sun will release over its entire
lifetime. They are the cataclysmic ends of certain types of stars, and
are responsible for seeding the universe with the material necessary
to form other stars, planets, and life itself. We owe our very
existence to generations of stars that lived and died billions of
years ago, before the formation of our Sun and solar system. The
processes of stellar evolution continue to occur today, with typical
galaxies like the Milky Way hosting several supernovae per century, on
average. The study of supernovae and the role they play in shaping the
evolution of star systems and galaxies is truly an exploration of our
own origins. Supernova remnants (SNRs), the expanding clouds of
material that remain after the explosion, spread elements over volumes
of thousands of cubic light-years, and heat the interstellar medium
through fast shock waves generated by the ejecta from the star.

\subsection{Stellar Evolution}

Any study of SNRs must begin with the processes which cause a star to
go supernova. Stars come in all sizes and colors (where the color of a
star is related to its temperature), but virtually all stars live out
their lives in a similar fashion. They spend the majority of their
lives fusing hydrogen in their cores into helium, a process which
releases energy. This process is not particularly efficient in stars;
nevertheless, the sheer mass of available material to burn means that
stars shine at a roughly constant brightness for millions, billions,
even tens of billions of years. The life expectancy of a star is a
rather sensitive function of its initial mass. This relationship is
perhaps counterintuitively inverted, such that the more massive a star
is, the shorter its lifetime. (This is due to the fact that massive
stars, while having much more fuel to burn, fuse it at a much faster
rate than do low-mass stars). Stars that are on the hydrogen burning
phase of their lives are said to be on the ``main sequence,'' a
reference to the Hertzsprung-Russell diagram seen in Figure 1.1.

This fusion process that takes place in the stellar core is one-half
of a balancing act, providing the internal pressure to support the
star against the other half, gravity. This cosmic ``tug-of-war'' can
only last as long as the star has fuel to burn; once the hydrogen fuel
in the core is depleted, gravity will pull the star in on itself
unless a new process can provide the necessary counter-balancing
force. The fate of a star at this point depends on its mass, with high
and low-mass stars following very different paths.

\subsubsection{Low-Mass Stars}

Stars that begin their main-sequence lives with a mass of less than
about 8 solar masses ($M_\odot$, where $M_\odot$ is $\sim 1.99 \times
10^{33}$ grams) will spend the majority of their lives in the hydrogen
burning phase. When their hydrogen runs out, they will swell into red
giants, increasing in volume by a factor of $>$ 1000. (The smallest of
stars, with masses $<$ 0.5 $M_\odot$, will not even have enough power
to reach this stage). The red giant phase is typically characterized
by a helium core surrounded by a hydrogen burning shell. When the core
contracts and heats to temperatures above $\sim 10^{8}$ K, helium
burning will begin, fusing helium to carbon via the triple-alpha
process. The core of the star that remains typically cannot fuse much
beyond carbon and oxygen, and collapses further when the helium supply
is used up. The outer layers of the star are ejected, creating the
misleadingly named ``planetary nebula.'' Collapse continues until the
degenerate pressure of electrons in the plasma is sufficient to
balance the gravitational forces, creating a ``white dwarf'' star. A
typical white dwarf has a mass of $\sim 0.6-0.7 M_\odot$ contained in
a volume about the size of the Earth (R $\sim 6000$ km), which leads
to an average density for a white dwarf of $\sim 10^{6}$ g
cm$^{-3}$. White dwarfs are typically very stable (although see
section 1.2), and will exist as burned-out remnants of once bright
stars indefinitely.

\subsubsection{High-Mass Stars}

Stars above $\sim 8$ $M_\odot$ are hot enough at their cores to fuse
hydrogen into helium, helium into carbon, carbon into oxygen, neon,
silicon, sulfur and other elements, stepping up the periodic
table. Once iron (element 26) is reached, though, it no longer becomes
energetically favorable for the fusion process to continue. The iron
core collapses in upon itself on timescales of the order of a second,
sidestepping electron degeneracy pressure by eliminating electrons,
combining them with protons to form neutrons. Once nuclear density
($\sim 10^{15}$ g cm$^{-3}$) is reached, the degeneracy pressure of
neutrons is sufficient to halt the collapse, and the core becomes a
proto-neutron star. The remaining infalling material from the core
bounces off of this now hard core, ejecting the outer layers of the
star in a fantastic explosion known as a ``core-collapse supernova.''
The proto-neutron star forms a neutron star, a stellar remnant of
order 1 $M_\odot$ and R $\sim 10$ km, with an average density of $\sim
10^{15}$ g cm$^{-3}$. For highly massive stars, it is possible that
even neutron degeneracy pressure cannot halt the collapse of the core,
and a stellar-mass black hole is formed.

\subsection{Supernova Classification} 

As in many sciences, observations of events or objects in astronomy
often precede theoretical explanations for said events. However,
unlike in most disciplines, astronomers normally do not have the means
to conduct laboratory tests of observed phenomena. This often leads to
a significant time delay between the observation of and the
theoretical description of a given event. As a result, the field is
riddled with examples of ``historical inaccuracies'' when it comes to
naming and classification schemes. A prime example of this is the
supernova classification scheme, where astronomers classify supernovae
as ``Type I'' or ``Type II'' based solely on the existence of hydrogen
lines in their spectra; Type I SNe show no hydrogen lines, Type II
do. It was only later realized that vastly different processes can be
responsible for this bit of observational data.

\subsubsection{Type Ia}

The vast majority of stars in the galaxy are $< 8 M_\odot$, meaning
(see Section 1.1.1) that they are destined to end their lives as white
dwarfs. However, many stars in the galaxy are also part of binary (or
even triple) systems. If a white dwarf is contained in a binary system
with a companion star that enters its red giant phase (or has even
slightly evolved off the main sequence), and the separation between
the stars is sufficiently close, the white dwarf can gravitationally
strip matter from its companion. Mass transfer occurs between the two
stars, with the white dwarf growing in mass via an accretion
disk. White dwarfs can only exist up to the ``Chandrasekhar limit,''
and a white dwarf pushed over this limit ($\sim 1.4 M_\odot$) will
become unstable, igniting a deflagration, or subsonic burning, of
material in the star. This deflagration will lead to a detonation, a
supersonic thermonuclear explosion of the entire white dwarf. The
entire 1.4 $M_\odot$ of material is ejected into the surrounding
medium at speeds of $\sim 10,000$ km s$^{-1}$, releasing about
10$^{51}$ ergs of kinetic energy. Type Ia SNe leave behind no compact
remnant, and their light curve, i.e. the brightness of the supernova
as a function of time, is primarily powered by the radioactive decay
chain of nickel-56 to cobalt-56 to iron-56. Their light curves peak a
few days after explosion, then slowly fade over the course of a few
months. Since all white dwarfs are thought to explode at an identical
mass via the same mechanism, their light curves are quite similar in
peak brightness, and can be used as a ``standard candle'' to determine
extragalactic distances. Type Ia spectra show no hydrogen because
white dwarfs are made mostly of carbon and oxygen, nearly all of which
is burned to nickel in the explosion.

\subsubsection{Type II}

Type II supernovae result from the deaths of massive stars, greater
than 8 $M_\odot$, but generally not more than $\sim 25-30
M_\odot$. These stars end their lives as core-collapse SNe (CCSNe),
ejecting their outer layers (5-25 $M_\odot$) at a velocity of $\sim
5,000-10,000$ km s$^{-1}$. Coincidentally, they yield roughly the same
amount of kinetic energy ($\sim 10^{51}$ ergs) as do type Ia SNe, but
their overall energetics are much greater. Ninety-nine percent of the
energy released in a CCSN is carried off by neutrinos. Type II SNe
leave behind a neutron star that is typically $\sim 1.4 M_\odot$ with
an initial temperature (kT) of several MeV. Type II SNe can be further
divided into subclasses based either on the shape of the light curve
or the spectrum (e.g. IIP, IIL, IIb, IIn, etc.). They show hydrogen
lines in their spectra because the outer atmosphere of the star at the
time of explosion still contained a significant amount of hydrogen.

\subsubsection{Type Ib and Ic}

Type Ib and Ic SNe are also the result of a core-collapse event,
differing from type II in that they do not show hydrogen lines in
their spectra. The physical reason behind this is that type Ib and Ic
SNe result from stars that have shed their outer layers via stellar
winds (``Wolf-Rayet'' stars) or gravitational interaction with a
companion star prior to explosion. Type Ib explosions result from
stars that have lost only their hydrogen envelope; type Ic from stars
that have lost both their hydrogen and helium envelopes. Type Ib and
Ic SNe are believed to be the result of stars with a progenitor mass
of $\gsim 30 M_\odot$, although this number could be lower in binary
systems. They leave behind neutron stars and stellar-mass black
holes. As with type II SNe, most of their energy is carried off in
neutrinos, and their light-curves, both in peak brightness and in
shape, can vary greatly.

\subsection{Astrophysical Shocks}

Shock waves, propagating supersonic disturbances, occur commonly in
all sorts of astro-environments throughout the universe, where
conditions are unlike those found on Earth. Densities in the
interstellar medium (ISM) are on the order of a few particles per
cubic centimeter, six orders of magnitude less than can be produced in
the best laboratory vacuum systems. Sound speeds in the ISM are
usually on the order of a few km s$^{-1}$, relatively slow by
astrophysical standards. The shock waves generated by a supernova are
an excellent example of a strong shock, with shock speeds often being
several thousand times the speed of sound in the ISM.  These shock
waves compress, sweep, and heat interstellar material. In fact,
supernova shock waves are one of the main sources of heating of the
interstellar medium, as well as the mechanism for distributing
material throughout the universe. 

The following is a mathematical description of a shock wave, beginning
with the Rankine-Hugoniot Conditions, given by

\begin{equation}
\rho_1v_1=\rho_2v_2,
\end{equation}

\begin{equation}
\rho_1v_1^2+p_1=\rho_2v_2^2+p_2,
\end{equation}

\begin{equation}
\frac{1}{2}v_1^2+E_1+\frac{p_1}{\rho_1}=\frac{1}{2}v_2^2+E_2+\frac{p_2}{\rho_2},
\end{equation}

\noindent
where $\rho$ is the density of the gas, $v$ is the velocity, $p$ is
the thermal pressure, and $E$ is the internal energy, given by
$dW=dE=-p$d$V$. Subscript 1 denotes pre-shock material (that is,
ambient interstellar material that is ahead of the expanding shock
wave) and 2 denotes post-shock gas. These equations are written in the
frame of reference of the shockwave (see Figure 1.2).  In particular,
this means that $v_1$ is not zero (in fact, $v_1 = -v_{shock}$) even
though there is no motion in front of the shock in the observer's
frame.

Equation (1) is the conservation of mass across the shock, while
Equation (2) shows that the sum of the ram pressure, $\rho v^2$, and
thermal pressure $p$ must be equal across the shock. This amounts to a
conservation of momentum. Equation (3) is the conservation of energy
(kinetic, internal, and thermal).

The volume of the gas is given by $V$, such that
d$V=$d$\frac{1}{\rho}$. Thus
$dE=-p$d$V=-K\rho^{\gamma}$d$\frac{1}{\rho}$, which can be integrated
to obtain $E=\frac{1}{\gamma-1}\frac{p}{\rho}$. The thermodynamic
relation between pressure and density is $p=K\rho^{\gamma}$ where $K$
is a constant (although it is not constant across the shock) and
$\gamma$ is the polytropic index of the gas. $\gamma$ varies depending
on the properties of the fluid and is given by: \\ $\gamma=5/3$ for
non-relativistic monoatomic gas (as is generally found in the ISM),\\
$\gamma=4/3$ for relativistic monoatomic gas,\\ $\gamma=7/5$ for
non-relativistic diatomic gas.\\

\noindent
Using this, eq. (3) becomes

\begin{equation}
\frac{1}{2}v_1^2+\frac{1}{\gamma-1}\frac{p_1}{\rho_1}=\frac{1}{2}v_2^2+\frac{1}{\gamma-1}\frac{p_2}{\rho_2}
\end{equation}

\noindent
Using the Rankine-Hugoniot conditions, it can be shown that

\begin{equation}
\frac{\rho_2}{\rho_1}=\frac{(\gamma+1)p_2+(\gamma-1)p_1}{(\gamma+1)p_1+(\gamma-1)p_2}=\frac{v_1}{v_2}.
\end{equation}

This is the shock jump condition for density, giving the {\bf
compression ratio}, which relates density (and thus fluid velocity)
ahead of the shock and behind the shock. In the limit of strong
shocks, that is, $p_2 \gg p_1$, or $M \gg 1$, where $M$ is the Mach
number (defined as $v_{shock}/v_{sound}$), we can neglect the
pre-shock pressure so that

\begin{equation}
\frac{\rho_2}{\rho_1}=\frac{\gamma+1}{\gamma-1}=4
\end{equation}

\noindent
for $\gamma=5/3$. In the limit of a strong shock, we also have

\begin{equation}
\frac{p_2}{p_1}=\frac{2\rho_1v_1^2}{p_1(\gamma+1)},
\end{equation}

\noindent
or

\begin{equation}
p_2=\frac{2\rho_1v_1^2}{\gamma+1}.
\end{equation}

\noindent
The gas is also governed by the ideal gas law such that
$p_{2}=n_{2}kT_{2}$, where k is Boltzmann's constant. So

\begin{equation}
p_2=\frac{2\rho_1v_1^2}{\gamma+1}=\frac{3}{4}\rho_1v_1^2=n_{2}kT_{2}, 
\end{equation}

\noindent
where $n=\frac{\rho}{m}$ = total particle number density, and $m$ is
the mean mass per particle. Thus,

\begin{equation} 
p_2=\frac{2\rho_1v_1^2}{\gamma+1}=\frac{3}{4}\rho_1v_1^2=\frac{\rho_{2}}{\mu
m_p}kT_{2},
\end{equation}

\noindent 
where $m_p$ is the mass of a proton, equal to $1.67 \times 10^{-24}$
grams. Further simplifying, we have

\begin{equation}
kT_{2}=\frac{3}{16} \mu m_pv_s^2,
\end{equation}

\noindent
where $\mu m_{p}$ is the mean mass per particle behind the shock (for
a fully ionized plasma of cosmic abundances, $\mu \sim 0.6$). This is
the temperature behind a shock, where $v_{s} = v_1$ is the shock
speed. $kT_{2}$ is the ``shock temperature,'' which is the average of
proton and electron temperatures. In the absence of heating of
electrons at the shock (and assuming no sharing of energy between
protons and alpha particles), the initial temperature ratio between
protons and electrons, $T_{p}/T_{e}$, is just the ratio of the masses
of protons and electrons, $m_{p}/m_{e}$ = 1836. Behind the shock,
these temperatures equilibrate to bring down $T_{p}$ and bring up
$T_{e}$, but the timescale for equilibration is long, and is a
function of the post-shock temperature and density. Observationally,
it is known \citep{ghavamian01} that an inverse correlation exists
between shock speed and degree of equilibration of protons and
electrons at the shock, with near full equilibration seen in old
remnants like the Cygnus Loop ($v_{s}$ $\sim 300-400$ km s$^{-1}$) and
little equilibration seen in younger remnants like Tycho ($v_{s}$
$\sim 2000$ km s$^{-1}$).

For supernova shock waves of order a few thousand km s$^{-1}$, this
shock temperature will be of order 10 million K. Thus, shocked gas
will radiate thermal emission at X-ray energies, however, other
emission mechanisms also operate in SNRs. Non-thermal emission from
synchrotron radiation is often seen near the forward edge of the shock
wave in young SNRs, and line emission from ionized elements is common
in the shocked ejecta. Shocks in SNRs are ``collisionless,'' in that
collisions between particles are extremely rare, meaning that particle
interactions are mediated by magnetic fields. This is a good
approximation when the Coulomb mean free path is much greater than the
gyroradius of the thermal particles.

\subsection{Supernova Remnants}

The expanding material ejected from the star, rich in heavy elements
like oxygen, silicon, and iron, as well as the shock wave which it
drives into the ISM is known as a supernova remnant (SNR). Figure 1.3
shows Cassiopeia A, an example of a young ($\sim 330$ yrs) SNR in our
own galaxy. Although Cassiopeia A is known to have resulted from a
CCSN, it is generally difficult to tell the type of SN only by looking
at the remnant. SNRs remain visible for thousands, often tens or
hundreds of thousands of years before dissipating their energy into
the ISM. The life of a SNR can be thought of as consisting of four
phases.

\subsubsection{Free Expansion Phase}

Immediately following the explosion, the ejecta from the supernova
race out into the ISM at speeds of $\sim 10,000$ km s$^{-1}$, driving
a strong shock at the leading edge. Since the ejected mass is
significantly greater than the mass of the rarefied medium it
encounters, there is no appreciable slowing of the ejecta by the
ambient medium. How long this phase lasts depends on both the ejected
mass of the SN and the density of the CSM in front of the shock, and
can be anywhere from a few days to years.

\subsubsection{Reverse-Shock Phase}

As the shock continues to expand and sweep material it encounters, the
accumulated mass becomes non-negligible, and gradually causes the
shock to slow. The ejecta behind the shock, however, are still
traveling at a free-expansion velocity, and slam into the decelerating
material ahead of it. This causes a reverse shock to form. Initially,
this ``reverse'' shock moves inward only in the Lagrangian shock
frame, and still moves outward in the observer's frame. The
reverse shock, however, eventually ``turns around'' and moves inward
in the frame of reference of the observer. It typically takes of order
hundreds of years for this transition to occur. When the ejecta are in
free-expansion cooling is almost entirely adiabatic. While this
adiabatic cooling is effective in lowering the temperature of the
ejecta, energy is still conserved for the system because the energy is
not radiated away. Upon encountering the reverse shock the ejecta are
heated, like the ISM at the forward shock, to very high temperatures,
thus radiating strongly in X-rays. This radiative cooling is still
relatively inefficient, and most of the energy of the SNR+ISM system
is conserved. The duration of the reverse shock phase can last tens to
thousands of years.

\subsubsection{Sedov-Taylor Phase}

Once the mass swept-up by the forward shock greatly exceeds the ejecta
mass, the remnant enters the Sedov-Taylor phase (often known simply as
the Sedov phase). The reverse shock has propagated all the way back
through the ejecta and dissipated, and the remnant can be described by
a self-similar solution (Sedov 1959). The similarity variable can be
derived by dimensional analysis, and is given by

\begin{equation}
\xi = R(\rho/Et^2)^{1/5}
\end{equation}

\noindent
where $R$ is the distance the blast wave has traveled from the
supernova, $E$ is the explosion energy of the initial event, $\rho$ is
the density of the ISM, and $t$ is time. $\xi$ is dimensionless, and
equation 12 can be used to show that the distance traveled by the
blast wave (for $\gamma = 5/3$) as a function of $E$, $\rho$, and $t$,
is given by

\begin{equation}
R \propto (\frac{E}{\rho})^\frac{1}{5} t^\frac{2}{5}.
\end{equation}

\noindent
It can be immediately seen from this that the shock velocity is given
by

\begin{equation}
V_s=\frac{dR}{dt}=\frac{2R}{5t}.
\end{equation}

\noindent
The Sedov phase lasts for thousands to tens of thousands of years
after the explosion.

\subsubsection{Radiative Phase}

The shock continues to sweep up material and decelerate, eventually
reaching a point where the forward shock speed is only a few hundred
km s$^{-1}$. At this point, the temperature of the post-shock gas
drops below 10$^{6}$ K, and radiative cooling of the gas becomes
important. Cooling of the gas is a runaway process, as the more it
cools, the more the cooling rate increases. As the gas temperature
drops further, the material once again becomes visible in optical
radiation. The forward shock is driven mostly by momentum conservation
at this point, and eventually will turn sub-sonic and dissipate into
the ISM.

\subsection{Radiation Mechanisms in SNRs}

SNRs radiate throughout the electromagnetic spectrum. In radio waves,
the emission is entirely non-thermal in origin, resulting from
synchrotron radiation from relativistic electrons spiraling around
magnetic fields. Synchrotron emission is characterized by a
featureless power-law spectrum, where the radio flux, $S_{\nu}$ is
given by $S_{\nu} \propto \nu^{-\alpha}$, where $\nu$ is the frequency
and $\alpha$ is the spectral index, which depends on the energy
distribution of the electron population.

The primary focus of this work is on infrared (IR) emission from SNRs,
the physical basis for which is described in Chapter 2. Briefly, IR
emission is dominated by thermal continuum from warm dust grains,
heated by collisions with the hot ions and electrons in the post-shock
region. Remnants in the radiative phase can show IR line emission as
well, from low to moderately ionized states of abundant heavy elements
like O, Ne, S, Si, Ar and Fe.

At optical wavelengths, radiation prior to the radiative phase comes
primarily from hydrogen Balmer lines (transitions from $n \ge 3 \to
2$), such as H$\alpha$, $\lambda$ = 656.3 nm, and H$\beta$, $\lambda$
= 486.1 nm. This requires the presence of neutral hydrogen ahead of
the shock, which is much more easily attained in the case of a type Ia
SN, since CC SNe generally ionize the surrounding medium, either with
ionizing radiation from the progenitor, or a flash of ultraviolet (UV)
radiation at the moment of explosion. Charge exchange between slow
neutral atoms and fast protons behind the shock produces fast-moving
neutral atoms, generating a broad H$\alpha$ line, with a narrow
component arising from stationary neutral atoms in the post-shock
medium. During the radiative phase, strong optical lines are seen from
a variety of atomic species, most strongly from H$\alpha$ and
singly-ionized sulfur ([S II]). Ultraviolet (UV) emission from SNRs is
also produced, generally from higher ionization states than in
optical.

Soft X-rays (0.1-2 keV) are generally thermal in origin, and in SNRs
are often dominated by line emission from highly ionized
elements. Typically, elements that emit X-ray line emission have been
stripped of all but one or two electrons, making them
``hydrogen-like'' or ``helium-like.'' As with optical and infrared
lines, downward transitions of electrons to lower energy levels causes
emission of a photon whose energy is equal to the transition energy
between the electron's bound states. For elements that still contain
multiple electrons, the transitions between energy states become more
complicated, and generally generate numerous lines which are smeared
together by current X-ray spectroscopic technology. Continuum
emission, detailed below, is also observed at these energies.

Hard X-rays (2-50 keV) in SNRs can be either thermal or non-thermal in
origin. Thermal X-rays are dominated by continuum emission, although
lines from K-shell electron transitions do exist beyond 2 keV for
elements such as Si, S, Ar, and Fe. Thermal continuum seen in X-ray
spectra is generally thermal Bremsstrahlung (also known as
``free-free'' emission), which occurs when electrons and protons
interact. This process causes the electron to slow down, and energy
conservation requires emission of a photon to account for the lost
kinetic energy of the particle. Free-bound emission (or radiative
recombination) can also take place when a proton or ion captures a
free electron, emitting in the process a photon whose energy depends
on both the free kinetic energy of the electron and the orbital to
which it is captured. 

Non-thermal emission in SNRs arises from synchrotron emission
identical to that seen in radio waves, but from much more energetic
electrons. The maximum photon energy in keV of an electron with energy
$E$ is given by

\begin{equation}
h \nu = 1.93 (E/100 TeV)^{2} (B/10\mu G) keV.
\end{equation}

\noindent
In order to produce synchrotron emission in the 2-10 keV range,
electrons with energies of 100-200 TeV are required. Because this is
well beyond the particle thermal energies for even a fast shock,
another process must accelerate particles to high energies in remnants
where non-thermal emission is conclusively identified. The origin of
these high-energy cosmic-ray electrons is discussed in the next
section.

At gamma-ray energies, emission can be produced by one of three
processes; two of which are leptonic in origin, one of which is
hadronic. Bremsstrahlung, both thermal and non-thermal in origin, can
account for photons of all energies, up to TeV
emission. Inverse-Compton scattering from relativistic electrons off
of cosmic microwave background or far-infrared photons can upscatter
the photons to very high energies. The only known hadronic source of
gamma-ray emission is the decay of neutral pions, or $\pi^{0}$
particles, into two gamma-ray photons. This process occurs 98.7\% of
the time in $\pi^{0}$ decays. The $\pi^{0}$ particles themselves are
produced in collisions between cosmic-ray protons and thermal protons,
as well as protons and alpha particles in the pre-shock gas. The
minimum pion energy required to produce a gamma-ray of energy
$E_{\gamma}$ is given by

\begin{equation}
E_{min}(\pi) = E_{\gamma} + (m_{\pi}^{2}c^{4}/4E_{\gamma})
\end{equation}

\noindent
To produce high energy gamma rays ($> 1$ GeV), the last term on the
right becomes negligible, and the minimum pion energy needed is
roughly equal to the gamma-ray energy observed.

It is likely that all of these processes play a role in the gamma-ray
emission observed from SNRs. For a thorough review of SNRs at high
energies, see Reynolds (2008).

\subsection{Cosmic-Ray Acceleration in SNRs}

Cosmic-rays are highly energetic particles, typically protons, alpha
particles, and nuclei of heavier elements, with a small percentage of
the population consisting of electrons, streaming through space at
relativistic speeds. They can either be detected directly (at lower
energies), or indirectly through interactions with atoms in the
Earth's upper atmosphere (at high energies). Upon the collision of a
cosmic-ray with our atmosphere, a shower of particles (mostly
containing pions) is produced. These pions decay further into
electrons, positrons, muons, neutrinos, and photons, and can be
detected from ground-based Cherenkov telescopes. Such telescopes can
even reconstruct the events to determine the location in the sky from
which the cosmic-ray came. Unfortunately, this location is not
indicative of the original source of the particle. Since cosmic-rays
are charged particles, they gyrate around the magnetic field lines of
our Galaxy, and are essentially randomized by the time they reach
Earth. This leads to a mystery: from where do cosmic-rays come?

\subsubsection{Cosmic-Ray Sources}

The fact that synchrotron emission is observed in radio waves for
every Galactic SNR known shows that, at the very least, electrons are
efficiently accelerated to energies of a few GeV. If electrons are
accelerated, protons and ions should be accelerated as well. This,
unfortunately, is a difficult thing to observationally verify, because
the synchrotron radiation from relativistic protons is orders of
magnitude weaker than from electrons spiraling around a magnetic
field. Gamma-ray production via the decay of $\pi^{0}$ particles,
discussed in the previous section, requires that protons at GeV
energies exist. Unambiguous detection of this hadronic gamma-ray
signal in SNRs would provide the observational confirmation that such
acceleration of ions is taking place. Searches for this signal are
currently underway.

Detection of non-thermal synchrotron emission at X-ray energies is a
clear indication that the shock is accelerating electrons beyond TeV
energies. If the shocks are equally as efficient at accelerating
protons, this could account for the galactic cosmic-ray spectrum to
energies up to $\sim 10^{15}$ eV. Cosmic-rays at much higher energies
have been detected, but it is difficult to produce them in large
quantities in SNR shocks. It is likely that these ultra-high energy
particles have an extra-galactic origin.

If supernova shock waves are efficiently accelerating cosmic rays,
then the equations detailed in Section 1.3 are no longer valid, since
escaping cosmic rays can rob the shock of energy. This leads to a
higher compression ratio (r $\equiv \rho_{2}/\rho_{1}$), and a lower
post-shock temperature. Even if no cosmic rays escape, the compression
ratio can still be increased if relativistic particles dominate, since
r $\to$ 7 as $\gamma \to \frac{4}{3}$.

\subsection{Summary}

Supernovae represent the end of a star's life, but in the process of
dying, elements that will go on to form future generations of stars
and planets are spread throughout the galaxy. The universe is nearly
14 billion years old, old enough that every cubic centimeter of a
galaxy like the Milky Way has been overrun numerous times by shock
waves produced by SNe. They represent one of the main feedback
mechanisms in the evolution of a galaxy, shaping and recycling
products in the ISM. 

SNRs are visible at all wavelengths of the electromagnetic spectrum,
though the physical processes responsible for emission at various
wavelengths differ. Nonetheless, these processes are often connected,
and a complete understanding of the dynamics of the remnant requires
connecting the physics behind these various faces of the remnant. 

\newpage

\begin{figure}
\figurenum{1.1}
\includegraphics[width=15cm]{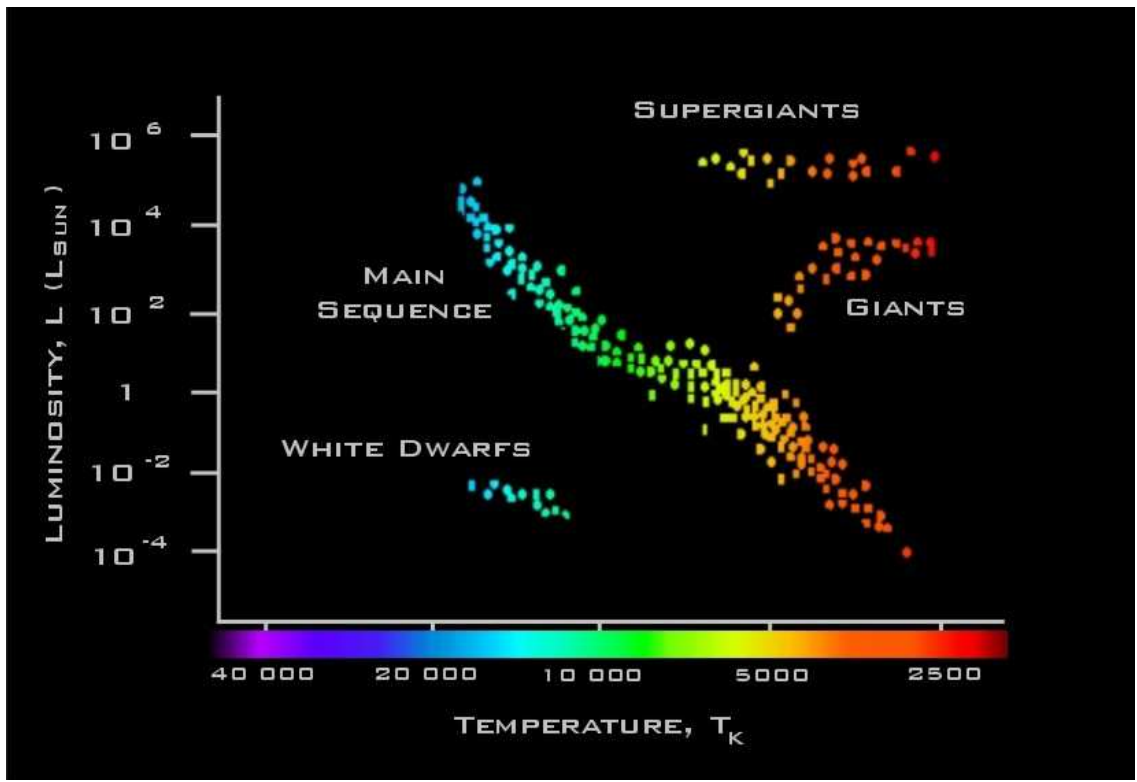}
\caption{Hertzsprung-Russell Diagram, showing the temperatures of
stars vs. luminosity. Taken from
{http://www.le.ac.uk/ph/faulkes/web/stars/o\_st\_overview.html}.
\label{hrdiagram}
}
\end{figure}

\newpage

\begin{figure}
\figurenum{1.2}
\includegraphics[width=17cm]{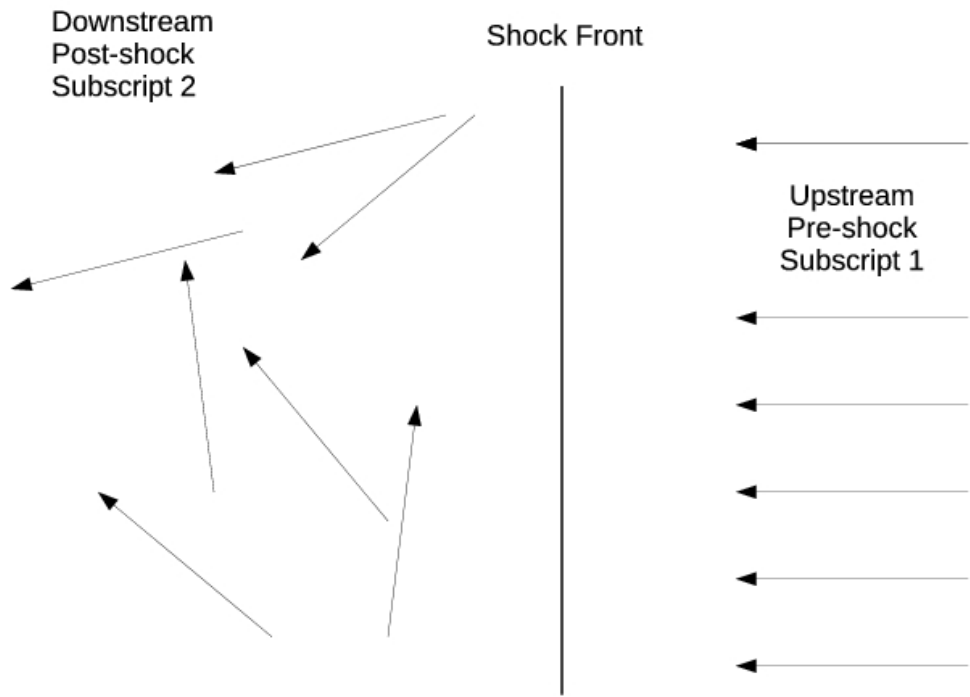}
\caption{Shockwave diagram, in the frame of reference of the
shock. Arrows represent the velocities of particles upstream and
downstream of the shock. In this frame, upstream material (at rest in
the observer's frame) races in towards the shock. The shock then
randomizes the velocities of the particles, and gives them a bulk
velocity downstream of 1/4 their initial velocity (for a strong shock
with $\gamma = 5/3$).
\label{shockframe}
}
\end{figure}

\newpage

\begin{figure}
\figurenum{1.3}
\includegraphics[width=15cm]{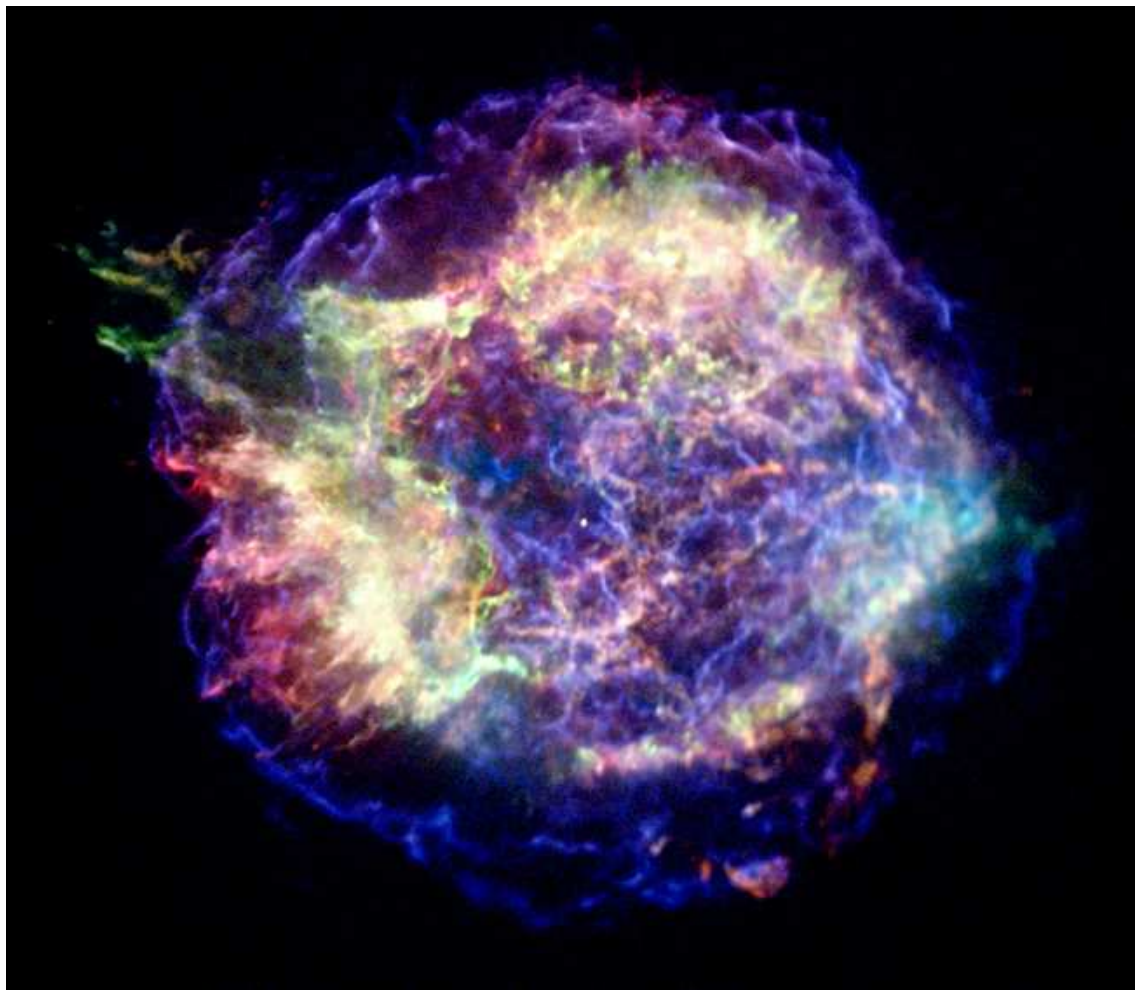}
\caption{SNR Cassiopeia A, seen in X-rays from the {\it Chandra X-ray
Observatory}. Image from {http://chandra.harvard.edu/photo/2006/casa/}
\label{casa}
}
\end{figure}

\newpage
\clearpage

\section{Infrared Emission from Young Supernova Remnants}

IR emission from SNRs is predominantly thermal emission from warm dust
grains, heated via collisions with hot electrons and ions in the
post-shock gas. Although IR line emission can become strong once a
shock reaches its radiative phase, it is virtually non-existent in
fast, non-radiative shocks, as Figure 2.1 shows. This work focuses on
emission from these non-radiative shocks, which typically persist for
hundreds or thousands of years in SNRs before becoming radiative.

\subsection{What is Dust?}

Dust grains in the ISM are not like the dust that accumulates on top
of TVs and countertops. ISM grains are microscopic, ranging in size
from molecules of a few atoms to small solid bodies, several microns
($\mu$m, where 1 $\mu$m = 10$^{-6}$ meters) in radius. Grains are made
up of various elements, most notably carbon (which can exist in either
crystalline forms like graphite or in amorphous forms), oxygen,
silicon, magnesium, and iron. Polycyclic aromatic hydrocarbons (PAHs)
have also been spectroscopically identified as residing in the
ISM. These molecules are similar to PAHs produced on Earth, typically
as byproducts of fuel burning. PAHs consist of aromatic rings of
carbon with hydrogen atoms at their edges. On average, about 0.1-1\%
of the mass of the ISM is contained in dust grains, with the remainder
being in the gaseous phase.

\subsection{Dust Formation Sites}

Dust plays an important role in both the evolution the ISM in galaxies
and the universe as a whole. It plays an important role in
star formation, acting as a catalyst for the formation of H$_{2}$
molecules, which are efficient coolants, giving dense clouds a chance
to contract and create new stars. Early observations of the disk of
the Milky Way galaxy showed dark lanes of dust that block starlight,
and high-redshift observations of galaxies in the early universe show
large quantities of dust present shortly after the Big Bang.

Dust condensation requires an abundance of heavy elements, a dense
environment where collisions between particles are frequent, and a low
temperature below the vaporization temperature for grains. These
conditions are not frequently found together in the universe, but two
sites are often suggested as potential hosts for dust nucleation:
atmospheres of AGB stars and supernovae. AGB stars are beyond the
scope of this work, but the amount of dust produced in supernovae can
be determined from observations of SNRs, and will be discussed at
length in a later chapter. Theoretical calculations of the amount of
dust produced in the ejecta of CC SNe can exceed several solar masses
(Nozawa et al. 2007). However, more recent work by
\citet{cherchneff10} has revised these estimates down by about a
factor of 5.

\subsection{Observing Dust Emission}

The majority of heating of grains in the ISM is done via radiative
heating by photons. This heating can be from stars, active galactic
nuclei, or the interstellar radiation field. For this work, however, I
focus on collisional heating by particles behind shock
waves. Collisionally heated grains in SNRs are typically warmed to
temperatures of 50-200 K, which is far too cold to be observed by
optical, ground-based instruments. Grains at this temperature radiate
in the mid-IR, with their spectra peaking anywhere from 20-100
$\mu$m. To effectively observe at these wavelengths, one needs to
travel outside the Earth's atmosphere, above the water vapor that
significantly absorbs mid and far-IR radiation. The majority of the
work described here is based on observations done by the {\it Spitzer
Space Telescope}.

\subsubsection{Spitzer}

NASA's {\it Spitzer Space Telescope} is the fourth and final mission
in the ``Great Observatories'' program, following the {\it Hubble
Space Telescope} (1990-present, optical wavelengths), the {\it Compton
Gamma-Ray Observatory} (1991-2000, gamma-rays), and the {\it Chandra
X-ray Observatory} (1999-present, X-rays). All four instruments were
large space-based observatories. {\it Spitzer} was launched in August
of 2003 and began full-time science operations in 2004. Unlike the
other observatories in the program, with orbits around the Earth, {\it
Spitzer} follows a heliocentric orbit that recedes away from Earth at
the rate of 0.1 astronomical units (AU) per year. Both the Earth and
the Moon are incredibly bright IR sources, and the telescope had to be
placed far away from both to achieve its desired sensitivity. The
spacecraft (see Figure 2.2) consists of an 85-centimeter telescope
outfitted with 3 separate instruments which can be placed in the field
of view at any given time. A sun-shield, which always faces the Sun,
protects the entire system, acting as the first line of defense
against photons that would warm the telescope and damage the
instruments. The telescope is cryogenically cooled, with the primary
coolant being liquid helium. This keeps the detectors at 4.2 K,
necessary for science observations in the mid and far-IR, but comes at
a price: the liquid helium is an expendable resource and cannot last
forever. The target lifetime for the ``cold mission'' (i.e., time
before the cryogen ran out) of {\it Spitzer} was 5 years; in reality,
it lasted over 5.5 years before running out in May of 2009. Thus began
the ``warm mission'' of {\it Spitzer}, involving only the shortest IR
wavelengths, which is expected to last until 2013.

\subsubsection{Spitzer's Instruments}

{\it Spitzer} has three instruments onboard, data from all of which is
featured in this work. The {\it Infrared Array Camera} (IRAC) provides
photometric (imaging) capabilities in the near and mid-IR, with four
channels covering the wavelength range of 3.3-8.5 $\mu$m. The {\it
Multi-band Imaging Photometer for Spitzer} (MIPS) contains three
broadband channels for photometric imaging, centered at 24, 70 and 160
$\mu$m for channels 1, 2, and 3, respectively. Spectroscopically, the
{\it Infrared Spectrograph} (IRS) provides both low and medium
spectral resolution data over the wavelength range of 5-40 $\mu$m. The
low-resolution spectrograph uses slit spectroscopy and is ideal for
continuum detection; its resolution, $\lambda/\delta\lambda$, is
64-128. The high-resolution module uses echelle spectrographs ideal
for observing lines, provides a resolution of $\lambda/\delta\lambda$
= 600. Both IRAC and MIPS provide diffraction limited optics, with the
angular resolution ranging from $\sim 1''$ for the 3.6 $\mu$m array to
$\sim 50''$ at 160 $\mu$m.

\subsection{Size Distribution of ISM Dust Grains}

Since dust can exist in all varieties and sizes, it is necessary to
have a more quantitative understanding of the distribution of these
various grains in the ISM. Numerous authors (Mathis et al. 1977,
Weingartner \& Draine 2001, hereinafter WD01, Zubko et al. 2004 and
more) have attempted to quantify the size distribution for both the
galaxy and the Magellanic Clouds (dwarf satellite galaxies of the
Milky Way). This work primarily uses the distributions of WD01, shown
in Figure 2.3, although alternative models are explored. As can be
seen in the figure, the size distribution of grains is steeply
weighted towards the small end. This is not unexpected, since it is
believed that grains coalesce in dense environments and grow in
size. Shattering of grains in grain-grain collisions may also play a
major role in establishing the ISM grain size distribution.

\subsection{Grain Heating and Cooling}

Dust grains in the ambient ISM are heated by the interstellar
radiation field, primarily by UV starlight. This radiation field can
heat dust to $\sim 10-20$ K, and warmer dust is often found in the
immediate vicinity of stars. In SNRs, however, the primary heating
mechanism for grains is collisional heating, where grains are warmed
by frequent collisions with the hot ($> 10^{6}$ K) electrons and ions
in the post-shock region behind the forward shock. The heating rate
for a grain immersed in a hot plasma is given by

\begin{equation}
H=\left(\frac{32}{\pi m}\right)^{1/2}\pi a^{2} n (kT)^{3/2} h(a,T),
\end{equation}

\noindent
where $m$ is the mass of the impinging particle (proton, electron,
etc.), $a$ is the radius of the grain, $n$ is the density of the gas,
$k$ is Boltzmann's constant, $T$ is the temperature of the gas, and
$h(a,T)$ is a function that describes the efficiency of the energy
deposition rate of a particle at a given $T$ for a grain with radius
$a$. It can be immediately seen from this equation that at a fixed
$T$, electrons will dominate the heating over protons, since their
mass is much smaller and they move much faster.

Since grains are virtually always smaller than the wavelength of light
they emit (i.e. $a << \lambda$), they will cool as modified
blackbodies (see Figure 2.4).  The cooling rate of a given grain at a
temperature $T_{d}$ is given by

\begin{equation}
{\cal L} = \int^{\infty}_{0} d\nu C_{abs}(\nu) 4\pi B_{\nu}(T_{d}),
\end{equation}

\noindent
where $\nu$ is the frequency, $C_{abs}$ is the absorption
cross section, and $B_{\nu}(T)$ is the Planck blackbody
function. Calculating the quantity $C_{abs}$ requires knowledge of the
dielectric function, $\epsilon$, of a given grain material. For an
excellent review, see Draine (2004). In equation (32) of that paper,
the absorption cross section for a sphere is given by (assuming $a <<
\lambda$)

\begin{equation}
C_{abs}= \frac{9\nu V}{c}
\frac{\epsilon_{2}}{(\epsilon_{1}+2)^{2}+\epsilon_{2}^{2}},
\end{equation} 

\noindent
where $c$ is the speed of light, $V$ is the grain volume,
$\epsilon_{1}$ is the real part of $\epsilon$, and $\epsilon_{2}$ is
the imaginary part. It is readily seen that in the limit of $a <<
\lambda$, $C_{abs}$ is proportional to grain volume. This is in
contrast to the heating rate, where heating was proportional to the
surface area of the grain. As a result, the equilibrium temperature
for a grain immersed in a plasma is a function of its size, even if
all other grain properties are identical. Additionally, small grains
(i.e. grains that have a sufficiently large surface-to-volume ratio)
may find collisions so infrequent and cooling times so rapid that they
never reach an equilibrium temperature, and instead constantly
fluctuate, spiking to high temperatures and emitting radiation much
more efficiently when at their maximum temperatures. See Figure 2.5
for plots of grain temperature versus time.

\subsection{Dust Grain Sputtering}

The same collisions that heat grains can also slowly destroy them via
sputtering. Sputtering is the ejection of atoms from the surface of a
grain during collisions with ions. This loss of material reduces the
size of the grain, and the ejected atoms are liberated back into the
gaseous phase. Thus, as a function of time, large grains are converted
into small grains, and small grains are completely destroyed in the
post-shock region of a SNR. This strongly modifies the grain size
distribution behind the shock. Nozawa et al. (2006) provides the
sputtering yield (number of particles ejected per collision) as a
function of impinging particle energy for various particles and dust
compositions (see Figure 2.6).

The equations given in Nozawa et al. are taken from Bodhansky (1984),
which calculates sputtering of solids with respect to industrial
applications, particularly that of building a nuclear reactor. For
these applications, the bulk approximation of a solid is acceptable,
as one never has to worry about the sides and back walls of the
reactor. For sufficiently small grains, however, these equations are
not sufficient, because sputtering can take place not only from the
front side of the grain where the initial impact occurs, but also from
the sides and back.  Specifically, Jurac et al. (1998) find that for
grains with $a$ $<$ 3R$_{P}$, where R$_{P}$ is the projected range of a
particle impacting a grain (where projected range is the average of
the depth to which a particle will penetrate the grain in the course
of slowing down), the sputtering yields are enhanced. For the smallest
of grains, this can lead to an order-of-magnitude increase in the
sputtering rate, as shown in Figure 2.7.

Finally, there is a competing effect for small grains. Sufficiently
fast impinging protons do not deposit all of their energy into a grain
when they collide; the rate at which they deposit energy is a function
both of the grain radius and the energy of the particle. As small
grains become transparent to protons, protons deposit less energy in
collisions with nuclei within grains, so sputtering rates are reduced
relative to Jurac et al. results \citep{serradiazcano08}. We therefore
scale sputtering yields for small grains in proportion to the
fractional energy deposited by the proton or alpha particle. Figure
2.8 shows the sputtering yield as a function of grain radius for a 10
keV proton.

\subsection{Modeling Grain Emission in SNR Shocks}

In order to create a model for collisionally heated dust emission in
the post-shock gas of an SNR, everything outlined above must be taken
into account. One must have an underlying model for the grain physics,
including the grain size distribution for each species of grain. In
this work, unless stated otherwise, we model dust in the ISM as
consisting of ``astronomical silicate'' (with predominantly
MgFeSiO$_{4}$ composition) (Draine \& Lee 1984) and graphite grains,
mixed in the proportions given in WD01. Optical constants
($\epsilon_{1}$ and $\epsilon_{2}$) are taken from Draine \& Lee
(1984), and sputtering yields are calculated as described above. We
use 100 grain sizes, logarithmically spaced from 1 nm to 1
$\mu$m. Although smaller grains are thought to be present in the ISM,
we assume that they are instantaneously destroyed in the shock and
contribute nothing to the emission seen by {\it Spitzer}
\citep{micelotta10}. Given {\it Spitzer's} limited spatial resolution
and the small number of remnants that are large enough to resolve the
immediate post-shock region, this is likely a good approximation.

The energy deposition rates of electrons and protons as a function of
energy and grain size must also be input to the code, as well as the
optical constants of various grain types. Since the heating rate of
grains depends on the density and temperature of different particle
species within the plasma, these must be included in the code. The
total sputtered number of atoms for a given grain depends on the time
it has been immersed in the plasma, i.e., the time since it was
shocked. This can be quantified by a parameter known as the
``sputtering timescale'', defined as $\tau = \int^{t}_{0} n_{p} dt$,
where $n_{p}$ is the post-shock proton density. This is similar to the
``ionization timescale'' found in X-ray analysis, defined as $\tau_{i}
= \int^{t}_{0} n_{e} dt$, where $n_{e}$ is electron density. In order
to model a region of any significant spatial width behind the shock,
it is necessary to create a shock model which superimposes regions of
different $\tau$ behind the shock. In the models described in this
paper, this is done by calculating the sputtering rate for all grains
in the distribution in each zone behind the shock, and adjusting the
grain size distribution accordingly. The final model appropriately
sums these post-shock distributions.

The output of such a model is the temperature of each grain in the
distribution, which is size-dependent. To account for stochastic
heating effects on small grains, we use a method devised by
Guhathakurta \& Draine (1989). Since the sputtering rates for grains
are calculated in the shock model, we can integrate them to obtain the
total amount of mass in grains that is destroyed. Of course, this mass
is not actually destroyed, merely converted back into the gaseous
phase. The thermal spectrum for each grain is calculated and summed
over the final distribution to create a single spectrum, which can be
compared directly to observations.

\subsection{Necessity for a Multi-Wavelength Approach}

In theory, one can tune any or all of the parameters in the model to
fit the observed data from {\it Spitzer} or other telescopes. In
practice, however, there are significant degeneracies present in the
model. Figure 2.9 shows an example of these degeneracies for only two
components, electron density and temperature. It is impossible, from
IR data alone, to eliminate the degeneracies; so we must use
information from other wavelengths as additional constraints on the
modeling.

\subsubsection{X-rays}

Thermal X-ray spectra are most sensitive to the electron temperature
of the shocked gas. Since both the shocked ambient medium and the
reverse-shocked ejecta are strong X-ray emitters, it is necessary to
separate (either spatially or spectroscopically) the components
belonging to each to get an accurate measure of the temperature in the
post-shock gas. As discussed in Chapters 3-5, we see very little
evidence for dust emission from ejecta in most SNRs, and thus are
typically only concerned about dust heated by the forward shock. A
shock model of an X-ray spectrum can also give the ionization
timescale. If the SNR is large enough and/or young enough, high
spatial resolution instruments like {\it Chandra} may be able to
resolve proper motion of the forward shock itself, yielding a shock
velocity. This is subject to uncertainties about the distance to the
object, which is often not well known.

\subsubsection{Optical/UV}

Optical emission from non-radiative shocks shows line emission from
both stationary atoms in the post-shock gas and fast-moving hydrogen
atoms, created by thermal protons, that have undergone charge-exchange
(i.e. the stealing of an electron) with slow neutrals entering the
shock. This requires at least partially-neutral material ahead of the
shock, and creates a fast-moving neutral atom moving with bulk
velocity $(3/4)v_{s}$ (for standard shock jump conditions), where
$v_{s}$ is the shock speed. The fast-moving neutrals are collisionally
excited by free electrons and protons, emitting radiation primarily
via the $n = 3 \rightarrow 2$ (Balmer-$\alpha$, or H$\alpha$ 656.3 nm
optical line) and the $n = 2 \rightarrow 1$ (Lyman-$\alpha$, 121.6 nm
ultraviolet line) transitions. This produces a ``broad'' hydrogen
line, where the broadening is a result of two (additive) phenomena:
thermal line broadening resulting from the random motions of the hot
neutrals in the shock frame, and Doppler broadening resulting from the
bulk velocities of the post-shock gas seen along a line-of-sight
through the front and back sides of the SNR. Hydrogen lines measured
directly on the limbs of the remnant show only thermal broadening,
while lines measured at any point interior to the outer shell show
additional broadening from the Doppler component. This broad line can
be used as a diagnostic of shock speed and proton temperature in the
post-shock gas.

A narrow component is also seen in optical/UV spectra of SNRs, arising
from collisional excitation of cold neutral atoms that survive passage
of the shock. The intensity ratio of the broad and neutral components
is sensitive to both the shock speed and the degree of equilibration
between electrons and protons at the shock front (Chevalier
1980). Excitations of H in collisions with protons and alpha particles
are most important at high shock speeds; electrons dominate at low
shock speeds. High spatial resolution optical images can also be used
to measure the proper motion of some SNR shocks.

\subsection{Density Diagnostics}

The density of the gas, either pre-shock or post-shock, is difficult
to determine from either X-ray or optical observations. X-rays can
give a measure of the root mean square (r.m.s.) post-shock density
through the emission measure, defined as $EM = \int^{V}_{0} dV f n_{e}
n_{H}$, where $n_{e}$ and $n_{H}$ are the post-shock electron and
proton densities, respectively, but this is dependent on $f$, the
filling fraction of the material in the volume considered. H$\alpha$
line strength measurements can yield the total number of hydrogen
atoms entering the shock at a given time, but only if the pre-shock
neutral fraction is known {\it a priori}. IR modeling of warm dust
emission provides an independent diagnostic that does not depend on
these uncertainties. If one knows the gas temperature and energy
deposition function, the heating of a grain is dependent only on the
grain size and gas density. Matching model results to observed IR
spectra, with density as a tunable parameter, gives a fit to the
post-shock density. 

Although IR modeling is not sensitive to pre-shock density, it is
nonetheless possible to use inferences derived from IR and X-ray fits
to constrain this quantity. X-ray spectral fitting provides the EM of
the gas, as defined above. This quantity can be rewritten (if $n_{e}$
is constant) as $EM \propto n_{e} M_{g}$, where $M_{g}$ is the mass in
gas that has been swept-up by the forward shock, defined by $M_{g}
\propto \int^{V}_{0} dV f n_{H}$. If the post-shock electron density
can be independently determined, as is the case in modeling IR
spectra, then this quantity can be divided out of the EM, leaving the
quantity of gas shocked by the remnant. This method is independent of
the filling fraction of the gas, since it merely measures the total
amount of swept-up gas, regardless of its distribution. If the
distance to the remnant is well known, this total amount of gas can be
divided by the volume enclosed by the forward shock to obtain the
average pre-shock density in the ambient ISM. The implicit assumption
in this method is that $\rho_{ISM}$ is constant.

\subsubsection{Errors on Density}

The shape of a dust spectrum is a fairly sensitive function of the gas
density. Although errors on derived quantities are model dependent, at
the very least one can make an estimate of the validity of results
reported in the following chapters from a purely statistical point of
view. Figure 2.10 shows a spectrum from a region of SNR 0509-67.5
(discussed at length in Chapter 8), overlaid with two models. These
models represent the 90\% confidence limits on the fits using
$\chi^{2}$ statistics, varying only the post-shock density. For a
given model fit to a dataset, $\chi^{2}$ is given by

\begin{equation}
\displaystyle\sum_{i=1}^n (\frac{X_{i}-\mu_{i}}{\sigma})^{2},
\end{equation}

\noindent
where $X_{i}$ is the value of the $i$th data point, $\mu_{i}$ the
value of the $i$th model point, and $\sigma$ the standard deviation of
the dataset. Once a best fit value for a given parameter is found by
minimizing the value of $\chi^{2}$, the 90\% confidence limits are
found by varying the parameter until $\delta\chi^{2}$ = 2.71. The best
fit was obtained with a density of $n_{p}$ = 0.88 cm$^{-3}$, and the
90\% error limits are 0.7 and 1.0 cm$^{-3}$. Thus, one can expect
errors of order 20\% in densities derived in IR fits, within the
framework of a given model.

\subsubsection{Application to Particle Acceleration in Shocks}

This multi-wavelength approach to determining both the pre- and
post-shock densities provides more robust estimates than analysis in
either wavelength could alone. Knowing both densities allows a direct
measurement of the compression ratio of the forward shock, defined as
$\rho_{2}/\rho_{1}$. Using the standard shock jump conditions found in
Chapter 1, this ratio for a strong shock should be 4. If, however,
cosmic rays are being accelerated in SNRs, the energy deposited into
these particles would have to come from somewhere. An alternative sink
for shock thermal energy is the turbulent magnetic fields found in the
immediate vicinity of the shock. Specifically, diffusive shock
acceleration (Bell 1978, Jones \& Ellison 1991), a process by which
the particles scatter back and forth across the shock off from
magnetic field irregularities, thus gaining a substantial amount of
energy, is widely believed to be the process which robs the shock of
energy. This process is believed to be capable of accelerating
particles up to the {\it knee} of the cosmic-ray spectrum, which
occurs at roughly $10^{15}$ eV. In fact, if supernova shocks are the
sole source of cosmic rays in the galaxy, accounting for the
cosmic-ray energy density observed requires that $\sim 10$\% of the
kinetic energy of a supernova explosion ($\sim 10^{51}$ ergs) must be
transferred to cosmic rays.

If such a process is happening, as appears the case with at least some
young SNRs (Abdo et al. 2010), the shock jump conditions found in
chapter 1, which ignore the contributions of cosmic rays or magnetic
fields, would no longer be strictly valid. The robbing of energy from
the post-shock gas to be injected into cosmic rays would increase the
magnetic field amplification at the shock, lower the temperature of
the post-shock gas, and increase the compression ratio of the
gas. Efforts have been made to measure the magnetic field
amplification (Uchiyama et al. 2007) and the correlation between shock
speed and post-shock gas temperature (Helder et al. 2009), but
observational confirmation of the increased compression ratio at the
forward shock is difficult to obtain. The method detailed above could
provide measurements, or at the very least, constraints, on this
number.

\newpage

\begin{figure}
\figurenum{2.1}
\includegraphics[width=15cm]{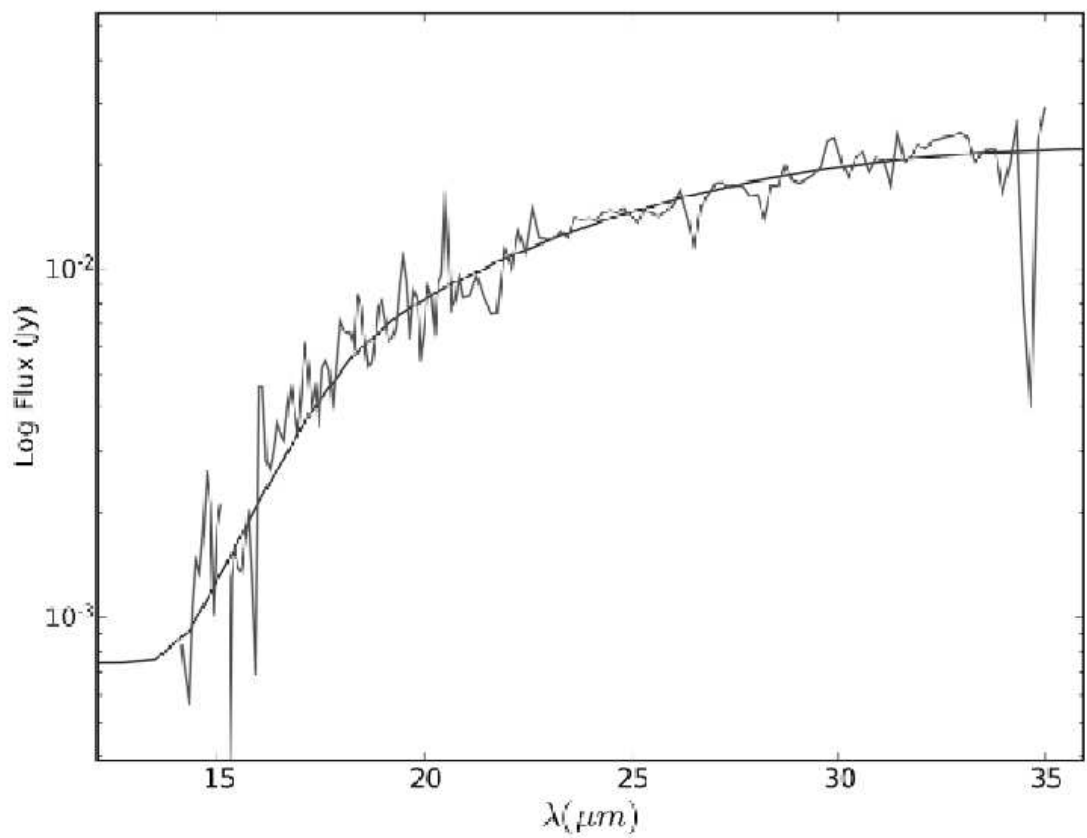}
\caption{Background-subtracted {\it Spitzer} IRS spectrum of warm dust
in SNR 0509-67.5, with model overlaid. The model will be discussed
further in chapter 7. Note absence of lines in the spectrum.
\label{0509spec}
}
\end{figure}

\newpage

\begin{figure}
\figurenum{2.2}
\includegraphics[width=13cm]{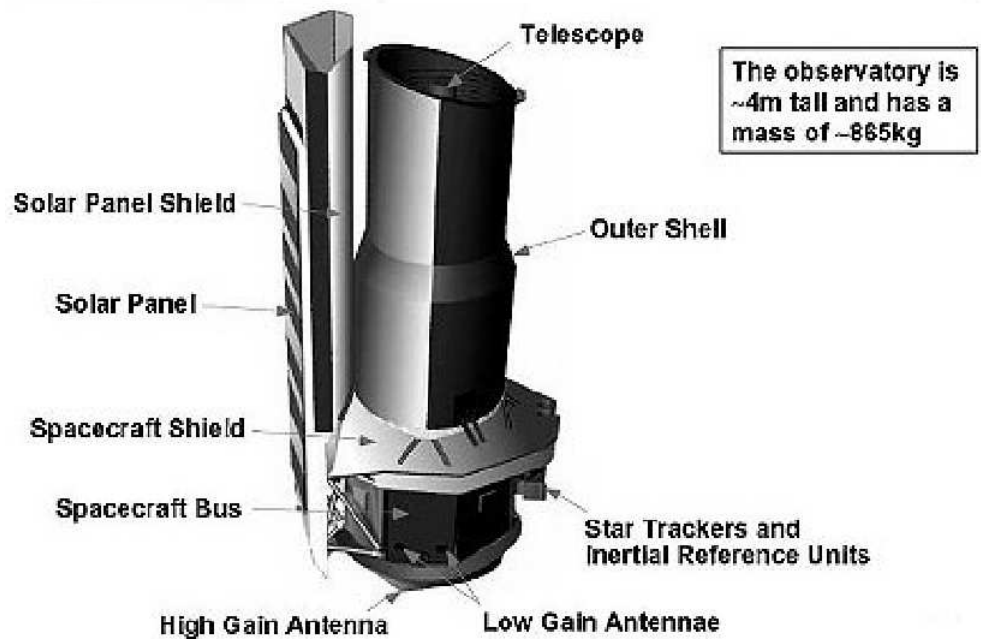}
\caption{Diagram of the {\it Spitzer Space Telescope}, taken from
{http://www.nasa.gov/missions/deepspace/f\_spitzerbirth\_prt.htm}
\label{spitzertelescope}
}
\end{figure}

\newpage

\begin{figure}
\figurenum{2.3}
\includegraphics[width=14cm]{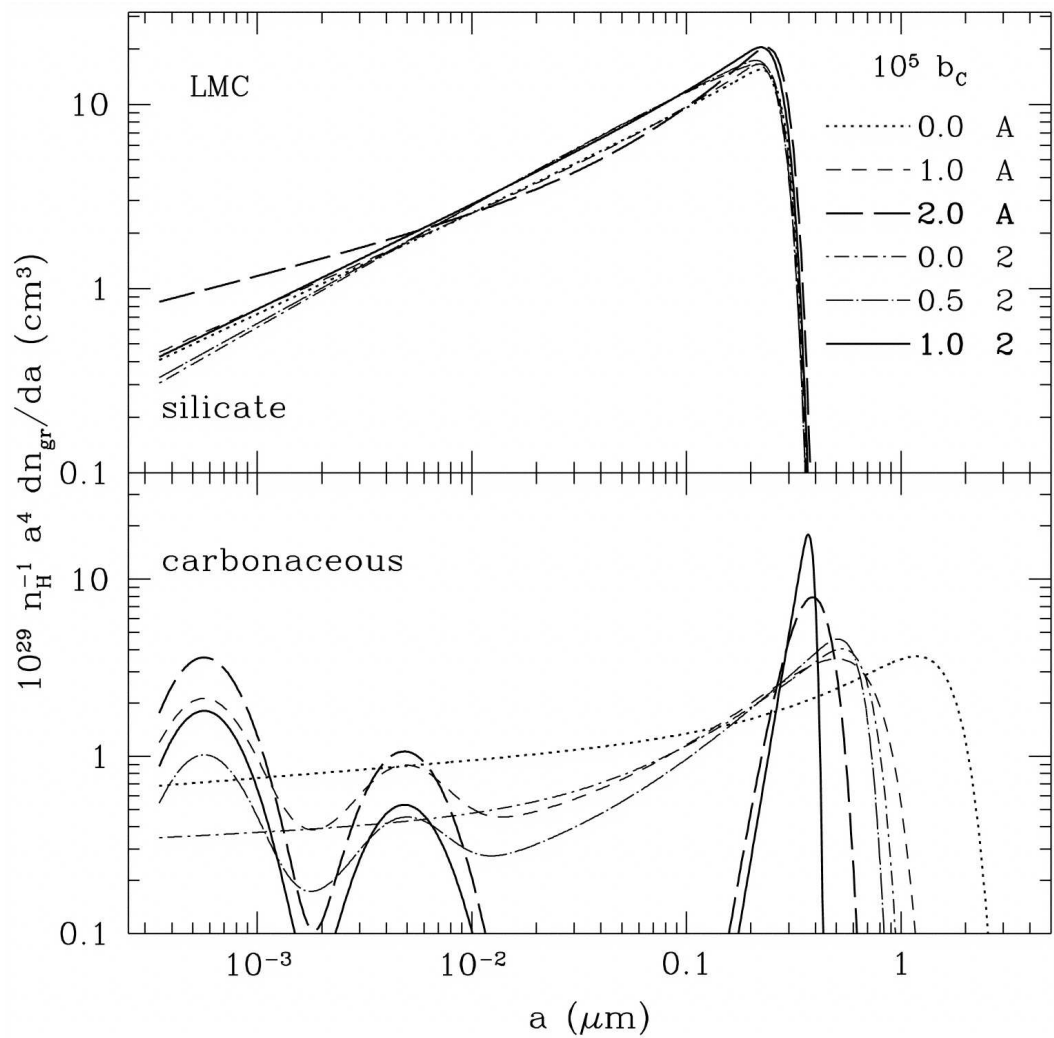}
\caption{Figure 18 from Weingartner and Draine (2001), showing the
size distribution of silicate (top) and carbonaceous grains (bottom)
for the Large Magellanic Cloud. For this work, we use their model
``2.0 A.''
\label{wd01}
}
\end{figure}

\newpage

\begin{figure}
\figurenum{2.4}
\includegraphics[width=15cm]{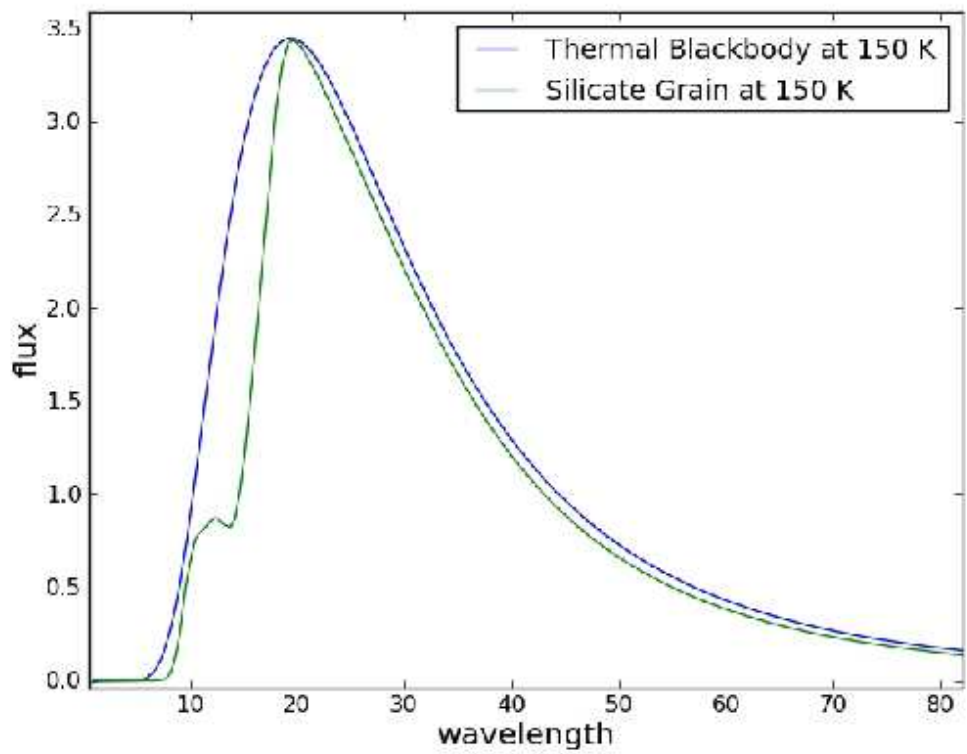}
\caption{Comparison of thermal blackbody at 150 K (blue) with single
silicate grain at 150 K (green).
\label{greybody}
}
\end{figure}

\newpage

\begin{figure}
\figurenum{2.5}
\includegraphics[width=13cm]{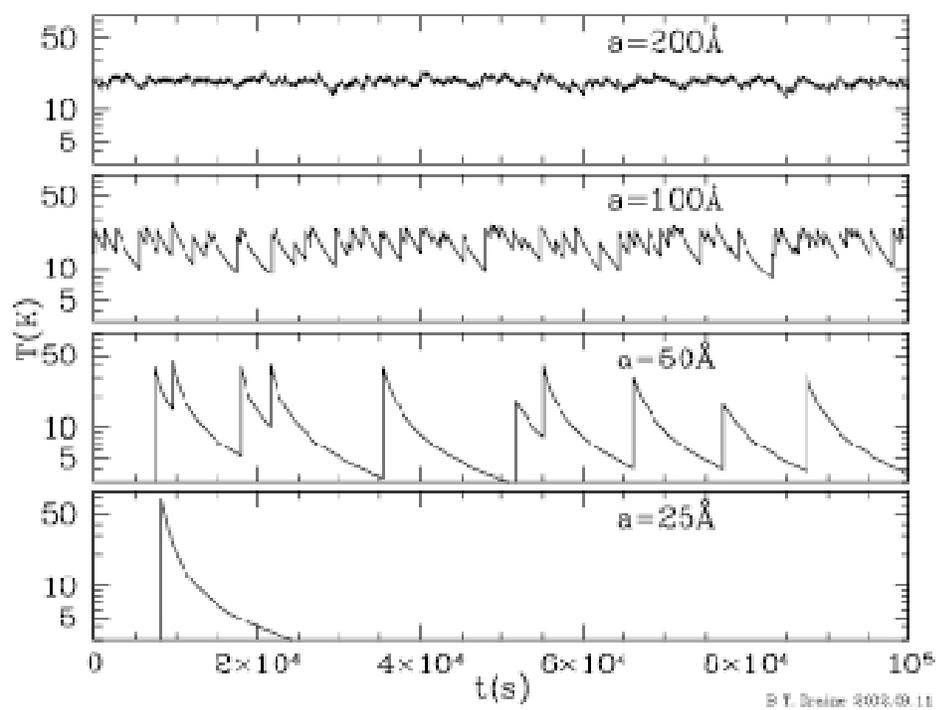}
\caption{Figure 17 from Draine (2003), showing grain temperature
vs. time for 4 different grain sizes, heated by photons from the
interstellar radiation field. A similar stochastic heating process
occurs for collisionally heated grains.
\label{stochastic}
}
\end{figure}

\newpage

\begin{figure}
\figurenum{2.6}
\includegraphics[width=14cm]{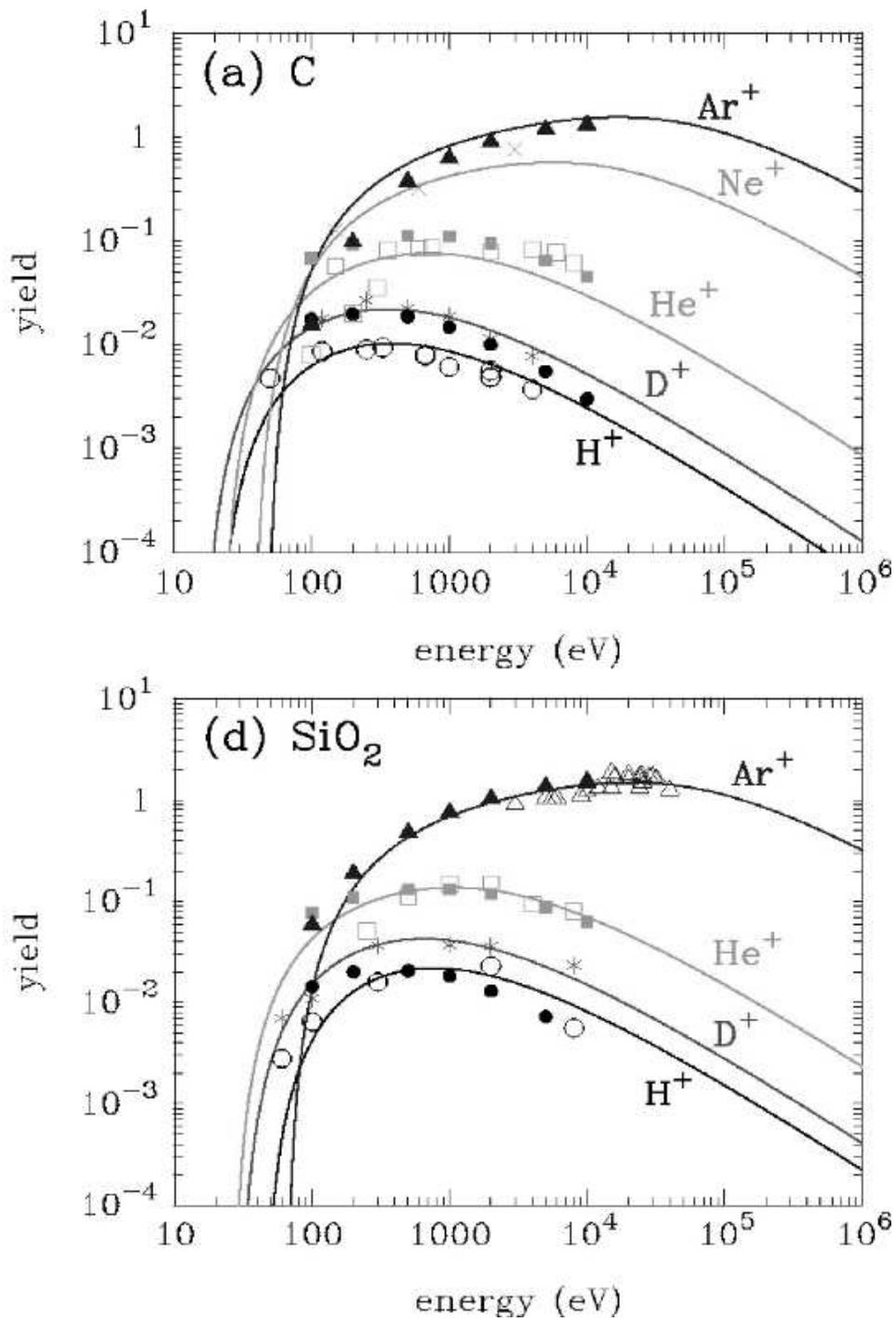}
\caption{Figure 1a and 1d of Nozawa et al. (2006), showing the
sputtering yield as a function of impinging particle energy for a
variety of ions into carbonaceous and SiO$_{2}$ grains.
\label{sputyield}
}
\end{figure}

\newpage

\begin{figure}
\figurenum{2.7}
\includegraphics[width=13cm]{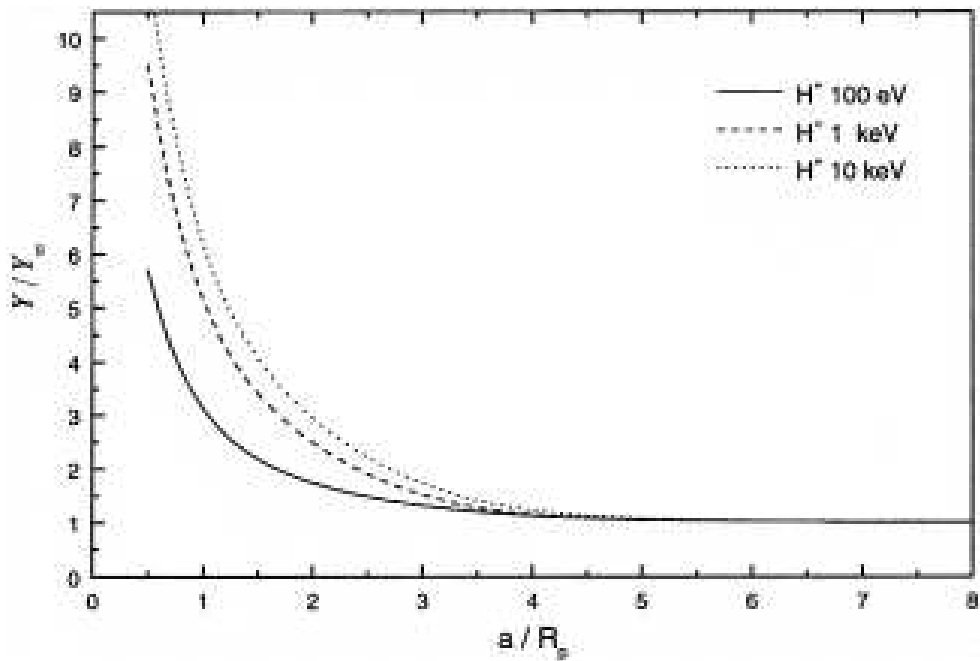}
\caption{Figure 3a from Jurac et al. (1998), showing the enhancement
in yield for protons of various energies, scaled to the yield for
isotropic bombardment of a flat surface (semi-infinite solid
approximation).
\label{jurac}
}
\end{figure}

\newpage

\begin{figure}
\figurenum{2.8}
\includegraphics[width=15cm]{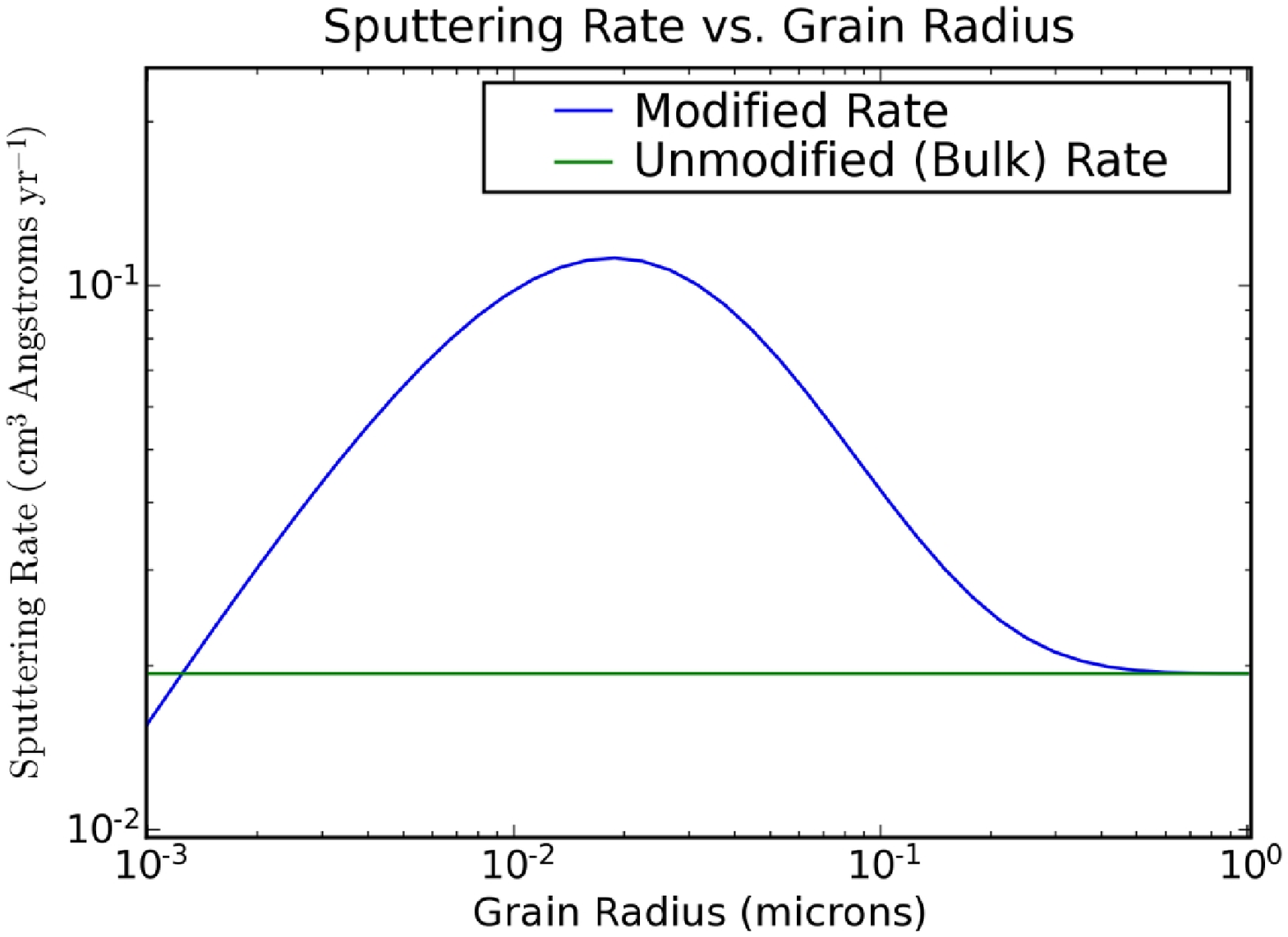}
\caption{Sputtering rate as a function of grain radius, where
sputtering is done by protons with energy 10 keV. Rates used in code
shown in blue, bulk solid approximation shown in green. Increase in
rate from $10^{-2}-10^{-1}$ $\mu$m due to enhancement of sputtering
for small grains, dropoff short of $\sim 10^{-2}$ $\mu$m due to
decreased energy deposition rate for energetic particles.
\label{modifiedyield}
}
\end{figure}

\newpage

\begin{figure}
\figurenum{2.9}
\includegraphics[width=16cm]{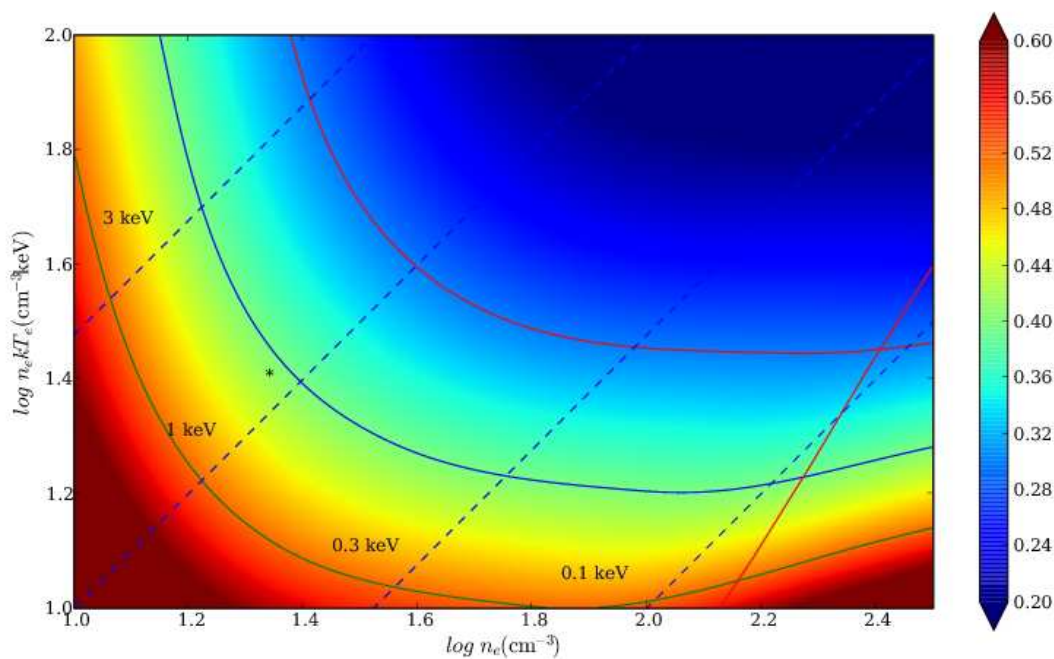}
\caption{The 70/24 $\mu$m MIPS flux ratio for Kepler's SNR (see
Chapter 6) as a function of electron density, $n_{e}$ and pressure,
$n_{e}kT_{e}$. The background color scale indicates 70/24 $\mu$m
ratio, as indicated by the color bar at right. Dashed lines are lines
of constant temperature. The solid magenta diagonal line at lower
right indicates where the modeled shocks would become radiative,
assuming solar abundance models and an age of 400 yr. Three solid
curves are lines of constant 70/24 $\mu$m MIPS flux ratios, 0.30,
0.40, and 0.52 (from top to bottom). Position of a Balmer-dominated
fast shock is marked by a star. There are many combinations of density
and temperature that can yield an identical 70/24 $\mu$m flux ratio.
\label{kep2dplot}
}
\end{figure}

\newpage
\clearpage

\begin{figure}
\figurenum{2.10}
\includegraphics[width=16cm]{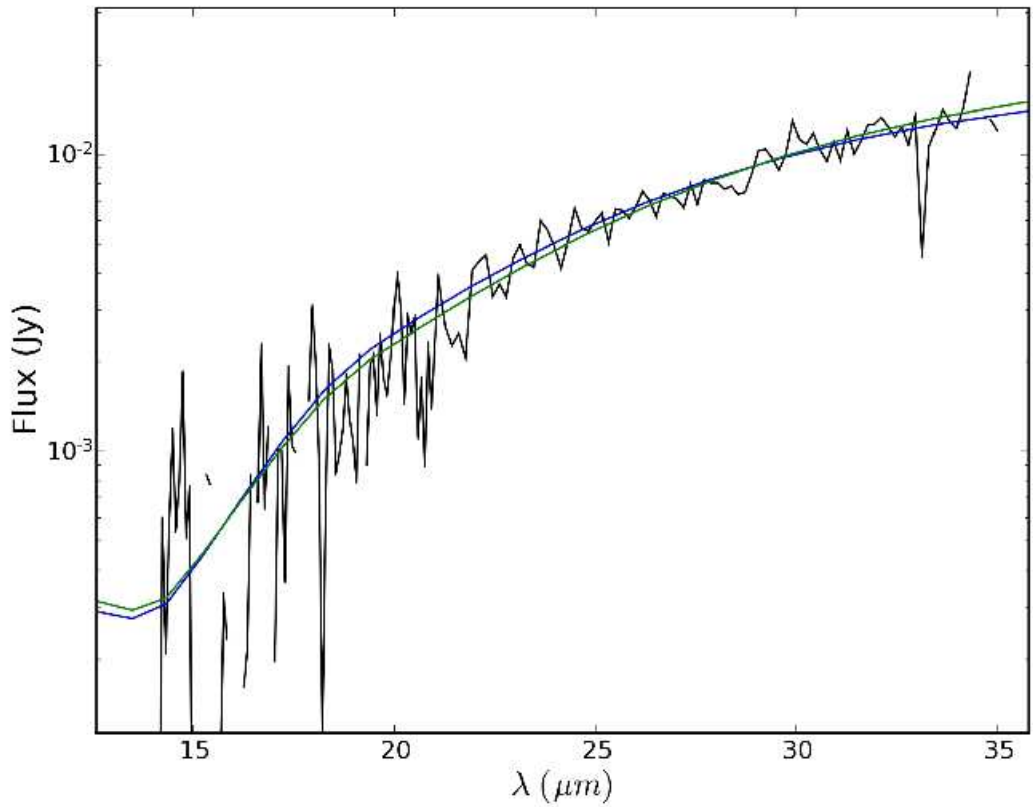}
\caption{{\it Spitzer} IRS data of ``faint'' region (see Chapter 8) of
0509-67.5, overlaid with 90\% confidence limits from model fit. The
best-fit density for this data was $n_{p}$ = 0.88 cm$^{-3}$ (not
shown), and the 90\% confidence lower and upper limits are 0.7 and 1.0
cm$^{-3}$, shown in green and blue, respectively. $\chi^{2}$ fits were
only performed in the region from 21-32 $\mu$m, where signal-to-noise
was strongest.
\label{conflimits}
}
\end{figure}

\newpage
\clearpage

\section{Dust Destruction in Type Ia Supernova Remnants in the Large Magellanic Cloud}

This chapter is reproduced in its entirety from Borkowski, K.J.,
Williams, B.J., Reynolds, S.P., Blair, W.P., Ghavamian, P., Sankrit,
R., Hendrick, S.P., Long, K.S., Raymond, J.C., Smith, R.C., Points,
S., \& Winkler, P.F. 2006, ApJ, 642, 141.

\subsection{Introduction}

The dust content in galaxies, dust composition, and grain size
distribution are determined by the balance between dust formation,
modification in the interstellar medium (ISM), and destruction
\citep{draine03}.  Some evidence exists for dust formation in the
ejecta of core-collapse supeyrnovae \citep[e.g., SN
1987A;][]{dekool98} but no reports exist for SNe Ia.  Dust {\it
destruction} is intrinsically linked to SN activity, through
sputtering in gas heated by energetic blast waves and through betatron
acceleration in radiative shocks \citep{jones04}.  Dust destruction in
SNRs can be studied by its strong influence on thermal IR emission
from collisionally heated dust.  The IRAS All Sky Survey provided
fundamental data on Galactic SNRs \citep{arendt89,saken92}. This
prompted extensive theoretical work on dust heating, emission, and
destruction within hot plasmas, summarized by
\citet{dwekarendt92}. Theory is broadly consistent with IRAS
observations, but limitations of those observations (low spatial and
spectral resolution and confusion with the Galactic IR background)
precluded any detailed comparisons. In particular, while it is clear
that thermal dust emission is prevalent in SNRs, our understanding of
dust destruction is quite poor.

To examine the nature of dust heating and destruction in the
interstellar medium, we conducted an imaging survey with the {\sl
Spitzer Space Telescope} (SST) of 39 SNRs in the Magellanic Clouds
(MCs).  We have selected a subset of our detections, four remnants of
Type Ia supernovae, to address questions of dust formation in Type Ia
ejecta, dust content of the diffuse ISM of the LMC, and dust
destruction in SNR shocks.  Both DEM L71
(0505-67.9; Rakowski, Ghavamian, \& Hughes 2003) and 
0548--70.4 (Hendrick, Borkowski, \& Reynolds 2003) show
X-ray evidence for iron-rich ejecta in the interior, and both have
well-studied Balmer emission from nonradiative shocks
\citep{ghavamian03,smith91}.  Two smaller remnants, 0509--67.5 and
0519--69.0, also show prominent H$\alpha$ and Ly$\beta$ emission from
nonradiative shocks (Tuohy et al.~1982; Smith et al.~1991; Ghavamian
et al. 2006, in preparation).  There appears to be little or no
optical contribution from radiative shocks.  Confusion in IR is
widespread in the LMC, but our remnants are less confused than
typical, easing the task of separating SNR emission from background.

\subsection{Observations and Data Reduction}

\begin{figure}
\figurenum{3.1}
\includegraphics[width=14cm]{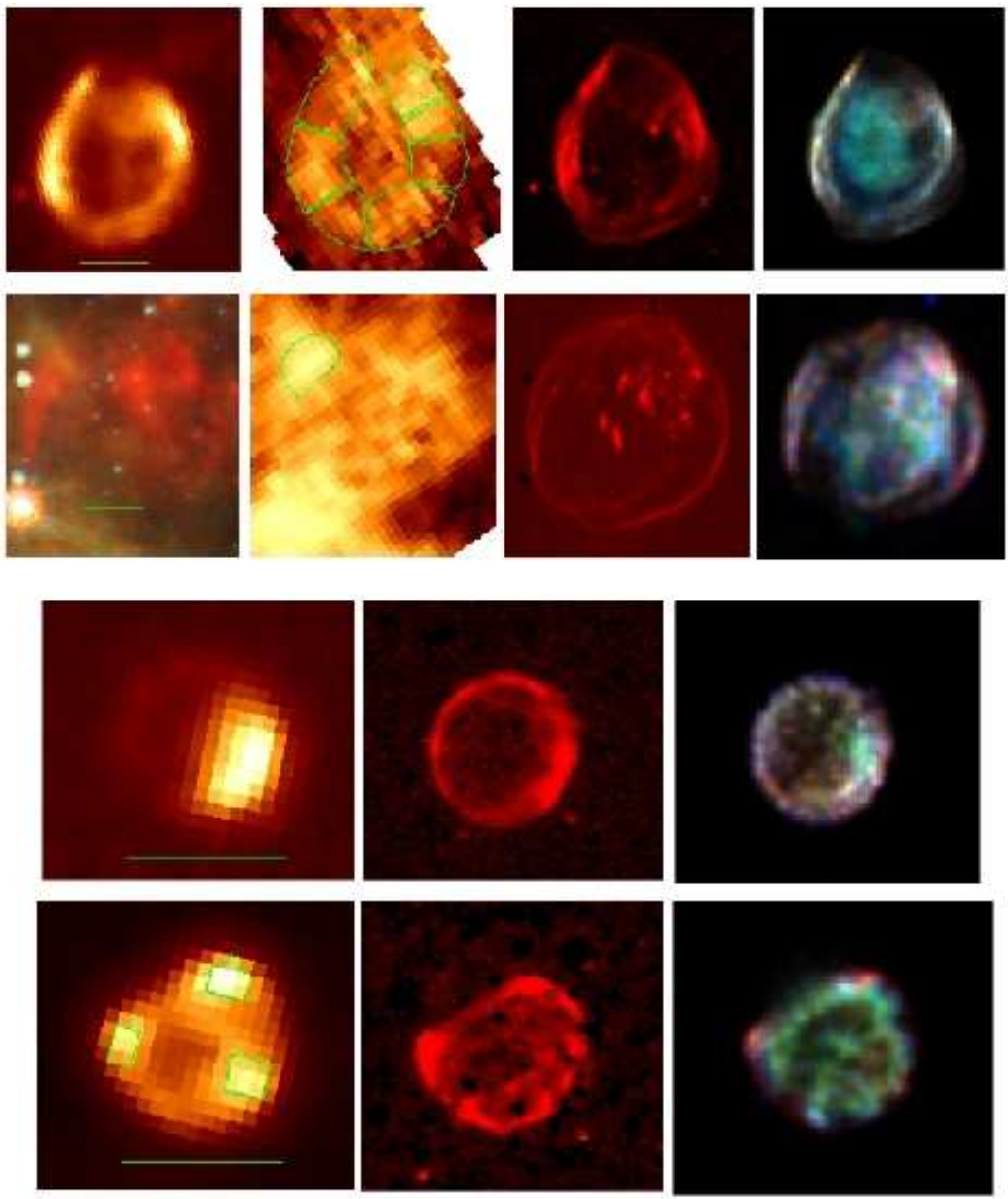}
\caption{Top row:  DEM L71 at 24 and 70 $\mu$m,
H$\alpha$, and X-ray (red, 0.3 -- 0.7 keV; green,
0.7 -- 1.0 keV; blue, 1.0 -- 3.5 keV; smoothed with
1 pixel Gaussian).
Second row: 0548--70.4 with red, 24 $\mu$m; green, IRAC 8.0 $\mu$m; 
blue $\mu$m, IRAC 5.8 $\mu$m; 70 $\mu$m; H$\alpha$, and X-ray image
as for DEM L71, smoothed with a 2 pixel Gaussian. 
Third row:  0509--67.5 at 24 $\mu$, 
H $\alpha$, and X-ray: red, 0.3 -- 0.7 keV; green, 0.7 -- 1.1 keV and blue, 
1.1 -- 7.0 keV, 
Bottom row:  0519--69.0, as in third row. Half-arcminute scales are shown for each SNR.
\label{imagesia}
}
\end{figure}

We observed all four objects in all four bands of the Infrared Array
Camera (IRAC), as well as with the Multiband Imaging Photometer for
Spitzer (MIPS) at 24 and 70 $\mu$m.  Each IRAC observation totaled 300
s (10 30-s frames); at 24 $\mu$m, 433 s total (14 frames); and at 70
$\mu$m, 986 s total (94 frames) for all but 0548--70.4, for which we
observed a total of 546 s in 52 frames.  The observations took place
between November 2004 and April 2005.  Images are shown in
Figure~\ref{imagesia}.  Confusion from widely distributed warm
dust made many 70 $\mu$m observations problematic, but we obtained
useful data on both DEM L71 and 0548--70.4.

MIPS images were processed from Basic Calibrated Data (BCD) to
Post-BCD (PBCD) by v.~11 of the SSC PBCD pipeline. For the 24 $\mu$m
images, we then re-mosaicked the stack of BCD images into a PBCD
mosaic using the SSC-provided software MOPEX, specifically the overlap
correction, to rid the images of artifacts.  For the 70 $\mu$m data,
we used the contributed software package GeRT, provided also by the
SSC, to remove some vertical streaking.  IRAC images were also
reprocessed using MOPEX to rid the image of artifacts caused by bright
stars.

All four remnants were clearly detected at 24 $\mu$m, with fluxes
from indicated regions
reported in Table~\ref{fluxtableia}.  As Figure~1 shows, emission is
clearly associated with the X-ray-delineated blast wave, though not
with interior X-ray emission.  Since we expect line emission from
fine-structure transitions of low-ionization material to be a
significant contributor only in cooler, denser regions identified by
radiative shocks, we conclude that the emission we detect is
predominantly from heated dust.  None of our objects was clearly
detected at 8 $\mu$m, with fairly stringent upper limits shown in
Table~\ref{fluxtableia}.

\subsection{Discussion}

We modeled the observed emission assuming collisionally heated dust
\citep[e.g.,][]{dwekarendt92}.  The models allow an arbitrary
grain-size distribution, and require as input parameters the hot gas
density $n$, electron temperature $T_e$, ion temperature $T_i$, and
shock sputtering age $\tau=\int_0^t n_p dt$.  The models use an
improved version of the code described by \citet{borkowski94},
including a method devised by \citet{guh89} to account for
transiently-heated grains, whose temperature fluctuates with time and
therefore radiate far more efficiently.  The energy deposition rates
by electrons and protons were calculated according to \citet{dwek87}
and \citet{dweksmith96}.  We used dust emissivities based on bulk
optical constants of \citet{draine84}.  Our non-detections in IRAC
bands showed that small grains are destroyed, so it was not necessary
to model emission features from small polycyclic aromatic hydrocarbon
(PAH) grains.  The preshock grain size distribution was taken from the
``provisional'' dust model of \citet{weingartnerdraine01}, consisting
of separate carbonaceous and silicate grain populations, in particular
their average LMC model with maximal amount of small carbonaceous
grains.  Sputtering rates are based on sputtering cross sections of
\citet{bianfer05}, augmented by calculations of an enhancement in
sputtering yields for small grains by \citet{jurac98}. We have modeled
1-D shocks, that is, superposed emission from regions of varying
sputtering age from zero up to a specified shock age \citep{dwek96}.

To estimate shock parameters, we used the non-radiative shock models
of Ghavamian et al.~(2001) to model the broad component H$\alpha$
widths and broad-to-narrow H$\alpha$ flux ratios measured by
\cite{tuohy82} and \cite{smith91} for the LMC SNRs.  Results for
electron and proton temperatures $T_e$ and $T_p$ are quoted in
Table~\ref{inputsia}.  For 0509--67.5, we assumed $T_e/T_p \le 0.1$ at
the shock front, consistent with the observed Ly$\beta$ FWHM of 3700
km s$^{-1}$ (Ghavamian et al. 2006, in preparation).  For Sedov
dynamics, the sputtering age $\tau$ (which is also the ionization
timescale) reaches a maximum of about $(1/3)n_p t$ where $t$ is the
true age of the blast wave \citep{borkowski01}.  Therefore we use an
``effective sputtering age'' of $n_p t/3$ when calculating effects of
sputtering.

\subsubsection{DEM L71 and 0548--70.4}

These two remnants have been well-studied in X-rays (DEM L71: Rakowski
et al.~2003; 0548--70.4, Hendrick et al.~2003).  They have ages of
4400 and 7100 yr, respectively, derived from Sedov models.  For DEM
L71, Ghavamian et al.~(2003) were able to infer shock velocities over
much of the periphery, ranging from 430 to 960 km s$^{-1}$, consistent
with X-ray inferences \citep{rakowski03}.

To model DEM L71, we used parameters deduced from {\sl Chandra}
observations \citep{rakowski03}, averaged over the entire blast wave
since different subregions were fairly similar.  We find a predicted
70/24 ratio in the absence of sputtering ($\tau = 0$) of about 2.3 
(including only grains larger than 0.001 $\mu$m in radius),
compared to the observed 5.1.  Using an effective age of 1/3 the Sedov
age gave a value of 5.1.  Table~\ref{resultsia} also gives the total
dust mass we derive, and the total IR luminosity produced by the
model.

For 0548--70.4, both the east and west limbs and some bright knots of
interior emission are visible at 24 $\mu$m, but only the north half of
the east limb is clearly detected at 70 $\mu$m. Only fluxes from this
region were measured; the results are summarized in
Table~\ref{fluxtableia}.  Using a 1-D model for Coulomb heating of
electrons by protons, we calculate a mean electron temperature in the
shock region of $T_e \sim 0.66$ keV.  A model using the postshock
density of 0.72 cm$^{-3}$ obtained by Hendrick et al.~(2003) for the
whole limb (including sputtering) gives too high a 70/24 $\mu$m
ratio. That ratio is very sensitive to density; we found that
increasing $n_p$ by a factor $< 2.5$ adequately reproduced the observed
ratio.  That fitted density appears in Table~\ref{inputsia} and the
corresponding results are in Table~\ref{resultsia}. Gas mass was derived
from the X-ray emission measure of the east limb \citep{hendrick03}, 
scaled to the region shown in Figure~1, and using electron density in 
Table~\ref{inputsia}.

\subsubsection{0509--67.5 and 0519--69.0}

Our other two objects are much smaller; X-ray data suggest young ages
\citep{warren04}.  Detections of light echoes \citep{rest05} indicate
an age of about 400 yr for 0509--67.5 and about 600 yr for 0519--69.0,
with $\sim 30$\% errors.  Much higher shock velocities inferred by
Ghavamian et al.~(2006, in preparation) mean that plasma heating
should be much more effective.  Higher dust temperatures, hence lower
70/24 $\mu$m ratios, should result.  In fact, we did not detect either
remnant at 70 $\mu$m, with upper limits on the ratio considerably
lower than the other two detections (Table~\ref{fluxtableia}).

In the case of 0509-67.5, optical-UV observations fix only $T_p$, so
we regarded the density $n_p$ as a free parameter, fixing $\tau$ at
$n_p t/3$ and finding $T_e$ assuming no collisionless heating.  Our 70
$\mu$m upper limit gives a lower limit on $n_p$, shown in
Table~\ref{inputsia}, as well as an upper limit on the total dust mass
(Table~\ref{resultsia}).

The analysis of 0519-69.0 was identical to that done for 0509-67.5. 
However, for 0519-69.0 we divided the remnant up
into two regions: the three bright knots (which we added together and
considered one region, accounting for 20\% of the total flux) and the
rest of the blast wave.  Optical spectroscopy (Ghavamian et al.~2006)
allowed determination of parameters separately for the knots and the
remainder.  Again regarding $n_p$, $T_e$ and $\tau$ as free parameters,
we place lower limits on the post-shock densities and $T_e$, and upper
limits on the amount of dust mass, including sputtering.  For both
remnants, density limits assume the effective sputtering age; if there
is no sputtering at all, we obtain firm lower limits on density lower
by less than a factor of 2.

\subsection{Results and Conclusions}

The IR emission in the Balmer-dominated SNRs in the LMC is spatially
coincident with the blast wave. It is produced within the shocked ISM
by the swept-up LMC dust heated in collisions with thermal electrons
and protons.  We find no evidence for infrared emission associated
with either shocked or unshocked ejecta of these thermonuclear
SNRs. While detailed modeling of small grains is required to make a
quantitative statement, apparently little or no dust forms in such
explosions, and any line emission produced by ejecta is below our
detection limit. This is consistent with observations of Type Ia SNe
where dust formation has never been observed.  It is also consistent
with the absence in meteorites of presolar grains formed in Type Ia
explosions \citep{clayton04}.

The measured 70/24 $\mu$m MIPS ratios in DEM L71 and 0548-70.4, and
the absence of detectable emission in the IRAC bands in all 4 SNRs,
can be accounted for with dust models which include destruction of
small grains. Without dust destruction, numerous small grains present
in the LMC ISM \citep[e.g.,][]{weingartnerdraine01} would produce too
much emission at short wavelengths when transiently heated to high
temperatures by energetic particles. Destruction of small grains is
required to reproduce the observed 70/24 $\mu$m MIPS ratios in DEM L71
and 0548-70.4: 90\%\ of the mass in grains smaller than 0.03--0.04
$\mu$m is destroyed in our models.  Even with this destruction, we
infer pre-sputtering dust/gas mass far smaller than the 0.25\% in the
Weingartner \& Draine model.

The two young remnants, 0509--67.5 and 0519--69.0, have been detected
only at 24 $\mu$m, but our rather stringent upper limits at 70
$\mu$m suggest the presence of much hotter dust than in the older
SNRs DEM L71 and 0548-70.4. Such hot dust is produced in our plane
shock models only if the postshock electron densities exceed $1.6$
cm$^{-3}$ and 3.4--7.7 cm$^{-3}$ in 0509--67.5 and 0519--69.0,
respectively (Table~\ref{inputsia}).  0509--67.5 is asymmetric, and the
quoted lower density limit needs to be reduced if an average postshock
electron density representative of the whole SNR is of interest. We
measure a flux ratio of 5 between the bright and faint hemispheres,
depending primarily on the gas density ratio between the hemispheres,
and on the ratio of swept-up ISM masses. For equal swept-up masses,
our models reproduce the observed ratio for a density contrast of 3 or
less; the actual density contrast is lower because more mass has been
swept up in the brighter hemisphere. The nearly circular shape of
0509--67.5 also favors a low density contrast.  The densities derived
here are several times higher than an upper limit to the postshock
density of $0.2$ cm$^{-3}$ obtained by \citet{warren04} who used
hydrodynamical models of \citet{dwarkadas98} to interpret {\it
Chandra} X-ray observations of this SNR. The origin of this
discrepancy is currently unknown.  Possible causes include: (1)
neglect of extreme temperature grain fluctuations in our dust models
for 0509--67.5, (2) modification of the blast wave by cosmic rays as
suggested for the Tycho SNR by \citet{warren05}, (3) contribution of
line emission in the 24 $\mu$m MIPS band. 
 
The measured 24 and 70 $\mu$m IR fluxes, in combination with estimates
of the swept-up gas from X-ray observations, imply a dust/gas ratio a
factor of several lower than typically assumed for the LMC. In order
to resolve this discrepancy, one needs much higher dust destruction
rates and/or a much lower dust/gas ratio in the pre-shock gas. Most
determinations of dust mass come from higher-density regions, but Type
Ia SNRs are generally located in the diffuse ISM, where densities are
low. Both the dust content and the grain size distribution might be
different in the diffuse ISM. In the Milky Way, the dust content is
lower in the more diffuse ISM \citep[e.g.,][]{savage96}, most likely
due to dust destruction by sputtering in fast SNR shocks (more
prevalent at low ISM densities) and by grain-grain collisions in
slower radiative shocks.  Grain-grain collisions are the more likely
destruction mechanism for large grains \citep{jones94,borkowski95}, so
such grains might be less common in the diffuse ISM.  Smaller grains
are more efficiently destroyed by sputtering in SNRs, so dust
destruction will be more efficient for a steeper preshock grain size
distribution (more weighted toward small grains).  This in combination
with the lower than average preshock dust content mostly likely
accounts for the observed deficit of dust in the Balmer-dominated SNRs
in the LMC. Apparently dust in the ambient medium near these SNRs has
been already affected (and partially destroyed) by shock waves prior
to its present encounter with fast SNR blast waves.  Spectroscopic
follow-up is required in order to confirm preliminary conclusions
presented in this work and learn more about dust and its destruction
in the diffuse ISM of the LMC.

We thank Joseph Weingartner and Karl Gordon for discussions about dust
in the LMC.

\newpage

\begin{deluxetable}{lccc}
\tablenum{3.1}
\vspace{-0.2truein}
\tablecolumns{4}
\tablewidth{0pc}
\tabletypesize{\footnotesize}
\tablecaption{Measured Fluxes and Upper Limits\tablenotemark{a}}
\tablehead{
\colhead{Object} & 8.0 $\mu$m & 24 $\mu$m & 70 $\mu$m}

\startdata
DEM L71 & $<1.06$ & 88.2 $\pm${8.8} & 455 $\pm${94}\\
0548-70.4 & $<3.82$ & 2.63 $\pm${0.30} & 19.9 $\pm${4.7}\\
0509-67.5 & $<0.2$ & 16.7 $\pm${1.7} & $<32.7$\\
0519-69.0 & $<0.9$ & 92.0 $\pm${9.2} & $<121$\\

\enddata

\tablenotetext{a}{All fluxes (not color-corrected) in mJy.  Limits are $3\sigma$.}
\label{fluxtableia}
\end{deluxetable}

\newpage
\clearpage

\begin{deluxetable}{lccccccc}
\tablenum{3.2}
\vspace{-0.3truein}

\tablecolumns{4}
\tablewidth{0pc}
\tabletypesize{\footnotesize}
\tablecaption{Model Input Parameters}
\tablehead{
\colhead{Object} & $T_e$ (keV) & $T_p$ (keV) & $n_p$ & $n_e$
& Age (yrs.)& $\tau (10^{10}$ cm$^{-3}$ s) & Ref.}

\startdata
DEM L71                    & 0.65    & 1.1 & 2.3    & 2.7    & 4400 & 11   & 1, 2\\
0548-70.4                  & 0.65    & 1.5 & 1.7    & 2.0    & 7100 & 12   & 3, 4\\
0509-67.5                  & 1.9     & 89  & $>1.4$ & $>1.6$ & 400  & 0.59 & 4, 5\\
0519-69.0\tablenotemark{a} & 2.1     & 36  & $>2.8$ & $>3.4$ & 600  & 1.8  & 4, 5\\
0519-69.0\tablenotemark{b} & 1.0     & 4.2 & $>6.4$ & $>7.7$ & 600  & 4.0  & 4, 5\\

\enddata

\tablenotetext{a}{Fainter portions of remnant}
\tablenotetext{b}{Three bright knots}

\tablecomments{Densities are post-shock.  References: (1) Rakowski et al 2003; 
(2) Ghavamian et al 2003; (3) Hendrick et al 2003; 
(4) Ghavamian et al 2006, in preparation; (5) Rest et al. 2005}
\label{inputsia}
\end{deluxetable}

\newpage
\clearpage

\begin{deluxetable}{lcccccccc}
\tablenum{3.3}
\vspace{-0.3truein}
\tablecolumns{4}
\tablewidth{0pc}
\tabletypesize{\footnotesize}
\tablecaption{Model Results}
\tablehead{
\colhead{Object} & 70/24 (0) & 70/24 sput. & 70/24 obs. & $T$(dust)(K) 
& Dust Mass 
& \% destr. 
& dust/gas 
& $L_{36}$ }

\startdata
DEM L71                    & 2.3       & 5.1         & 5.1        & 55--65   
   & 0.034                   & 35                & 4.2$\times 10^{-4}$    & 12\\
0548-70.4                  & 2.7       & 7.6         & 7.6        & 53--62   
   & 0.0018                & 40                & 7.5$\times 10^{-4}$  & 2.1\\
0509-67.5                  & $<2.0$    & $<2.0$      & $<2.0$     & 66--70   
   & $<1.1 \times 10^{-3}$    & $>18$                & ... & ...\\
0519-69.0\tablenotemark{a} & $<1.3$    & $<1.3 $     & $<1.3$     & 72--77  
   & $<2.7 \times 10^{-3}$    & $>34$                & ... & ...\\
0519-69.0\tablenotemark{b} & $<1.3$    & $<1.3 $     & $<1.3$     & 73--86  
   & $<6.4 \times 10^{-4}$    & $>38$                & ...                   & ...\\

\enddata

\tablenotetext{a}{Fainter portions of remnant}
\tablenotetext{b}{Three bright knots}
\tablecomments{Column 2: model prediction without sputtering; column 3, including
sputtering with $\tau = n_p t/3$; column 4, observations; column 5, for 0.02--0.1 $\mu$m grains; column 6, mass of dust
currently observed (after sputtering), in $M_\odot$; column 7, percentage of original
dust destroyed; column 8, ratio of swept-up dust to gas masses; column 9,
$L_{36} \equiv L_{IR}/10^{36}$ erg s$^{-1}$.}

\label{resultsia}
\end{deluxetable}

\newpage
\clearpage

\newpage
\clearpage

\section{Dust Destruction in Fast Shocks of Core-Collapse Supernova Remnants in the Large Magellanic Cloud}

This chapter is reproduced in its entirety from Williams, B.J.,
Borkowski, K.J., Reynolds, S.P., Blair, W.P., Ghavamian, P., Hendrick,
S.P., Long, K.S., Points, S., Raymond, J.C., Sankrit, R., Smith, R.C.,
\& Winkler, P.F. 2006, ApJ, 652, 33.

\subsection{Introduction}

Dust plays an important role in all stages of galaxy evolution. The
life-cycle of dust grains and the amount and relative abundances
present in the interstellar medium (ISM) are determined by the balance
between dust formation, grain modification, and dust destruction
\citep{draine03}.  Dust destruction is known to occur in both fast and
slow shocks in SNRs \citep{jones04}. We focus here on dust destruction
via sputtering by high energy ions in fast (non-radiative) shocks in
SNRs.

SNRs make excellent probes of the dust content of the diffuse ISM in
galaxies, since their shock waves create X-ray plasmas that heat dust
in their vicinities.  Modeling of the X-ray emission provides the
basis for understanding the dust emission.  Inferences of dust content
and properties from SNR observations are thus complementary to
UV-absorption studies, which rely on the fortuitous locations of
background UV-bright stars \citep{jenkins84}.  The combination of
these two types of investigation may lead to significant advances in
our understanding of dust properties, with potential repercussions for
gas-phase abundance determinations, theories of chemical evolution of
galaxies, and dust-catalyzed cosmochemistry.

To examine the nature of dust heating and destruction in the ISM, we
conducted an imaging survey with the {\it Spitzer Space Telescope} of
39 SNRs in the Magellanic Clouds.  We detected at least 17 remnants at
one or more wavelengths.  In a previous paper (Borkowski et al.~2006;
Paper I) we analyzed IR emission from SNRs resulting from Type Ia
SNe. We found that the dust-to-gas mass ratio was lower than the
expected value of 0.25\% \citep{weingartnerdraine01} (hereafter WD)
for the surrounding ISM, a result that we attributed to Type Ia SNe
exploding in lower density media. Since core-collapse SNe are thought
to occur in more dense areas, we focus here on performing a similar
analysis on remnants resulting from core-collapse SNe, to determine
whether our hypothesis is correct.

We have chosen four remnants of core-collapse supernovae from our
sample to further address the question of dust formation in SNe
ejecta, dust destruction in SNR shocks, and dust abundance in the
diffuse ISM of the LMC.  We base our inference of the core-collapse
nature of our objects on the presence of a pulsar-wind nebula in
0453--68.5 \citep{gaensler03}, a central compact object in N23 (Hayato
et al.~2006, ApJ, submitted; Hughes et al.~2006), the O-rich
classification of N132D \citep{lasker78}, and the presence of a large
mass of Mg in N49B \citep{park03}.

All four of these remnants were detected in both the 24 and 70 $\mu$m
bands of the Multiband Imaging Photometer for {\it Spitzer} (MIPS).
Morphological comparisons with optical images show that there is
apparently little contribution from radiative shocks.  As was the case
with the Type Ia remnants (Paper I), we see no clear indication of IR
emission coming from remnant interiors. Rather, we find a similar
result as we found for the Type Ia remnants in Paper I: the
dust-to-gas ratios we infer are lower by a factor of $\sim 4$ than
what is expected for the LMC. In section 4.3, we discuss
possible reasons for this apparent dust deficiency.

\subsection{Observations and Data Reduction}

All four objects were observed with the 24 and 70 $\mu$m arrays, and
N49B with all four channels of the Infrared Array Camera (IRAC).  N49B
was not detected in any IRAC channel, so only the 24 and 70 $\mu$m
observations will be discussed here. At 24 $\mu$m, we mapped each
remnant in our survey and the surrounding background with 14 frames of
30.93 seconds each, for a total exposure time of 433 s. At 70 $\mu$m,
we observed a total of 545 s in 52 frames for all but N132D, which we
observed for 986 s in 94 frames.

MIPS images were processed from Basic Calibrated Data (BCD) to
Post-BCD (PBCD) at the {\it Spitzer} Science Center (SSC) by version
13.2 of the PBCD pipeline. For the 70 $\mu$m images, we used the
contributed software package GeRT to reprocess the raw telescope
images into BCD images, then reprocessed the BCD images using the SSC
software package MOPEX. GeRT was useful in removing some of the
artifacts, such as vertical stripes, from the 70 $\mu$m data.

Figure 4.1 shows the 24 and 70 $\mu$m images as well as X-ray images
from {\it Chandra} archival data and optical images from the
Magellanic Cloud Emission-Line Survey (MCELS; Smith et
al.~2005). Because line emission from low-ionization gases should be a
significant contributor only in slower, radiative shocks, we conclude
that we are seeing thermal IR emission primarily resulting from dust.
Our measured fluxes are presented in Table~\ref{fluxtablecc}.

\subsection{Modeling}

Because of the morphological similarities between the IR emission and
the blast wave seen in X-rays, we believe the dust present in the ISM
is being collisionally heated by electrons and protons in the outward
moving shock wave \citep{dwekarendt92}. The modeling of dust emission
for these remnants is similar to the modeling done on the four Type Ia
remnants in Paper I, where it is described more fully. Our model uses
a one-dimensional plane-shock approximation.  The sputtering
timescale, $\tau_p$, is equal to $\int_0^t n_p dt$, and is one of the
inputs to the code. The other inputs are electron temperature $T_e$,
ion temperature $T_i$, gas density $n$, grain size distribution, and
grain composition and relative abundances. For the composition,
abundances, and distribution of dust grains, we follow
\citet{weingartnerdraine01}.

X-ray analysis provides estimates of the electron temperature,
ionization timescale $\tau_i \equiv \int_0^t n_e dt$, and emission
measure of the plasma. We used archival {\it Chandra} data for our
analysis, and fit X-ray spectra using Sedov non-equilibrium ionization
(NEI) thermal models in XSPEC
\citep{arnaud96,borkowski01}. Emission-measure averaged $T_e$ and
$T_i$ from these models and the reduced $\tau_i$ ($1/3$ of the Sedov
$\tau_{ised}$, defined as the product of the postshock electron
density and the SNR age) were then used as inputs to our plane shock
model.  (The approximate factor 1/3 arises from applying results of a
spherical model to plane-shock calculations; see Fig.~4 of Borkowski
et al.~2001).  Since the $T_i/T_e$ ratio is close to 1 in these
models, we set $T_i = T_e$.

We employed two methods to model dust emission. In the first, we fix
$\tau_p$, $T_e$, and $T_p$ as derived from X-rays (taking $\tau_p =
\tau_i/1.2$), leaving only the density and total dust mass as free
parameters. In the second, we use the dynamical age of the remnant, as
derived from optical or global X-ray studies, leaving the shock age
and density as free, but correlated, parameters. We find that these
two different methods produce very similar results, within a factor of
$\sim 25\%$. We thus report only the inputs and results from the first
method. Model input parameters are given in Table~\ref{inputscc}.  The
density of the gas is then adjusted to reproduce observed 70/24 $\mu$m
flux ratio, and 24 $\mu$m flux is normalized to the observed value to
provide a total dust mass in the region of interest. The emission
measure divided by the electron density gives an estimate for the
amount of gas swept up by the blast wave, and dividing the dust mass
by the gas mass gives us a dust-to-gas mass ratio for the shocked
ISM. Results are summarized in Table~\ref{resultscc}.

\subsubsection{N132D}

SNR N132D has been well studied in optical wavelengths
\citep{morse96}, and is one of the brightest remnants in the
Magellanic Clouds at X-ray wavelengths.  The remnant is
extraordinarily bright at 24 $\mu$m, with a total flux of $\sim 3$
Jy. It is the only remnant in our sample that is brighter at 24 $\mu$m
than at 70 $\mu$m. This implies warm dust, and therefore a dense
environment (to provide the inferred heating).  We derive a plasma
temperature from X-rays of 0.6 keV for the NW, and 1 keV for the
south. The dense environment is consistent with the remnant being very
bright in X-rays. \citet{morse96} estimate a preshock hydrogen density
for this remnant of 3 cm$^{-3}$ based on modeling of the photoionized
shock precursor.  We analyzed the NW rim and the bright southern rim
separately. We find high densities, in good agreement with the
expected postshock proton density $n_p$ of 12 cm$^{-3}$. Densities are
higher by a factor of $\sim 2$ in the NW, and thus (since the emission
measure is fixed), comparatively less mass in gas in that region. The
mass in dust, however, was comparable to what was found in the south,
adjusting for the different sizes of the regions.

\subsubsection{N49B}

N49B is believed to be older than N132D, perhaps as old as 10,000 yrs.
The ionization age is lower than this, however, perhaps due to an
explosion into a pre-existing cavity \citep{hughes98}. This remnant is
much fainter than N132D at all wavelengths, but especially so at 24
$\mu$m, and the flux ratio is lower by more than order of magnitude in
N49B compared to N132D. This can be explained by the much lower
density we find in N49B. We find a density of $n_{p}$ = 1.1 cm$^{-3}$
from our dust models, and a density of $n_{p}$ = 2.1 cm$^{-3}$ from a
Sedov analysis of X-ray data. Since dust heating rates are strongly
dependent on density, a low-density environment will result in much
cooler dust, whose spectrum peaks at longer wavelengths. We find a
mean plasma temperature in this remnant of 0.36 keV. Despite the
contrasts in density and flux ratio of more than an order of
magnitude, the dust-to-gas ratio in this remnant is only twice as low
as that found in N132D.

\subsubsection{N23}

At 70 $\mu$m, the brightest parts of the shell of N23 are clearly
visible. The region selected for analysis completely enclosed the
emission visible at 70 $\mu$m, which contained about 75\% of the
emission visible at 24 $\mu$m. We derive a mean plasma temperature of
0.56 keV, and again a lower dust-to-gas ratio than is expected. Dust
modeling yields $n_{p}$ = 5.8 cm$^{-3}$. \citet{hughes06} found
densities of 10 and 23 cm$^{-3}$ in a couple of X-ray bright shocks in
this region of the remnant.

\subsubsection{0453--68.5}

Although 0453--68.5 had the highest 70/24 $\mu$m ratio of any remnant
in this sample, only the north rim of the shell was clearly separable
from the background at 70 $\mu$m. This is due to the overall faintness
of the remnant, the high levels of background confusion in the region,
and the lower sensitivity of the 70 $\mu$m array. Nonetheless, there
was sufficient S/N to analyze the north rim of the remnant. We
obtained a temperature of 0.29 keV, the lowest in our sample. The
density found from modeling dust emission, $n_{p}$ = 0.63 cm$^{-3}$,
was also the lowest of these remnants. Sedov modeling of X-rays
yielded a value of $n_{p}$ = 1.1 cm$^{-3}$. Both the temperature and
density are consistent with the observations of overall faintness at
24 $\mu$m and a high 70/24 $\mu$m ratio.

\subsection{Discussion and Conclusions}

We find excellent (generally within a factor of 2) agreement between
densities derived from dust modeling and estimated from optical and
X-ray data. As has been proposed in the past
\citep[e.g.,][]{dwekarendt92}, dust emission is indeed a valuable
density diagnostic of X-ray emitting plasmas.

Our basic quantitative results are contained in Table~\ref{resultscc}.
First, we note that the total mass in dust we infer is quite small, of
order $10^{-2} - 10^{-1} \msun$, for all objects.  Morphological
evidence suggests that all this is associated with the blast wave, but
these values also serve as restrictive upper limits for the amount of
ejecta dust that can have been produced in these core-collapse
supernovae. They are, for the most part, less than the $0.08 \msun
\lesssim M_{d} \lesssim 0.3 \msun$ calculated by \citep{todini01}.
Substantial amounts of dust will have been destroyed when the ejecta
are reverse-shocked, but the ejecta dust mass ultimately delivered to
the interstellar medium appears to be quite small.

We find that on average, about 40\% of the mass in dust grains,
including $\sim 90\%$ of the mass in grains smaller than 0.04 $\mu$m,
has been destroyed via sputtering in these remnants.  After accounting
for this sputtering, we find an initial dust-to-gas ratio that is
lower by a factor of roughly four than what is generally expected for
the LMC.  The values in Table~\ref{resultscc} are similar to the
dust-to-gas ratios we found in remnants from Type Ia SNe, and they are
still well below the value reported in WD of $\sim 2.5 \times
10^{-3}$. In Paper I we speculated that low dust/gas ratios might be
expected for Type Ia remnants expanding into lower-than-average
density material, in which previous supernovae might have destroyed
some grains.  Since we now have a similar deficit for core-collapse
remnants expanding into denser media, this explanation is unlikely.

The overall lower dust/gas ratios we infer will require another
explanation: either our destruction rates have been underestimated, or
dust masses deduced from UV and optical absorption studies have been
overestimated.  We examine these possibilities in turn.  First,
sputtering might have been underestimated, by underestimating either
sputtering timescales or rates.  Our models indicate that sputtering
timescales would have to be increased by an order of magnitude, on
average, to match the expected dust-to-gas ratio in the WD model, an
unlikely possibility.  Various factors could increase sputtering
rates. Sputtering can be enhanced by increased relative velocities
between ions and grains, either caused by grain motions (not included
in our models) or by higher than assumed ion temperatures. In order to
estimate importance of such effects, we doubled ion temperatures in
our shock models. This rather large enhancement in ion temperatures
resulted in only 10--15\%\ increase in the fraction of the original
dust destroyed for relatively fast shocks considered here.  Sputtering
is expected to be enhanced for nonspherical grains because of the
increased surface-to-volume ratio $S/V$.  For a prolate ellipsoidal
grain with an axial ratio of 2, $S/V$ is twice as large as for a
spherical grain, which enhances sputtering rates by a modest factor of
2. There is a possibility that sputtering yields may also be enhanced
by the same factor because of rough grain surfaces, although both
enhancement and reduction in sputtering yields may occur depending on
various parameters charaterizing the surface roughness
\citep[e.g.,][]{makeev04}. We find it unlikely that these modest
enhancements in sputtering rates can account for the lower than
expected dust-to-gas mass ratios.

It is also possible that the dust mass in the WD dust model (and more
generally in most dust models considered in the past) may be
overestimated. The total grain volume per H atom in the Milky Way
deduced from depletion of elements onto dust in the ISM
\citep[assuming solar abundances;][]{jenkins04} is only $\sim 60\%$ of
the total grain volume in the WD model \citep{draine04}. This
discrepancy becomes less severe if oversolar abundances are assumed
for the ISM. For example, \citet{cartledge06} favor oversolar
abundances based on observations of young F and G stars and the
observed depletion patterns.

A more interesting explanation for our discrepancy could both reduce
inferred dust/gas ratios from absorption studies, and increase
sputtering rates.  Porous (fluffy) grains \citep{mathis96} can be
characterized by a porosity (fraction of grain volume devoid of
material) $\cal P$.  Such grains have an extinction per unit mass
generally larger than for spherical compact
grains. \citet{voshchinnikov06} modeled visual and UV absorption
towards $\zeta$ Oph and $\sigma$ Sco with a mixture of highly porous
(${\cal P} \gapprox 0.9$) and compact grains with dust mass of only
70\%\ and 44\%\ of the dust mass in the WD model.

We expect enhanced sputtering rates for porous grains compared with
spherical compact grains. A highly porous (${\cal P}=0.9$) grain has
surface area 4.6 times larger than a compact grain of the same
mass. If sputtering yields were the same for porous and compact
grains, an enhancement of sputtering rates by a factor of 4.6 is
expected. In addition, protons, $\alpha$ particles, and heavier ions
can penetrate much deeper into porous grains, because their range
within the solid is inversely proportional to the mean grain
density. A proton with energy $E$ will penetrate $0.09 (E/1~{\rm
keV})$ $\mu$m into a porous silicate grain with ${\cal P}=0.9$ (we
used eq.~18 of Draine \&\ Salpeter 1979 to estimate the proton
projected range $R_p$). \citet{jurac98} found that sputtering yields
are enhanced if grain radius $a$ becomes less than $3R_p$ because of
additional sputtering from the back and sides of the grain. Grains as
large as $0.25 (E/1~{\rm keV})$ $\mu$m are then destroyed more
efficiently.  It is also possible that the rough surface geometry
expected for porous grains could result in a further enhancement of
sputtering rates, as the protruding grain extensions comparable in
size to the proton or $\alpha$ particle projected ranges could be more
readily sputtered \citep{jurac98}. An order of magnitude enhancement
in sputtering rates is possible for highly porous grains.

The reduced preshock dust content and substantial enhancement in
sputtering rates that result from porous grain models provide a
promising explanation for the apparent dust deficit in LMC SNRs. These
effects are partially offset by the increased IR emissivity of porous
grains, as less dust is then required to account for the observed IR
emission. Modeling of {\it Spitzer} observations with porous grain
models is required to assess whether these models can resolve the
apparent dust deficit in LMC SNRs. Only further study in this area
will resolve the question of whether our results are a deficiency in
our knowledge of how to model dust emission or a real indicator of
conditions in the LMC. Spectroscopic follow-up on these remnants in
the far-IR is required to confirm and strengthen our results, and
resolve the issues of dust content in the ISM.  Implications of a
resolution go beyond these particular SNRs or even the LMC as a whole
and could help bring about a significant improvement in our
understanding of interstellar dust in general.

\newpage
\clearpage

\begin{deluxetable}{lccc}
\tablenum{4.1}

\tablecolumns{4}
\tablewidth{0pc}
\tabletypesize{\footnotesize}
\tablecaption{Measured Fluxes}
\tablehead{
\colhead{Object} & 24 $\mu$m & 70 $\mu$m & 70/24}

\startdata
N132D NW   & 730 $\pm${73} & 430 $\pm${96} & 0.59 $\pm${0.13}\\
N132 S     & 1000 $\pm${100} & 770 $\pm${170} & 0.76 $\pm${0.17}\\
N49B       & 43 $\pm${4.3} & 395 $\pm${79} & 9.1 $\pm${2.0}\\
0453--68.5 & 13 $\pm${1.3} & 250 $\pm${50} & 19 $\pm${4.2}\\
N23        & 100 $\pm${10} & 240 $\pm${50} & 2.4 $\pm${0.5}\\

\enddata

\tablenotetext{a}{All fluxes given in millijanskys}

\label{fluxtablecc}
\end{deluxetable}

\newpage
\clearpage

\begin{deluxetable}{lccc}
\tablenum{4.2}
\tablecolumns{6}
\tablewidth{0pc}
\tabletypesize{\footnotesize}
\tablecaption{Model Inputs}
\tablehead{
\colhead{Object} & $T_e(=T_p)$  & $\tau_{p}$ & Age \\
& (keV) & ($10^{11}$ cm$^{-3}$ s) &(yr) 
} 

\startdata
N132D NW   & 0.61 & 5.3 & 2500{\tablenotemark{a}}\\
N132D S    & 1.02 & 1.0 & 2500\\
N49B       & 0.36 & 0.76 & 10,900{\tablenotemark{b}}\\
0453--68.5 & 0.29 & 2.7 & 8700{\tablenotemark{b}}\\
N23        & 0.56 & 2.1 & 4600{\tablenotemark{c}}\\

\enddata

\tablenotetext{a}{\citet{morse96}}
\tablenotetext{b}{\citet{hughes98}}
\tablenotetext{c}{\citet{hughes06}}

\label{inputscc}
\end{deluxetable}

\newpage
\clearpage

\begin{deluxetable}{lcccccc}
\tablenum{4.3}
\tablecolumns{7}
\tablewidth{0pc}
\tabletypesize{\footnotesize}
\tablecaption{Model Results}
\tablehead{
\colhead{Object} & $n_{p}$ (cm$^{-3})$ & T(dust) (K) & M$_{d}$ ($M_\odot$) (curr.) & Dust/gas (curr.) & \% dest. & Dust/Gas (orig.)}

\startdata
N132D NW & 34 & 95-120 & 0.0075 & $4.6 \times 10^{-4}$ & 50 & $9.2 \times 10^{-4}$\\
N132D S & 14 & 85-105 & 0.015 & $2.5 \times 10^{-4}$ & 38 & $4.0 \times 10^{-4}$\\
N49B & 1.1 & 50-60 & 0.08 & $1.8 \times 10^{-4}$ & 27 & $2.4 \times 10^{-4}$\\
0453--68.5 & 0.63 & 40-55 & 0.1 & $6.1 \times 10^{-4}$ & 33 & $9.8 \times 10^{-4}$\\
N23 & 5.8 & 70-85 & 0.011 & $3.0 \times 10^{-4}$ & 39 & $4.9 \times 10^{-4}$\\

\enddata

\tablecomments{Col (2): postshock proton density; col. (3): for
0.02-0.1 $\mu$m grains; col (4): mass of dust currently observed
(after sputtering), in $\msun$; col (5): ratio of current dust mass to
swept-up gas mass; col (6): percentage of dust destroyed via
sputtering; col (7): ratio of dust mass originally present to swept-up
gas mass.}

\label{resultscc}
\end{deluxetable}

\newpage
\clearpage

\begin{figure}
\figurenum{4.1}
\plotone{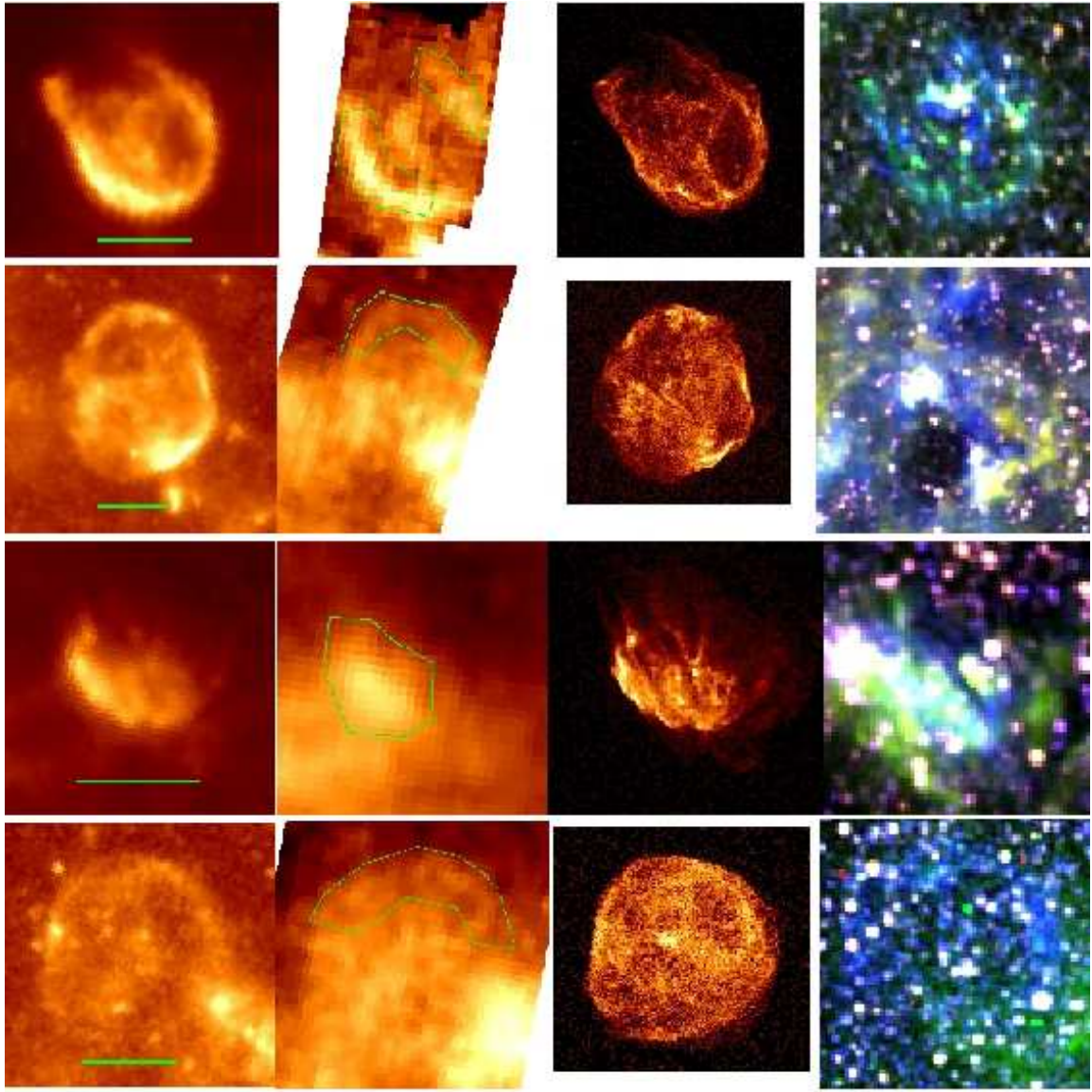}
\caption{Top row, from left to right: N132D at 24 and 70 microns (the region of
interest at 70 microns is marked on the image), in the X-rays (broadband,
Chandra image) and in the optical (overlay of MCELS images, with [S II],
Halpha and [O III] marked in red, green and blue, respectively).   Second,
third and fourth rows show the same sequence for N49B, N23 and 0453-68.5,
respectively. One arc-minute scales are shown on the 24 $\mu$m images.
\label{images}
}
\end{figure}

\newpage
\clearpage

\section{{\it Spitzer Space Telescope} Observations of Kepler's Supernova Remnant: A Detailed Look at the Circumstellar Dust Component}

This chapter is reproduced in its entirety from Blair, W.P.,
Ghavamian, P., Long, K.S., Williams, B.J., Borkowski, K.J., Reynolds,
S.P., \& Sankrit, R. 2007, ApJ, 662, 998.

\subsection{Introduction}

Each of the remnants of historical galactic supernovae (SNe) provides
a unique and important perspective to our understanding of young
supernova remnants (SNRs) and their interaction with the interstellar
medium (ISM).  SN~1604 is the second youngest galactic SNR (behind Cas
A), and was first sighted in October 1604 by Johannes Kepler and
others.  Kepler was not the first to see the SN, but he published the
most detailed account of the SN light curve (Kepler 1606), and largely
because of this the SNR has come to be known as Kepler's SNR.  Baade
(1943) was the first to recover the optical SNR. It has been well
observed at all wavelengths (X-ray: Cassam-Chena\"i et al. 2004,
Hughes 1999; Radio: DeLaney et al. 2002; Optical: Blair et al. 1991,
Sollerman et al. 2003; NIR/IR: Gerardy \& Fesen 2001, Douvion et
al. 2001; sub-mm: Morgan et al. 2003; and numerous earlier papers
referenced therein).

The distance to Kepler's SNR has been uncertain, with most
investigators adopting a value of near 5 kpc, as discussed by Reynoso
\& Goss (1999).  However, Kepler's SNR lies almost directly toward the
galactic center, making the use of H~I velocities and a galactic
rotation curve very unreliable.  Using an HST/ACS image compared to
ground-based data, Sankrit et al. (2005) have recently determined a
distance of 3.9 (+1.9, -1.4) kpc to Kepler using the proper motion of
a filament whose velocity is known with reasonable accuracy (Blair et
al. 1991).  Since the uncertainty is still rather large, we will adopt
a value of 4 kpc in this paper and scale other parameters to this
value.  The SNR lies $6.8^{\circ}$ off the galactic plane (473 $\rm
d_4$ pc).  With an angular diameter of $\sim$200\arcsec, the radius is
1.93 $\rm d_4$ pc and the mean expansion velocity has been $\rm
\sim4720 ~ d_4 ~ km~s^{-1}$.  Since the current velocity of the
primary blast wave is $\sim$1660 $\rm \pm 120 ~km~s^{-1}$ (Blair et
al. 1991; Sankrit et al. 2005), the primary shock has been
significantly decelerated over 400 years.

The brightest optical emission is from a region in the WNW, as shown
in the images of Blair et al. (1991).  Bright [S~II] emission
indicates slower, denser, radiative shocks.  However, more extensive
nonradiative emission is present where smoother H$\alpha$ filaments
are not accompanied by [S~II], for instance, across the N rim and in
portions of two centrally-projected regions.  We shall refer
frequently below to the different character of the radiative and
nonradiative emission regions.

Douvion et al.~(2001) observed Kepler with the ISOCAM instrument
aboard the Infrared Space Observatory (ISO), with an angular
resolution of 6\arcsec, in a band from 10.7 -- 12.0 $\mu$m.  Spectra
were also obtained between 6.5 and 16 $\mu$m.  They found the emission
to have a similar morphology to the H$\alpha$ image, but made no
distinction between radiative and non-radiative regions.  They found
almost no line emission to be present and were able to fit their
spectra with collisionally heated dust at a fixed temperature, using
``astronomical silicate" composition (Draine \& Lee 1984), assuming
that dust was heated in regions with densities and temperatures
consistent with radiative shock emission ($n_e \sim (2 - 10) \times
10^3$ cm$^{-3}$, $T_e \sim 10^4$ K.  They reported a total dust mass
of order $10^{-4}$ $M_\odot$.

Kepler's SNR is the only of the historical SNe from the last
millennium whose progenitor type is a matter of serious debate (Blair
2005).  The early claim by Baade (1943) of consistency of the
historical light curve with a Type Ia SN has been questioned by
Doggett \& Branch (1985) and Schaefer (1996) among others.  The
presence of dense, (apparently) N-rich circumstellar material (CSM)
surrounding the SNR so high off the galactic plane was taken to
indicate pre-SN mass loss from a massive star and hence a core
collapse SN (Bandiera 1987).  The progenitor could have been a runaway
star, which is consistent with the observed morphology (Borkowski et
al.  1992; 1994; Vel\'{a}zquez et al. 2006).  On the other hand,
analyses of X-ray spectra from Exosat (Decourchelle \& Ballet 1994),
ASCA (Kinugasa \& Tsunemi 1999, 2000) and more recently XMM-Newton
(Cassam-Chena\"i et al. 2004) indicate an overabundance of Fe and Si
enhanced ejecta, which is not expected from the explosion of a massive
star but instead suggests a Type Ia designation.

At some point in the evolution of a young SNR, some of the SN ejecta
may form into interstellar dust of various types, depending on the SN
type (e.g.  Arendt et al. 1999). Evidence exists for the formation of
dust on 1-2 year time scales in core-collapse SN (SN 1987A--McCray
1993 and Bouchet et al. 2004; SN 1999em--Elmhamdi et al.~2003).  The
large mass of iron produced by Type Ia SNe suggests that grains may
form there as well, although direct evidence for this process is scant
and timescales are not well known.  Dunne et al. (2003) have reported
the detection of a large mass of very cold dust in Cas A based on
SCUBA (sub-mm) observations, although these conclusions have been
called into question (Dwek 2004a,b; Krause et al. 2004).  Morgan et
al. (2003) have reported SCUBA observations of Kepler's SNR and
claimed a large mass of cold (17 K) dust.  A SN blast wave sweeps up
gas and dust as it expands, and so any circumstellar or interstellar
dust should also be heated, and ultimately sputtered and destroyed by
collisions in the post shock region.  Hence, young SNRs can in
principle be dust-processing laboratories.  With the advent of the
{\it Spitzer Space Telescope}, it is possible to study these effects
in unprecedented detail.

{\it Spitzer} observations of Magellanic Cloud SNRs are shedding some
light on these issues.  Borkowski et al. (2006) looked exclusively at
four remnants of Type Ia SNe.  They found no evidence of dust from the
ejecta, but rather emission consistent with heating of ISM dust by the
expanding blast wave.  B. Williams et al. (2006) selected a group of
suspected remnants of core collaspe SNe in the LMC and have found
similar results: no obvious dust emission associated with regions of
ejecta but dust emission associated with the primary blast wave was
prominent.  R. Williams et al. (2006) report {\it Spitzer} observations
of a different sample of LMC SNRs and suggest that, at least in some
objects, line emission may dominate over the dust continuum.  However,
even in these objects, the emission arises at positions consistent
with shocked ISM.

In this paper, we report infrared imaging and spectroscopy of Kepler's
SNR using the MIPS and IRAC instruments on {\it Spitzer}, obtained as
part of a Cycle 1 Guest Observer program (\#3413). We find evidence
for swept-up dust heated by the primary blast wave, but do not see
evidence for the cold dust component reported by Morgan et al. (2003).
Modeling of the dust and comparison to observed ratios are used to
constrain the allowed plasma conditions in the shocked gas.

\subsection{Observations and Data Processing}

Below we describe the {\it Spitzer} MIPS (Rieke et al. 2004) and IRAC
(Fazio et al. 2004) imaging and present the data.  The MIPS instrument
also has a low resolution spectral energy distribution (SED)
spectroscopy mode, and the data obtained on Kepler are described in
the subsequent sub-section.

\subsubsection{{\it Spitzer} Imaging}

We observed Kepler's SNR using MIPS imaging at 24, 70 and 160 $\mu$m,
and with all four bands of IRAC (3.6, 4.5, 5.8 and 8.0 $\mu$m).  All
MIPS photometric imaging used the `small' field size.  The 160 $\mu$m
band should be most sensitive to the distribution of cold dust, being
near the peak of the curve fit by Morgan et al.  (2003) to the SCUBA
data.  Based on the ISO measurements of Douvion et al.  (2001), the 24
and 70 $\mu$m MIPS bands and 8 $\mu$m IRAC band should assess the
warmer dust component near 120 K.

At 24 $\mu$m we used one cycle and no mapping with a 10 s exposure
time.  The 70 $\mu$m band used 10 s exposure times, 1 cycle and a 3x1
map for two cycles (step size at 1/2 array for columns and full array
for rows).  This results in 252 sec integration time and covers the
object with room to spare to assess background, and enough overlap to
account for the side B problem.  The smaller FOV of the 160 $\mu$m
band (along with the anomalous block of 5-pixels) required a greater
number of positions to cover the desired region. We chose 10 sec
exposures and 4 cycles.  We then chose mapping with 1x3 for 3 map
cycles.  This results in an integration time of 252 sec.

We worked with the standard post-BCD reduction data sets for the MIPS
imaging data, as retrieved with {\tt Leopard} from the $\spitzer$
archive.  The data have been reprocessed several times over the course
of this work, and here we report the S11.0.2 version of the processed
data.  Figure 1 shows the full field of view MIPS 24 $\mu$m image,
which is by far the most sensitive and detailed of the $\spitzer$
images obtained.

Kepler fits within a single IRAC field. Our AOR used the 12 position
Reuleaux dither pattern and 2-100 sec frames per pointing to to
achieve a 2400 sec integration time in all four bands.  We again
worked with the post-BCD data sets, which were processed with version
S11.4.0.  While some SNR emission is apparent in the 8 $\mu$m image,
it is difficult to assess the shorter wavelength IRAC images for SNR
emission because of stellar contamination.

Hence, to better investigate the extent of SNR emission in the IRAC
bands, we have displayed these data in two ways.  First, we made
three-color images.  Figure 2a shows a three-color IRAC image with 8
$\mu$m (red), 5.6 $\mu$m (green), and 3.6 $\mu$m (blue).  Stars are
white or bluish and the SNR emissions appear orange and yellow.
Orange filaments show emission at both 8.0 and 5.6 $\mu$m, but are
dominated by the longer wavelength band.  Yellow filaments appear to
correspond closely with the brighter radiative filaments and indicate
relatively stronger emission at 5.6 $\mu$m. Any SNR emission at 3.6
$\mu$m is completely dominated by the longer wavelengths. Figure 2b
shows a similar comparison, but using 5.6 $\mu$m as red, 4.5 $\mu$m in
green and again 3.6 $\mu$m in blue. Even more so than Fig. 2a, this
combination shows only emission from the densest (radiative) optical
filaments, which appear as orange (indicating emission in both the 4.5
and 5.6 $\mu$m bands).  The two patches of green emission in this
Figure adjacent to bright stars indicate instrumental effects in the
4.5 $\mu$m image.

Secondly, we have made difference images of adjacent bands to minimize
the effects of stellar contamination.  Since SNR emission likely
occurs in the band being used for subtraction, these images are not
useful for quantitative analysis, but rather for highlighting the
overall extent of detected emission.  Fig. 2c shows the residual
emission when the 5.6 $\mu$m band is subtracted from the 8 $\mu$m
image, and Figure 2d shows 4.5 $\mu$m minus 3.6 $\mu$m.  While some
stellar residuals remain, these two panels confirm the discussion from
the three-color Figures above: fairly extensive emission is present at
8$\mu$m, and only a few of the brightest clumpy filaments remain
visible at 4.5 $\mu$m.

For comparison, we show a medium scaling of the MIPS 24 $\mu$m image
in Figure 2e, and a 0.3 - 0.6 keV energy cut of archival {\it Chandra}
soft X-ray data in Figure 2f. Comparing Figure 2c and 2e shows that
the region visible at 8 $\mu$m corresponds very closely to the
brightest regions in the 24 $\mu$m image. There is also tremendous
similarity between the softest X-rays and the 8 $\mu$m image. Since
the densest regions should be coolest, this confirms to first order
that the brightest regions in the 8 and 24 $\mu$m images are largely
due to higher densities at these locations.

In Figure 3, we show the MIPS data and some additional comparisons,
all on the same scale as the panels of Figure 2. Figure 3a and 3b show
hard and soft stretches of the 24 $\mu$m data, respectively,
highlighting the faintest and brightest regions detected at 24 $\mu$m.
The MIPS 70 $\mu$m data were reprocessed using the GeRT software
available from the $\spitzer$ Science Center web site in an attempt to
minimize the obvious striping that traverses these data from NNE to
SSW.  This made a modest improvement in the cosmetic appearance of the
70 $\mu$m image, but the original data set was used for all
measurements reported in this paper.  The corrected data are shown in
Figure 3c. Although the resolution is lower at 70 $\mu$m, it is quite
clear that the regions brightest at 70 $\mu$m correspond closely to
the brightest emission at 24 $\mu$m.  Figure 3d shows the
star-subtracted H$\alpha$ image from Blair et al.  (1991). The overall
extent of the H$\alpha$ emission is quite silimar to the brightest
infrared regions, although differences are apparent in the brightest
region of radiative filaments in the WNW.  Figure 3e shows the MIPS
160 $\mu$m image of the same region as the other panels.  No emission
is detected above the complex and variable background at 160 $\mu$m.
Finally, Figure 3f shows a three color rendition of the {\it Chandra}
X-ray data. The red shows the same energy cut as in Figure 2f (0.3 -
0.6 keV), green is the energy band 0.75 - 1.2 keV, and blue is a band
from 1.64 - 2.02 keV.

\subsubsection{MIPS SED Spectroscopy}

The MIPS SED spectroscopy mode covers the spectral region from 55
$\mu$m to 95 $\mu$m with a resolution of 15 - 25.  The SED aperture
covers a region 3.8\arcmin\ $\times$ 0.32\arcmin\ and produces a
two-dimensional spectral output file with 16 pixels in the spatial
direction and 32 pixels in the dispersion direction.  A grid of
aperture positions covering Kepler's SNR and an adjacent sky region
were observed on 25 Sep. 2005.  The spectral grid used a 3\arcmin\
chop distance in the dispersion direction and an overlapping ``a" and
``b" position in the spatial direction, covering the northern 2/3rds
of the shell (mapping with two overlapping columns and 7 rows).  With
five cycles, this provided $\sim$300 sec integration times per
position.

The primary purpose of the SED spectra was to search for evidence that
strong emission lines might be contaminating the broad band imaging,
and thus affect their intercomparison. The individual spectra are low
in signal and so we chose to sum the data covering the bright NW
quadrant of the remnant (grid positions 2b, 3b, 4b, and 5b), using the
corresponding sky positions that were furthest removed from SNR
emission (sky positions s4b, s5b, s6b and s7b). The aperture positions
used are shown in Figure 4, projected onto the 24 $\mu$m image.  After
summing the object data and subtracting sky, it was clear from
displaying the two-dimensional data that the strongest signal was in
columns 8 - 12.  We collapsed these four columns into the spectrum
shown in Fig. 5, which represents the bright NW radiative filaments
and a portion of the NW shell.  Because of the relatively crude state
of SED calibration and software for handling these data, we do not
attempt quantitative fits to the resulting spectrum, but rather rely
on it to provide a more general description of the mid-IR spectrum.

The SED spectrum is dominated by the tail of the warm dust continuum
as it fades to longer wavelengths.  There is no evidence at this
spectral resolution for the [O~I] 63.2 $\mu$m line predicted to be
strong in radiative shock models.  There may be some indication of the
88.3 $\mu$m line of [O~III], but it does not dominate the spectrum.  A
similar bump near 75 $\mu$m may be due to [N~II] 76.5 $\mu$m, which
might be consistent with the N overabundance of the shell material as
judged from optical spectra, although the wavelength agreement is not
very good. In any event, since lines do not dominate in this spectral
region where the dust continuum is relatively faint, it is unlikely
that line contamination has a signficant effect on the observed 24
$\mu$m flux where the dust emission is much stronger.

This point is strengthened by an IRS spectrum of the NW region by
Roellig \& Onaka (2004).  This spectrum shows a modest emission
feature near 26 $\mu$m that they mark as [Fe II] but which may be a
blend of this line with [O IV] 25.9 $\mu$m.  However, this emission
feature makes only a small contribution to the total flux in the 24
$\mu$m band.  The IRS spectrum extends down to the IRAC range, showing
the dust continuum almost disappearing.  In the IRAC 8 $\mu$m
bandpass, a moderately strong [Ar II] line appears at 7.0 $\mu$m,
accounting for $\sim$ 20\% of the total flux.  Hence, this is
potentially important in considering ratios between the 8 $\mu$m image
and other bands, at least for the NW filaments where bright radiative
emission is dominating (e.g. Blair et al. 1991).

\subsection{Analysis and Modeling}

\subsubsection{Morphological Comparisons}

What is the structure of Kepler's SNR in the infrared and how does it
compare to images at other wavelengths?  Although the effective
resolutions in the various data sets are different, they are close
enough to allow some meaningful intercomparisons.  The 24 $\mu$m data
are the deepest of the $\spitzer$ data. Scaling to show only the
highest surface brightness features, the similarity of Fig. 3b to the
IRAC 8 $\mu$m (Fig. 2c) and 70 $\mu$m (Fig. 3c) indicates to first
order that the same regions dominate all three bands.  The further
similarity to the optical H$\alpha$ image (Fig. 3d) and the softest
band of X-ray emission (Fig. 2f and red band in Fig. 3f) is striking.
Since the majority of the optical emission especially across the
northern limb arises from nonradiative shocks associated with the
primary blast wave, it is clear that the brightest 8 $\mu$m, 24
$\mu$m, and 70 $\mu$m emissions are associated with this same
component.  Hence, heating of dust in dense, swept up CSM/ISM is the
dominant mechanism operating in Kepler's SNR.

The morphology of the WNW region including the primary region of
bright radiative filaments contains some subtle but important effects.
Careful comparison of Fig. 3b and Fig. 3d shows differences between
the brightest 24 $\mu$m morphology and the chaotic, more extended
emission seen in the H$\alpha$ image. We suggest the 24 $\mu$m image
is dominated by emission from the primary blast wave in the moderate
density interclump gas, and not the slower radiative shocks in the
very dense radiative knots, which are dominated by [S~II], [N~II], and
[O~III] line emission (e.g. Blair et al. 1991).  The 8 $\mu$m image
(Fig. 2c) looks intermediate between the 24 $\mu$m and H$\alpha$
appearance, which is another indicator that line emission from the
radiative shocks contributes somewhat at this wavelength, as indicated
above in the discussion of the IRS spectrum.

The 24 $\mu$m image scaled to highlight the lowest surface brightness
structures (Fig. 3a) is the only of the IR images with sufficient
signal to show the entire outer shell of the SNR.  A general
similarity is seen with the 3-color X-ray image in Fig. 3f, with the
exception of the extended `ears' to the WNW and ESE.  Similarities to
the 6 cm VLA data are apparent (e.g.  DeLaney et al. 2002), in
particular in the south and along the eastern rim.  However
differences are also apparent, including a relatively higher IR
surface brightness in the NW, along the northern rim, and across the
projected middle.  The `ears' mentioned above are most apparent in the
radio data.

The disagreement in appearance in the projected interior of the shell
deserves specific comment.  Here an important clue comes from the
optical spectroscopy of these central filaments by Blair et
al. (1991).  The SE central grouping of filaments are significantly
red-shifted and thus are a portion of the receding shell of the SNR.
In comparison, the NW central filaments are blue-shifted and are part
of the approaching shell of the SNR. These filaments have direct
counterparts in the soft X-ray band, but the association with the
higher energy X-ray bands is less clear. The approaching filaments
have no counterpart in the radio data, and while there is faint radio
emission at the projected position of the SE central grouping of
filaments, the morphologies are different.  Hence, it is not clear
whether any of the central radio emission is correlated with either
the optical or bright IR emissions in these regions.

In Figure 6, we show a different kind of morphological
comparsion. This color composite, star-subtracted image shows Kepler's
SNR as viewed by NASA's three Great Observatories, with $\spitzer$ 24
$\mu$m data in red, $\hubble$ H$\alpha$ in yellow, $\chandra$ 0.3 -
1.0 keV emission in green, and harder 2 - 10 keV $\chandra$ emission
in blue.  This image provides a combination of physical information
and some subtle affects due to differing spatial resolution of the
data sets used.  The $\hubble$ data have the highest spatial
resolution, but highlight the fact that the optical emission arises in
knotty, dense structures being encountered by the primary blast wave.
These filaments are bathed in the glow of the IR component, which
arises from dust heated by the blast wave.  The X-ray emission in the
north extends out to the optical limb and stops.  The red rim around
the top is largely an artifact of the lower resolution of the 24
$\mu$m data and the stretch applied to this component of the
image. (As we indicated in the discussion above, the brightest 24
$\mu$m emission is coincident with the shock front position.)
However, the larger extension of the red emission in the NW appears to
be a real effect.

The distribution of the two X-ray components in Fig. 6 is quite
interesting.  The harder X-rays (blue) primarily arise in sychrotron
emission at the shock front, and are seen most clearly along the
southern and eastern limbs.  The softer X-ray component is dominated
by thermal emission from Si and Fe rich ejecta (e.g. Cassam-Chena\"i
et al. 2004). In the south, this component lies directly interior to
the blue component.  In the north, this component largely fills in
between the northern limb emission and the centrally-projected
emission regions. Note the distinct absence of red emission in the
green regions, which indicates {\it no significant warm dust emission}
coincident with regions of ejecta.

\subsubsection{Derivation of IR Fluxes and Ratios}

We now turn to more quantitative information, concentrating first on
the 24 $\mu$m and 70 $\mu$m data sets, where contamination of the SNR
emission by stars is not a significant problem.  We derived the total
fluxes at these two wavelengths using the following method.  Because
the post-BCD data are in units of MJy $\rm sr^{-1}$, we extract
regions of pixels corresponding to the object, and use the known post
BCD pixel sizes (from the file headers) and number of pixels included
in each region to scale appropriately to total fluxes.  We similarly
extract representative regions of background surrounding the SNR
region (as allowed by field coverage) and average these to determine
the most appropriate overall background levels to subtract.  Using
this technique, we obtain total fluxes at 70 and 24 $\mu$m of 4.90 Jy
and 9.5 Jy, respectively, and thus a ratio of 70/24 $\mu$m of 0.52.

Our total flux at 70 $\mu$m appears to disagree with published {\it
IRAS} fluxes at 60 $\mu$m (range from 7.1 - 10.5 Jy), summarized by
Saken et al.  (1992, their Table 6), although our derived 24 $\mu$m
flux lies within the range listed (8.1 - 11.7 Jy) at IRAS 25 $\mu$m.
The apparent disagreement arises simply from the differing bandpasses
used.  Bandpasses for {\it IRAS} are broader than for {\it Spitzer},
and thus would encompass more flux at a given wavelength.  To confirm
this, we integrated the output spectrum from the dust model described
below for the entire remnant over the {\it IRAS} 60 $\mu$m bandpass
and obtained an ``expected" 60 $\mu$m flux of $\sim$8 Jy.  A
corresponding exercise for the {\it Spitzer} 70 $\mu$m band predicts a
flux of ~5 Jy.

In addition to the total fluxes, we have determined the 24 and 70
$\mu$m fluxes for several sub-regions of the remnant to search for
variations in these ratios that might arise due to different shock or
other parameters.  We have also measured fluxes from relevant regions
of the IRAC 8 $\mu$m image, although stellar contamination is a more
significant problem for these data.  Figure 7 shows the extraction
regions projected onto the relevant images.  These figures also define
the nomenclature we will use below to reference the regions.

Because {\it Spitzer}'s optics provide images at or near the
diffraction limit for all wavelengths, the resolution in images from
various instruments differs by roughly the ratio of wavelengths.  The
angular resolutions of various images range from 2\arcsec\ at 8
$\mu$m, to 6.2\arcsec\ at 24 $\mu$m, 18\arcsec\ at 70 $\mu$m, and
41\arcsec\ at 160 $\mu$m.  For determination of flux ratios from
various regions of the SNR, it is necessary to convolve the
higher-resolution image of interest to the resolution of the
lower-resolution image.  For the 8/24 $\mu$m ratios, we convolve the 8
$\mu$m image to the resolution of that at 24 $\mu$m, and for the 24/70
$\mu$m ratios, we convolve the 24 $\mu$m image to the resolution of
that at 70 $\mu$m.  We used contributed software by K.~Gordon
(U.~Arizona) distributed by the SSC to convolve the images, using
kernels for convolving 24 $\mu$m images to the same PSF as 70 $\mu$m,
and similarly for 8 $\mu$m to 24 $\mu$m.  These kernels are slightly
temperature-dependent; we used versions appropriate for 100 K.  Though
we do not expect blackbody spectral shapes, this value is close to the
appproximate grain temperatures reported by Douvion et al. (2001).  In
each case, the higher-resolution images were resampled onto a grid
identical to that of the lower resolution image, using AIPS
(Astronomical Image Processing System)\footnote{ AIPS is produced and
supported by the National Radio Astronomy Observatory, operated by
Associated Universities, Inc., under contract with the National
Science Foundation.}  task HGEOM.  These are the images shown in
Figure 7.

The extraction regions were defined to select physically associated
regions of emission so that the results could be compared.  We
displayed and aligned the 24 and 70 $\mu$m images using the display
tool {\tt ds9}. \footnote{See {\tt http://hea-www.harvard.edu/RD/ds9/}
.}  We also displayed the optical H$\alpha$ and other images on the
same scale for reference and comparison while defining regions.  The
defined 24 and 70 $\mu$m regions are shown in Figure 7a-c, and details
are given at the top of Table 1.  Object position O1 enclosed the
region of bright radiative emission in the NW, position O2 enclosed
the entire bright NW rim at 24 $\mu$m. Note that by differencing these
two positions, the area corresponding to just the relatively bright
nonradiative portion of the NW rim can be extracted.  O3 samples the
northern (mainly) nonradiative rim, and O4 and O5 enclose the two
centrally-projected regions of knotty filaments.

Background regions were identified to the north and south of the
remnant to account for the gradient observed in the background from
north to south.  Some backgrounds were defined specifically for use
with the 70 $\mu$m image in an attempt to better account for the
significant striping in this image. For each object position, the most
appropriate backgrounds were summed and scaled to the size of the
object region being measured.  Specifically, we used the following
regions as background at 70 $\mu$m: For positions O1 and O2 in Figure
7a-c, background regions B1 and B5 were averaged and used. For
positions O3 and O5, background regions B2 and B4 were used. And for
position O4, background regions B2 and B3 were used.  At 24 $\mu$m we
simply used regions B1, B2 and B4, which do not overlap SNR emission.

The signal in each sampled region was measured using the {\tt
FUNtools} package that interfaces with {\tt ds9}. \footnote{{\tt
http://hea-www.harvard.edu/RD/funtools/help.html} .}  As with the
total flux estimates, it is necessary to multiply the total signal
above background measured in a region by the pixel scale (sr $\rm
pixel^{-1}$).  Thus, ratios use the same spatial object regions, even
though the pixel sizes varied. The region fluxes and ratios between 24
and 70 $\mu$m are provided in Table 2.  Only small differences are
obtained.  In particular, the similarity in ratio between O1 and O3,
or O1 and the difference between O2-O1, indicates that there is no
significant difference in 70/24 $\mu$m ratio between bright radiative
and nonradiative regions.  This is in keeping with the earlier
discussion about the IRS and SED spectra, where these emissions are
dominated by the main blast wave.

We have also derived ratios of regions between 8 $\mu$m and 24 $\mu$m
using a similar technique, although for somewhat modified spatial
regions than used with the 70/24 $\mu$m ratios.  The defined 8 and 24
$\mu$m regions are shown in Figure 7d-f, and details are given at the
bottom of Table 1.  This comparison is complicated by the faintness of
the 8 $\mu$m emission, by the presence of many more stars at 8 $\mu$m,
and by the difference in resolution between the images.  Because of
potential contamination of the 8 $\mu$m image by the [Ar~II] 7.0
$\mu$m line seen by Roellig \& Onaka (2004), the derived ratios may be
skewed toward slightly higher 8/24 $\mu$m ratios than true.  On the
other hand, any relative changes in the ratio should be real.

For the 8 $\mu$m to 24 $\mu$m comparison, positions O1 and O2 isolate
the two brightest regions of radiative optical filaments in the NW.
Position O3 samples the so-called `bump' region, which includes both
radiative and nonradiative optical emission in the NNW.  O4 provides
the cleanest sampling of nonradiative shock emission on the northern
rim.  O5 and O6 sample two regions of central emission knots, but are
smaller regions than measured above for the 70/24 $\mu$m ratio. Again,
optical data suggests a mixture of radiative and nonradiative shocks
in these regions.  Stars have largely been avoided, with the exception
of position O6, for which it was impossible to totally exclude stars
from the selected region.  At 8 and 24 $\mu$m, we simply averaged all
three of the background regions shown and applied this as
representative to all positions.

We have applied recommended photometric corrections for diffuse
sources appropriate for the 8 $\mu$m band, as described in the IRAC
section of the SSC website\footnote{See {\tt
http://scc.spitzer.caltech.edu/irac/} .}.  This correction for sources
large compared to the calibration aperture for point sources
(12\arcsec) is a factor of 0.74, which has been applied to measured
fluxes and backgrounds for this band.  However, this correction is
sufficiently poorly known that an additional uncertainty of order 10\%
is introduced. Any relative changes in the 8/24 $\mu$m ratio should
remain valid, however.

The derived 8/24 $\mu$m ratios are summarized in Table 3.  Comparing
the measurements at positions O1 and O2 to O4, we see higher ratios at
the strongly radiative positions, indicative of contamination of the 8
$\mu$m image by line emission (likely [Ar II], as indicated in the IRS
spectrum discussion above).  Hence, the O4 measurement likely
represents the most accurate assessment of the 8/24 $\mu$m dust
continuum. It is unclear without IRS data whether the higher ratios
observed at O3, O5, and O6 are due to variations in the dust continuum
or due to significant IR line emission from the optically faint
radiative filaments in these regions.

We have not color-corrected the observed fluxes reported in the
Tables.  Our model fluxes have been produced by integration of the
calculated spectrum over the {\it Spitzer} bandpasses at 24 and 70
$\mu$m.  Since {\it Spitzer} fluxes are calibrated by comparison with
stars whose spectra at 24 $\mu$m and longward are well-approximated by
the Rayleigh-Jeans limit of a blackbody, we have assumed a
$\lambda^{-2}$ spectrum and calculated what {\it Spitzer} would report
for the flux at the nominal frequency for each band.  That is, we have
converted our model fluxes into ``{\it Spitzer} space'' before
computing model ratios, rather than color-correcting observed
fluxes. However, such corrections would not be large in any case.  Our
shock models that reproduce the observed flux ratios give grain
temperatures between 75 and 95 K.  While the spectra are not exact
blackbodies, color corrections for a 70 K blackbody are less than 10\%
at both 24 and 70 $\mu$m, as reported in the MIPS Data Handbook (v3.0,
p.~29); at 100 K, they are less than 7\% at both wavelengths.  For
IRAC, color corrections are reported in the IRAC Data Handbook only
down to blackbody temperatures of 200 K.  For that temperature,
corrections are less than 20\% for both Channels 3 and 4.  We believe
that overall calibration errors, estimated in the {\it Spitzer}
Observing Manual, Sec. 8.3.3 (p. 355) to be 10\% for extended sources
at 24 $\mu$m and 15--20\% at 70$\mu$m, will dominate the errors at
those bands.  We can also estimate internal statistical errors from
the pixel-to-pixel dispersion in the background; these errors are
negligibly small compared to those due to calibration.  We adopt
conservative overall error estimates of 10\% at 24 $\mu$m and 20\% at
70 $\mu$m.

In addition to deriving region ratios as described above, we also
created a ratio map from the convolved 24 $\mu$m image and the 70
$\mu$m image as follows.  First, we measured a mean background from
most of the region on the image not occupied by the source (avoiding
obvious point sources), and subtracted that value, at each wavelength.
Then we blanked the convolved 24 $\mu$m image below 10 MJy/sr, a level
corresponding to about 20\% of peak, which left all obvious structure
intact.  (The off-source rms fluctuation level was 0.6 MJy/sr.)  At 70
$\mu$m, a much noisier image, the off-source rms was about 1.6 MJy/sr;
we blanked below 5 MJy/sr, about three times this value.  The point of
the blanking is to make sure that only pixels whose measured fluxes
are highly significant at both wavelengths are used to compute ratios.
We then generated the ratio image shown in Fig.~8.

Figure 8 shows several features of interest.  First, a general
anticorrelation of ratio with 24 $\mu$m brightness indicates that
hotter (lower 70/24 $\mu$m ratio) regions are brighter.  Second, the
minima in the ratio are actually offset from the 24 $\mu$m brightness
peaks slightly.  Third, the western region of bright radiative shocks
appears to have somewhat different ratio.  The range of pixels in the
image is about 0.3 -- 0.8, with a broad maximum around 0.45 and most
pixels between 0.35 and 0.5, consistent with the values measured in
regions shown in Table~2. Higher ratios in the fainter regions is the
primary reason the total 70/24 $\mu$m = 0.52 even though most of the
brighter regions have lower ratios.

\subsubsection{Grain Emission Modeling}

The morphological comparisons across different wavelengths show a
clear correlation between the soft X-ray images and IR images at all
three MIPS wavelengths, 8, 24 and 70 $\mu$m. We thus attribute
emission from Kepler in these bands to shocked interstellar and
circumstellar dust, heated by the hot, X-ray emitting plasma in the
primary blast wave (Dwek \& Arendt 1992).  Collisions with energetic
electrons and ions heat dust grains to $\sim 100 K$, where they emit
thermal radiation visible to {\it Spitzer's} mid-IR instruments. In
addition to heating, the ions in the plasma sputter dust grains,
rearranging the grain-size distribution by destroying small grains and
sputtering material off of large grains. We employ computer models of
collisionally heated dust to explain what is seen in Kepler, identical
to what was done for Type Ia SNRs (Borkowski et al. 2006) and
core-collapse SNRs (B. Williams et al. 2006) for SNRs in the Large
Magellanic Cloud.  Our models use as input an arbitrary grain-size
distribution, grain type (astronomical silicates, carbonaceous, etc.),
proton and electron density $n_p$ and $n_e$, ion and electron
temperature $T_i$ and $T_e$, and shock age (or sputtering time scale)
$\tau_p=\int_0^t n_p dt$. The model is based on the code described by
Borkowski et al. (1994) in the context of photon-heated dust in
planetary nebula Abell 30, and augmented to allow for heating by
energetic particles in hot plasmas.

Because little is known about the surroundings of Kepler, we adopt a
power-law grain size distribution with index $\alpha$ = -3.5 and an
exponential cutoff.  We use a range of 100 grain sizes from 1 nm to
0.5 $\mu$m. We also use only astronomical silicates.  This is
consistent with past efforts to model dust emission from Kepler
(Douvion et al. 2001).  We use bulk optical constants for astronomical
silicates from Draine \& Lee (1984). Energy deposition rates for
electrons and protons were calculated according to Dwek (1987) and
Dwek \& Smith (1996). Because small grains are stochastically heated
and have large temperature fluctuations as a function of time, we must
account for the increased radiation produced by such transient
heating. We use the method described by Guhathakurta \& Draine (1989)
for this purpose.

It is also necessary to model sputtering for all grains, since
sputtering alters the grain size distribution downstream of the shock.
Sputtering rates for grains in a hot plasma are taken from Bianchi et
al. (2005). Small grains can actually experience an enhancement in
sputtering due to the ion knocking off atoms not only from the front
of the grain, but also from the sides or the back. We have included
such enhancements in our models, with enhanced sputtering yields
described by Jurac et al. (1998).  Sputtering is very important in the
dense CSM environment of Kepler, and results in the efficient
destruction of small grains in the postshock gas and very significant
modification of the preshock grain size distribution.  Even MIPS
fluxes and their ratios depend on the shock sputtering age, but
because emission in these bands is mostly produced by relatively large
grains with moderate temperature fluctuations, sputtering effects are
less extreme than at shorter wavelengths. We consider our dust models
reliable for modeling MIPS fluxes and their ratios. Thermal
fluctuations are particularly important in very small grains, which
reradiate their energy at short wavelengths, so sputtering
dramatically reduces the amount of radiation in the IRAC bands.
 
Contrary to expectations, the 8 $\mu$m emission is much stronger than
predicted by our models.  While we do not include PAHs in our modeling
code, they are not likely to be present in large quantities given the
absence of their distinctive spectral features in IRS spectra (Roellig
\& Onaka 2004).  However, the model does not account for several
physical effects likely to be important for very small grains, such as
discrete heating (i.e., discrete energy losses as an electron or
proton traverses a grain), and corrections to heating and sputtering
rates required when particle mean free paths are much greater than
grain radius.  For these reasons, we have not attempted to model the
8/24 $\mu$m flux ratios shown in Table~3. A great deal of theoretical
work remains to be done before the wealth of data produced by {\it
Spitzer} can be used effectively to understand the properties of small
grains.

For modeling an outward moving shock wave, we have used a
one-dimensional plane-shock approximation. The plane-shock model
assumes a constant temperature, but superimposes regions of varying
sputtering timescale from zero up to a specified shock age (Dwek et
al. 1996). The shock model effectively varies the product of density
and time behind the shock, allowing us to account for material that
has just been shocked and material that was shocked long ago, since
these will experience different amounts of sputtering. The output of
our models is a single spectrum, which is produced by superimposing
spectra of many different grain sizes.  Since we do not model observed
spectra directly, we can only fit flux ratios from photometric
measurements. We focus here on reproducing the observed 70/24 $\mu$m
ratios.

We constructed a grid of dust models to explore the parameter spaces
of temperature and density, covering electron temperatures $kT_e$ from
0.03 to 10 keV and electron densities $n_e$ ranging from 3 to $10^{3}$
cm$^{-3}$.  We assumed that ion and electron temperatures are
equal. The output of the grid was $\sim 300$ separate models with
varying $n$, $T$, and $\tau_p$, the shock sputtering age. (Since
$\tau_p$ inherently contains $n_p$, and $n_p$ is related by a constant
factor to $n_e$, $\tau_p$ is varied from model to model.  See
Borkowski et al. 2006.)  We divide the actual age of Kepler by 3 to
approximate an ``effective shock age" for Kepler, which was used in
the models. The factor of $1/3$ arises from applying results of a
spherical blast wave model to the plane-shock calculation (Borkowski
et al. 2001).

For each of the 300 models, a 70/24 $\mu$m ratio was calculated from
the output spectrum, and the value of the ratio was plotted on a
two-dimensional color-coded plot as a function of electron density and
pressure (see Figure 9). We then added contours to the plot which
correspond to measured values of the 70/24 $\mu$m ratio from the MIPS
images. The three contours are the highest and lowest measured region
values (regions O3 and O5, 0.40 and 0.30, respectively; see Table 2),
and the value of 0.52 appropriate for the spatially-integrated
fluxes. The plasma conditions for the nonradiative regions of Kepler's
SNR can be contained for the most part between these contours. Our
nonradiative shock models are not applicable at high densities and low
temperatures (toward the lower right corner of Fig. 9) because of the
onset of radiative cooling. Shocks with an age of 400 yr and solar
abundances. for instance, become radiative to the right of the line
shown in the lower right corner of Figure 9.  We used postshock
cooling ages tabulated by Hartigan et al. (1987) to draw this line in
the electron temperature--pressure plane.

At low densities (near the left boundary of Fig. 9), our models with
equal ion and electron temperatures overestimate the 70/24 $\mu$m flux
ratio.  In this region of plasma parameters, shocks are fast and ion
temperatures are generally larger than electron temperatures. Heating
of grains by ions becomes relatively more important, resulting in
increased grain temperatures and lower 70/24 $\mu$m flux ratios
relative to shocks with equal ion and electron temperatures.  We
estimated an electron temperature for the northwest portion of the
remnant, where Sankrit et al. (2005) determined a shock speed of 1660
$\rm km~s^{-1}$. We used a simple model for the ion-electron
equilibration through Coulomb collisions behind the shock. From this
we derived a $T_e$ of 1.2 keV (in the absence of significant
collisionless electron heating and assuming a shock age of $\sim 150$
yr).  A shock model with a postshock electron density of 22 cm$^{-3}$
reproduces the measured ratio of 0.40; we mark its position by a star
in Figure 9. (Without sputtering, this ratio would have been equal to
0.31.) A noticeable displacement of this model from the middle contour
is caused by additional heating by ions in the more sophisticated
model with unequal ion and electron temperatures used for the
Balmer-dominated shock in the north. Grain temperatures in this model
vary from 75 K to 95 K.

Our estimate of a typical electron density in dust-emitting regions of
$n_e \sim 20$ cm$^{-3}$ is in reasonable agreement with other
estimates, such as the estimate of $7 - 12$ cm$^{-3}$ in the central
optical knots (Blair et al.~1991).  It should characterize the bulk of
the shocked CSM around Kepler, although higher postshock electron
densities and lower temperatures (and hence lower shock speeds) are
also possible because contours of constant 70/24 $\mu$m ratio
approximately coincide with lines of constant pressure on the right
hand side of Figure 9.  While nonradiative Balmer-dominated shocks
with speeds less than 1660 km s$^{-1}$ have not been measured to date
in Kepler, the highly inhomogeneous optical and X-ray morphologies
suggest that plasma conditions in the shocked CSM may vary greatly
with position within the remnant. It is likely that shocks are present
with velocities less than seen in the Balmer-dominated shocks but more
than seen in radiative shocks. Shocks with such intermediate
velocities are best studied at X-ray wavelengths, and a future
analysis of the CSM in Kepler based on a new long {\it Chandra}
observation is in progress (Reynolds et al. 2006).  It is also likely
that fast nonradiative shocks in the south and east travel through gas
with densities much less than 20 cm$^{-3}$; the current MIPS 70 $\mu$m
data and optical observations are not sensitive enough to study these
shocks in much detail.

Our derived value of $n_e T_e = 3 \times 10^{8}$ cm$^{-3}$ K ($n_e
kT_e = 26$ cm$^{-3}$ keV) is in the pressure range considered by
Douvion et al. (2001). However, based on the morphological resemblance
between the 24 $\mu$m image and the nonradiative optical emission, a
lower density and higher temperature better characterize the typical
dust emitting regions.

\subsubsection{Total Dust Mass and Dust/Gas Ratio}

The spatially-integrated IR spectrum of Kepler is produced by grains
of widely varying sizes immersed in inhomogeneous plasmas. This
results in a wide range of grain temperatures, making determination of
a total dust mass model dependent because of the extreme sensitivity
of the radiated IR power to the grain temperature. In Appendix A, we
estimate the shocked CSM X-ray emission measure in Kepler using a
simple plane shock model with a mean temperature of 5 keV and
ionization age of $10^{11}$ cm$^{-3}$ s, without any collisionless
heating at the shock front but allowing for energy transfer from ions
to electrons through Coulomb collisions. Emission measure-averaged ion
and electron temperatures are 8.9 keV and 1.4 keV in this shock
model. The only remaining free parameters in the model are a postshock
electron density $n_e$ and a total dust mass. We obtain $n_e = 13$
cm$^{-3}$ by matching the measured spatially-integrated 70/24 $\mu$m
flux ratio of 0.52.  (There is a nonnegligible contribution to grain
heating from hot ions in this fast shock; as can be inferred from
Figure 9, simpler shock models with equal ion and electron temperature
of 1.4 keV predict a slighty higher 70/24 $\mu$m ratio of 0.58.) We
then derive a total dust mass of $5.4 \times 10^{-4}$ $M_\odot$ from
the measured spatially-integrated MIPS fluxes.

We obtain nearly the same mass when we use instead plasma conditions
assumed by Douvion et al. (2001), $n_e = 6000$ cm$^{-3}$ and $kT_e =
0.0043$ keV, corresponding to a model spectrum shown by a solid line
in their Figure 3, and based on fits to IRAS and ISO
observations. (Douvion et al. 2001 quote a smaller dust mass of $1-2
\times 10^{-4}$ $M_\odot$, appropriate for their simple model with hot
silicate dust at temperature of 107.5 K, significantly hotter than
dust in our models.) This agreement between two different mass
determinations based on independent data and very different assumed
plasma conditions is encouraging; a future more detailed
spatially-resolved joint study of {\it Spitzer} and {\it Chandra} data
should provide us with a refined dust mass determination. The mass of
$5.4 \times 10^{-4}$ $M_\odot$ refers to dust currently present in the
shocked CSM. Most (78\%) of dust was destroyed in our fast shock
model, implying an initial (preshock) dust mass of 0.0024
$M_\odot$. Because dust destruction rates depend sensitively on the
assumed shock speed and its age, our current estimate of the preshock
dust mass is rather uncertain.

The total IR flux in the plane shock model discussed above is $1.4
\times 10^{-9}$ ergs s$^{-1}$ cm$^{-2}$, in good agreement with
IRAS-based fluxes of $1.3 \times 10^{-9}$ ergs s$^{-1}$ cm$^{-2}$ and
$1.6 \times 10^{-9}$ ergs s$^{-1}$ cm$^{-2}$ listed by Dwek (1987) and
Arendt (1989), respectively. The IR luminosity is $2.8 \times 10^{36}$
ergs s$^{-1}$. Kepler is a low-luminosity object when compared with
SNRs for which this type of analysis has been done with IRAS (e.g.,
Saken et al. 1992). More recently, Borkowski et al. (2006) derived
total luminosities for two of the four SNRs from Type Ia SNe in their
study, and B. Williams et al. (2006) derived luminosities for all four
of their sample of SNRs from core-collapse SNe in the Large Magellanic
Cloud. Of these six remnants, only SNR 0548-70.4 has a luminosity
lower than Kepler, and even it is comparable at $2 \times 10^{36}$
ergs s$^{-1}$. This SNR is, however, several times larger (and thus
likely older) than Kepler.  Indeed, most of the remnants in these two
studies have luminosities much higher than Kepler.

A specific object of interest for comparison is Tycho, a Type Ia SNR
of comparable age located at a 2 kpc distance. The IR flux of Tycho is
$5 \times 10^{-9}$ ergs cm$^{-2}$ s$^{-1}$ (Arendt 1989), so its IR
luminosity is comparable to Kepler. Dust in Tycho is much cooler than
in Kepler (Arendt 1989, Saken et al. 1992), consistent with a much
lower ISM density around Tycho. SNR 0509-67.5, one of the remnants
studied by Borkowski et al. (2006) in the LMC, appears very similar to
Tycho.  While observations of light echoes (Rest et al. 2005) have
placed the age of this SNR at $\sim$ 400 yr, a high shock speed
deduced from optical and UV observations of Balmer-dominated shocks
(Ghavamian et al.  2007) implies a low ambient ISM density. Its MIPS
24 $\mu$m flux is 16.7 mJy, about 2--4 times lower than what Tycho or
Kepler would have at a distance of 50 kpc. An upper limit to its dust
mass is 0.0011 $M_\odot$, and SNR 0509-67.5 is likely less luminous in
the IR than Tycho or Kepler.

Determination of the dust/gas mass ratio requires knowledge of the
shocked CSM gas mass. Using an X-ray emission measure of 10 $M_\odot$
cm$^{-3}$ (Appendix A) and $n_e = 13$ cm$^{-3}$, we derive a total
shocked CSM mass in Kepler of 0.77 $M_\odot$. This is in good
agreement with a shocked CSM mass of 0.95 $M_\odot$ derived from {\it
ASCA} observations (Kinugasa \& Tsunemi 1999), after scaling their
results to the 4 kpc distance used here and using an electron density
of 13 cm$^{-3}$. The hydrodynamical model of Borkowski et al. (1992,
1994), based on the massive core-collapse runaway progenitor scenario
of Bandiera (1987), requires $\sim 1 ~ M_\odot$, also in reasonable
agreement with the present mass estimate. Recent hydrodynamical models
of Vel\'{a}zquez et al. (2006), based on a Type Ia progenitor
scenario, require several $M_\odot$ of shocked CSM.  But Vel\'{a}zquez
et al. (2006) used collisional equilibrium ionization X-ray spectral
models, which underestimate X-ray emission by an order of magnitude or
more for young SNRs (Hamilton et al. 1983; see also Appendix A),
resulting in an excessive estimate of the shocked mass by a comparable
factor.

Combined with the dust mass from above of 0.0024 $M_\odot$, we arrive
at a dust-to-gas mass ratio for the CSM surrounding Kepler of
0.003. This is lower than the generally accepted figure for the Galaxy
(e.g., Weingartner \& Draine 2001) by a factor of several.  Since the
dust in Kepler appears to have originated in some kind of a stellar
outflow, the lower dust content may be related to this and not a
characteristic of the general ISM in the region. Kepler's dust/gas
ratio is higher by a factor of several than found for SNRs in the
Magellanic Clouds (Borkowski et al.~2006; B.~Williams et al.~2006;
Bouchet et al. 2006), most likely because of the high (near or above
solar) metallicity of its SN progenitor compared with the $\sim$0.4
solar LMC abundance.

\subsubsection{North-South Density Gradient}

A number of indicators imply a potential difference in preshock
density from north to south across the Kepler region.  The difference
in the surface brightness of the north rim compared to the south rim
is one indicator, and the effect may be visible directly from the
background in Fig. 1 and the appearance of the 160 $\mu$m image in
Fig. 2e. The south rim is only faintly visible at 24 $\mu$m, while the
north rim is extremely bright. We believe the 24 $\mu$m is dominated
by dust emission, although some contribution from emission lines
cannot be ruled out without spectra.  Under this assumption, we took a
small region of the southern rim and an equal-sized region of the
northern rim that is dominated by nonradiative shocks and measured the
flux from both regions at 24 $\mu$m. The regions were arcs following
the shape of the rim, approximately 23\arcsec\ in thickness and
75\arcsec\ in length.  The south region follows the faint southern rim
at a declination of approximately -21:31:04 (J2000).  We found a
north/south ratio of 24 $\mu$m fluxes of $\sim$30. A possible
explanation for this contrast is that the shocks are encountering
regions of different density.

We have made several estimates of the density contrast necessary to
produce the observed intensity ratio.  First, we assume that the
dust-to-gas mass ratio in the two regions is the same (although we
make no {\it a priori} assumptions about what that ratio must be). We
do not assume equal amounts of swept-up material in the two regions,
but rather simply vary the gas density (and dust mass accordingly) in
the models and assume the same shock speed (and thus the same
temperature) for both regions.  This method reproduces the observed
flux difference with a modest density contrast of only a factor of
$\sim$4.5. It should be noted that these are all post-shock densities.

As an alternative approach to this problem, one can relax the
requirement that the shock speeds be equivalent in the two regions,
since a variation in density should cause variations in shock speed.
We repeated the calculations for several different shock speeds in the
south region, while keeping the north region constant at 1660 km
s$^{-1}$, and a density of $n_{e} = 19.5$ cm$^{-3}$.  We calculated
electron and proton temperatures for the south region with these
different shock speeds assuming no equilibration at the shock front,
and used these different temperatures to predict dust emission at 24
$\mu$m.  In these models, the density contrast required is only weakly
dependent on shock speed.  While grains are heated to hotter
temperatures in faster shocks, the amount of destruction of grains
also increases.  Overall, we find that for a reasonable range of shock
speeds up to 3000 km s$^{-1}$, a post-shock density contrast between 4
and 7 times lower in the south (and accordingly 4--7 times less dust
present) can reproduce the observed ratio. The density contrast
increases with faster shock speeds.

As a third approach, we consider the remnant to be in pressure
equilibrium.  Keeping the density and shock speed in the northern
region constant with the values mentioned above, we varied the shock
speed in the south and derived density from the pressure equilibrium
expression $n_{N}V_{N}^{2} \equiv n_{S}V_{S}^{2}$, where subscripts N
and S refer to the north and south regions, respectively.  Under this
constraint, we determined the shock speed required to reproduce the
observed 24 $\mu$m flux ratio, again assuming a constant dust-to-gas
mass ratio. We find that a shock with speed $\sim$5000 km s$^{-1}$,
and thus a density of $\sim$ 2.1 cm$^{-3}$ can account for the
difference in fluxes coming from the two regions. This is a density
contrast of 9.1.  Because 5000 km s$^{-1}$ is much higher than
inferred from other indicators, this contrast in density is considered
an upper limit. The predicted 70/24 $\mu$m flux ratio in the south
region is $\sim$ 1.2, which is significantly higher than in the
brighter regions of the remnant, but consistent with the idea that the
dust in that region is cooler.  It is also consistent in a general
sense with the higher overall average 70/24 $\mu$m ratio of 0.52.
Since the these models are only constrained by the fluxes observed at
24 $\mu$m in the south, these results should be considered tentative.
Detailed mid-infrared spectroscopy will be required to investigate
this issue further.

\subsubsection{Whither the Cold Dust Component?}

Morgan et al. (2003; hereafter M03) have recently reported excess
emission in Kepler using SCUBA 450 and 850 $\mu$m observations.  They
model this as a cold astronomical silicate dust component from ejecta
($T_d$ = 17 K), inferring as much as a solar mass of material in this
component.  Dwek (2004a) proposes a differing interpretation, arguing
instead for a component at $T_d$ = 8 K and a much lower mass of dust
in the form of Fe needles.  He supports the general possibility of a
cold dust component related to the ejecta, pointing to a similar
component identified in Cas A (Dwek 2004b), but the amount of mass
involved would be much lower than inferred by M03.  Douvion et
al. (1991) would not have been able to detect this cold component.

Our MIPS 160 $\mu$m image shown in Figure 3 is directly relevant to
this discussion since this wavelength is close to the peak of the cold
component curve predicted by M03 (see their Figure 2).  No emission
related to the SNR shell or interior is detected at 160 $\mu$m,
although the background is patchy and there appears to be a general
gradient of intensity from NE toward the SW in the image.

To derive an upper limit at 160 $\mu$m, we extracted emission from the
region corresponding to the SNR location (judging from an aligned
overlay of the 24$\mu$m image) and from nearby background regions, as
described above for the 24 and 70 $\mu$m images.  Background levels to
the north and east of the SNR are at a higher level than the average
from the object location, while the background level in the south is
slightly below the object region average.  The observed field of view
is very close to the western edge of the SNR, but the background in
that region appears to be comparable to that in the south.  To be
conservative, we apply the southern background level to the entire
object, which will over-estimate any contribution from the SNR.  The
resulting upper limit we derive is 0.8 Jy at 160 $\mu$m, which is
approximately a factor of 10 below the value predicted by M03 from the
SCUBA-based model. Hence, our $\spitzer$ 160 $\mu$m data do not
confirm the M03 cold-dust picture. This fairly conservative upper
limit does not rule out the idea of Fe needles as discussed by Dwek
(2004a), but the presence of any such component in Kepler has been
argued against on theoretical grounds (Gomez et al.  2005).  Even if
present, any such component in Kepler's SNR would contain well below
0.1 solar mass, and would not require a massive precursor star as in
the M03 interpretation.

\subsubsection{Synchrotron Emission}

X-ray synchrotron emission has been reported for Kepler: thin
filaments seen in the eastern ``ear" extension (Bamba et al.~2005),
broader emission from the SE quadrant (Cassam-Chena\"i et al.~2004),
and an extension of the integrated flux to hard X-rays (Allen,
Gotthelf, \& Petre 1999).  The X-ray fluxes demand a steepening of the
extrapolated radio spectrum; if that steepening occurs at shorter than
IR wavelengths, IR synchrotron emission should be present with a
morphology identical to the radio and at readily extrapolated
brightness levels.  While the average radio spectral index of Kepler
is --0.71 ($S_\nu \propto \nu^\alpha$) (DeLaney et al.~2002),
higher-frequency archival radio data imply a value of --0.59 (Reynolds
\& Ellison 1992).  This concave-up curvature, or hardening to shorter
wavelengths, can occur for electrons accelerated in a shock modified
by the pressure of accelerated ions (Ellison \& Reynolds 1991).  If we
extrapolate from images at cm wavelengths (e.g., DeLaney et al.~2002)
with this value, we predict surface brightnesses at 3.6 $\mu$m in the
range 0.6 -- 1.1 $\mu$Jy arcsec$^{-2}$ (or $(2.5 - 4.8) \times
10^{-2}$ MJy sr$^{-1}$).  Our 3.6 $\mu$m image shows only a faint hint
of any emission associated with Kepler.  In particular, there is no
apparent emission in the SE quadrant.  Typical brightness levels are
0.4 -- 0.5 MJy sr$^{-1}$, or 10 -- 20 times the extrapolated
synchrotron flux.  Even with very significant hardening of the
spectrum above 10 GHz, the 3.6 $\mu$m data are not constraining the
presence of synchrotron emission.  In particular, the regions
identified by Bamba et al. (2005) and Cassam-Chena\"i et al.~(2004),
the SE quadrant and the eastern ``ear," are very faint in IR in
general.  The western ``ear" may also harbor X-ray synchrotron
emission, and it too is fainter at 24 $\mu$m relative to the rest of
the shell emission. Far more sensitive IR observations would be
necessary to detect synchrotron emission.

\subsection{Discussion}

Of the historical SNe, Kepler's remains enigmatic because a clear
determination of the SN type (and thus precursor star) has proven
elusive.  Claims of evidence supporting both a massive precursor (core
collapse SN type) or a white dwarf precursor (Type Ia SN) abound in
the literature over the last decade or more.  In this section, we
discuss this issue and highlight new insights that may help resolve
this dichotomy.

Firstly, since not all SN ejecta can form into dust, the claim by M03
of a solar mass or more of cold dust in Kepler was an indication of
even more ejecta mass and hence a massive precursor star. The negative
detection at 160 $\mu$m here points to an apparent problem with the
SCUBA result, presumably due to the complex and variable background
observed in this region.  Hence, a massive precursor is not required
by the IR data.

Another indication pointing toward a massive precursor star has been
the overabundance of nitrogen in the optically-emitting filaments,
which are thought to represent dense knots of CSM being overrun by the
blast wave.  We have run a small grid of shock models using the
current version of the shock code described by Raymond (1979) and
Hartigan et al. (1987) for comparison with the optical spectrum of
knot `D3,' a bright, radiative shock knot presented by Blair et
al. (1991, their Table 2).  We find that the main features of this
spectrum can be matched with a model similar to model E100 of Hartigan
et al. (1987), but with the N abundance increased by 0.5 dex.  This is
indeed a significant enhancement over solar abundance.

However, solar abundances are the wrong reference point.  At a
distance of $\sim$4 kpc and galactic coordinate G4.5+6.8, Kepler's SNR
is nearly half way to the galactic center.  Rudolph et al. (2006)
provide a summary of galactic abundance gradients, and for nitrogen
find an increase of $\sim$0.3 dex for the assumed distance of Kepler's
SNR.  Additionally, inspection of their Figure 4 shows that the
scatter in observed points around the best fit line for the gradient
readily encompasses the required value of 0.5 dex enhancement in
nitrogen at Kepler's distance from the galactic center.  While the
nitrogen abundance around Kepler's SNR may be enhanced relative to its
local surroundings, it is not required, and certainly the magnitude of
any enhancement is considerably less than has been recognized
previously.

We note, however, that this in no way negates the fact that the
presence of dense material surrounding Kepler's SN is surprising for
an object nearly 500 pc off the galactic plane. This material must
have its origin in either the precursor star, or the precursor system
(if a binary of some type was involved).  Canonical wisdom says that
core collapse precursors are massive stars, and massive stars shed
material via stellar winds prior to exploding.  Type Ia SNe are
thought to arise from white dwarf stars that are pushed over the
Chandrasekhar limit, although the exact details of the precursor
system that gives arise to this are still widely debated (e.g.  Livio
\& Riess 2003, and references therein).  The absence of hydrogen lines
in the spectra of Type Ia SNe is taken as evidence for CO white dwarfs
with little or no photospheric hydrogen, and certainly little if any
CSM.

Improved statistics on extragalactic SNe, however, are finding
exceptions to this general scenario.  A small but growing class of
bona fide (confirmed with spectra) Type Ia SNe has been found that
show narrow hydrogen lines at late times, indicating the presence of a
CSM component close to these objects. SN 2002ic is a recent example of
this phenomenon.  Some authors dub these objects Type Ia/IIn (e.g., a
Type Ia with narrow hydrogen lines; Kotak et al. 2004) while others
denote such objects as a new SN type, IIa (i.e., a Type Ia with
hydrogen lines; Deng et al. 2004).  Whatever the designation, these
objects demonstrate that some SNIa explosions can occur in regions
with significant CSM, albeit at distances closer in to the SN than
inferred for Kepler.  Any such close in component near Kepler's SN, if
present, would have long since been overrun by the blast wave.

Direct evidence for the presence of (presumably) CSM material around
another Type Ia SN was provided by {\it Hubble} Space Telescope
images of light echoes around SN 1998bu (Garnavich et al. 2001; Patat
2005).  Among several light echoes detected in this SN, there is an
echo generated by scattering off dust located closer that 10 pc to
the SN.  At the same time, SN 1998bu has one of the most stringent
upper limits on the density of the stellar wind (Panagia et
al. 2006). The presence of a detached CSM (or perhaps an ISM shell
swept-up by winds of the SN progenitor) is likely in this nearby SN.
These and similar observations open up the possibility that Kepler's
SNR represents a local example of this phenomenon.

Finally, it should be noted that recent detailed X-ray observations
and modeling (Kinugasa \& Tsunemi 1999; Cassam-Chena\"i et al. 2004)
are consistent with Si and Fe-rich ejecta, but show no evidence for
other enhancements seen in core collapse SNRs such as Cas A (Hughes et
al.  2000; Hwang \& Laming 2003).  The temperature structure inferred
within the X-ray ejecta, with Fe K peaking interior to Fe L is
reminscent of Type Ia SNRs such as SN1006 and Tycho (Hwang \& Gotthelf
1997; DeCourchelle et al. 2001).  These early results are strengthened
almost to the point of certainty by a recent deep {\it Chandra} X-ray
exposure on Kepler that will allow a detailed assessment of abundances
within the ejecta of Kepler (Reynolds et al. 2006). The preponderance
of evidence from the X-rays now points toward a Type Ia origin for the
precursor of Kepler's SNR.

If Kepler represents a SNIa explosion in a region with significant
CSM, it would be important from two directions: If plausible models
are put forward that can explain a Type Ia with significant CSM,
applying them to Kepler's SNR may provide a stringent test because of
its proximity and wealth of supporting observational data. On the
other hand, Kepler is only one object, and observations of additional
extragalactic examples of SNIa's with CSM in various forms may provide
important clues about the frequency and/or the progenitor population
of such explosions (e.g. Mannucci 2005).

\subsection{Conclusions}

We have presented {\it Spitzer} imaging of Kepler's supernova
remnant at 3.6, 4.5, 5.8, 8, 24, 70, and 160 $\mu$m wavelengths
and compared with data from other wave bands.
Emission associated with the remnant is obvious at
all except 160 $\mu$m, but emission in the two shortest-wavelength IRAC
bands is visible only at the locations of bright optical 
radiative shocks in the WNW.  However, at 24 $\mu$m, the entire periphery of 
the remnant can be seen, along with emission seen, at least in projection,
toward the interior.

To summarize, we find:

\begin{itemize}

\item The 24 $\mu$m emission is well correlated with the outer blast
wave as delineated by soft X-ray emission and by nonradiative
(Balmer-dominated) shocks seen in the optical. This is clearly
emission from dust heated by collisions in the X-ray emitting
material.  It is not well correlated either with ejecta emission
(shown by strong Fe L-shell emission in X-rays) or with dense regions
containing radiative shocks.  In particular, we find no evidence for
dust newly formed within the ejecta material.

\item The emission at 8 $\mu$m largely resembles that of the brightest
regions at 24 $\mu$m, although contamination by line emission in the 8
$\mu$m is apparent in the radiative shock regions.  This similarity
indicates that even short-wavelength emission originates from the same
grain population as that at longer wavelengths.  However, current
models are not yet able to describe emission from the implied
population of small grains.

\item The emission at 70 $\mu$m is similar to that at 24 $\mu$m, but a
higher background makes it difficult to discern the fainter southern
half of the remnant.  Where both 70 $\mu$m and 24 $\mu$m emission can
be seen, 70/24 $\mu$m flux ratios for discrete regions range from 0.37
to 0.49.  Lower values, implying higher temperatures, are correlated
with brighter regions. A total object average ratio of 0.52 implies
that the fainter regions tend toward higher values of this ratio.

\item The SED spectrum indicates that lines make at most a small
contribution between 55 and 95 $\mu$m.  The spectrum, along with the
absence of emission at 160 $\mu$m, rules out the presence of large
amounts of cold dust.

\item Dust models using a power-law grain size distribution and
including grain heating and sputtering by X-ray emitting gas can
explain observed 70/24 $\mu$m flux ratios with sensible parameters.
The models give gas densities of 10 -- 20 cm$^{-3}$.  The range of
observed ratios can be explained by ranges of temperature and density
of different regions in rough pressure equilibrium.

\item We find a total dust mass of about $5.4 \times 10^{-4}\ M_\odot$
after sputtering, and infer an original mass of about $2.4 \times
10^{-3} \ M_\odot$.  With an estimate of shocked gas mass from X-ray
data, we infer an original dust/gas ratio of about $3 \times 10^{-3}$,
lower by a factor of several than normally assumed for the Galaxy, as
has been found for several other supernova remnants.

\item We find that a moderate density contrast in the range of $\sim$4
-- 9 is required to explain the brightness variations observed between
the north and south rims of the remnant at 24 $\mu$m, depending
somewhat on the assumptions and models applied.

\item We suggest the preponderance of current evidence from optical,
X-ray, and infrared data and modelling now points toward a Type Ia
supernova, albeit in a region of significant surrounding CSM/ISM,
especially for an object so far off the galactic plane.  A possible
similarity to several extragalactic Type Ia supernovas with narrow
hydrogen lines at late times is pointed out.

\end{itemize}

Data from the {\it Spitzer Space Telescope} clearly demand more
sophisticated grain modeling.  The absence of large quantities of
newly formed dust challenges models hypothesizing such dust
formation in the SN ejecta.  However, the heated dust from the 
circumstellar medium heated by the blast wave can provide useful 
diagnostics of plasma conditions.

\acknowledgments

It is a pleasure to thank the $\spitzer$ operations team at 
JPL for their efforts in obtaining these data.  We also thank the
public relations staffs at the $\spitzer$, STScI, and $\chandra$ operations
centers for producing the color image in Figure 6, which was part of a 
photo release for the 400th anniversary of SN1604 in October 2004. 
This research has made use of SAOimage {\tt ds9}, developed by the
Smithsonian Astrophysical Observatory.  This work is supported
by JPL grant JPL-1264303 to the Johns Hopkins University.

\newpage
\clearpage

\begin{deluxetable}{lccccc}
\tablenum{5.1}
\tabletypesize{\footnotesize}
\tablecaption{Aperture Parameters for Region Extractions}
\tablehead{\colhead{Region} &\colhead{$\alpha$ (2000) } &
\colhead{$\delta$ (2000) } &\colhead{Semi-Major } &\colhead{Semi-Minor  } &
\colhead{PA ($^{\circ}$) } \\
\colhead{ } & \colhead{ } & \colhead{ } &  
\colhead{Axis (\arcsec)} & \colhead{Axis (\arcsec)} &  \colhead{ }}
\startdata
{\bf 24 $\micron$ to 70 $\micron$: }  & &          &         &     &   \\
Object 1 & 17:30:35.76 & -21:28:44.5  &  21.4  & 30.0  & 192.9  \\
Object 2 & 17:30:36.73 & -21:28:27.2  &  20.8  & 50.2  & 218.9  \\
Object 3 & 17:30:41.81 & -21:27:54.6  &  39.8  & 18.4  & 182.9  \\
Object 4 & 17:30:43.51 & -21:29:48.0  &  31.2  & 20.2  & 182.9  \\
Object 5 & 17:30:39.93 & -21:29:17.9  &  28.8  & 17.2  & 182.9  \\
Background 1 & 17:30:36.82 & -21:27:05.2  &  18.5  & 27.9  & 186.3  \\
Background 2 & 17:30:42.96 & -21:27:03.2  &  39.8  & 18.4  & 182.9  \\
Background 3 & 17:30:41.91 & -21:31:45.7  &  34.0  & 21.6  & 186.3  \\
Background 4 & 17:30:38.24 & -21:32:15.7  &  35.1  & 17.2  & 182.9  \\
Background 5 & 17:30:34.05 & -21:31:40.2  &  22.3  & 48.7  & 186.3  \\
{\bf 8 $\micron$ to 24 $\micron$: }  & &          &         &     &   \\
Object 1 & 17:30:35.81 & -21:28:55.0  &  7.2  & 7.8  & 93.2  \\
Object 2 & 17:30:36.28 & -21:28:33.7  &  10.2  & 10.2  & 0  \\
Object 3 & 17:30:38.14 & -21:27:59.7  &  7.2  & 7.2  & 0  \\
Object 4 & 17:30:42.61 & -21:27:51.8  &  9.6  & 15.6  & 103.2  \\
Object 5 & 17:30:42.61 & -21:27:51.8  &  9.6  & 15.6  & 103.2  \\
Object 6 & 17:30:40.09 & -21:29:21.8  &  9.0  & 14.4  & 103.2  \\
Background 1 & 17:30:35.92 & -21:27:34.5  &  29.4  & 14.4  & 138.2  \\
Background 2 & 17:30:42.07 & -21:27:12.3  &  10.2  & 20.4  & 43.2  \\
Background 3 & 17:30:40.28 & -21:31:23.9  &  12.0  & 17.4  & 93.2 \\ 

\enddata

\end{deluxetable} 

\newpage
\clearpage

\begin{table}
\tablenum{5.2}
\begin{center}
{Table 5.2 -- MIPS 70/24 $\mu$m Regions Summary}
\begin{tabular}{lccc}
Region$^{a}$ & Net 70 $\mu$m$^{b}$ Flux & Net 24 $\mu$m Flux$^{b}$ & 70/24 Ratio \\
       & mJy & mJy &       \\
\hline
O1  & 778   &  2133  & 0.36 \\
O2  & 1171   &  3125 & 0.37 \\
O2 $-$ O1  & 394   &  992 & 0.40 \\
O3  & 412   &  1032 & 0.40 \\
O4  & 349   &  886  & 0.39 \\
O5  & 242   &  766  & 0.30 \\
\hline
\end{tabular}
\end{center}

Notes:  (a) Refer to Figure 7a-c for region definitions.
(b) Net fluxes have been background-subtracted as described in the text.
\end{table}

\newpage
\clearpage

\begin{table}
\tablenum{5.3}
\begin{center}
{Table 3 -- MIPS 8/24 $\mu$m Regions Summary}
\begin{tabular}{lccc}
Region$^{a}$ & Net 8 $\mu$m$^{b}$ Flux & Net 24 $\mu$m Flux$^{b}$ & 8/24 Ratio \\
       & mJy & mJy &       \\
\hline
O1  & 10.4   &  409   & 0.025 \\
O2  & 16.5   &  571   & 0.029 \\
O3  & 6.9    &  242   & 0.029 \\
O4  & 6.9    &  314   & 0.022 \\
O5  & 5.4    &  228   & 0.024 \\
O6  & 11.7   &  339   & 0.035 \\
\hline
\end{tabular}
\end{center}

Notes:  (a) Refer to Figure 7d-f for region definitions.
(b) Net fluxes have been background-subtracted as described in the text.
\end{table}

\newpage
\clearpage

\begin{figure}
\figurenum{5.1}
\includegraphics[width=14cm]{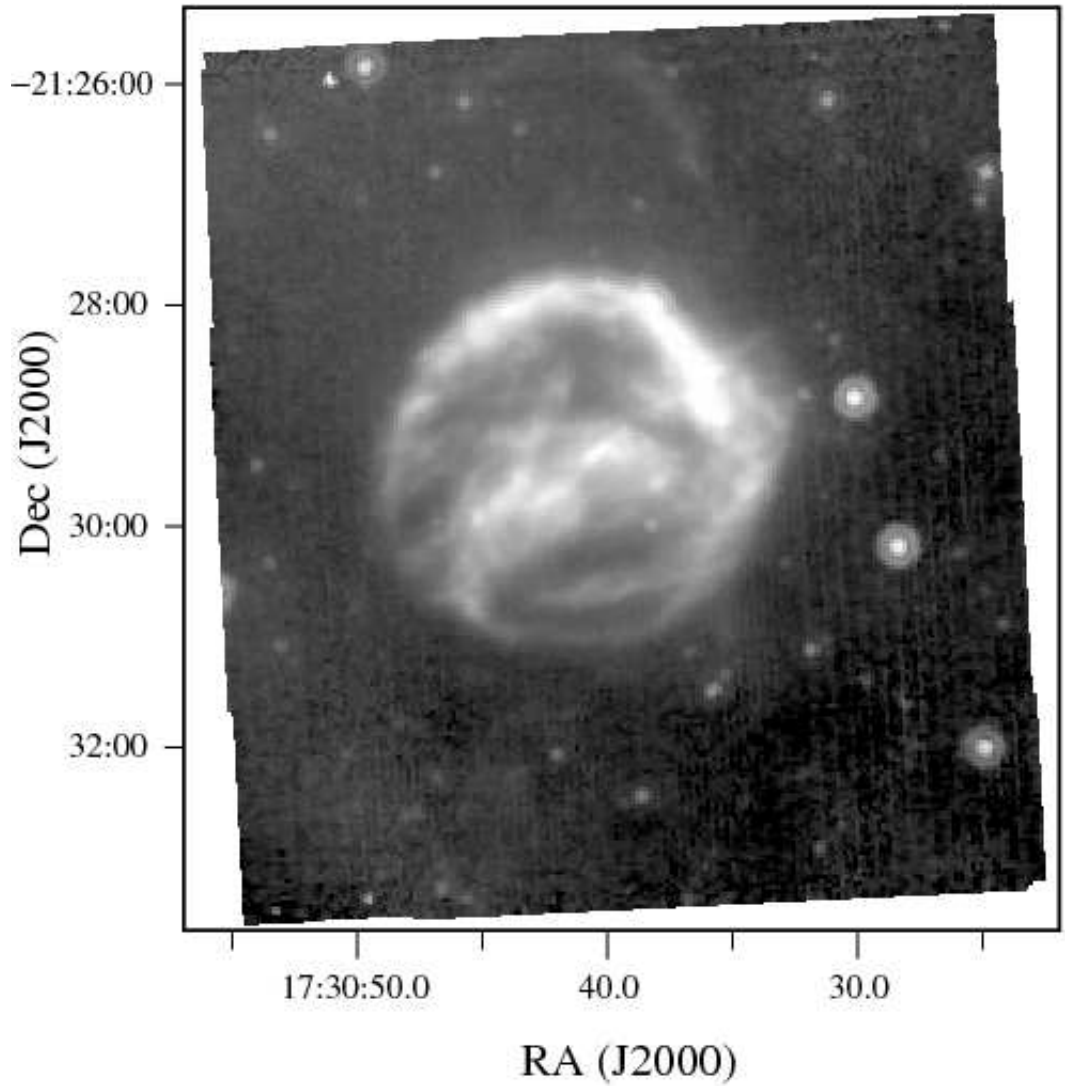}
\caption{Full field of view MIPS 24 $\mu$m image of Kepler's SNR.
Scaling is set to a compromise level to show the overall structure
to best advantage. The 24 $\mu$m image is by far the deepest and most
detailed of the {\it Spitzer} images.
\label{5-1}
}
\end{figure}

\newpage

\begin{figure}
\figurenum{5.2}
\includegraphics[width=10cm]{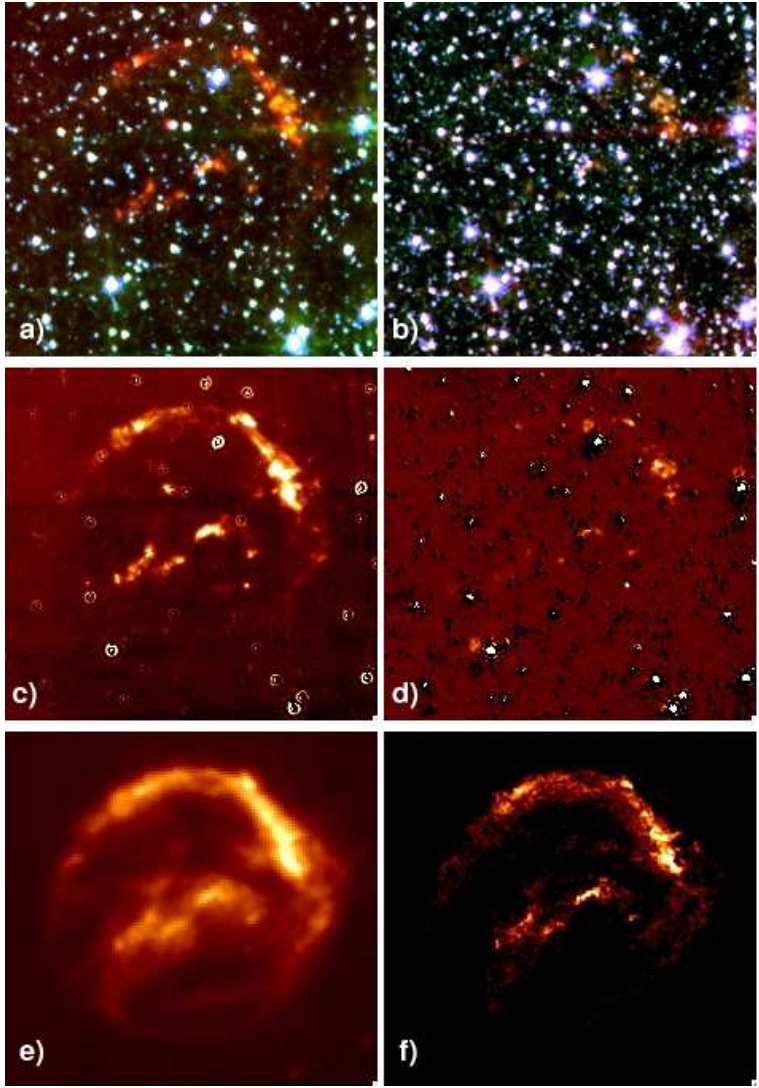}
\caption{A six-panel color figure concentrating on the IRAC images and 
their comparison to other wavelength bands. Panel a shows a three-color
IRAC image with 8 $\mu$m in red, 5.6 $\mu$m in green and 3.6 $\mu$m 
in blue.  Panel b is similar, but for 5.6 $\mu$m in red, 4.5 $\mu$m 
in green, and 3.6 $\mu$m in blue.  The orange color of the SNR filaments 
indicates emission in both 4.5 and 5.6 $\mu$m, but only from the 
brightest filaments seen at 8 $\mu$m.  Panel c is a difference image 
of 8 $\mu$m minus 5.6 $\mu$m, scaled to show the extent of faint 
emission at 8 $\mu$m.  Panel d shows the 4.5 $\mu$m minus 3.6 
$\mu$m difference image.  Panel e shows the 24 $\mu$m MIPS 
image from Fig. 1 to the same scale as the other images.  The
8 $\mu$m image closely tracks the brightest regions at 24 $\mu$m.
Panel f shows the soft band (0.3 - 0.6 keV) {\it Chandra} image from 
archival data, which again looks astonishingly like the 8 $\mu$m 
image in panel c.  All images are aligned and scaled exactly the same.
\label{5-2}
}
\end{figure}

\newpage

\begin{figure}
\figurenum{5.3}
\includegraphics[width=11cm]{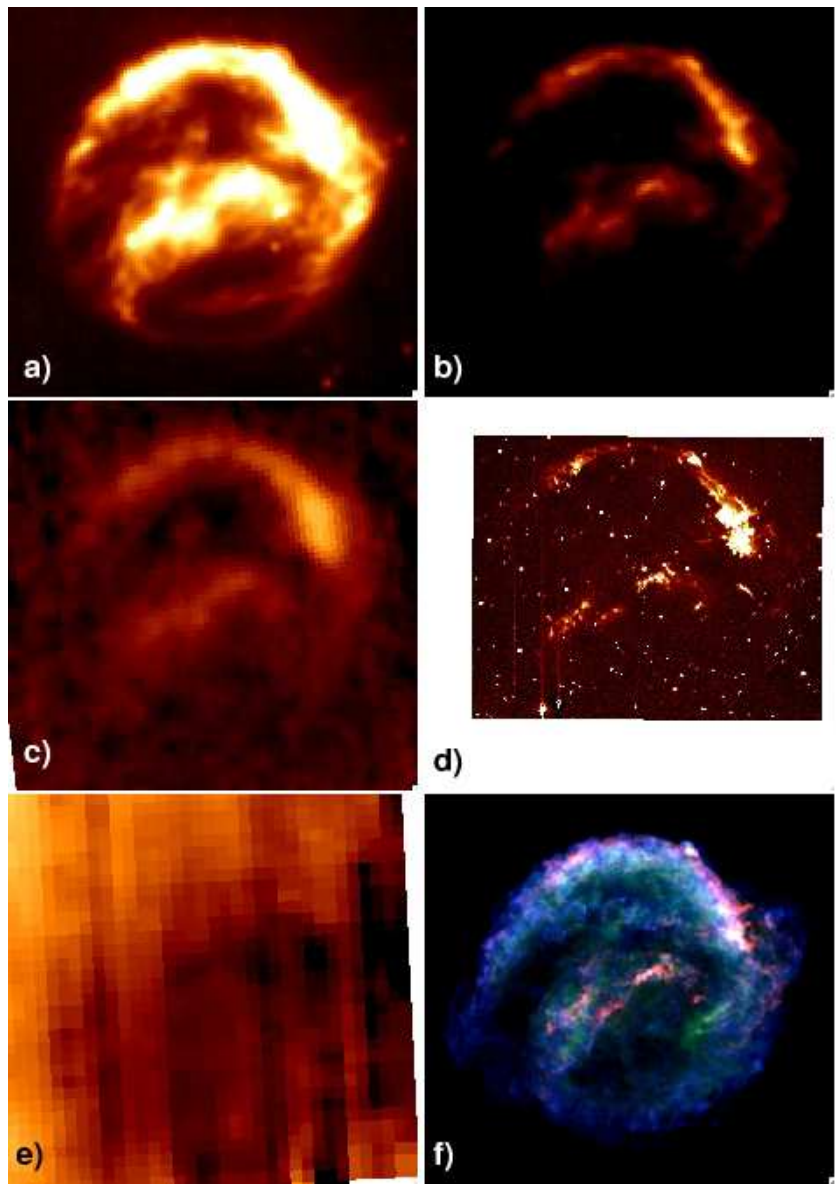}
\caption{A six-panel color figure concentrating on the MIPS images and 
their comparison to other wavelength bands. Panels a and b show the MIPS 24 
$\mu$m data with a hard stretch and a soft stretch, respectively, to 
show the full dynamic range of these data.  Panel c shows the MIPS 70 
$\mu$m data after running through the GeRT software to improve the 
appearance of the background.  Panel e shows the MIPS 160 $\mu$m data 
for the same field of view, although no SNR emission is actually seen 
at this wavelength. Panel d shows the star-subtracted H$\alpha$ 
image from Blair et al. (1991).  Panel f shows a three-color 
representation of the {\it Chandra} data for Kepler, with the red being 
0.3 - 0.6 keV (as in Fig. 2f), green being 0.75 - 1.2 keV, and blue being
1.64 - 2.02 keV.
\label{5-3}
}
\end{figure}

\newpage

\begin{figure}
\figurenum{5.4}
\includegraphics[width=14cm]{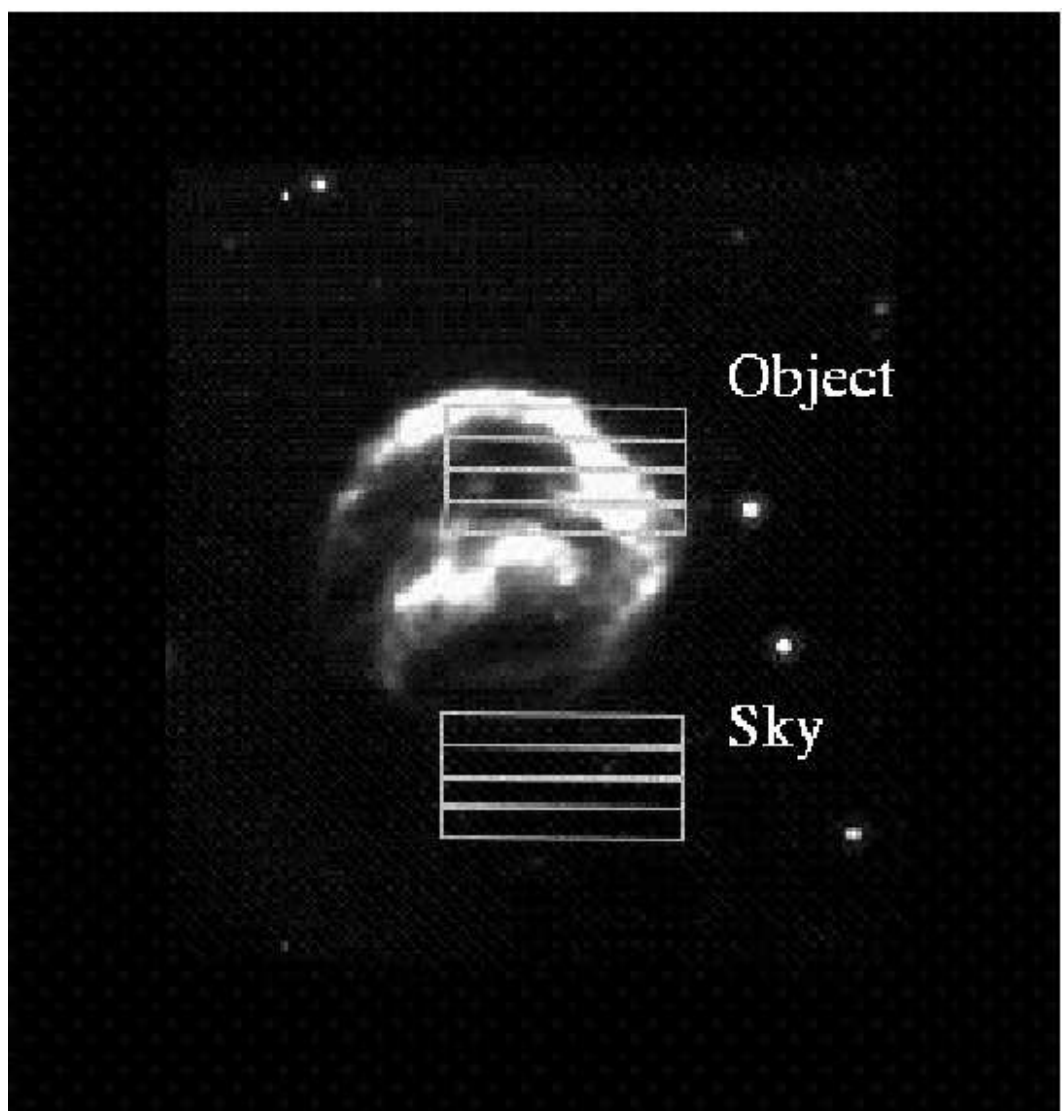}
\caption{SED apertures selected for assessing the bright NW
radiative emission and the sky background, projected
on the 24 $\mu$m image.
\label{5-4}
}
\end{figure}

\newpage

\begin{figure}
\figurenum{5.5}
\includegraphics[width=14cm]{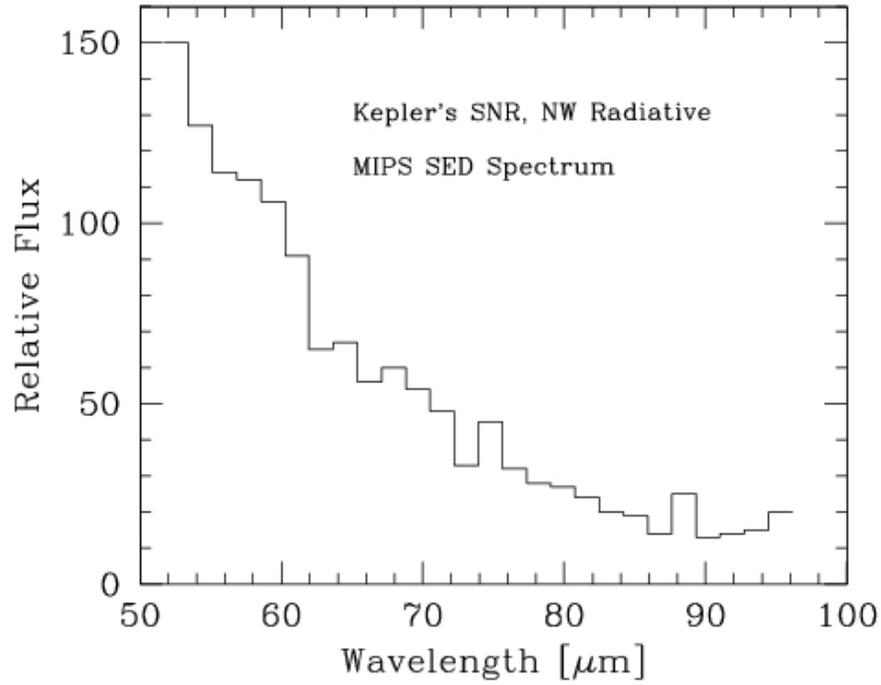}
\caption{Background subtracted SED 55 - 95 $\mu$m spectrum of the NW region
of Kepler's SNR, as indicated in Figure 4.
\label{5-5}
}
\end{figure}

\newpage

\begin{figure}
\figurenum{5.6}
\includegraphics[width=14cm]{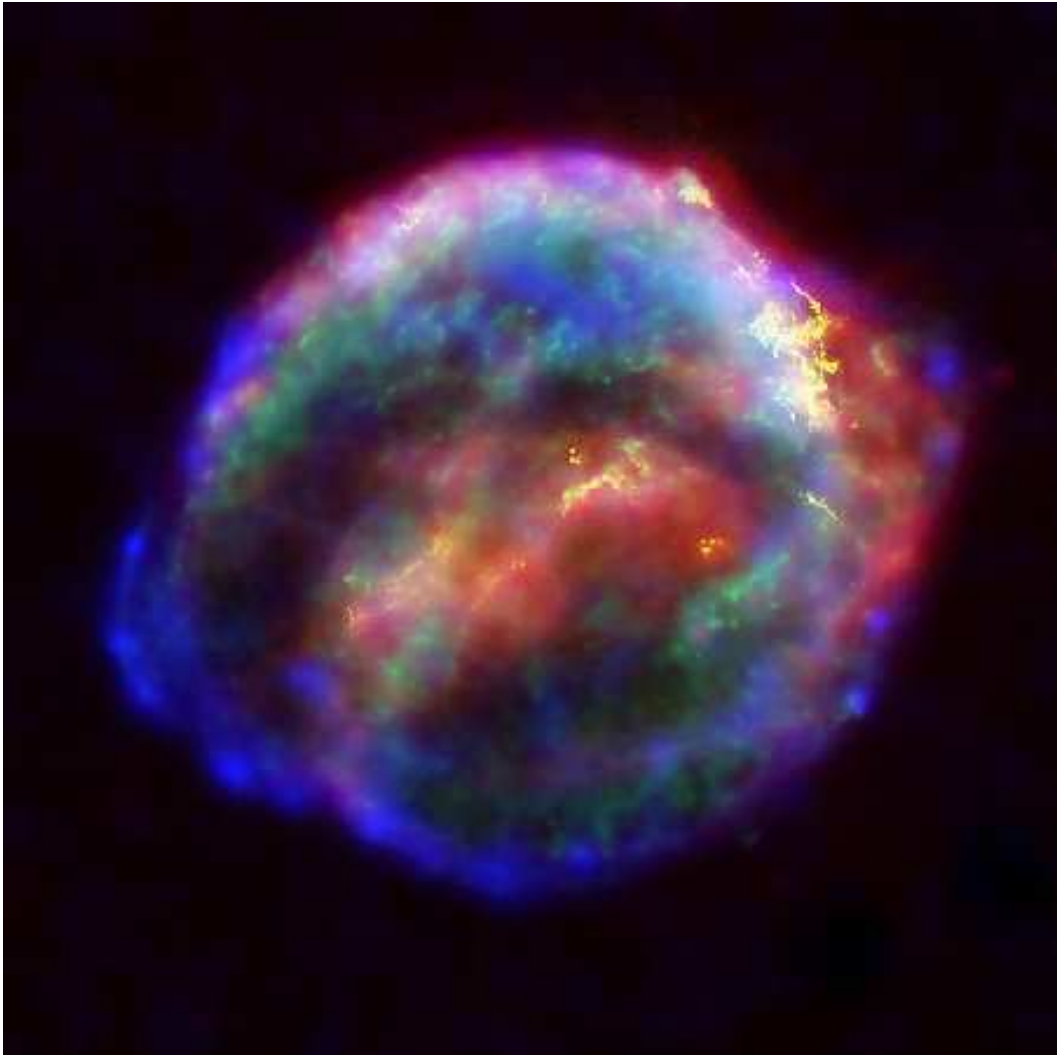}
\caption{A color view of Kepler showing data from $\spitzer$ 24 $\mu$m 
(red), $\hubble$ ACS H$\alpha$ (yellow), and $\chandra$ medium (blue) and 
soft (green) X-ray emission bands.  Despite the differing intrinsic 
resolutions of the various data sets, they have been carefully 
coaligned.  A $\hubble$ continuum-band image was used to subtract the stars 
from this image.  See text for details.
\label{5-6}
}
\end{figure}

\newpage

\begin{figure}
\figurenum{5.7}
\includegraphics[width=15cm]{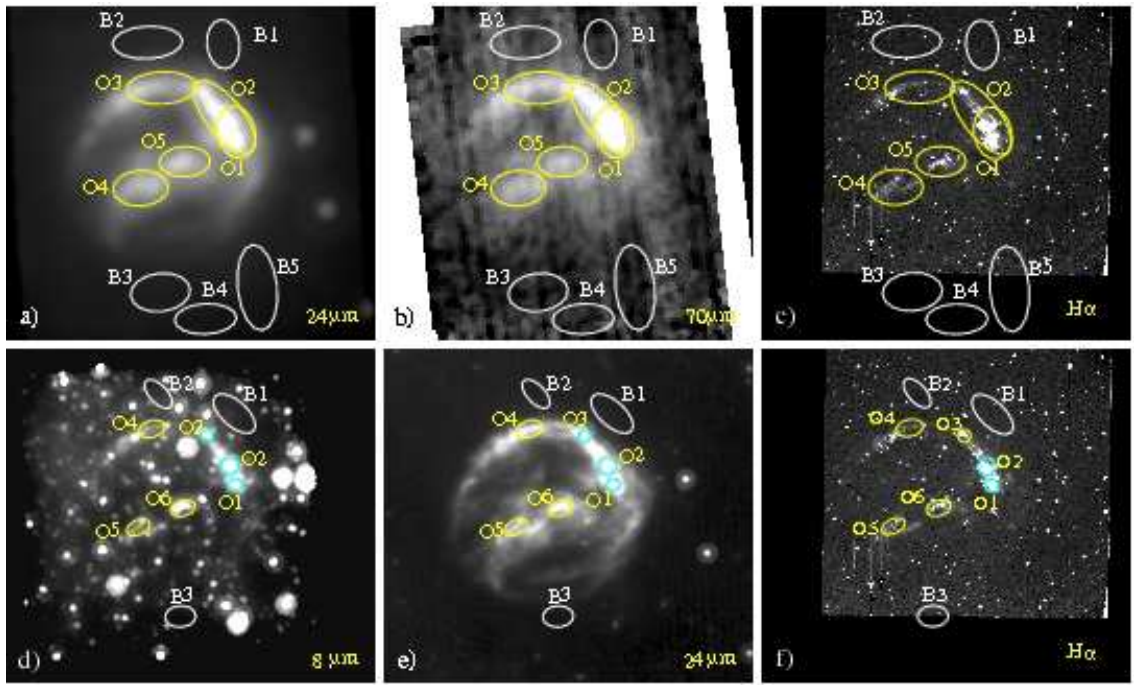}
\caption{This figure shows the object and background extraction
regions selected for determining ratios between 24 and 70 $\mu$m
(panels a and b) and between 8 and 24 $\mu$m (panels d and e).  For
reference, the regions are also projected onto the optical H$\alpha$
image from Blair et al. (1991) in panels c and f. Note that the images
shown in panel a and d are the versions that have been convolved to
the lower resolution image.  Also, panel b shows the original
(non-GeRT-corrected) 70 $\mu$m image.  The labels are used in the text
and Tables 5-2 and 5-3.
\label{5-7}
}
\end{figure}

\newpage

\begin{figure}
\figurenum{5.8}
\includegraphics[width=15cm]{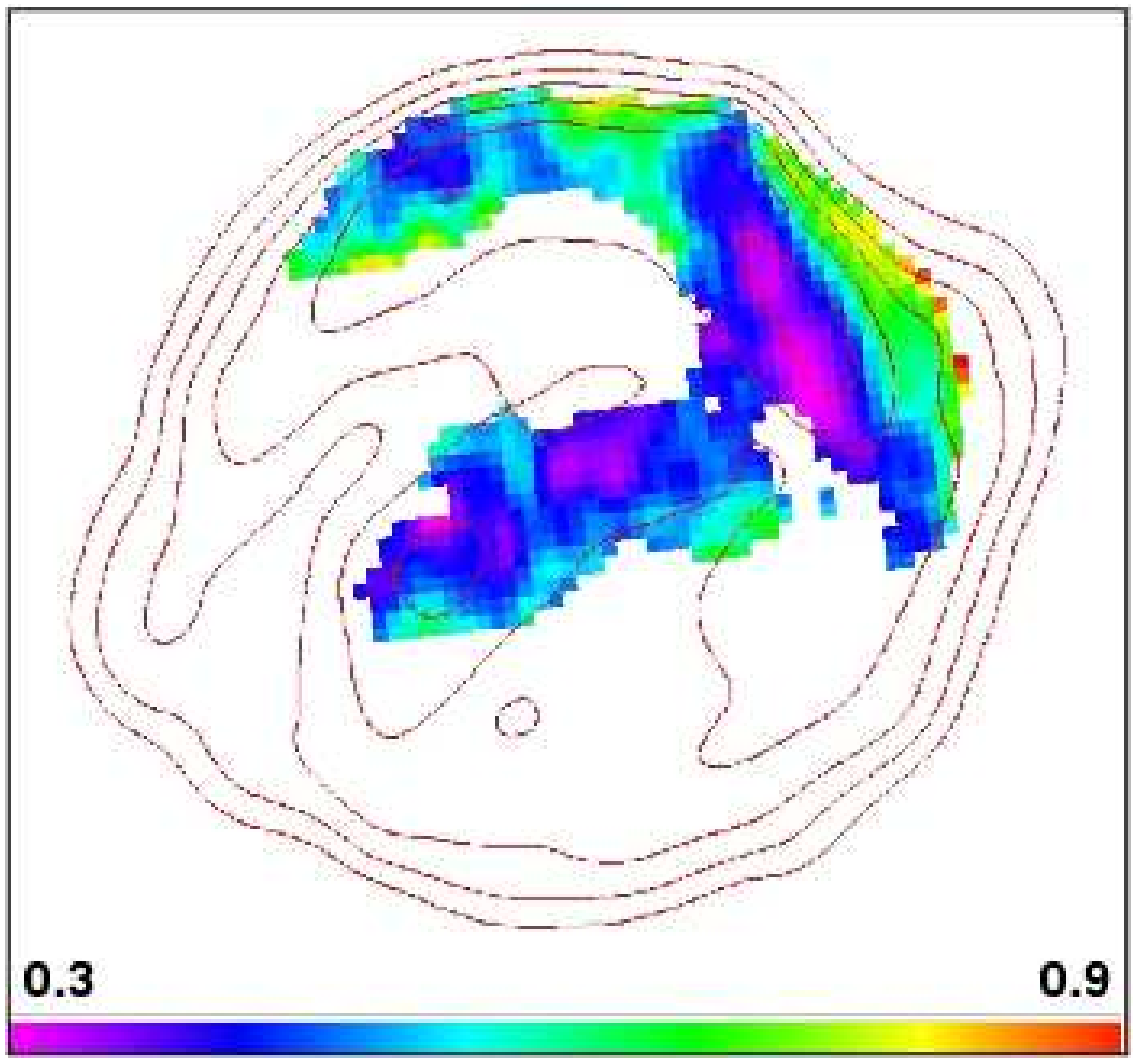}
\caption{A ratio image of the 70 $\mu$m and 24 $\mu$m data, where only
the regions with significant signal have been kept.  The color bar
provides an indication of the measured ratio, ranging from 0.28 to
0.9. A simple contour from the 24 $\mu$m image is shown for
comparison.  Note the lower values of the ratio in the regions of
brightest emission, indicating they are somewhat warmer.
\label{5-8}
}
\end{figure}

\newpage

\begin{figure}
\figurenum{5.9}
\includegraphics[width=16cm]{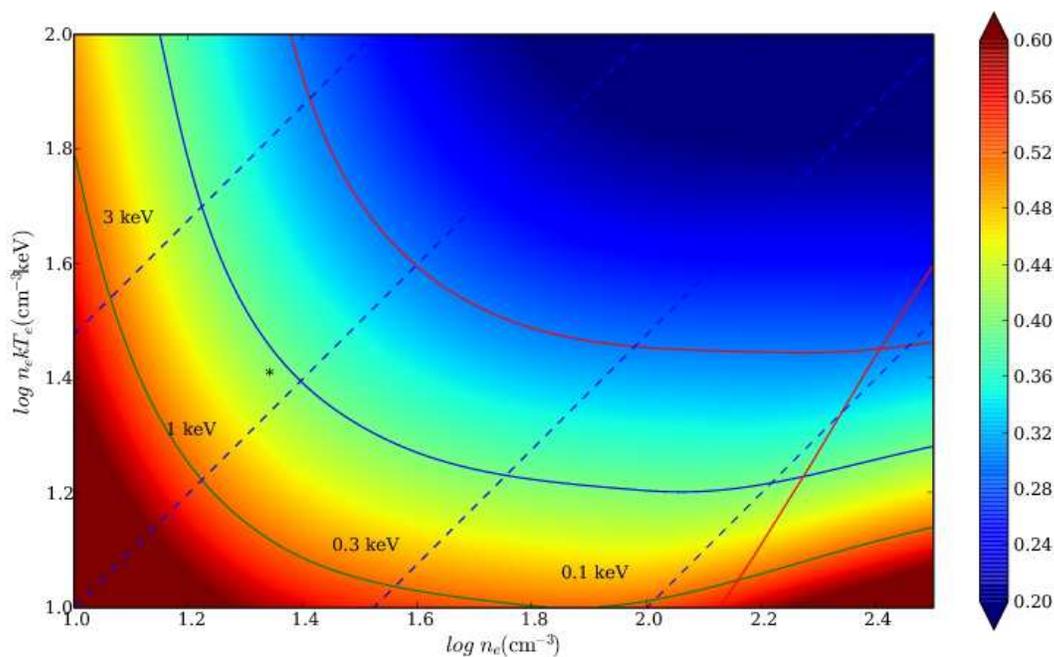}
\caption{The 70/24 $\mu$m MIPS flux ratio as a function of electron
density and pressure for plane shock dust models discussed in the
text.  The background color scale indicates 70/24 $\mu$m ratio, as
indicated by the color bar at right. Dashed lines are lines of
constant temperature.  The solid magenta diagonal line at lower right
indicates where the modeled shocks would become radiative, assuming
solar abundance models and an age of 400 years.  Three {\it solid
curves} are lines of constant 70/24 MIPS flux ratios, 0.30, 0.40, and
0.52 (from top to bottom), encompassing measured ratios listed in
Table 2 and the spatially-integrated ratio.  Position of a
Balmer-dominated fast (1660 km s$^{-1}$) shock in the north is marked
by a star.
\label{5-9}
}
\end{figure}

\newpage
\clearpage

\section{Ejecta, Dust, and Synchrotron Radiation in SNR B0540-69.3: A More Crab-like Remnant than the Crab}

This chapter is reproduced in its entirety from Williams, B.J.,
Borkowski, K.J., Reynolds, S.P., Raymond, J.C., Long, K.S., Morse,
J.A., Blair, W.P., Ghavamian, P., Sankrit, R., Hendrick, S.P., Smith,
R.C., Points, S., \& Winkler, P.F. 2008, ApJ, 687, 1054.

\subsection{Introduction}

Many core-collapse supernovae (SNe) leave behind a neutron star as a
compact remnant.  Some of these neutron stars are active pulsars which
inflate a bubble of relativistic particles and magnetic fields
confined by the ejecta or interstellar medium (ISM), known as a
pulsar-wind nebula.  The combination of a shell supernova remnant
(SNR) and associated pulsar-wind nebula can allow the investigation of
various issues of importance in supernova and pulsar physics,
including pulsar kicks, ejecta structure and composition, and particle
acceleration at relativistic shocks.  Pulsar-wind nebulae serve as
calorimeters for pulsar spindown energy loss, and as test systems to
study the behavior of relativistic shocks where the pulsar wind is
thermalized. We know of few cases of a ``normal'' radio and X-ray
shell supernova remnant containing an active pulsar and synchrotron
nebula.  Probably the best known such ideal case is the Large
Magellanic Cloud remnant B0540-69.3 (or ``0540'' for short). 0540 is
also one of a highly exclusive group of ``oxygen-rich'' SNRs, a group
that includes Cas A, Puppis A, G292+1.8, 1E0102-72.3, and N132D.

Theoretical studies of PWNe have either concentrated on the gross
evolution, assuming a homogeneous nebula
\citep{rees74,pacini73,reynolds84} or the detailed spatial structure,
neglecting evolution \citep{kennel84}.  Since the advent of the
new generation of X-ray observatories, the study of PWNe has
accelerated, with the identification of many new objects and more
detailed information on known ones (see Gaensler \& Slane 2006 for a
recent review).  \cite{chevalier05} modeled PWNe for different
assumptions about the ejecta profiles into which they expand, to
relate properties of supernovae to those of the PWNe.
 
PWNe produce extremely broad-band spectral-energy distributions
(SEDs), well described in various frequency regimes with power laws.
Most PWNe are observed in radio and X-rays; only a few are detected
optically (here as in many other ways the Crab Nebula is an
exception), and almost nothing is known about infrared or ultraviolet
spectra.  Typical radio spectra are featureless, and are well described by
power-laws with spectral indices $\alpha < 0.3$ ($S_\nu \propto
\nu^{-\alpha}$), with X-ray indices steeper by 0.5 -- 1.3 (see data in
Chevalier 2005).  Since simple models of synchrotron losses predict a
steepening of exactly 0.5, they lack some essential physics, which may
be constrained if the complete spectrum is known.  Galactic PWNe are
all found close to the Galactic plane, where they suffer from
extinction in optical and UV and confusion in IR.  Filling in the SED
between radio and X-rays can best be done with a high-latitude object.
For this reason as for many others, 0540 is an interesting target.

0540 was first catalogued as a radio source of unknown nature, a minor
feature on a 408 MHz map of the 30 Dor region made with the Molonglo
telescope \citep{lemarne68}.  \cite{mathewson73} first classified it
as a supernova remnant on the basis of its steep radio spectrum, although
their optical survey did not detect it.  Early reports associated 0540 with 
the H$\alpha$ emission nebula N 158A 
\citep{henize56}, though that object is $3'$ from the centroid of the
early radio positions (which could be localized to better than
$10''$).  The absence of strong H$\alpha$ emission from 0540 further
demonstrates that the association with N 158A is erroneous.
Subsequent radio observations \citep{milne80} gave an improved
spectral index of $-0.44$, typical for a shell supernova remnant.  The
first indication of something unusual was the X-ray detection
\citep{long79} with the {\sl Einstein} Observatory, in which 0540 was
the third brightest X-ray remnant in the LMC.  The X-ray spectrum was
shown to be featureless by \cite{clark82} with the {\sl Einstein}
Solid-State Spectrometer.  The first optical detection was reported by
Mathewson et al.~(1980), motivated by pre-publication reports of the
observations of Clark et al~(1982).  Mathewson et al.~did not see
H$\alpha$ but instead a spectacular ring in [O III] of $8''$ diameter,
with a smaller ring in fainter [N II] emission ($4''$ diameter) and no
appreciable Balmer emission. In addition to classifying 0540 as an
``oxygen-rich'' SNR, \cite{mathewson80} also reported spectroscopic
observations indicating expansion speeds of order 1500 km s$^{-1}$.
The discovery of the 50 ms X-ray pulsar \citep{seward84} and optical
synchrotron nebula \citep{chanan84} added to the complexity and
interest of the system.  The optical emission was shown definitively
to be synchrotron by the discovery of polarization \citep{chanan90}.
The pulsar spindown timescale $P/2 {\dot P}$ is about 1660 yr
\citep{seward84}, somewhat longer than the kinematic age estimate
resulting from dividing the radius ($4'' = 1$ pc at our assumed
distance of 50 kpc) by the expansion speed of about 1500 km s$^{-1}$,
which yields a value of $\sim 700$ yr.  The pulsar spindown luminosity
is $1.5 \times 10^{38}$ erg s$^{-1}$.

\cite{reynolds85} modeled 0540 with the formalism of
\cite{reynolds84}, with the pulsar driving an accelerating synchrotron
nebula into the inner edge of expanding ejecta.  At that time, there
were no more than hints of extended structure that could be identified
with the outer blast wave.  \cite{reynolds85} found that the current
radio, optical, and X-ray observations could be explained without
requiring extreme values for the pulsar initial energy or other
parameters.  He deduced an initial pulsar period of about $30 \pm 8$
ms, that is, relatively slow, and concluded that the true age of 0540
was between 800 and 1100 yr, somewhat longer than the kinematic age
due to the pulsar-driven acceleration.

Up to this time, all observations were consistent with 0540 being a
standard Crab-like remnant (i.e., a nonthermal center-brightened radio
and X-ray nebula surrounding a pulsar), except for the hint of
larger-scale structure from radio images and from X-ray observations
\citep{seward84}.  Definitive information on the structure came from
higher-resolution radio observations with the Australia Telescope
\citep{manchester93} which showed a clear radio shell with diameter
about $65''$ surrounding a radio nebula with size (about $5''$ FWHM)
comparable to the bright X-ray nebula and [O III] ring.  The shell has
a radio spectral index $\alpha$ of about $-0.4$, while the central
nebula has $\alpha = -0.25$.  At this point it was clear that 0540 is
even more Crab-like than the Crab, as it possesses a clear outer blast
wave interacting with surrounding material, so that we could be sure
that the interior PWN is interacting with the inner SN ejecta as in
Reynolds \& Chevalier (1984).  X-ray emission from the blast wave was
confirmed with {\it Chandra} observations \citep{hwang01}; the
emission is brightest in the W and SW, like the radio.  Spectral fits
indicated abundances typical of the LMC, with a temperature of order 4
keV (for a Sedov blast wave model) and ionization timescale $\tau
\equiv n_e t = 3.7 \times 10^{10}$ cm$^{-3}$ s, though spectral
differences are apparent in different regions, and a hard component
may be called for.

The most thorough optical spectroscopic study to date was reported by
\cite{kirshner89}.  They confirmed the high velocities (FWZI $\sim
2800$ km s$^{-1}$), and reported weak H$\alpha$ emission.  The average
centroid of SNR lines (as opposed to narrower lines from a nearby H II
region) was shifted by $+370$ km s$^{-1}$.  No [Ne III] was reported ($<
1.5$ \% of [O III]); they concluded that this was a real abundance
deficit rather than a temperature or density effect.  A detailed study
by \cite{serafimovich04}, focusing on the optical nonthermal
continuum, revised the reddening and optical slope to give a power-law
index in the optical of $\alpha_o = -1.07.$ Recent observations by
\cite{morse06} report the discovery of faint [O III] emission
extending to a radius of $8''$, with a velocity of 1650 km s$^{-1}$.
They find the centroid of this velocity component to be the same as
that of the LMC, so that a large peculiar velocity of the system is
not required.

\cite{chevalier05} modeled 0540, along with several other PWNe, with
the goal of learning more about the SN explosion.  He obtained several
results for a simple dynamical model of a PWN expanding into ejecta of
various density profiles driven by a pulsar of given power. He
interpreted 0540 as the result of a SN Ib/c, an exploding Wolf-Rayet
star, with the prediction of a lack of significant emission from
hydrogen. However, recent optical observations by
\citet{serafimovich04} and \citet{morse06} have detected hydrogen. In
light of this, it is now believed \citep{chevalier06} that 0540 is the
result of a type IIP supernova.

The infrared observations of 0540, which was detected by the {\it
Infrared Space Observatory} (ISO) \citep{gallant99}, presented in this
paper promise to advance our understanding on several fronts. The
outline of our paper is as follows: In section 6.2, we describe the
observations and data reduction, and results are given in section
6.3. In section 6.4.1, we discuss a general picture of the PWN, and
sections 6.4.2 and 6.4.3 discuss in detail the line emission and dust
continuum emission, respectively. In section 6.4.4, we discuss the
origin of the O-rich clumps, whose existence we posit in section
4.2. Section 6.5 serves as a summary of our findings.

\subsection{Observations and Data Reduction}

During Cycle 1 of {\it Spitzer} observations, we obtained pointed
observations of 0540 with the Infrared Array Camera (IRAC) and the
Multiband Imaging Photometer for Spitzer (MIPS) as part of a survey of
$\sim 40$ known supernova remnants in the Large and Small Magellanic
Clouds \citep{borkowski06,williams06}. Our IRAC observations (28
November 2004) consisted of a dither pattern of 5 pointings with a
frame time of 30 seconds for each frame. This pattern was used for all
4 IRAC channels. Our MIPS observations (7 March 2005) differed based
on the module used. At 24 $\mu$m, we mapped the region with 42
overlapping pointings of 10 seconds each. At 70 $\mu$m, we mapped the
remnant with 94 pointings of 10 seconds each. At 160 $\mu$m, we mapped
the region with 252 pointings of 3.15 seconds. Since 0540 was not
detected at 160 $\mu$m, we do not discuss 160 $\mu$m data here. Both
the Basic Calibrated Data (BCD) and Post-BCD products were processed
with version S14.4 of the PBCD pipeline. We then used the {\it Spitzer
Science Center} (SSC) contributed software package MOPEX to ``clean
up'' the images, although the improvements were minimal. MOPEX was
able to remove some of the streaks caused by bright stars in the IRAC
images of the region.

\begin{figure}
\figurenum{6.1}
\plotone{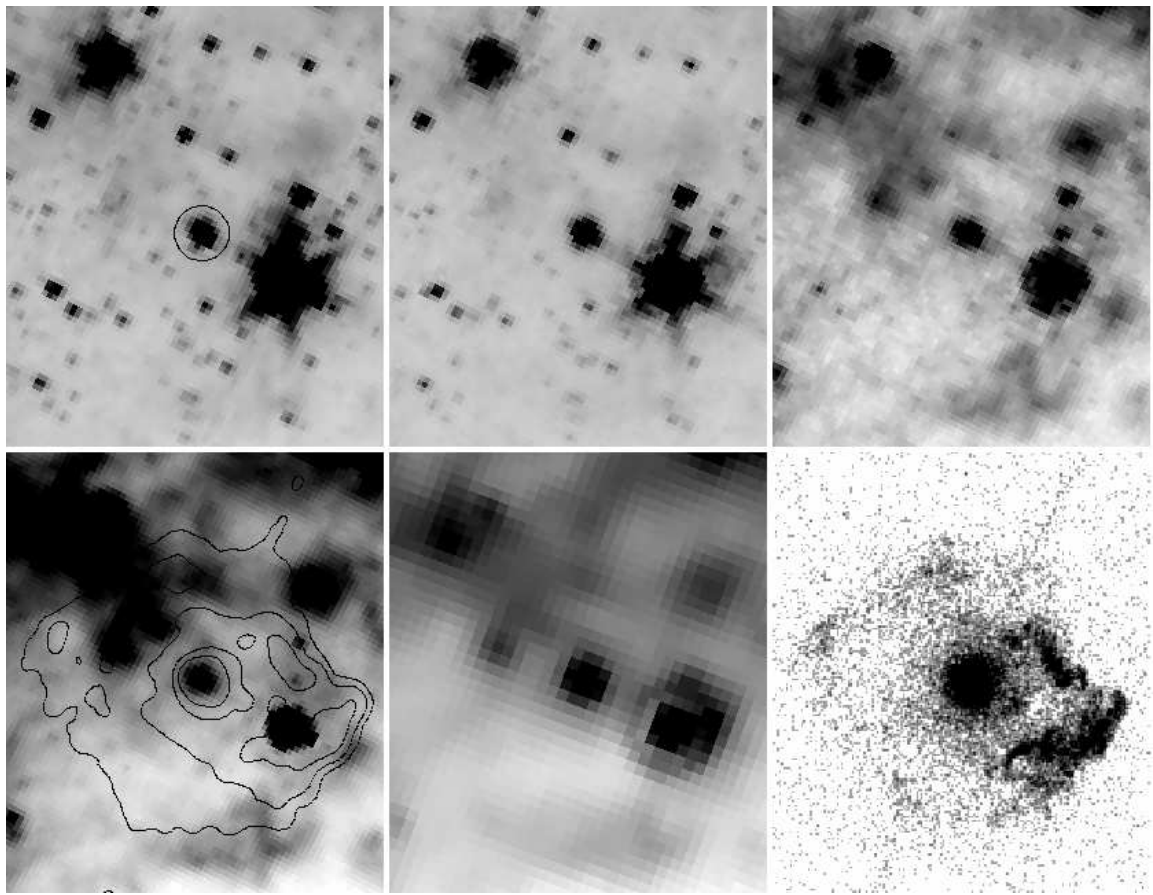}
\caption{Images of PWN 0540-69.3. Each image is approximately 100
arcseconds across. Left to Right, Top to Bottom: IRAC Chs. 1-4 (3.6,
4.5, 5.6, and 8.0 $\mu$m, respectively), MIPS 24 $\mu$m, {\it Chandra}
broadband X-ray image. The location of the PWN is marked with a circle
in the IRAC Ch.1 image, and X-ray contours are overlaid on the IRAC
Ch.4 image.}
\label{fig6-1}
\end{figure}

Our images of the source are shown in Figure~\ref{fig6-1}. With a
radius of $\sim 4''$, the PWN is resolved by {\it Spitzer}, and it
clearly stands out from the background in IRAC and MIPS 24 $\mu$m
bands. In IRAC ch. 3 \& 4 (5.8 \& 8.0 $\mu$m), as well as MIPS 24
$\mu$m, there is a hint of a shell around the nebula, at approximately
$30''$. We considered the possibility that this shell is related to the
SNR, perhaps the collisionally heated dust from the outer blast wave,
as we have observed in several other SNRs. However, the morphology of
the IR shell does not correspond with any features in the X-ray or
radio shell. Spectroscopy of the shell shows it to be virtually
identical to the surrounding background unrelated to the remnant, so we
are forced to conclude that its apparent relation to the SNR is
coincidental.

In Cycle 2, we obtained spectroscopic pointings for 0540 using all
four instruments of the Infrared Spectrograph (IRS). Our observations
were done between 8-10 July 2005. We used the spectral mapping mode
for the low-resolution modules, and staring mode for the
high-resolution echelle modules. Figure ~\ref{fig6-2} shows our
coverage of the PWN with IRS overlaid on our MIPS 24 $\mu$m image. For
the short-wavelength, low-resolution module (SL) we obtained 5
parallel pointings with each of the two orders, with a step direction
of $3.5"$ perpendicular to the dispersion direction of the slit. A
total of 480 seconds (2 cycles of 240 s) was obtained for each slit
position. For the long-wavelength, low-resolution (LL) module, we
obtained 3 parallel pointings for each order with a step direction of
$10.5"$ perpendicular to the dispersion direction. A single cycle of
120 seconds was used for LL. Since we are primarily interested in
determining the shape of the continuum from the low resolution
spectra, it was important to obtain spectra of the local background as
well as the source. Figure~\ref{fig6-1} illustrates the complex nature
of the local background and the difficulty of accurate background
subtraction. Because our source is only $\sim 8''$ in diameter, we
were able to extract background spectra from not only the parallel
slit pointings, but also from different parts of the slit containing
the source. We downloaded the Post-BCD data, pipeline version S15,
from the SSC. We used the Spitzer IRS Custom Extraction (SPICE)
software provided by the SSC to extract our spectra. Although the PWN
is slightly extended, it is close enough to a point source, especially
at longer wavelengths, to use the point source extraction mode in
SPICE.

In order to determine line profiles and strengths from the source
itself, we also obtained pointings with both the short-wavelength,
high-resolution (SH) and long-wavelength, high-resolution (LH) modules
in staring mode.  For these pointings, we centered the echelle
spectrographs only on the source, without a dedicated
background pointing. Since staring mode automatically provides 2 nod
positions for each pointing, we averaged the two to obtain a single
spectrum for SH and a single spectrum for LH. For SH, we used 3 ramp
cycles of 480 seconds each, for a total of 1440 s. The same total
integration time was obtained for LH, but was broken up into 6 cycles
of 240 s each.

\subsection{Results}

\subsubsection{Flux Extraction}

0540-69.3 is located close to the 30 Doradus region of the LMC, and
thus is in a region of high infrared background. For our IRAC and MIPS
images, we simply used an annular background region to subtract off
the background flux from the PWN. Because the nebula is only about $4''$
in radius, we used an on-source region of $\sim 6''$ radius to be sure
to capture all of the flux from the object, and a background annulus
between $6''$ and $10''$ radius. At 70 $\mu$m, we derive an upper limit to the
flux that could be contained in the region based purely on error
analysis of the pixels. Our results, with a 3$\sigma$ upper limit at
70 $\mu$m, are given in Table 6.1.

\subsubsection{Spectral Extraction}

For each of the SL orders, our procedure was as follows. First, we
extracted spectra from three different, non-overlapping positions on
each of the 5 slits. The positions corresponded to the middle and the
ends of each slit. This gave us a total of 15 different spectra. Given
the spatially varying background in the vicinity of the PWN, we
elected to use only the background regions that were closest to the
source. Thus, we excluded the 4 ``corner" regions, leaving us with 11
total regions. We considered the middle 3 regions to be our
``on-source" region (since flux from the PWN was extended into all 3)
and added them together. We then used the remaining regions as
background. As a check of this method, we integrated the background
subtracted spectra over the appropriate wavelengths corresponding to
the 5.8 and 8.0 $\mu$m IRAC channels, factoring in the spectral
response curves. Within errors, we obtained the same flux here as we
did using aperture photometry on the IRAC images. In
Figure~\ref{fig6-3}, we show the short-low spectrum of the PWN with
the both the original source spectrum and background spectrum
overlaid.

For the LL slits, we followed a similar procedure. Because we only had
3 parallel slit positions, we had 9 total spectra extracted from
spatially different areas. Because the LL slit is wider than SL (about
$10.5''$), we only considered the middle region of the middle slit
to be the on-source region. In keeping with our policy of only using
the closest background regions, we again excluded the 4 corner
regions, and only used the 4 regions corresponding to the 2 middle
regions of the parallel slit pointings, and the 2 extractions from the
ends of the middle slit. We then averaged the 4 background spectra and
subtracted the result from the on source spectrum to get a background
subtracted spectrum. Again, as a check on this method, we integrated
the resulting spectrum over the appropriate wavelengths, and with the
appropriate spectral response curves, calculated a 24 $\mu$m flux
that could be compared with that derived from aperture photometry on
the MIPS image. Within errors, there was excellent agreement between
these two methods. Figure~\ref{fig6-4} shows the long-low spectrum of
the PWN.

\begin{figure}
\figurenum{6.2}
\plotone{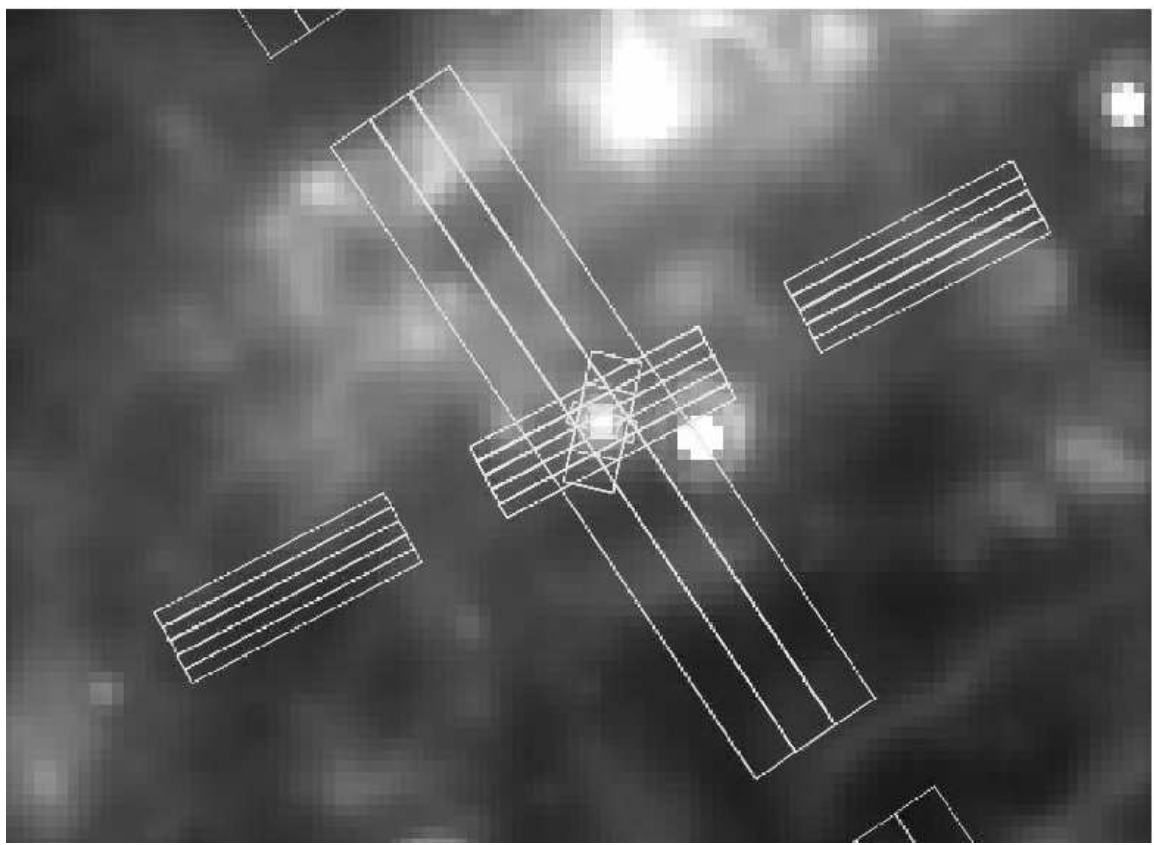}
\caption{Coverage of IRS slits overlaid on MIPS 24 $\mu$m image.}
\label{fig6-2}

\end{figure}

\begin{figure}
\figurenum{6.3}
\plotone{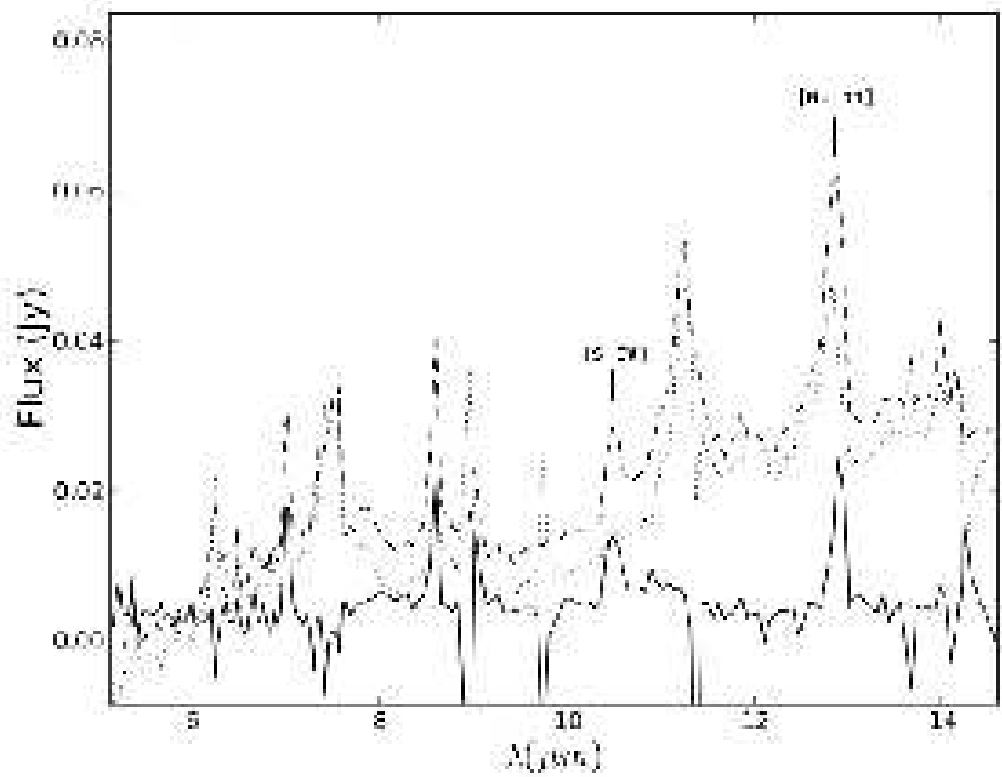}
\caption{The short-wavelength, low-resolution spectrum of the
PWN. Local background has been subtracted as described in the
text. Dashed line is source $+$ background; dotted line is background;
solid line is the spectrum of the source only.}
\label{fig6-3}
\end{figure}

\begin{figure}
\figurenum{6.4}
\plotone{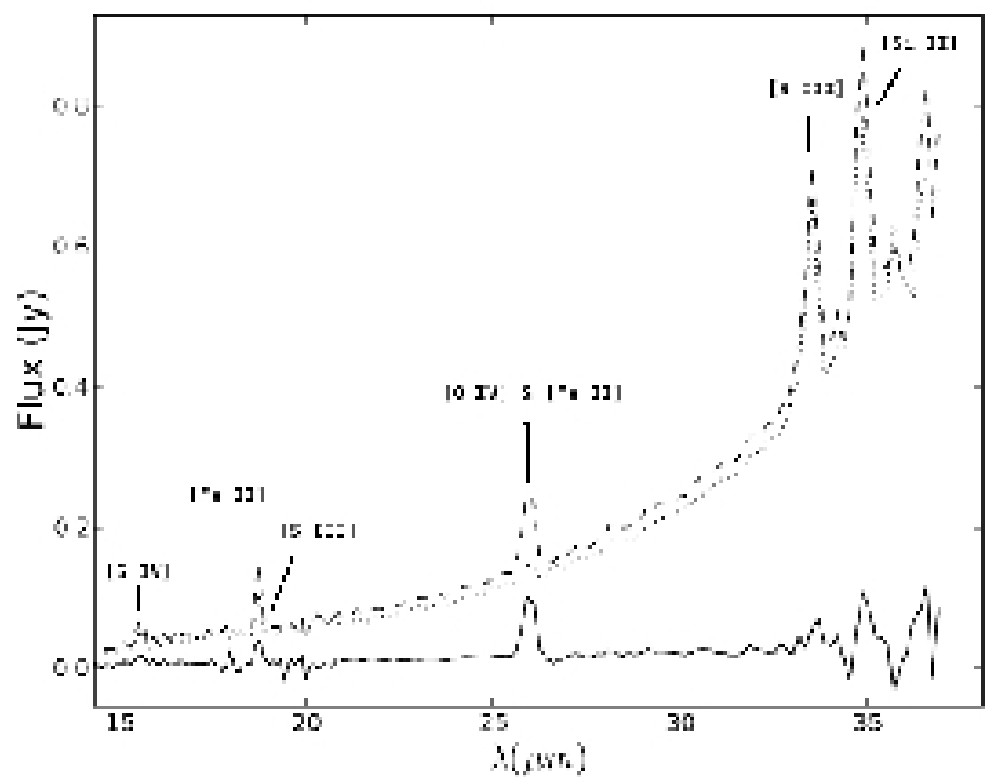}
\caption{The long-wavelength, low-resolution spectrum of the
PWN. Lines are the same as in Figure 6.3.}
\label{fig6-4}
\end{figure}

In order to examine the shape of the synchrotron continuum from the
low-resolution data, it was necessary to remove the lines from the
spectra, as well as artifacts produced by obviously bad
pixels. Although we detected the PWN at all wavelengths, the spatially
varying background made our background subtraction procedure somewhat
uncertain. Using different background subtraction regions does produce
different results for the final spectrum, mainly due to the steep
north-south gradient in the infrared background in the region of 0540
(see Figure~\ref{fig6-2}). We believe our approach of defining an
``annulus'' region and averaging the backgrounds around the source is
the best solution to this problem. However, it is not without
significant uncertainties. We tried several variations of different
background regions to see what the effects were. The largest
differences came in comparing the two same-slit background positions
with the two parallel slit background positions. We found variations
in the absolute flux level between these two choices to be on the
order of 40\%. Because we have no reason to favor one over the other,
we averaged them together with equal weights, thus creating our
annulus. Because of this, we have used caution in interpreting the
results of the extraction of the continuum. We also considered the
possibility that a large number of weak lines could be interpreted as
continuum. We reject this hypothesis for two reasons. First, there are
only a handful of IR lines predicted in this wavelength range, and
with the exception of [Ar III] at 8.99 $\mu$m, none of them even come
close to the detection limit based on our line models. Second, we have
high-resolution spectra of this region, and there is no evidence of
lines that would be unresolved in the low-resolution data.

\subsubsection{Line Fitting}

\begin{figure}
\figurenum{6.5}
\plotone{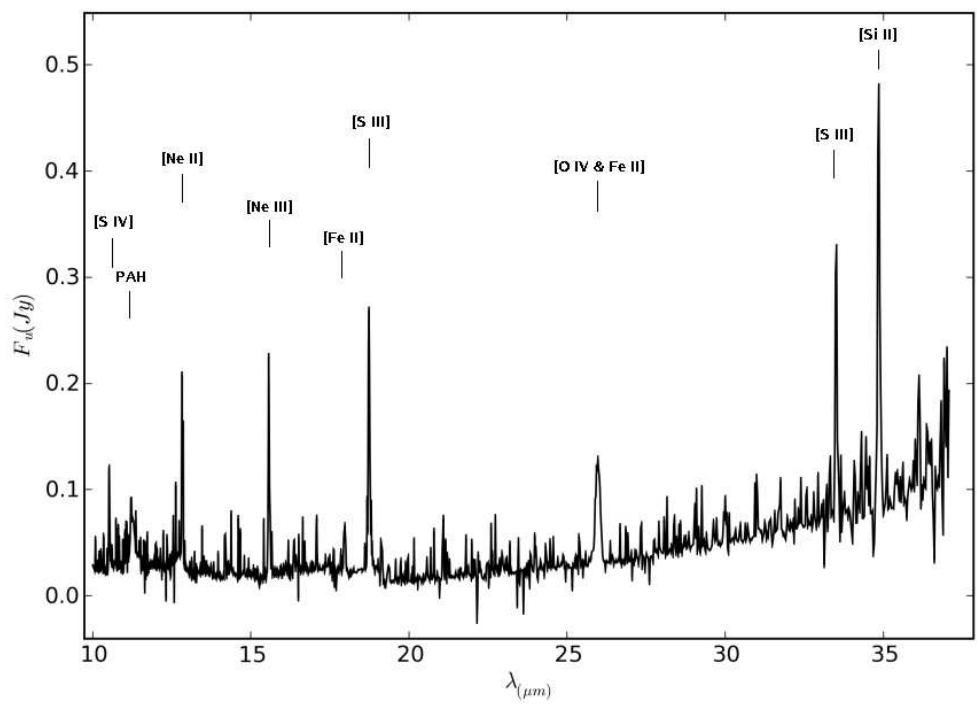}
\caption{The high-resolution spectrum of the PWN, with no background
subtraction. Measured lines are marked, along with a dust feature at
${\sim 11} \mu$m.}
\label{fig6-5}
\end{figure}

We used SPICE to extract spectra from the high-resolution data as
well. In order to fit the lines, we used the open-source software
Peak-O-Mat, which runs on SciPy (Scientific Python) and is available
from http://lorentz.sourceforge.net/. Peak-O-Mat is an interactive
program that is designed to fit curves using a least-squares algorithm
to a user-specified function. Because our extraction region contains
not only the entire expanding shell of the PWN, but also the
foreground and background emission from the surrounding ISM, we
expected to see both broad and narrow components for most of the lines
detected, as has been seen in optical spectroscopy of the nebula. We
assumed Gaussian profiles for both the broad and narrow components,
and fit these on top of a linear background. We manually removed
artifacts that were clearly caused by bad pixels, as determined by
examining the 2-D dispersed image. We also clipped bad pixels from the
backgrounds in the vicinity of each line, in order to make the fitting
of the actual lines easier with a longer tail for the Gaussian. We did
not remove or alter any of the pixels that were contained in the line
itself, except in the case of the [Ne II] line at 12.8 $\mu$m. There
was an obvious bad pixel that was contaminating the line structure at
around 12.86 $\mu$m. In order to correct for this, we interpolated the
strength of that pixel based on the strengths of neighboring
wavelength pixels.  Line profiles and strengths are discussed in
section 6.4. The complete high-resolution spectrum of the PWN
is shown in Figure~\ref{fig6-5}.

\begin{figure}
\figurenum{6.6}
\plotone{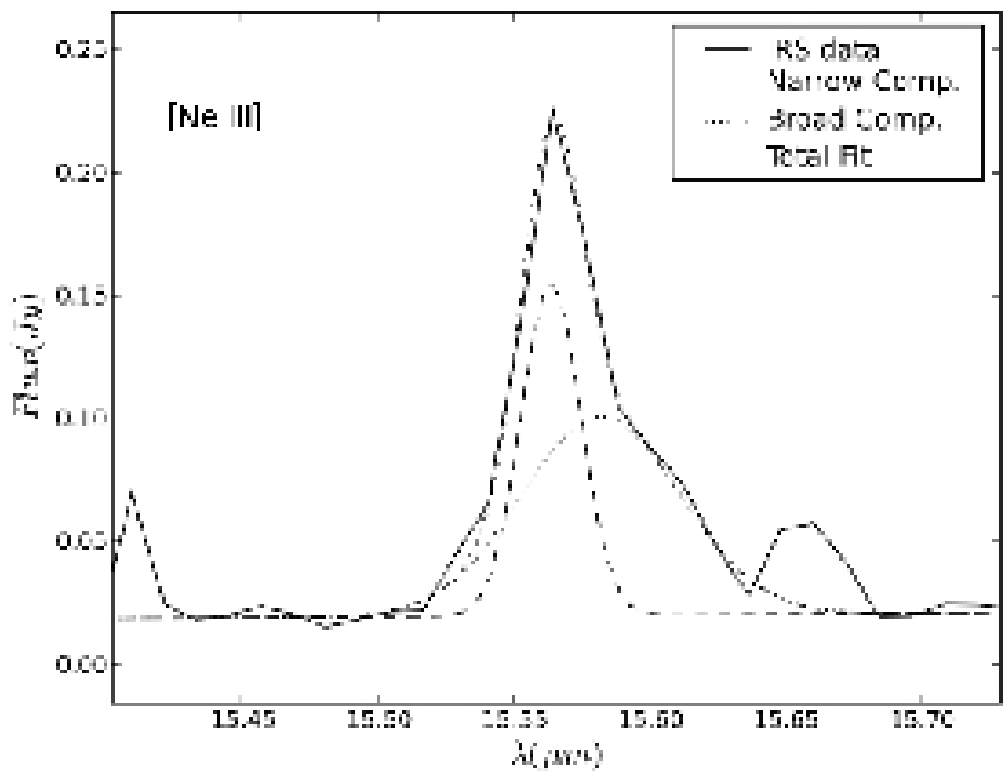}
\caption{An example of our two-component fit to the lines identified
in the high-resolution spectrum of the PWN. [Ne III] is clearly seen
to have two components. Noisy pixels were clipped out for the fitting,
but were left in this image to show their relative level of
contribution.}
\label{fig6-6}
\end{figure}

We find that nearly all of the lines in the spectrum have a
two-component nature, with a narrow component we attribute to the
surrounding H II region, and a broad component coming from the
PWN. Figure~\ref{fig6-6} shows an example of a two-component fit to
a line, in this case [Ne III], at 15.5 $\mu$m. The spectral resolution
of both SH and LH is $\lambda/\Delta\lambda$ $\sim 600$, which
corresponds to a minimum FWHM of 500 km s$^{-1}$. Since we do not expect the
narrow component widths to be wider than this, we fixed the narrow
component widths to this value. Furthermore, the LMC has an overall
recession velocity relative to the Sun of +270 km s$^{-1}$, so all narrow
components should be redshifted by this amount. However, when we fixed
the centroid of the Gaussian for the narrow component to this
velocity, the fits were unacceptably poor. According to the { \it
Spitzer Observer's Manual}, the wavelength calibration in IRS is 1/5
of a resolution element, which for the high-resolution module
corresponds to 0.003-0.011 $\mu$m, or 100 km s$^{-1}$. Since we found that
all the narrow components seem to be off by a comparable systematic
shift, we believe that the uncertainties in wavelength calibration are
responsible. Thus, we measured the shift for each narrow component and
averaged them to obtain a value to which we would fix each narrow
component. We considered SH and LH separately, and calculated that
each narrow component was redshifted on average (relative to its rest
wavelength) 171 km s$^{-1}$ for SH and 230.5 km s$^{-1}$ for LH. Fixing the
centroids of the narrow components to these values returned much more
acceptable fits.

After we used Peak-O-Mat to determine the best values for the
parameters of either one or two Gaussians, we then used our own
least-squares algorithm to obtain errors. The errors listed on the
parameters in Table 6.2 are 90\% confidence limits,
corresponding to a rise in $\chi^{2}$ of 2.706 from its minimum
value. This procedure was repeated for each parameter
separately. Errors on line fluxes were obtained through the standard
error propagation formula.

\subsection{Discussion}

\subsubsection{General Picture}

We aim at a self-consistent, semi-quantitative picture of the PWN that
accounts for the presence of lines (optical and IR), the extent of the
synchrotron nebula, and the source of the [O III] emission at $8''$
radius. We find it useful to first point out some contrasts between
0540 and the most widely-known object of its class, the Crab
Nebula. Although 0540 has been referred to as ``The Crab's Twin,'' the
two differ in some important ways. The most obvious difference is the
lack of an outer shell in the Crab, while 0540's $30''$ shell has been
seen in both radio and X-ray observations. For the purposes of this
paper, however, the important differences lie in the PWN. In the Crab,
the size of the nebula decreases with increasing frequency, so that
the radio nebula is larger than the optical, which is larger than the
X-ray, etc. In 0540, the synchrotron nebula is approximately identical
in extent throughout all wavelengths, around $5''$. The other
fundamental difference is the presence in 0540 of emission located
beyond the synchrotron nebula (the [O III] halo). There is nothing
like this seen in the Crab, where the radio synchrotron emission
extends to the outer boundary of anything known to be associated with
the nebula.

In modeling the Crab Nebula, \cite{sankrit97} considered two models,
one a pure shock model and the other a pure photoionization model to
explain the optical emission. They concluded that shocks from an
expanding shell were more likely. In the case of 0540, however, a pure
shock model cannot reproduce the [O III] extended emission. We
therefore propose an extension to their models that incorporates both
a global shock {\it and} photoionization. The specifics of our model
will be described more fully in the sections below, but our general
picture of the nebula is as follows.  It is based on the dynamical
picture of Chevalier (2005; C05).

Approximately a millenium ago, a star exploded via the core-collapse
mechanism, leaving behind a pulsar, and sending a shock wave out into
the interstellar medium. The outer boundary of this forward shock is
now about 8 pc (angular distance of about $30''$) from the pulsar, and
the reverse shock into the ejecta is somewhere between $10'' - 30''$,
having not yet reached back to the inner ejecta. The pulsar has since
formed a pulsar-wind nebula, which itself is driving a shock into the
inner edge of the surrounding ejecta, which are in free expansion. The
shock wave heats the inner ejecta and sweeps them into a thin
shell. Since the shell of material is being continuously injected with
energy from the pulsar, it is accelerating and overtaking less dense
material as it expands. The shock speed relative to upstream material,
however, reaches a maximum and then begins to drop since the
free-expansion speed of the ejecta material is also higher at larger
radii. There is no reason, however, to expect the ejecta to be
completely homogeneous. The $^{56}$Ni synthesized in the explosion
will have heated the central ejecta by radioactive decay, causing them
to expand in an ``iron-nickel bubble'' \citep{li93}, and compressing
intermediate-mass ejecta into a denser surrounding shell.

We propose that the PWN shock has reached a radius of about 1.2 pc
from the pulsar, which corresponds to a size of $\sim 5''$, the size
of the nebula as determined by X-ray observations. The layer of
shocked ejecta is geometrically thin, bounded on the inside by a
contact discontinuity separating it from the the PWN proper, which is
the shocked pulsar wind. The shock has already encountered and
propagated through the low-density iron-nickel bubble and its
surrounding shell. That shell is likely to be highly clumpy
\citep{basko94}; shocks driven into the clumps of heavy-element ejecta
will be slow. Finally, at a sub-arcsecond radius we expect the
inward-facing pulsar wind shock where the relativistic pulsar wind is
thermalized. Interior to the shock driven into the ejecta, emission in
optical/IR is both thermal and non-thermal, with the dominant
component being synchrotron continuum emission from the relativistic
electrons.  However, multiple emission lines are clearly detected from
dense clumps and filaments of thermal gas.  In addition to this, we
identify a rising continuum in the mid-infrared above the synchrotron
continuum that we interpret as a small amount of warm dust,
collisionally heated by electrons heated by the shock. Most lines seen
in optical and infrared then come from dense clumps of ejecta, where
the shock wave has slowed significantly and become highly radiative.

What remains is to explain the faint [O III] emission seen at
$8''$. We propose that this is material that is still in free
expansion, i.e. unshocked, that has been photoionized by ultraviolet
photons emitted from the shockwave. The source of photoionization is
two-fold; ultraviolet photons from the synchrotron nebula and those
produced in fast radiative shocks both contribute appreciable amounts
of ionizing radiation. We show below that to within a factor of 2,
there are enough ionizing photons produced to account for the [O III]
halo at $8''$.

\begin{figure}
\figurenum{6.7}
\plotone{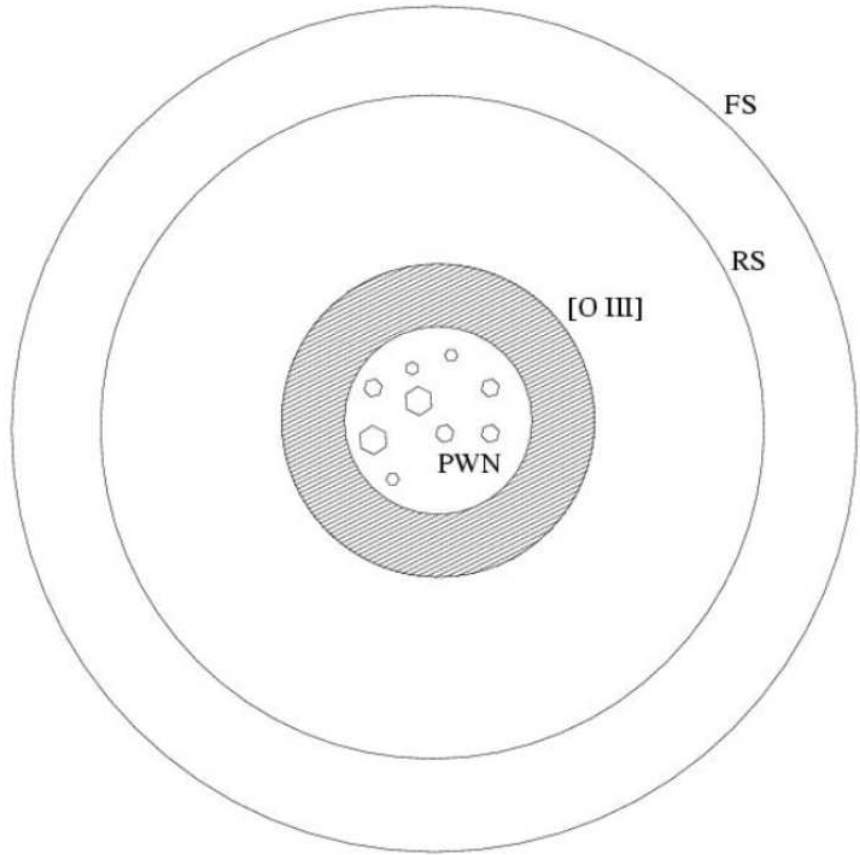}
\caption{A cartoon sketch of our general picture discussed in section
  4.1. Not to scale. FS refers to the forward shock from the SN blast
  wave, at a radius of $\sim 30''$. RS refers to the reverse shock,
  which has not yet been observed, and is at an unknown position
  between 10 and 30$''$ from the pulsar. [O III] refers to the extent
  of the halo of material that has been photoionized, and is seen in
  optical images to extend to 8$''$. PWN refers to the edge of the
  shock driven by the pulsar wind, and is located at a radius of
  5$''$. Interior to this shock, ejecta material has fragmented into
  clumps. The PWN as a whole is observed to have a redshifted velocity
  as reported in previous optical observations, possibly resulting
  from a pulsar kick. This is also the region where relativistic
  particles from the pulsar create observed synchrotron emission; see
  discussion in text.}
\label{fig6-7}
\end{figure}

We have included, in Figure~\ref{fig6-7}, a cartoon sketch of this
picture, which will be further discussed in the following sections. A
factor of a few is all we expect to be able to accomplish in modeling
the nebula, due to the large uncertainties involved.  These
uncertainties include, but are not limited to; nature of the
progenitor star (which affects the post-explosion density distribution
of the ejecta), heavy element abundances in the ejecta, degree of
clumping of the ejecta, etc. We have endeavored in the following
sections to point out places where uncertainties arise, and where
possible, to assign quantitative values to them.

\paragraph{PWN Model}

C05 discusses a model, based on a thin-shell approximation, for a
pulsar wind nebula interacting with an inner supernova ejecta density
profile. We have used this model along with our observations to
determine various quantities about 0540, including how much hot gas
should be present. Observable quantities for the pulsar include period
($P$), period derivative ($\dot P$), luminosity ($\dot E$) and for the
nebula, size ($R$) and shell velocity ($V_{sh}$). While the quantities
for the pulsar are fairly well established by previous observations,
those for the PWN are much more uncertain. Previous optical studies of
the remnant \citep{mathewson80, kirshner89}, as well as radio
observations \citep{dickel02} interpreted the PWN as a bubble of
radius $\sim 4''$, and the optical observations gave expansion
velocities less than 1400 km s$^{-1}$. Based on {\it Chandra}
observations, \citet{petre07} concluded the nebula was slightly
larger, with a radius of $\sim 5''$. We shall adopt $5''$ (1.2 pc) as
an estimate of the location of the ejecta shock.

However, \citet{morse06} detected faint [O III] emission in images
extending out to a radius of $\sim 8''$. Based on similarities to the
Crab Nebula, they interpreted this [O III] halo as being the outer
edge of the shock from the pulsar wind overtaking the slower moving
ejecta. Here we present an alternative interpretation of this [O III]
halo emission. We propose that it is undecelerated ejecta that have
been photoionized, rather than shock-ionized. The FWZI of the [O III]
emission from Morse et al. was 3300 km s$^{-1}$, which, given the
extent of 1.8 pc and our interpretation of this as undecelerated
ejecta, provides the remnant age of 1140 years. While this is somewhat
longer than the favored model of \citet{reynolds85}, it is at least
reasonable given other age estimates made by previous studies of the
object. Photoionization calculations are discussed in Appendix B.

As a first attempt to model the observations, we considered the case
of a spherically symmetric shock wave driven by the energy input from
the pulsar expanding into a medium with density profile described by
$\rho_{SN} = At^{-3}(r/t)^{-m}$. We considered different values of the
parameter $m$, as dynamical mixing between the ejecta and surrounding
medium would produce a complicated density structure. The swept-up
mass does not exceed $1 \msun$ in this model. Although this model did
a reasonably good job at producing shock speeds high enough to account
for the necessary dust grain heating rate, a spherically symmetric
model does not adequately reproduce line radiation observed in both
optical and IR. A slow shock into dense material is required to
explain these lines, and the spherical model cannot account for this,
since presence of lines requires a departure from the overall
homogeneous density profile. We present the spherically symmetric
calculations in Appendix C. A more robust model is required to explain
both the slow shocks required for lines and the faster shocks required
for dust emission. We will return to this picture at the end of the
following section, but we must first describe our line observations in
detail.

\subsubsection{Lines}

Eight emission lines are detected in the Spitzer spectrum.  They
provide constraints on the density and temperature of the emitting
gas, and perhaps more importantly on the elemental abundances.  They
complement the optical spectra published by \citet{kirshner89} (K89),
\citet{morse06} (M06) and \citet{serafimovich05} (S05). We first
summarize the implications of the optical spectra, then consider shock
wave models for the combined optical and IR emission.

Several temperature estimates are available from the optical spectra.
The [O III] line ratio I(4363)/I(5007) gives temperatures of about
24,000 K according to S05 or 34,000 K (K89).  According to the CHIANTI
database \citep{landi06}, the ratio given by K89 corresponds to 50,000
K, while that given in M06 implies 24,000 K.  K89 also find
temperatures $>$30,000 K from the [O II] I(7325)/I(3727) ratio and
$<$10,000 K from [S II] I(4072)/I(6723).  The [S II] ratio of M06
implies $T$ = 14,000 K.  Assuming a temperature of 10,000 K, S05 find
a density of 1400 -- 4300 $\rm cm^{-3}$, and at 14,000 K the range
would be 1700 -- 5000 $\rm cm^{-3}$.  The differences among the
various temperature estimates may result partly from different
reddening corrections and different slit positions, but it is clear
that the [O III] lines are formed in hotter gas than the [S II] lines.
The Spitzer data include only one pair of lines from a single ion, [Fe
II] I(17.9$\mu$)/I(26.0$\mu$), which is constrained to be larger than
1.13.  Again using CHIANTI, this requires a density above about 5000
$\rm cm^{-3}$ and a temperature above 4000 K.  However, the ratio
depends upon the deblending of the [Fe II] and [O IV] lines at
26$\mu$m, and the uncertainty may be larger than the formal value.
The density contrast between the optically emitting material and the
mean post-shock density from the global model indicates that as in the
Crab Nebula, optically emitting material is concentrated in dense
knots and/or filaments.

The next step in interpreting the spectra and constructing models is
to estimate the relative importance of photoionization and shock
heating.  In the Crab nebula, photoionization dominates, though shocks
are important for the UV lines produced at higher temperatures and for
compressing the gas to increase the optical emissivity
\citep{sankrit97}.  In the oxygen-rich SNRs, such as N132D and
1E0102-7219, shock heating dominates \citep{blair00}.  0540
shows both synchrotron emission reminiscent of the Crab and extreme
heavy element enhancement.  A pure photoionization model with strongly
enhanced abundances and the observed density gives too low a
temperature to account for the [O II], [O III] and [S II] line ratios,
while shock models cool so rapidly that they produce little [O I] or
[S II] unless they produce no [O III] at all. Therefore, it seems
likely that a model of a shock including the PWN ionizing radiation is
needed.

We have computed models with the shock model described in
\citet{blair00} illuminated by the power law continuum described by
\citet{serafimovich04}.  Briefly, the code is similar to that of
\citet{raymond79} and \citet{cox85}, but it has been modified to
describe SNR ejecta with little or no hydrogen.  The most important
difference is that the cooling rate is enormous, so that the electron
temperature is well below the ion temperature in the hotter parts of
the flow.  The model is similar to those of \citet{itoh81} and
\citet{sutherland95}.  Unlike those models, we do not include the
photoionization precursor of the shock, because the ionizing emission
from the shock is considerably weaker than the ambient synchrotron
radiation.  In comparison with the spectra of Cas A, N123D and
1E0102-7219, shock models have the problems that no single shock model
produces the observed range of ionization states, and that they tend
to predict too much emission in the O I 7774 \AA\/ recombination line
unless the cooling region is somewhat arbitrarily truncated (Itoh
1988).  However, they do predict reasonable relative intensities from
the UV to the near IR for O III and O II. Below, we attribute the
truncation to mixing with hotter, lower density gas.

We assume a 20 $\rm km~s^{-1}$ shock with a pre-shock density of 30
$\rm cm^{-3}$, which produces a density of around 5000 $\rm cm^{-3}$
where the [S II] lines are formed.  The elemental abundances are O:
Ne: Mg: Si: S: Ar: Ca: Fe = 1: 0.2: 0.1: 0.1: 0.1: 0.1: 0.1: 0.1 by
number.  H, He and N are not included in the model, because it seems
likely that the lines from these elements arise in some other gas,
like either the quasi-stationary flocculi or the outer shell of ejecta
seen as very fast knots in Cas A \citep{kirshner77,fesen01}.  The
normalization of the power law flux assumes that the shocked gas is 1
pc from the center of the PWN.

The list of caveats is long.  There is undoubtedly a range of shock
speeds and pre-shock densities.  The shocked gas is unlikely to be a
uniform mixture of the various elements, and large variations in the
composition among different clumps, as observed in Cas A, are likely.
There may well be a significant contribution from unshocked
photoionized gas for some lines \citep[e.g.,][]{blair89}, as we shall
argue below for the [O III] halo.  The shock models are plane
parallel, with the power law illumination incident from the PWN,
while the X-rays are more likely to illuminate the shocked gas from
behind.  The models terminate somewhat arbitrarily at 250 K because of
numerical limitations.  This will affect the IR lines and the O I
recombination line at 7774 \AA .  Also, as a compromise between energy
resolution and energy range of the ionizing radiation, the power law
only extends to 2 keV.  This means that the inner shell ionization and
Auger ionization of S and Fe is not included.  Finally, the atomic
data in the code are somewhat out of date and need to be updated.
Nevertheless, the code gives a reasonable idea of the relative line
intensities.

To compare this model with our observed IR spectra, we must place the
Spitzer spectrum on the same scale as the optical spectra.  We
normalize the IR lines to [O III] 5007 = 100 by dividing the Spitzer
intensities by 4 times the [O III] 5007 intensity given by M06.  The
factor of 4 is meant to account for the fact that the $2''$ slit used
by M06 covers only about 1/4 of the remnant.  This is obviously not a
very accurate correction, but it is probably good to a factor of 2.
Since the relative fluxes of many optical lines differ by a factor of
2 between M06 and K89, this is unfortunately the best we can do until
an optical spectrum of the entire remnant becomes available.

The result is shown in Table 3.  Overall, the agreement is
astonishingly good for such a simple model.  Several of the low
ionization lines, [O I], [Ne II] and [Si II] are underpredicted,
though the [Si II] line could be increased simply by increasing the
silicon abundance.  The ratio of the [S IV] to [S III] IR lines is too
low, but inclusion of the harder part of the power law spectrum would
improve that.  Inclusion of the harder X-rays would also increase the
intensity of the [Fe VII] line, though a lower pre-shock density or a
higher shock speed would have the same effect. The oxygen column
density of the model is only about $10^{14}~\rm cm^{-2}$, and the
thickness of the emitting region is only $6 \times 10^{11}$ cm.  If
the thickness were large enough to allow the remaining O$^+$ to
recombine, the predicted O I recombination line, which is comfortably
lower than the weakest detected lines, would increase to about 4 times
the apparent detection limit of K89.  The agreement would improve if
the argon abundance were cut in half, but otherwise the abundances
appear to match the observations.

The shock model shown in Table 2 produces $1.1 \times 10^{-14} ~\rm
erg~cm^{-2}~s^{-1}$ in the [Ne III] 15.5 micron emission line, so the
flux shown in Table 1 would come from a region with surface area of $2
\times 10^{38}$ cm$^{2}$. This area is roughly equal to the area of a
sphere of $5''$ radius but the emission could come from many smaller
volumes with a total filling fraction of a few percent. Heavy-element
ejecta enter these slow shocks at a rate of $\sim 0.01 \msun$
yr$^{-1}$.

We conclude that, as in the Crab \citep{sankrit97}, the observations
can be explained by shocks that heat and compress the gas in the
radiation field of the PWN.  The shock heating seems needed to reach
the high temperatures seen in some line ratios and to provide the high
densities observed, while the photoionization heating strengthens the
low and moderate ionization lines.  Oxygen is about ten times as
abundant as the other elements.  The shock speed and pre-shock density
are not very well constrained, but a shock as fast as 80 $\rm
km~s^{-1}$ requires a low pre-shock density to match the observed
density, and that in turn implies a very high pre-shock ionization
state, overly strong [O IV] and overly weak [O II] emission.

In the last 100 years alone, about 1 $\msun$ of heavy-element ejecta
have been shocked, more than the total mass of the swept-up ejecta in
the global model described in Appendix B. This casts some doubt on the
validity of the global, spherically symmetric model, where density
within freely expanding ejecta was assumed to be a smooth
power-law. It is possible that the innermost ejecta have been swept-up
by an iron-nickel bubble, as inferred for SN 1987A by \cite{li93} and
modeled by \cite{basko94} and \cite{wang05} (see also brief
discussion in C05). We explored the possibility that the global shock
could be contained within the shell swept-up by the iron-nickel
bubble. In this one-dimensional picture, the shock passed through the
inner, low-density region in $\sim 50$ years, and has since been
contained within the high-density ($n \sim 30$ cm$^{-3}$) bubble
wall. We varied parameters of the model until the shock speed in the
bubble was approximately 20 km s$^{-1}$, as required by line
models. We find, however, that the mass flux of material entering the
shock throughout the remnant's entire lifetime has been unreasonably
high for this model, approximately $0.01 (t/1140$ yr$)^{-1/2} \msun$
yr$^{-1}$. In addition, a 20 km s$^{-1}$ shock, even at such density,
would not adequately heat dust grains to temperatures observed. Dust
heating is discussed in more detail in section 6.4.3.3.

\begin{figure}
\figurenum{6.8}
\plotone{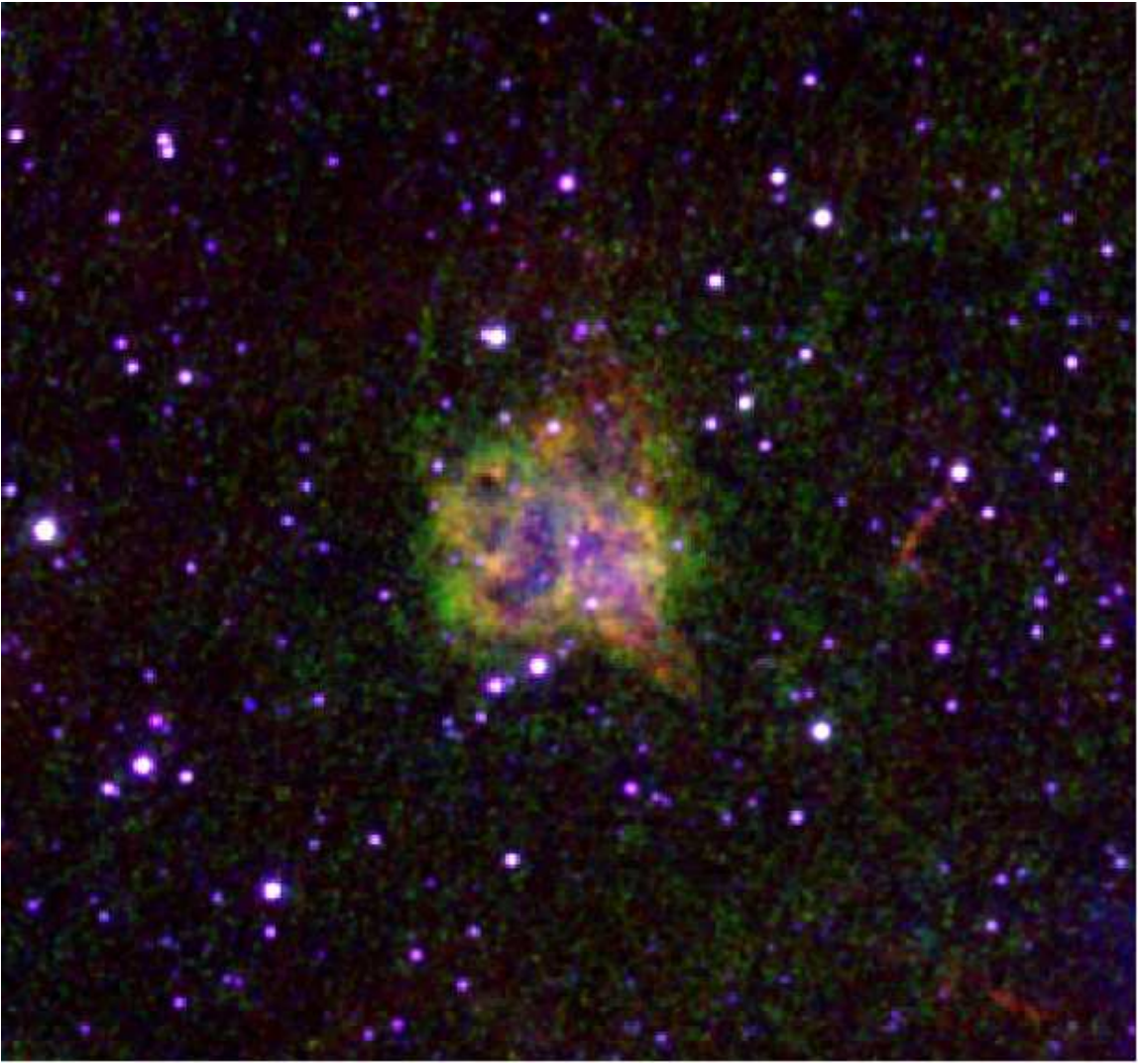}
\caption{{\it Hubble Space Telescope} WFPC2 image of PWN 0540-69.3, from
\cite{morse06}. Colors are: Blue - F791W continuum; Green - F502N [O
III]; Red - F673N [S II]}
\label{fig6-8}
\end{figure}

We are forced to consider inhomogeneous ejecta with a fast global
shock to heat dust to observed temperatures, and slower shocks
producing observed line emission. We propose the following picture:
The shock swept through the low-density iron-nickel bubble interior
early in the life of the SNR. It then encountered the dense, clumpy
shell of the bubble, slowing down and further fragmenting the shell
into dense clumps, which are still being overrun by slow shocks,
currently 20 km s$^{-1}$. The global shock has now exited the
iron-nickel bubble shell, and is propagating through the ejecta with
relatively low ambient density. The speed of this shock is not well
known, but 250 km s$^{-1}$ would be sufficient to heat the dust to the
observed temperature of around 50 K (see below). Assuming pressure
equilibrium between the dense clumps and the ambient ejecta, we derive
a density contrast, given the difference in shock velocities, of $\sim
150$. Support for this model can be inferred from {\it HST} images of
the nebula, as seen in Figure~\ref{fig6-8}, which shows [O III]
filaments in the interior, not just in a shell. [O III] line profiles
(M06) also do not match the shape that would be expected from a
spherically symmetric expanding shell, i.e. a flat top. The slow
shocks driven into the dense clumps are in rough pressure equilibrium
with the fast shock driven into the less dense ejecta.

\paragraph{Progenitor Mass}

We have compared the abundances of heavy elements listed in
section 6.4.2 with the predicted abundances of \cite{woosley95},
who consider abundance yields from core-collapse SNe ranging in mass
from 11-40 $M_\odot$ and metallicities between zero and solar. We
consider models with both solar and 0.1 solar metallicity, as this
range is most likely to reflect a massive star in the LMC. Our
abundances listed are somewhat uncertain, and result from fits to
optical and infrared line strengths. We considered the ratios of O to
Ne, Mg, Si, and Fe. The data do not single out a particular model from
\cite{woosley95}, but ratios of heavy elements to oxygen do favor a
low-to-medium-mass progenitor. High-mass progenitors ($\gapprox$ 30
$M_\odot$) are less favored, since they produce larger amounts of
oxygen relative to other elements. This interpretation is consistent
with that of \cite{chevalier06}, who favored a type IIP explosion for
this object based on observations of hydrogen in the spectrum. This is
also consistent with the idea that type IIP SNe should result from the
explosion of a single star of 8-25 $M_\odot$ \citep{woosley02}.

It is possible to quantify these results even further. If we assume a
constant heavy-element mass flux through the radiative shocks for
$10^{3}$ yr of 0.01 $M_\odot$ yr$^{-1}$ with our abundances listed
above, we get a total ejected mass in oxygen of $\sim 3.5 M_\odot$,
though this number should only be considered accurate to a factor of a
few, and is likely an upper limit. When compared with predictions from
models, this value favors stars in the range of 20-25
$M_\odot$. \cite{maeder92} gives slightly different abundance yields
for SNe, with lower overall oxygen abundances produced. In his model,
high-mass stars ($\gapprox 25 M_\odot$) actually produce less oxygen
than their lower-mass counterparts, due to mass-loss of outer layers
and inability to synthesize O from He and C. However, these massive
stars would be Wolf-Rayet stars, and can be ruled out based on the
detection of hydrogen in optical spectra.

\subsubsection{Dust}

\begin{figure}
\figurenum{6.9}
\plotone{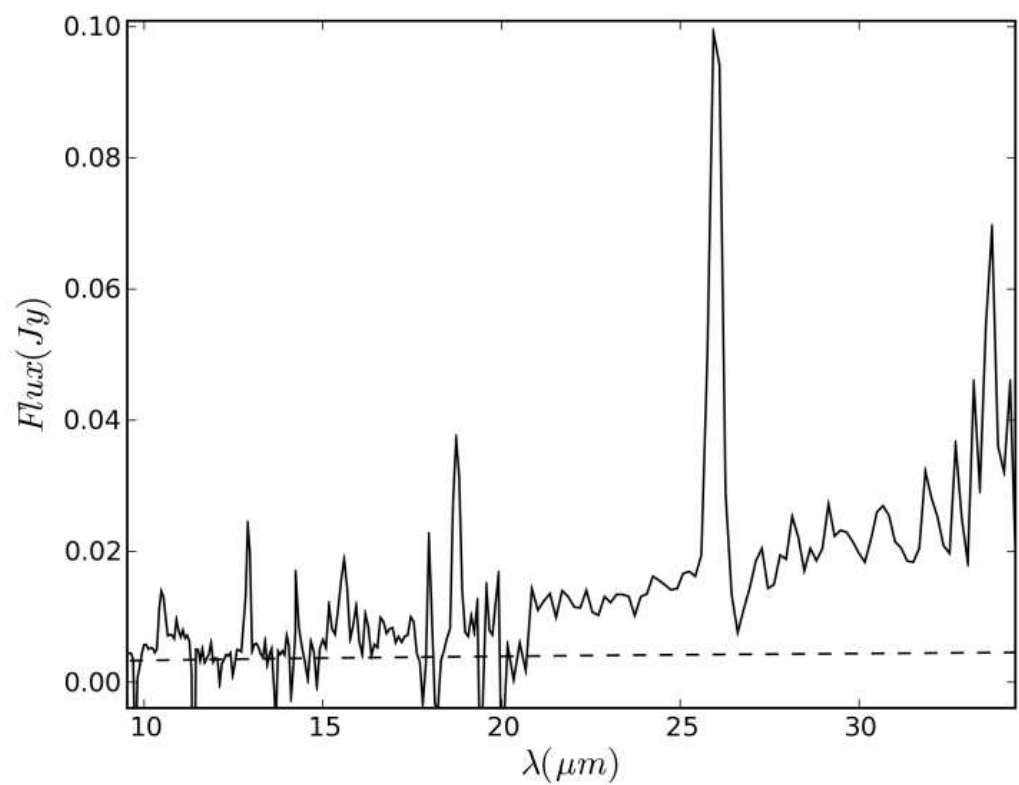}
\caption{The background-subtracted low-resolution spectrum of the PWN
is plotted as the solid line, with the radio synchrotron component shown as
a dashed line. A clear rising excess can be seen longward of 20 $\mu$m.}
\label{fig6-9}
\end{figure}

One of the more obvious features of the continuum in 0540 as seen in
Figure~\ref{fig6-9} is the excess of emission above the extrapolated
radio synchrotron spectrum at longer wavelengths. A similar excess has
been observed in the Crab \citep{temim06}, and has been attributed to
warm dust. We have inferred the temperature and the amount of dust
present, and have examined several possible mechanisms for grain
heating.

In order to fit the long-wavelength excess above the continuum, it was
necessary to remove contributions from emission lines and the
underlying synchrotron continuum. The flux contributed by the lines is
negligible, but their presence makes fitting of a model dust spectrum
more difficult. We have thus clipped obvious emission lines and bad
pixels out of the spectrum for this analysis.

\paragraph{Synchrotron Component}

\begin{figure}
\figurenum{6.10}
\plotone{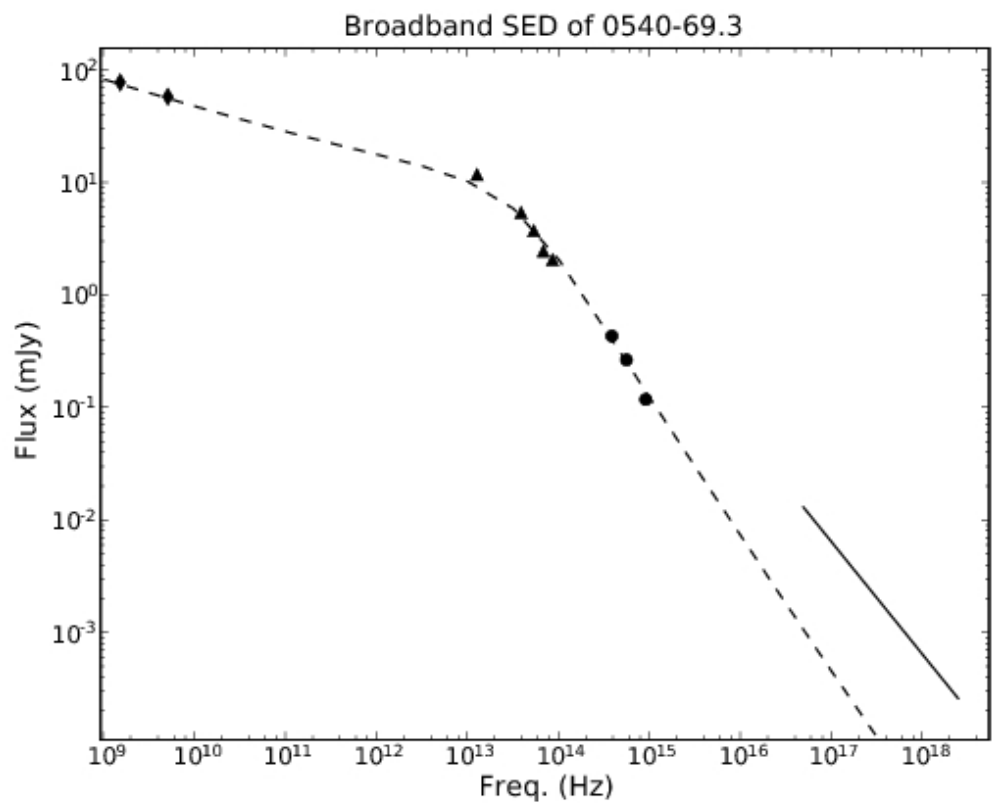}
\caption{Broadband spectrum of 0540.  Radio points (diamonds):
Manchester et al.~(1993).  IR points (triangles): our MIPS and IRAC fluxes.
Optical points (circles): Serafimovich et al.  ~(2004).  X-rays (solid line):
{\it Chandra} (Kaaret et al.~2001).  Dashed line: model described in
text.}
\label{fig6-10}
\end{figure}

In order to subtract the synchrotron component, it was necessary to
produce a model synchrotron spectrum that includes the break in
power-law indices from optical to radio. The synchrotron model used
here is one of a class of simple outflow models in which various quantities
are allowed to have power-law dependencies on radius: flow-tube width,
flow velocity, gas density (where mass loading might allow a range of
possibilities), and magnetic-field strength (Reynolds, in
preparation).  Such models can produce synchrotron-loss steepening in
spectral index both steeper and flatter than the homogeneous-source
value 0.5 \citep{reynolds06}.  Here the model, used for illustrative
purposes, invokes a simple outflow geometry with conical flow tubes
(width $w \propto r$), mass increasing as radius (due presumably to
mass loading), flow velocity decreasing as $r^{-2}$ roughly, and
magnetic field as $r^{-1}$.  The initial magnetic field at the
injection radius is $B_0 = 2.5 \times 10^{-4}$ G.  This model predicts
a decrease in size with frequency as $\theta \propto \nu^{-0.34}$,
which might be slow enough to be consistent with observations,
especially as it might take place along the line of sight.  While this
is not meant as a definitive model for 0540, it describes the data
well as shown on Figure~\ref{fig6-10} and was used to estimate the
synchrotron contribution. 

The result of radiative losses on electrons above the break energy in
a flat ($N(E) \propto E^{-s}$ with $s < 2$) energy distribution is for
such electrons to move to just below the break energy, where they can
produce a perceptible ``bump''.  However, the ``bump'' is almost
undetectable unless $s$ is very close to 0; for the 0540 value $s =
1.5$, there is almost no departure from the power-law below the break
frequency.  The model in Figure~\ref{fig6-10} was calculated
including the redistribution of electron energies, and it can be seen
that the excess we observe below 24 $\mu$m cannot be attributed to
this cause.

\paragraph{Fitting the Dust Component}

This left us with a residual rising continuum that we then fit with a
model dust spectrum. Since we presume that the dust present in 0540
would be newly formed ejecta dust, as seen in SN 1987A
\citep{ercolano07}, we have little {\it a priori} knowledge about the
grain-size distribution. However, since the wavelength of IR radiation
is much larger than typical ISM grain sizes, we adopt a model with a
single grain size, arbitrarily chosen to be $a=0.05$ $\mu$m in
radius. In any case, in the limit of $a$ $\ll \lambda$, the results
are independent of the choice of grain radius. We also do not know the
grain composition, as general results from the LMC should not apply to
ejecta dust. We thus consider two models; a graphite dust model and
the ``astronomical silicates'' model of \citet{draine84}. We calculate
the dust grain absorption cross section for both as a function of
wavelength. We then fit a simple modified blackbody model
(incorporating the grain absorption cross-section) to the data using a
least-squares algorithm designed for this model. We obtain a dust mass
of $\sim 3 \times 10^{-3}$ $\msun$ at a temperature of $50 \pm 8$ K
for silicate dust, while the resulting fit to the temperature with
graphite grains was slightly higher, $\sim 65$ K, and the required
dust mass was lower, $\sim 1 \times 10^{-3}$. 

The errors on the dust temperature are estimates based on using
different methods of removing lines and subtracting the background and
the underlying synchrotron spectrum. The resultant dust spectrum is
sensitive to these details. The dust mass should be considered
uncertain, and is probably only accurate to within a factor of a few,
as evidenced by the difference between derived masses for graphite and
silicate grains. Our data do not allow us to distinguish between
various dust compositions. It should also be noted that we are only
sensitive to dust that has been warmed by the shock wave from the
pulsar wind, and that there could be more dust that has not yet been
shocked, and is still too cool to be detected. Thus, our mass estimate
should be considered a lower limit.

\paragraph{Grain Heating Mechanisms}
\label{grains}

We now turn our attention to heating mechanisms for this dust. We
first consider heating by the synchrotron radiation field from the
PWN. Since the spectrum of the synchrotron radiation is known in the
optical/UV portion of the spectrum and grain absorption cross-sections
can be calculated as a function of wavelength, it is possible to
estimate whether there is enough radiation to heat the dust to
temperatures observed. We calculate the optical depth of the dust
around the PWN, and integrate over all wavelengths from radio up 1
keV. Although the flux from the PWN is higher at longer wavelengths,
most of the absorption occurs in the optical/UV portion of the
spectrum, due to the steeply rising absorption cross-sections in this
regime. We compare this number to the luminosity in dust derived from
our dust model, $\sim 5 \times 10^{35}$ ergs s$^{-1}$. A simple
calculation showed that the radiation available falls short by several
orders of magnitude of what is necessary.

However, this method tells us nothing about the total amount of dust
that could be present to absorb the synchrotron radiation. Thus, to
further test this hypothesis, we calculated the temperature to which
dust would be heated if it were exposed to such an ultraviolet
radiation field. We find that dust would only be heated to $\sim{20}$
K. If this were the source of the emission seen in IRS, it would
predict a 70 $\mu$m flux that is several orders of magnitude higher
than the upper limit we have placed on emission there. Given that
these order of magnitude estimates are drastically inconsistent with
our observations, we consider heating by photons from the PWN to be
ruled out.

We then considered the somewhat more exotic possibility of the
observed excess arising from a protoplanetary disk around the pulsar,
unassociated with the nebula. It has long been known that planets can
form around pulsars \citep{wolszczan92}, and the supposition has been
that these planets arise from a protoplanetary disk around the pulsar,
the source of which has been attributed to several mechnisms
\citep{bryden06}. Various surveys of known pulsars have been made in
infrared and submillimeter wavelengths, but for the most part these
surveys have only produced upper limits on the dust emission present.

However, \citet{wang06} conducted a survey of neutron stars with IRAC
and found a debris disk around the young isolated neutron star 4U
0142+61. The authors suggest that the age of the debris disk compared
to the spin-down age of the pulsar favors a supernova fallback
origin. The IRAC observations combined with K-band Keck-I observations
suggest a multi-temperature thermal model with temperatures ranging
from 700-1200 K, where the disk has inner and outer radii of 2.9 and
9.7 $\rsun$, respectively. Using the same model the authors use
\citep{vrtilek90}, we calculate the necessary radius to reproduce
observed fluxes for 0540 for a disk with temperature $\sim{50}$ K to
be on the order of $10^{4}$ $\rsun$. A survey of disks around
Anomalous X-ray Pulsars (AXPs) \citep{durant05} found several
candidates for fallback disks which consistently had IR(K-band)/X-ray
flux ratios of order $10^{-4}$. Although we were not able to find any
archival near-infrared observations of the PWN, we can make an
estimate of this ratio by looking at the overall spectrum of the IRAC
and optical points. An estimate of $5 \times 10^{-2}$ is reasonable
for this ratio in 0540, significantly different than that found in the
AXPs.  Additionally, \citet{wang07} observed 3 known AXPs with
Spitzer, and found no mid-IR counterpart to any of them. Given these
discrepancies between these cases and that of 0540, we do not believe
that a protoplanetary disk around the pulsar is the origin of the
far-IR excess.

What then is the cause? Collisional heating by hot gas heated by
shocks driven into the ejecta can provide both a qualitative and
quantitative explanation for the dust present. Grain heating
rate, $\cal{H}$, goes as

\begin{equation}
\mathcal{H} \propto n_ev_{e}T_{e} \propto PT_{e}^{1/2},
\end{equation}
where $n_{e}$, $v_{e}$, and $T_{e}$ are electron density, velocity,
and temperature, and $P$ is the pressure, $nT$. In the PWN, P
is constant throughout the bubble, so that grain heating is more
efficient in hotter gas. We find that the slow, radiative shocks are
incapable of heating dust grains to temperatures much above $\sim 25$
K. Faster shocks, and thus higher temperatures, are required to heat
grains to observed temperatures.

To determine whether this is plausible, given the conditions in the
object, it is necessary to make an estimate of the amount of gas that
is still hot, i.e. capable of heating dust grains through collisions
with electrons. The shock cooling time \citep{mckee87} is

\begin{equation}
t_{cool} = 2.5 \times 10^{10} v^{3}_{s7}/\alpha \rho_{0},
\end{equation}
where $v_{s7}$ is the shock speed in units of $10^{7}$ cm s$^{-1}$,
$\rho_{0}$ is the pre-shock density in amu cm$^{-3}$, and $\alpha \ge
1$ is a parameter for the enhancement of cooling due to higher metal
content. We find that a shock with velocity $\sim 250$ km s$^{-1}$
would effectively heat dust to 50 K, with a pre-shock density of $\sim
8$ amu cm$^{-3}$, assuming the same pressure as in slow shocks. If the
dust component is composed of graphite grains at $\sim 65$ K, a
slightly faster shock of 325 km s$^{-1}$ is required. Using the above
equation, we find that the amount of hot gas is on the order of a few
tenths of a solar mass. This yields dust-to-gas ratios of a few
percent. Given the unknown dust content within the inner ejecta of a
supernova, we believe this is a reasonable explanation.

As a check on the constraints of such a fast shock, we calculated the
expected X-ray emission from such a shock, and found it to be below
the upper limits of thermal X-ray emission observed from the PWN,
except for very metal-rich ejecta.

\subsubsection{Origin of O-rich Clumps}
\label{clumps}

\cite{matzner99} considered a spherically-symmetric explosion of a 15
$M_\odot$ RSG, and found that its He core and heavy element ejecta
formed an approximately constant density, freely expanding ejecta
core.  C05 rescaled their results to other values of ejecta mass
$M_{ej}$ and kinetic energy $E_{51}$, arriving at the core density of

\begin{equation}
\rho_ct^{3} = 2.4 \times 10^{9} (M_{ej}/15 M_\odot)^{5/2}
E_{51}^{-3/2} {\rm g\ cm^{-3}\ s^{3}}.
\end{equation}
An additional compression is expected from the iron-nickel bubble
effect. For the centrally-located Ni with mass $M_{Ni}$, the adjacent
ejecta are expected to be swept up into a shell with velocity 

\begin{equation}
V_1 = 975 (M_{Ni}/0.1 M_\odot)^{1/5} (\rho_c t^{3}/10^{9} {\rm g\
cm}^{-3}\ {\rm s}^{3})^{-1/5} {\rm km\ s^{-1}}.
\end{equation}
The compression within the Fe-Ni bubble shell is at least by a factor
of 7, expected in strong, radiation dominated shocks with $\gamma =
4/3$. The shell density increases inward from a shock front to a
contact discontinuity separating the shocked ejecta from the Fe-Ni
bubble. In one dimensional hydrodynamical simulations, Wang (2005)
finds an average shell compression by a factor of 20. The average
shell density is then 

\begin{equation}
\rho_1 t^{3} = 4.8 \times 10^{10} (M_{ej}/15 M_\odot)^{5/2}
E_{51}^{-3/2} {\rm g\ cm^{-3}\ s^3}.
\end{equation}
(Diffusion of radiation might reduce this compression by a modest
factor of $\le 2$ -- Wang 2005.) At the current remnant's age of 1140
yr, the shell density is 

\begin{equation}
\rho_1 = 1.0 \times 10^{-21} (M_{ej}/15 M_\odot)^{5/2}
E_{51}^{-3/2} {\rm g\ cm^{-3}}.
\end{equation}
Because the dense iron-nickel bubble shell has been accelerated by
low-density gas within the bubble, the shell is subjected to the
Rayleigh-Taylor instability, and we expect it to fragment into
clumps. Within a factor of 2, their expected density is equal to the
preshock density for the O-rich clumps in 0540. We conclude that these
clumps are remnants of the iron-nickel bubble shell.

\cite{matzner99} found a sharp density drop by a factor of 10
at the interface between the He core and the H envelope, with an
approximately constant density through much of the H envelope. The
envelope density $\rho_{env}$ is then 

\begin{equation}
\rho_{env} t^{3} = 2.4 \times 10^{8} (M_{ej}/15 M_\odot)^{5/2}
E_{51}^{-3/2} {\rm g\ cm^{-3}\ s^3},
\end{equation}
200 times less dense than the iron-nickel bubble shell. This density
contrast is similar to the density contrast inferred between the
O-rich clumps and the more tenuous inter-clump gas. It is likely that
the PWN nebula expands now into the H envelope. Because the dense He
core has been decelerated by the less dense H envelope during the SN
explosion, the ensuing Rayleigh-Taylor instability led to a
large-scale macroscopic mixing between them. As a result, we expect a
two-phase medium ahead of the PWN shell, consisting of more tenuous
H-rich gas and denser He-rich gas. It is possible that shocks driven
into the He-rich gas became radiative; that could explain the presence
of H and He recombination lines in optical spectra of 0540.

The dense iron-nickel bubble shell should contain not only O-rich
ejecta, substantial amounts of He-rich gas are also expected. Slow (20
km s$^{-1}$) shocks driven into the dense He-rich gas may also become
radiative; if so, they could produce strong lines of low ionization
species. This could explain excess emission seen in optical and IR
spectra for low ionization species (see discussion in \S~4.2). More
detailed shock models are necessary to determine whether or not our
picture is consistent with observations.

The identification of dense O-rich clumps in 0540 with a compressed
and fragmented shell swept up by the iron-nickel bubble has important
implications for ejecta detection in SNRs. Dense O-rich clumps are
expected to produce strong optical or X-ray emission, once shocked and
heated by the reverse shock. The optical emission should be most
prominent for remnants with a particularly dense ambient medium,
either of circumstellar (e.g., Cas A) or interstellar (N132D)
origin. The entire class of optically emitting O-rich remnants may owe
its existence to the iron-nickel bubble effect. For ejecta expanding
into less dense ambient medium, X-ray emission is expected instead
since clumps will be reverse-shocked much later when their densities
dropped significantly because of free expansion. The O-rich clumps
such as seen in 0540, even when shocked to X-ray emitting temperatures
10,000 yr after the explosion, will have substantial ($\sim 1$
cm$^{-3}$) electron densities and emission measures. Even old remnants
should show O-rich ejecta in their interiors, in agreement with the
accumulating evidence gathered by modern X-ray satellites. A good
example is a 14,000 yr old SNR 0049 $-$73.6 in the SMC, where {\it
Chandra} imaging and spectroscopy revealed the presence of a clumpy
O-rich ring in its interior \citep{hendrick05}. Hendrick et
al. interpreted this ring as the shell swept up by the iron-nickel
bubble, based on mostly theoretical arguments. Observational evidence
for the iron-nickel bubble effect in 0540 strengthens this
interpretation for 0049$-$73.6, and possibly for many more mature SNRs
with detected ejecta emission in their interiors.

Dust formation is most likely to occur where ejecta density is the
highest. The dense O-rich clumps likely contain dust; this dust may
survive the passage through the radiative shock. If it were mixed into
the much hotter ambient medium, this surviving dust may be the source
of the observed infrared emission.

\subsection{Summary}

We have observed the supernova remnant B0540-69.3 with all three
instruments aboard the {\it Spitzer Space Telescope}. We detected the
PWN in all 4 IRAC bands, as well as the 24 $\mu$m MIPS band. We did
not detect any emission from the PWN at 70 $\mu$m, though the upper
limit is rather unconstraining. We found no hint of detection at any
wavelength of the $\sim 30''$ shell surrounding the PWN, as seen in
X-rays and radio. Both the IRAC and the MIPS 24 $\mu$m photometric
fluxes are consistent with the emission being primarily dominated by
synchrotron emission, as synchrotron models extended both down from
the radio and up from optical wavelengths roughly reproduce the flux
seen in infrared. There is a change in slope of the overall
synchrotron spectrum taking place in mid-infrared wavelengths, as is
required to match the radio synchrotron power-law with the optical
power-law.

The IRS spectra in the 10-37 $\mu$m region show a clear excess of
infrared emission that cannot reasonably be attributed to any
synchrotron radiation. We conclude that this excess emission is coming
from a small amount ($\sim 1-3 \times 10^{-3} \msun$) of warm dust that
has been formed in the expanding ejecta from the SN. We consider
multiple heating mechanisms for this dust, ruling out both a fallback
disk around the neutron star and heating by the synchrotron radiation
from the PWN itself. We conclude that the dust is being heated by
shocks being driven into the ejecta by the energy input from the
pulsar. We derive a dust-to-gas mass ratio of the order of a few
percent, which is reasonable given how little is known about dust
content in the inner ejecta of SNe.

We consider the extended ($8''$ in radius) [O III] emission discovered
by Morse et al. in HST images of the nebula, and attribute this to
undecelerated ejecta that have been photoionized by photons from both
the radiative pulsar wind shocks and the synchrotron radiation from
the nebula. While there are not enough ionizing photons to do this
assuming solar abundances, we show that realistic assumptions about
the heavy element abundances in the ejecta, which are almost certainly
not solar, provide a plausible explanation of the [O III] halo.

We also detect a number of lines coming from both the ejecta in the
PWN and the background/foreground H II region. Most of the line
structures contained both a broad and a narrow component, blended
together due to the modest spectral resolution of the instrument. We
performed multi-gaussian fits to the line structures to identify both
of these components separately. The widths of the lines, as well as
their redshift from their rest wavelength, are broadly consistent with
previous line studies done in optical wavelengths. We find line widths
of order 1000-1300 km s$^{-1}$, and shifts between broad and narrow
components of lines of order 300-400 km s$^{-1}$. We model these
lines, as well as those found in optical wavelengths, and conclude
that slow ($\sim 20$ km s$^{-1}$) shocks driven into dense ($\sim 30$
cm$^{-3}$), O-rich clumps of material provide the most satisfactory
agreement with measured intensities. We find a preshock density
contrast of $\sim 100$ between the dense, optically-emitting clumps
and the rarefied gas behind the global shock, assuming rough pressure
balance between the two phases.

Our global picture of the pulsar-wind nebula consists of several
elements. An expanding, accelerating shell of material is driven into
the inner ejecta from the supernova, passing through the iron-nickel
bubble and the dense, clumpy shell, into which shocks are being driven
at 20 km s$^{-1}$. The fast, global shock has exited the bubble walls,
and has now reached a radius of about 1.2 pc. Beyond this shock, out
to a radius of 1.9 pc, material has been photoionized by UV photons
from both the shock and the synchrotron nebula, and this photoionized
material is observed in the form of an [O III] halo. This picture is
able to account for observations in the broad wavelength range from
radio to X-rays.

Future, high-resolution observations of this object in infrared
wavelengths, such as those which will be possible with the {\it James
Webb Space Telescope}, will serve to further its understanding. Just a
few of the possibilities that could be studied with such observations
are: spatial identification of the location of infrared lines, further
search for an infrared shell at $30''$ corresponding morphologically
with the X-ray shell, and identification of the spatial location of
the dust in the PWN. Further spectroscopy on the warm dust component
could potentially constrain the composition of dust formed out of
ejecta. The global shock is just one possible location for the hot gas
capable of heating dust grains, it is also possible that the shocked
and dusty heavy-element ejecta have been reheated in the turbulent and
hot PWN interior. The order of magnitude increase in the spatial
resolution of JWST can shed light on our hypothesis of the global
picture of the PWN. In addition, deep ground-based spectra of the [O
III] halo can confirm or refute the photoionization origin we have
suggested here.

\newpage
\clearpage

\begin{deluxetable}{lc}
\tablenum{6.1}
\tablecolumns{2}
\tablewidth{0pc}
\tabletypesize{\footnotesize}
\tablecaption{Measured Fluxes}
\tablehead{
\colhead{Channel} & Flux}

\startdata
IRAC Ch.1 (3.6 $\mu$m) & 1.77 $\pm${0.23}\\
IRAC Ch.2 (4.5 $\mu$m) & 2.19 $\pm${0.27}\\
IRAC Ch.3 (5.8 $\mu$m) & 3.61 $\pm${0.46}\\
IRAC Ch.4 (8.0 $\mu$m) & 5.10 $\pm${0.74}\\
MIPS Ch.1 (24 $\mu$m) & 13.19 $\pm${3.95}\\
MIPS Ch.2 (70 $\mu$m) & $<366$\\

\enddata

\tablenotetext{a}{All fluxes given in milliJanskys}

\label{fluxtable}
\end{deluxetable}

\newpage
\clearpage

\def\res#1#2#3{$#1^{+#2}_{-#3}$}
\begin{deluxetable}{lcccccccc}
\tablenum{6.2}
\tabletypesize{\scriptsize}
\tablecaption{Line Fits\label{linetable}}
\tablewidth{0pt}
\tablehead{\colhead{}  & \multicolumn{3}{c}{Narrow Component} & \multicolumn{5}{c}{ Broad Component} \\

\colhead{Line} & $\lambda$ ($\mu$m) & Flux\tablenotemark{a} &
FWHM\tablenotemark{b} ($\mu$m) & $\lambda$ &
Flux\tablenotemark{a} & FWHM\tablenotemark{b} & FWHM & Shift
}

\startdata

[S IV] (10.5105) & 10.5165 & \res{2.12}{0.53}{0.53} & 1.75 & \res{10.5261}{0.0017}{0.0015} & \res{7.32}{1.0}{1.0} & \res{3.89}{0.32}{0.21} & \res{1110}{91}{60} & \res{+274}{49}{43} \\

[Ne II] (12.8135) & 12.8208 & \res{5.86}{0.50}{0.50} & 2.14 & \res{12.8436}{0.0036}{0.0034} & \res{4.98}{0.85}{0.85} & \res{4.28}{0.61}{0.66} & \res{1000}{72}{154} & \res{+534}{84}{80} \\

[Ne III] (15.5551) & 15.5639 & \res{4.59}{0.29}{0.29} & 2.59 & \res{15.5823}{0.0018}{0.0019} & \res{7.29}{0.56}{0.56} & \res{6.86}{0.31}{0.32} & \res{1320}{62}{62} & \res{+354}{35}{37} \\

[Fe II] (17.9359) & - & - & - & \res{17.9663}{0.0025}{0.0025} & \res{3.01}{0.38}{0.38} & \res{6.84}{0.54}{0.65} & \res{1140}{90}{109} & - \\

[S III] (18.7130) & 18.7236 & \res{2.22}{0.40}{0.40} & 3.12 & \res{18.7407}{0.0012}{0.0011} & \res{10.18}{0.59}{0.59} & \res{6.07}{0.18}{0.16} & \res{972}{28}{26} & \res{+274}{19}{19} \\

[O IV] (25.8903) & - & - & - & \res{25.9454}{0.025}{0.0062} & \res{5.32}{1.7}{1.7} & \res{13.39}{3.8}{1.4} & \res{1650}{300}{180} & - \\

[Fe II] (25.9883) & - & - & - & 26.0375 & \res{1.71}{0.61}{1.7} & 10.03 & 1140 & - \\

[Si II] (34.8152) & 34.8419 & \res{5.13}{0.27}{0.27} & 5.81 & \res{34.8875}{0.0018}{0.0061} & \res{2.75}{0.31}{0.31} & \res{8.42}{0.25}{0.79} & \res{724}{22}{68} & \res{+393}{16}{53} \\

\enddata

\tablenotetext{a}{Flux in units of 10$^{-14}$ ergs cm$^{-2}$ s$^{-1}$}
\tablenotetext{b}{FWHM in units of 10$^{-2} \mu$m}

\tablecomments{Centroid position and FWHM of narrow components fixed
to values specified in the text. [Fe II] at 26 $\mu$m fixed to
redshift and FWHM of [Fe II] 17.9 $\mu$m. Col. (8): FWHM of broad line
in km s$^{-1}$. Col. (9): Shift of broad line relative to narrow line,
in km s$^{-1}$.}

\end{deluxetable}

\newpage
\clearpage

\begin{table}
\tablenum{6.3}
\begin{center}
\centerline{Table 3}

\vspace*{2mm}
\centerline{Normalized Emission Line Fluxes}

\vspace{4mm}
\begin{tabular}{| lrrrr |}

\hline \hline
Line            & M06  & K89   & Spitzer & Model \\
\hline
 O II  3727     & 46.  & 39.   & - & 52.8  \\ 
 Ne III  3869   &  7.2 & $<$1.5 & - &  9.3 \\
 S II    4072   &  3.6 &  3.   & - & 4.9  \\
 O III   4363   &  3.3 &  7.   & - & 4.2 \\
 O III   5007   & 100. & 100.  & - & 100. \\
 Fe VII  6085   &  -   &  2.:  & - & 0.02 \\
 O I     6303   &  3.3 &  5.   & - & 0.9 \\
 S II    6722   & 33.8 & 67.   & - & 36.1 \\
 Ar III  7136   &  -   &  8.   & - & 19.2 \\
 Ca II   7291   &  -   &$<$2.  & - & 0.6 \\
 O II    7325   &  -   &  6.   & - & 3.6 \\
 Fe II   8617   &  -   &  2.   & - & 5.7 \\
 S III   9532   &  -   &  34.  & - & 30.0 \\
 O I     7774   &  -   &  -    & - & 0.01 \\
 S IV    10$\mu$m&  -   &  -    & 26. & 7.4 \\
 Ne II   12$\mu$m&  -   &  -    & 31. & 3.2 \\
 Ne III  15$\mu$m&  -   &  -    & 33. & 29.0 \\
 Fe II   17$\mu$m&  -   &  -    &  8.4&  7.6 \\
 S III   18$\mu$m&  -   &  -    & 35. & 20.5 \\
 O IV    26$\mu$m&  -   &  -    & 15. & 11.6 \\
 Fe II   26$\mu$m&  -   &  -    &  4.8& 12.8 \\
 Si II   35$\mu$m&  -   &  -    & 22. & 7.9 \\
\hline

\end{tabular}
\label{linemodel}
\end{center}

\end{table}

\newpage
\clearpage

\newpage

\newpage

\newpage

\newpage

\newpage

\newpage

\newpage

\newpage

\newpage

\newpage
\clearpage

\section{Further Advances in Grain Modeling}

Portions of this chapter regarding work on the Cygnus Loop are from
Sankrit, R., Williams, B.J., Borkowski, K.J., Gaetz, T.J., Raymond,
J.C., Blair, W.P., Ghavamian, P., Long, K.S., \& Reynolds, S.P. 2010,
ApJ, 712, 1092.

\subsection{Introduction}

In this chapter, I discuss work I have done in advancing the
theoretical models of collisionally heated grain emission. In order to
account for dust observed in SNRs, models must include the proper
physics for heating, cooling, and sputtering of grains. The following
is a list of what I consider to be the primary issues in modeling
grains: \\ \\ - Grain porosity/compositeness\\ - Energy deposition
rates for ions/electrons\\ - Ion heating of grains in fast shocks ($>$
2000 km s$^{-1}$)\\ - Non-thermal sputtering of grains due to relative
gas-grain motions\\ - Sputtering rates for small grains\\ - Liberation
of elements into gaseous phase\\

Certainly, most of these have been addressed in some form or another
by various authors, but what the field still lacks, and what my work
here will address, is a self-consistent model that takes all of these
issues into account.

\subsection{Porous Grains}

Typical models of ISM grains assume solid particles composed of a
uniform material. This is probably unrealistic, as real grains almost
certainly show some degree of porosity (i.e. part of their volume is
taken up by vacuum) \citep{okamoto94}. Porosity, $\cal P,$ is defined
as the fraction of the volume of the grain that is occupied by
vacuum. It is also likely that grain formation mechanisms allow for
composite grains, i.e. grains made up of multiple types of materials
\citep{shen08}. Grains that are porous, composite, or both have
properties that are different from their non-porous, or ``compact''
counterparts. Their optical constants are different, and must be
approximated with either an effective medium theory (EMT), where
different optical constants are added together by various means (see
Bohren \& Huffman, 1983, for a review of various EMTs), or something
like the multilayer sphere (MLS) approach developed by
\citet{voshchinnikov99}, or the discrete dipole approximation
developed by \citet{draine94}. I have found an EMT to be the most
straightforward way to approximate the dielectric constants for
arbitrary materials. In particular, I follow the approach laid out by
\citet{bruggeman35}, where the average dielectric function is given by
solving the following equation for $\epsilon_{av}$:

\begin{equation}
\sum_{i=1}^n f_{i}
\frac{\epsilon_{i}-\epsilon_{av}}{\epsilon_{i}+2\epsilon_{av}} = 0,
\end{equation}

\noindent
where $n$ is the number of constituent materials (where vacuum is
considered a material), $f$ is the relative filling fraction of each
material (by volume), and $\epsilon$ is the complex dielectric
function of each material. 

I have developed code that is capable of calculating the optical
constants and absorption coefficients for grains of any arbitrary
combination of astronomical silicates \citep{draine84}, graphite,
amorphous carbon, and vacuum. Because the graphite crystalline
structure is composed of layered sheets of graphene, the components of
the dielectric constant for the electric field parallel,
$\epsilon_{\parallel}$, and perpendicular, $\epsilon_{\perp}$ to the
graphene axis will be different. To account for this, I adopt the
``$\frac{1}{3} - \frac{2}{3}$'' approximation of \citet{draine93},
where the effective optical constant for graphite comes $\frac{2}{3}$
from the perpendicular component and $\frac{1}{3}$ from the parallel.

Heating of grains depends on the enthalpies of the particles that
compose the grain. For composite grains, we approximate the total
enthalpy of the grain to be a weighted average based on the number of
particles of a given material relative to the total number of
particles in the grain. The enthalpy per atom is a function of
temperature, and I use the same enthalpy for both forms of carbon
(graphite and amorphous).

Perhaps the most significant effect of introducing porosity to grains
is that the grain volume per unit mass increases, since the overall
density of the grain is lowered; see Section 7.2.2.

\subsubsection{Size Distribution of Porous Grains}

The size distribution of porous grains is unknown, because most
interstellar extinction studies have been interpreted with compact
grain models \citep{weingartnerdraine01}. There are only a few porous
grain models available in the literature, and there are often
significant discrepancies between them. For this work, I show results
for two models. The primary model is a slightly modified version of
that from \citet{clayton03}, shown in Figure 7.1. This model consists
of grains in a distribution from 0.0025 to 1.5 $\mu$m consisting of
50\% vacuum, 33.5\% astronomical silicate (MgFeSiO$_{4}$, whereas the
original Clayton et al. (2003) model uses 28.5\% pyroxene
([Fe$_{0.6}$Mg$_{0.4}$]SiO$_{3}$]) and 5\% oxide
([Fe$_{0.4}$Mg$_{0.6}$]O)), and 16.5\% amorphous carbon. I also show
results in Chapter 8 from a second model, taken from \citet{mathis96},
which contains 15\% of the mass in grains contained in a small
population ($<$ 0.01 $\mu$m) of compact silicate grains. For the other
85\%, I use a composite grain consisting of 50\% vacuum and 50\%
solid material, where the solid material is made up of silicate and
graphite in the approximate fractions in which they are found in the
ISM of the Milky Way or LMC, as appropriate. The size distribution of
porous grains in this model is shown in Figure 7.2. 50\% vacuum is
chosen for two reasons: 1) for ease of comparison to the Clayton model
above; and 2) it likely represents the extreme \citep{heng09} of
interstellar grain porosities (although \citet{voshchinnikov06}
reports that grains may be as high as 90\% porous, using an MLS
approach). The parameter space one can explore here is vast, and
composite grains need not even be limited to the astronomical
silicates and carbonaceous grains I have listed above. Whatever the
distribution, compositeness, and porosity of interstellar grains is, I
have written the code in such a way that any changes can be easily
incorporated.

\subsubsection{Collisional Heating and Sputtering of Porous Grains}

In addition to having different optical properties, porous grains are
heated and sputtered differently than their solid counterparts. As a
simple example, consider a given mass $M$ of dust grains in two
populations: one population of solid (compact) silicate grains, all of
radius $a = 0.05\ \mu$m, and another of 50\% porous silicate grains
(with the other 50\% vacuum), also all of radius $a = 0.05\
\mu$m. Because half of the volume of porous grains is occupied by
vacuum, the mass per grain is half that of the compact grains. In
order to contain the same amount of total mass, $M$, there must be in
total twice as many porous grains as compact. Now consider how these
grains would be heated by a plasma. For this example, I choose a
plasma of number density $n_{H}$ = 2.0, $T_{e}$ = 1 keV, and a total
dust mass of $10^{-3}\ \msun$. I ignore heating by protons and
consider only heating of grains by electrons (for all but the youngest
SNRs, this is a reasonable approximation, see Figure 7.3). Sputtering
is neglected.

Solid grains are heated to a higher temperature when subjected to the
same plasma conditions. In a numerical model of the above example,
0.05 $\mu$m solid grains are heated to a temperature, $T_{dust}$, of
$\sim 62.3$ K and have a total IR luminosity of $8.85 \times 10^{35}\
\lsun$, while porous grains are heated to $\sim 57.3$ K and have a
total IR luminosity of $1.15 \times 10^{36}\ \lsun$, for a model
luminosity ratio $L_{porous}/L_{compact} = 1.3$.

As an alternative example, I modify the porous grain population above
to now contain the same {\it mass} per grain as the solid grains. For
a porosity of 50\%, this means that their radius must be larger by a
factor of $(1 - {\cal P})^{-1/3} = 1.26,$ so the porous grains now have
radius $a = 0.63\ \mu$m. In this example, the number, $N_{gr}$, of
radiating grains is identical. Under the same plasma conditions from
above, this porous grain population is heated to $T_{dust} = 56.5$ K,
and has a total luminosity of $1.09 \times 10^{36}\ \lsun$ in the
model, for a luminosity ratio $L_{porous}/L_{compact} = 1.23$.

As a result of the equilibrium temperature being lower for porous
grains, more heating is necessary to fit the same spectrum with a
porous grain model than a solid grain model. For a fixed plasma
temperature, this can only be achieved by increasing the post-shock
density. When using dust as a density diagnostic, this has significant
effects on inferences for the pre-shock density and compression ratio
of the shock. Raising the post-shock density for a given X-ray
emission measure must lower the pre-shock density (and thus, the gas
mass swept-up by the forward shock). A porous grain model fit to the
same spectrum will require a lower dust mass and gas mass. This is
explored further with respect to model spectral data from SNRs in
Chapter 8. In Figure 7.4, I show model spectra for compact, 10\%,
25\%, and 50\% porous silicate grains heated under the conditions
assumed above. These spectra are calculated over the entire grain size
distribution of \citet{clayton03}.

To include the effects of sputtering for porous grains, we assume that
the sputtered yield from a porous grain is equal to that of a compact
grain. For composite grains, a weighted harmonic mean of the
sputtering rate for each grain component is used.

\subsection{Energy Deposition Rates for Ions}

If electrons and ions are at the same temperature, grains will be
heated predominantly by electrons, since their velocities are much
higher. In fact, in many SNRs heating by protons can be virtually
neglected as a heating source for grains. However, it is often not the
case in SNRs that there is a temperature equilibration between these
particle species. In the simplest model of a collisionless shock,
there is a significant temperature difference between protons and
electrons, due to the fact that protons have a much higher mass. In
reality, the situation is much more complicated than that, with
contributions from collisionless heating at the shock and heating in
the precursor. Nonetheless, observations show that, for fast shocks,
protons are hotter than electrons. Recent evidence has shown that the
post-shock ion/electron temperature ratio increases as $v_{s}^{2}$
\citep{ghavamian07}. In the case of fast shocks (i.e. shocks of a few
thousand km s$^{-1}$ or more), heating by protons can have a
significant effect, as shown in Figure 7.3.

If heating by protons is to be considered, then energy deposition
rates must be properly accounted for in a heating model. Generally
speaking, when dealing with gas at high temperatures, both protons and
electrons deposit only a fraction of their energy into a grain, a
fraction which depends on both the grain radius and the temperature of
the impinging particle. This deposition function is represented by
$h(a,T)$ in eq. 17. Aside from its dependence on $a$, the grain
radius, and $T$, the temperature (or energy) of the particle, the
amount of energy deposited by an impinging particle depends on the
stopping power of the grain material. From the stopping power, one can
calculate the projected range of an impinging particle into the
grain. If this projected range is larger than the path traveled
through the grain material, then the particle will exit the other side
of the grain, having only deposited some of its energy. If the
projected range is shorter than the path traveled, the particle will
be stopped, and will deposit all of its energy into the
grain. (Actually, for electrons, the total energy deposited only
reaches about 0.875 of the initial energy, due to reflection from the
surface of the grain \citep{dwek87}.) As an approximation to the total
length traversed through the grain, I follow the standard in the field
of approximating the average distance through a grain as being $R=
\frac{4}{3} a$, which is the average path length through a sphere.

The stopping power (energy loss per unit path length) and projected
range (average length traveled by a particle into a material) for both
electrons and protons into a given grain are generally given by
analytic expressions approximating experimental data. For electrons,
these expressions fit the experimental data to within 15\% for the
energy range of 20 eV to 1 MeV \citep{dwek87}. For protons and alpha
particles, the approximate projected range, based on experimental data
from \citet{andersen77} and \citet{ziegler77}, is given by
\citet{draine79} as

\begin{equation}
R_{H} = 3 \times 10^{-6} (E/keV)\ {\rm g\ cm}^{-2}, R_{He} = 0.6R_{H},
\end{equation}

\noindent
where $\rho$ is the mass density of the grain. The authors caution
that this expression is valid only for $E < 100$ keV. I raise here two
questions: 1) Are these approximations appropriate for energies at or
exceeding the 100 keV threshold (as may well be the case in very young
SNRs with shock speeds exceeding 6000 km s$^{-1}$)? 2) These
approximations are based on data over 30 years old; are they still
valid?

\subsection{Ion Heating of Grains in Fast Shocks}

To answer both of these questions, I turn to more recent laboratory
data on the stopping power of protons and alpha particles in various
materials, obtained from the online {\it PSTAR} and {\it ASTAR}
databases published and maintained by the National Institute of
Standards and Technology. These databases are freely available online. In
Figure 7.5, I show the stopping power of a proton in silicon dioxide
as a function of energy. The stopping power turns over at $\sim 100$
keV. Since, for an extremely fast shock, there will be a significant
population of protons with energies in this range, this turnover
should be properly accounted for in calculating the projected range of
the particle.

Interstellar grains are more complicated than silicon dioxide
(although it is possible that SiO$_{2}$ is a minor component of ISM
dust). Since the NIST databases only have information available for
actual materials that can be measured in the lab, materials like
``astronomical silicate (MgFeSiO$_{4}$)'' are not available. The
individual stopping powers for the constituent elements are available,
however, and can be combined to approximate the stopping power for a
grain of arbitrary composition. To do this, I use ``Bragg's Rule''
\citep{bragg05}, given by

\begin{equation}
S(A_{m}B_{n}) = m \cdot S(A) + n \cdot S(B),
\end{equation}

\noindent
where $S(A_{m}B_{n})$ is the total stopping power of a molecule
$A_{m}B_{n}$, and $S(A)$ and $S(B)$ are the stopping powers of the
individual constituents. Since Mg is not included in the NIST
database, I use the results of \citet{fischer96}. Thus, for
MgFeSiO$_{4}$, one has

\begin{equation}
S(MgFeSiO_{4}) = S(Mg) + S(Fe) + S(Si) + 4\cdot S(O).
\end{equation}

In Figure 7.6, I show the stopping power of MgFeSiO$_{4}$, derived via
Bragg's rule. The calculated projected range for protons based on this
stopping power is shown in Figure 7.7, along with the approximation
from eq. 27 above. The projected range curves I show are given by the
following polynomial expressions in logarithmic space, where $E$ is
the energy of the impinging particle in keV:

\underline{Protons}

\begin{equation}
\log\ (R_{H,silicate}) = 0.053\ \log^{3}\ (E)-0.202\ \log^{2}\ (E)+1.21\ \log\ (E)-5.68
\end{equation}

\begin{equation}
\log\ (R_{H,graphite}) = 0.053\ \log^{3}\ (E)-0.158\ \log^{2}\ (E)+0.961\ \log\ (E)-5.46
\end{equation} 

\underline{Alpha Particles}

\begin{eqnarray*}
\log\ (R_{He,silicate}) &=& 3.78 \times 10^{-3}\ \log^{5}\ (E)-8.79 \times 10^{-3}\ \log^{4}\ (E)-4.25 \times 10^{-2}\ \\
& & \log^{3}\ (E)+6.25 \times 10^{-2}\ \log^{2}\ (E) + 1.04\ \log\ (E)-5.79
\end{eqnarray*}

\begin{eqnarray*}
\log\ (R_{He,graphite}) &=& 3.33 \times 10^{-3}\ \log^{5}\ (E)-6.13 \times 10^{-3}\ \log^{4}\ (E)-3.46 \times 10^{-2}\ \\
& & \log^{3}\ (E)+3.03 \times 10^{-2}\ \log^{2}\ (E) + 0.966\ \log\ (E) - 5.76.
\end{eqnarray*}

Ultimately, the purpose of calculating the projected range of a
particle into a grain is to determine how much energy is deposited
into the grain by an impinging particle. If a particle exits the grain
with energy $E^{\prime}$, then we can define $\Delta E$ =
$E-E^{\prime}$, and the fractional energy deposited as $\Delta E /
E$. This function is generally referred to as $\zeta(a,E)$, and the
grain heating efficiency function, $h(a,T)$, is given by
\citet{dwek81} as

\begin{equation}
h(a,T) = \frac{1}{2} \int_{0}^\infty x^{2} \zeta(a,E) e^{-x} dx,
\end{equation}

\noindent
where $x = E/kT$. In Figure 7.8, I show calculations of
$\zeta_{p}(a,E)$ for several grain sizes, and in Figure 7.9, I show
$\zeta(E)$ for a single grain of radius 0.05 $\mu$m compared with the
analytic approximation from \citet{dwek81}, given by

\begin{equation}
\zeta(a,E) = E^{*}/E \mbox{ for}\ E>E^{*},
\end{equation}

\noindent
where $E^{*}$ is given by \citet{draine79} as $E^{*}$(keV) = 133$a$
and 222$a$ ($a$ in $\mu$m) for protons and alpha particles,
respectively. Of course, the ultimate question is how much these
improved energy deposition rates matter. In Figure 7.10, I show the
ratio of the heating rate, $H(a,T)$, calculated from the NIST energy
deposition rates and that calculated from the analytical
approximations of \citet{draine79} and \citet{dwek81}, for a proton
temperature of 100 keV, which is relevant for fast shocks such as
those seen in SNR 0509-67.5 (see Chapter 8). Figure 7.11 shows a
comparison of the spectra of grains heated behind such a shock, for
both the modified energy deposition rates described above and the
analytical approximations. Figure 7.12 shows the ratio of the
spectra. The main difference is in the 18-40 $\mu$m range, but the
maximum difference is only $\sim 15$\%, which is roughly comparable to
the errors in fitting models to data (see Section 2.9.1). The use of
analytical approximations to proton heating rates should be valid for
most cases. Nonetheless, I use the modified rates in reporting results
in Chapter 8.

\subsection{Non-thermal Sputtering of Grains Due To Gas-Grain Motions}

On the other end of the SNR age spectrum are old remnants whose shock
speeds are much lower, and the above issues with proton heating are
irrelevant. In fast shocks, sputtering is mostly due to the thermal
motions of the particles in the gas bombarding the grain. For slow
shocks, however, an additional sputtering component is present that
arises from the relative gas-grain motions in the post-shock
region. Dust grains in the ISM typically have a low charge-to-mass
ratio when compared with ions. This means that they are largely
unaffected by the passage of the shock front over them. In their frame
of reference, though, they are suddenly thrust into a plasma with a
bulk-velocity of $\frac{3}{4}v_{s}$, where $v_{s}$ is the shock
speed. They are eventually brought to a rest in the frame of the gas
due to the drag force from collisions with ions, but this takes a
non-negligible amount of time, during which their sputtering rate is
increased. This sputtering is referred to as ``non-thermal''
sputtering, to contrast it with the case in which the sputtering is
purely ``thermal.'' (This should not be confused with other uses of
the phrase ``non-thermal'' in this thesis, which typically refer to
the emission produced by a relativistic population of electrons.)

A particular case where this is relevant is in the Cygnus Loop, a
10,000-20,000 year old supernova remnant at a distance of $\sim
500-600$ pc. Shock speeds in the Cygnus Loop are 300-500 km s$^{-1}$,
and the remnant as a whole is roughly 3 degrees in diameter. The
entire shell was detected in soft X-ray bands by the {\it ROSAT}
All-Sky Survey. {\it Spitzer} observations of a non-radiative shock on
the northeast limb of the Cygnus Loop were obtained in 2005 December
as part of Cycle 2 Guest Observer program 20743 (R. Sankrit, P.I.).
$^{\dagger}$\footnote{The text from here to $^{\ddagger}$ is taken
verbatim from Sankrit et al. (2010)} The target field was centered on
$\alpha_{\rm{J2000}}$ = 20\fh\ 54\fm\ 35\fs, $\delta_{\rm{J2000}}$ =
+32\arcdeg\ 17\arcmin\ 30\arcsec.  The $24\,\mu$m observations
included 12 frames and totalled 375 s.  The field-of-view of the
shocked region is about 5.7\arcmin\ $\times$ 5.7\arcmin\@.  The
$70\,\mu$m images consisted of a $3\times 3$ raster obtained in the
narrow field-of-view mode.  For each position, 40 frames were obtained
with a total exposure time of about 340~s.  The overall field-of-view
of the $70\,\mu$m image is about 8\arcmin\ $\times$ 13\arcmin, and
completely overlaps the $24\,\mu$m image field-of-view.

A three-color image of the non-radiative shock is shown in Figure
7.13. Red is an H$\alpha$ image obtained at the Mt.  Hopkins Whipple
Observatory 1.2m telescope, green is the $\spitzer$ $24\,\mu$m image,
and blue is an exposure-corrected 0.35--2.0~keV \textit{Chandra} image
made from the ACIS-S3 data (ObsID 2821) using CIAO 3.4.  The H$\alpha$
emission traces a narrow zone ($\sim 10^{14}$~cm thickness) just
behind the shock front.  The infrared emission arises in a more
extended zone behind the shock front.  The X-ray emission comes from
the entire post-shock region where the gas initially is heated to a
temperature of a few times $10^6$~K.  As the gas ionizes, the X-ray
emission is dominated by different species: Si and Fe make the major
contribution just behind the shock, and O and Ne further downstream.
A fainter shock is clearly seen ahead of the main shock, near the left
edge of the $24\,\mu$m image field of view.  The three images are
shown separately along the bottom of the figure.

The $\spitzer$ $24\,\mu$m and $70\,\mu$m images are shown in Figure
7.14.  Fluxes were measured along each of the three strips shown on
these images.  The strips were chosen to be perpendicular to the shock
front as seen in the $24\,\mu$m image.  Each strip is about 60\arcsec\
across and each box is 20.33\arcsec\ wide.  This corresponds to 22
pixels and 8 pixels, respectively in the $24\,\mu$m image and about 14
pixels and 5 pixels in the $70\,\mu$m image.

The total fluxes within each box were calculated for the MIPS images
by summing the data values (in MJy/sr) for pixels contained in the
box multiplied by the solid angle subtended.  These were used to
construct flux profiles along each strip.  The background in the
$24\,\mu$m image is about 3~mJy/pixel, and fairly uniform over most
of the detector.  We subtract this value off each pixel to obtain
background subtracted profiles.  The uncertainty in the background
value is $\sim 0.003$~mJy/pixel, or $\sim 0.53$~mJy in each
box.  This uncertainty translates to about 5\% error in the flux
measurement near the peak of the $24\, \mu$m emission.  The exception
to the near-uniform background is in the region around the Northern
corner of the FOV, where the background is lower and more variable.
There is no such difference associated with that part of the sky
in the $70\,\mu$m image.  This suggests that the anomaly is due to
the $24\,\mu$m detector and does not have any astrophysical
significance.  Therefore, we avoid the region in our analysis.

The background in the $70\,\mu$m image varies across the field of
view.  For each profile, we fit a low-order polynomial to the boxes
near the ends and used that for the background.  We varied the
background fit in several reasonable ways - using different numbers
of points, different order polynomials (including constant value
backgrounds) - to see how these changes affected the measured fluxes.
We find that uncertainties in the background contribute an error
of about 5\% to the flux measurements near the peaks of the profiles.

To model non-thermal sputtering, we use formulae for the gas drag
force found in \citet{draine79}, where we neglect the Coulomb drag, as
this should be negligible in a hot, X-ray emitting plasma. As a grain
slows down, sputtering causes its mass and radius to decrease. The
sputtering rate is given by \citet{nozawa06}. The coupled equations
for the grain radius and velocity are solved with standard numerical
ordinary differential equation solvers. Since grain gyroradii are much
smaller than the spatial scales of interest, we assume that dust and
gas motions are tightly coupled, i.e. grains gyrate with small (and
negligible) gyroradii around magnetic fields frozen into and moving
with the shocked gas. As in previous work on Galactic and LMC
remnants, we use a plane-parallel approximation for the shock
structure. The most important shock parameters are the pre-shock
density and the shock speed, $v_{s}$, which determine the post-shock
temperature and density through the standard strong-shock jump
conditions. The electron and ion temperatures in the post-shock region
are expected to be equal for the relatively slow shock being studied
\citep{ghavamian01}. In contrast with previous work, the shock
structure here is resolved, and each of the spatial bins can be
conveniently described by $\tau_{l}$ and $\tau_{u}$ ($\tau \equiv
\int_{0}^{\tau} n_{p} dt$), the lower and upper ``sputtering'' ages
bracketing its position.

At the relatively low plasma temperatures that we obtain in the Cygnus
Loop shocks, non-thermal sputtering is expected to contribute
significantly to the dust destruction \citep[e.g.][]{dwek96}.  In
order to check this, we ran a grid of models that included only
thermal sputtering.  The $70\mu$m to $24\mu$m flux ratios predicted in
these models were about 40\% lower than in the original set, and
failed to reproduce the observed ratios.  Our data clearly show that
non-thermal sputtering is an important process for dust destruction in
shocks with speeds $\lesssim$500~\kms.$^{\ddagger}$ In Figure 7.15, I
show a plot of the 70/24 $\mu$m ratio behind the shock for identical
conditions, with non-thermal sputtering turned on and off.

\subsection{Sputtering Rates for Small Grains}

The sputtering rate for small grains is much more uncertain that that
for large grains. For large grains, the ``semi-infinite solid''
approximation can be used, which is what most laboratory experiments
can measure. This approximation (which is typically fine for
industrial applications) assumes that all sputtering comes from the
surface of the grain where the impact takes place. With dust, however,
this is not always a valid approximation. If the grain is small
enough, sputtering can occur from not only the front side of the
grain, but the back and sides as well. To account for this, we use the
``enhanced'' sputtering rates of \citet{jurac98}. Even this
enhancement may not paint the full picture though, as for really small
grains (where a grain is considered ``small'' if $R_{p} >
\frac{4}{3}a\rho$), the sputtering yield may again be decreased due to
the incomplete deposition of energy from protons into the grain
\citep{serradiazcano08}. To account for this, we make the assumption
that, to first order, the sputtering rate is directly proportional to
the energy deposited into the grain, which drops with decreasing grain
size. An example of the sputtering rate as a function of grain size,
taking all three of these effects into account, is shown in Figure
7.16.

\subsection{Liberation of Elements into Gaseous Phase}

As dust is eroded via sputtering, the atoms ejected from the grain do
not simply disappear. They are ``liberated'' back into the gaseous
phase and are equilibrated with their surroundings in the hot
post-shock region. The refractory elements that make up dust grains
(C, O, Mg, Si, Fe, etc.) are the same elements whose lines are
commonly seen in X-ray spectra of SNRs. If sufficient amounts of these
atoms are injected into the post-shock gas from grain sputtering, it
is possible to see the effect on X-ray spectra as a function of
distance behind the shock. In Figures 7.17 and 7.18, I show the amount
of material sputtered for silicate and graphite grains as a function
of the sputtering timescale (which, in a plane-shock model, is
proportional to the distance behind the shock or the time since
shocked) for a range of post-shock proton temperatures. The sputtered
fraction listed in the figures is dependent on the grain size
distribution assumed. It is given by mass, and can thus be converted
to absolute masses of various elements, if the total dust mass needed
to fit the observed fluxes or spectra can be determined.

\subsection{Summary}

Since SNRs were first detected in the mid-IR with {\it IRAS} more than
25 years ago, numerous authors have worked to model the emission seen
with models of collisionally heated dust. Despite this work, there is
still much that is unknown about interstellar dust, and a lot of
physics that is known that is still not properly incorporated into
such models. I have built on previous work in this field by extending
models \citep{borkowski94} to include the effects listed above. 

Specifically, I have developed a code to approximate the optical
constants for a grain composed of arbitrary amounts of astronomical
silicate, graphite, amorphous carbon, and vacuum. This code could be
extended in the future, if necessary, to incorporate other grain
materials. The code also calculates modified heating and sputtering
rates for these composite/porous grains. I have also incorporated more
up-to-date energy deposition rates for ions, particularly important in
the case of very fast shocks, such as those seen in Cas A or SNR
0509-67.5. For slow shocks, such as those in the Cygnus Loop, I have
extended the code to include effects of non-thermal sputtering caused
by relative motions between gas and dust grains. The sputtering rate
for small grains has been updated to reflect the amount of energy
deposited into a grain. Liberation of elements from grains can be
followed in the code as a function of sputtering timescale. If some
parameters of the SNR are known, such as the distance to the object,
this sputtering could be followed in terms of a post-shock distance
and/or absolute mass scale. This injection of elements into the
gaseous phase could be compared directly with X-ray fits for
post-shock regions in sufficiently well-resolved SNRs.

Current and future generations of IR telescopes, such as the {\it
Stratospheric Observatory for Infrared Astronomy (SOFIA)}, the {\it
Wide-field Infrared Survey Explorer (WISE)}, the {\it Herschel Space
Observatory}, and the {\it James Webb Space Telescope}, as well as
X-ray telescopes like the {\it International X-ray Observatory}, will
allow testing of these models, as will archival study of {\it
Spitzer}, {\it Chandra}, and {\it XMM-Newton} data.

\newpage
\clearpage

\begin{figure}
\figurenum{7.1}
\includegraphics[width=16cm]{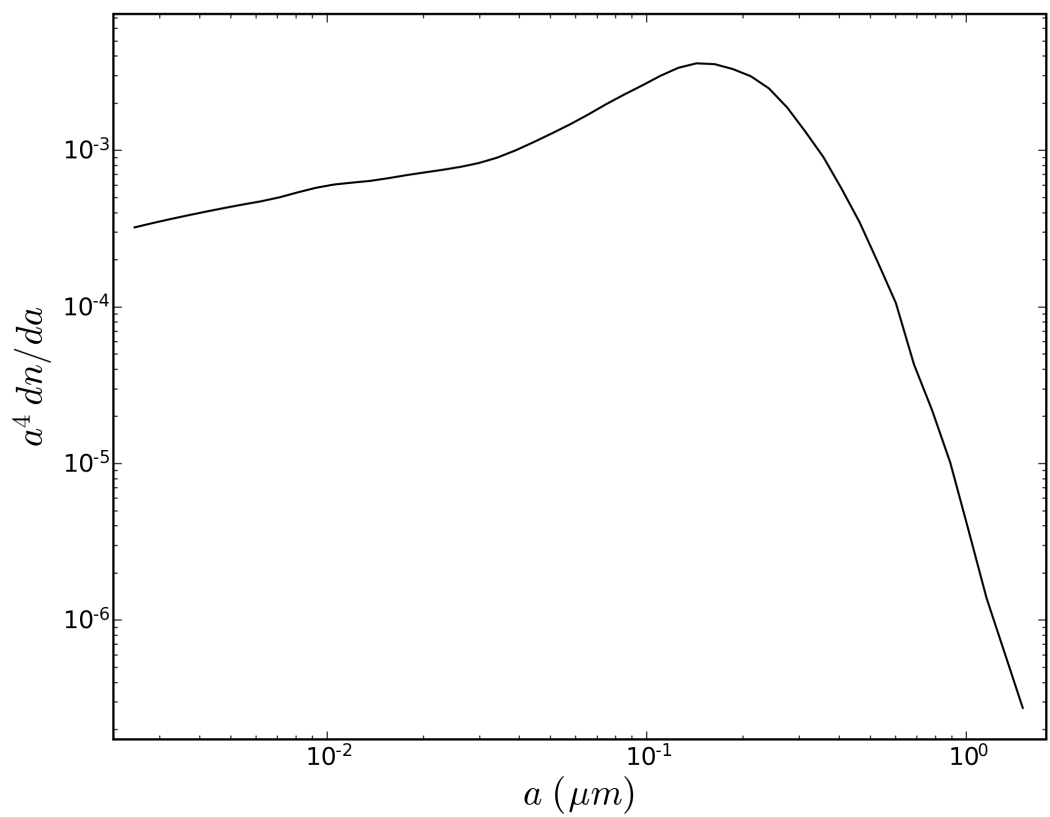}
\caption{Size distribution of porous grains in the ISM, from
\citet{clayton03}. Compact grain distributions of
\citet{weingartnerdraine01} are shown in Figure 2.3.
\label{clayton}
}
\end{figure}

\newpage
\clearpage

\begin{figure}
\figurenum{7.2}
\includegraphics[width=16cm]{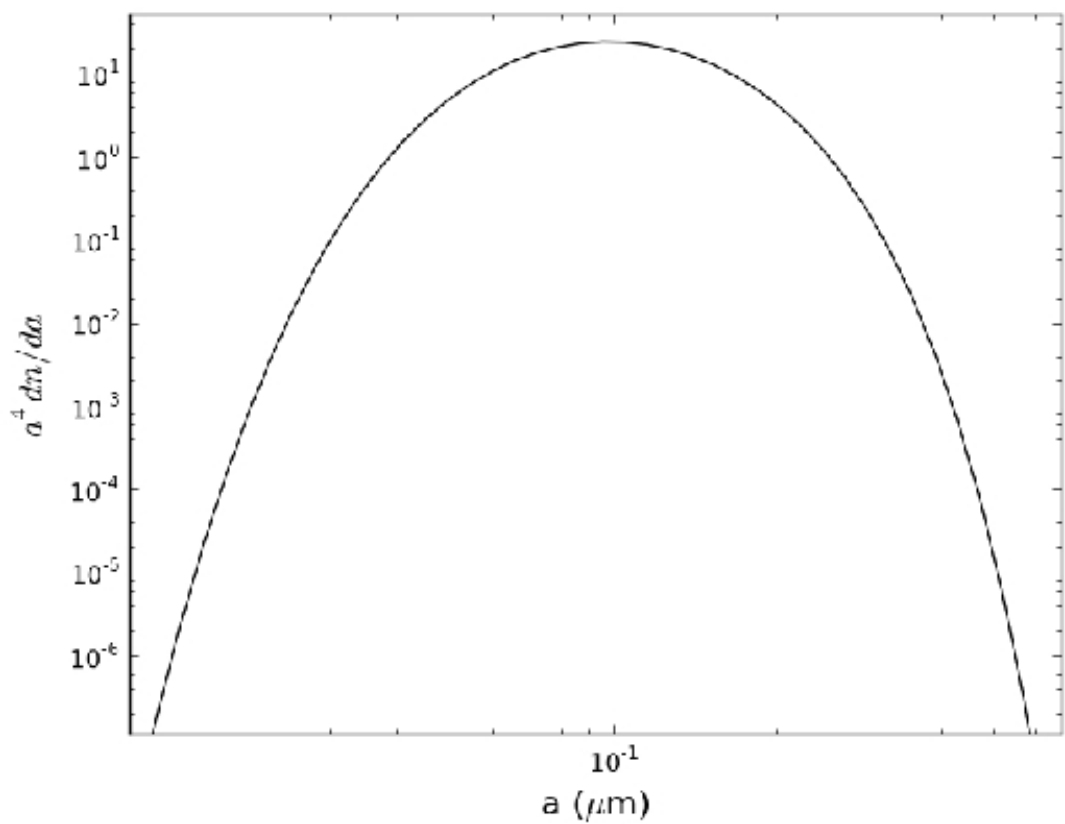}
\caption{Size distribution of porous grains in the ISM, from Mathis
  (1996). Eq. (3) of his paper gives the functional form of the
  distribution as $dn/da \propto a^{-\alpha_{0}}exp[-(\alpha_{1} a +
    \alpha_{2} /a +\alpha_{3} a^{2})]$, where $\alpha_{0}$ = 3.5,
  $\alpha_{1}$ = 0.00333 $\mu$m$^{-1}$, $\alpha_{2}$ = 0.437 $\mu$m,
  and $\alpha_{3}$ = 50 $\mu$m$^{-2}$. 
\label{mathis}
}
\end{figure}

\newpage
\clearpage

\begin{figure}
\figurenum{7.3}
\includegraphics[width=11cm]{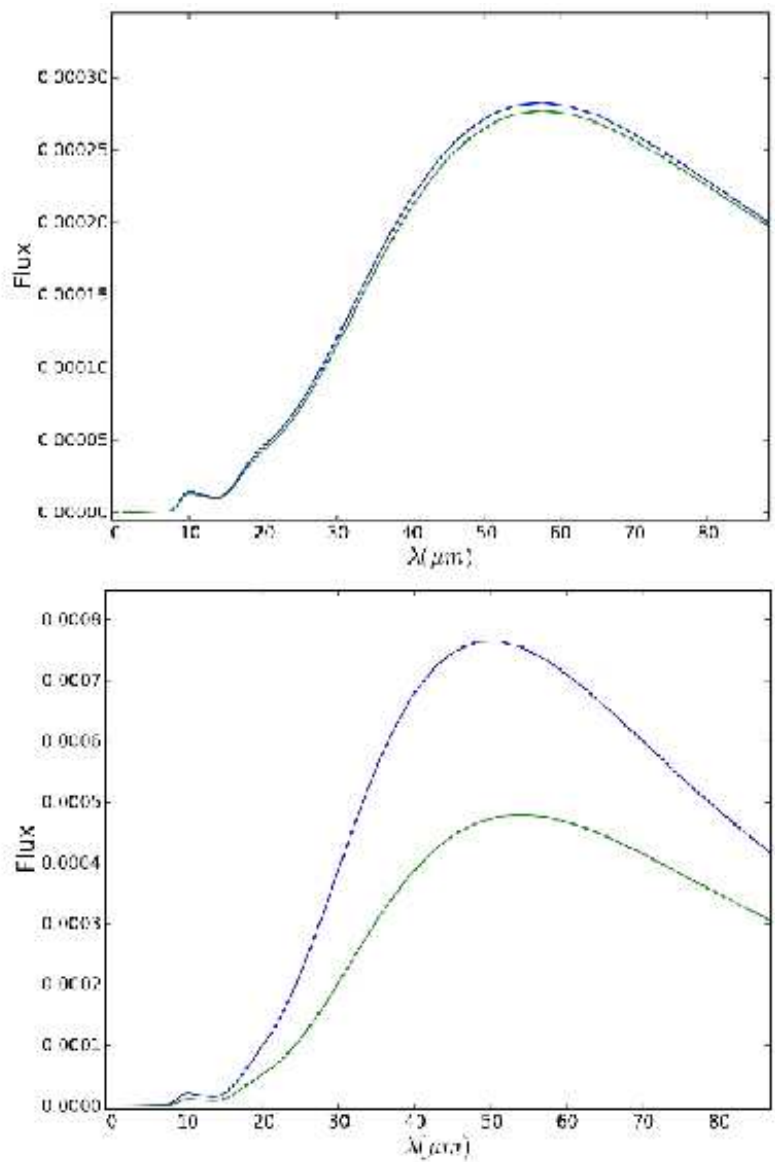}
\caption{The impact of proton heating of grains for slow ($\sim 1000$)
and fast ($\sim 3000$ km s$^{-1}$) shocks. For slow shocks, the
difference between proton heating and no proton heating is hardly
noticeable, but for fast shocks it becomes quite significant. Top:
Thermal spectrum of warm grains heated behind slow non-radiative
shock. Blue curve shows spectrum with $T_{p}$ = 1 keV, $T_{e}$ = 0.5
keV, Green curve shows $T_{p}$ = 0 keV, $T_{e}$ = 0.5 keV. Bottom:
Same as top, but for fast non-radiative shock. Blue curve shows
spectrum with $T_{p}$ = 20 keV, $T_{e}$ = 1 keV, Green curve shows
$T_{p}$ = 0 keV, $T_{e}$ = 1 keV.
\label{protheating}
}
\end{figure}

\newpage
\clearpage

\begin{figure}
\figurenum{7.4}
\includegraphics[width=16cm]{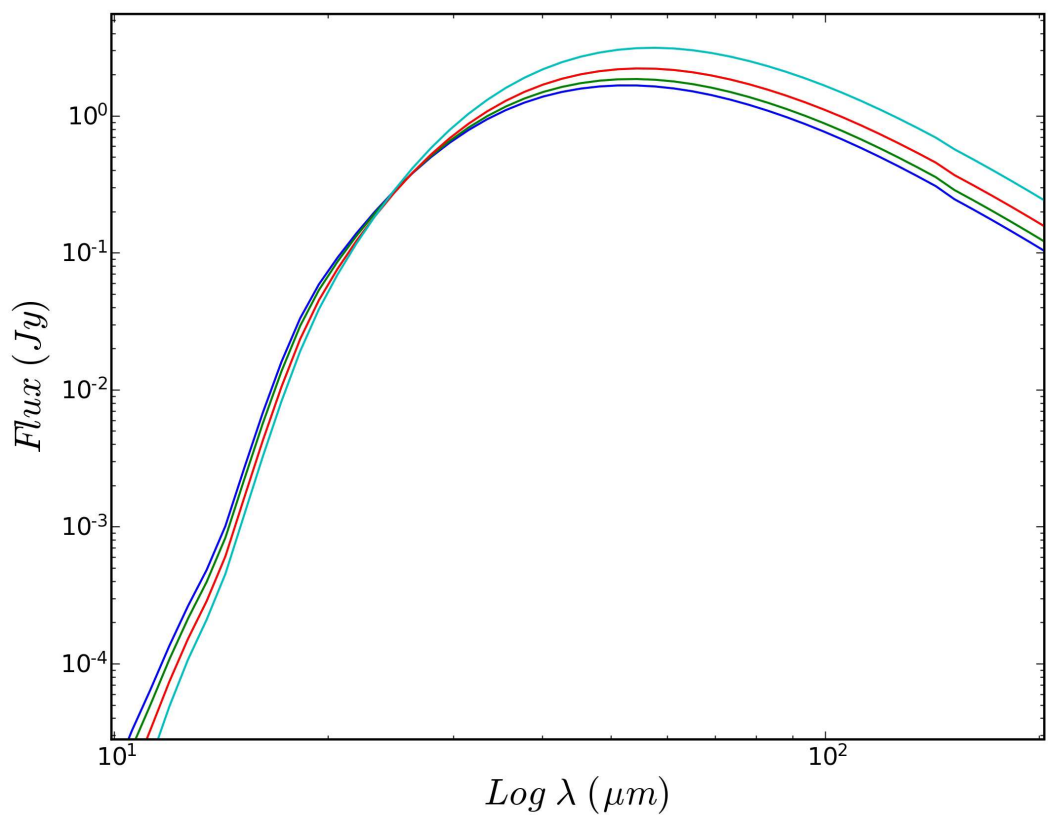}
\caption{Comparison of model spectra from 10-200 $\mu$m for silicate
grain size distributions of varying porosity. Blue: Compact silicate
grains; Green: 10\% porous silicate grains; Red: 25\% porous; Cyan:
50\% porous. Size distribution of \citet{clayton03} assumed, heating
model (identical plasma conditions for all grain populations)
described in text. Sputtering is neglected to highlight differences by
heating only.
\label{none}
}
\end{figure}

\newpage
\clearpage

\begin{figure}
\figurenum{7.5}
\includegraphics[width=15cm]{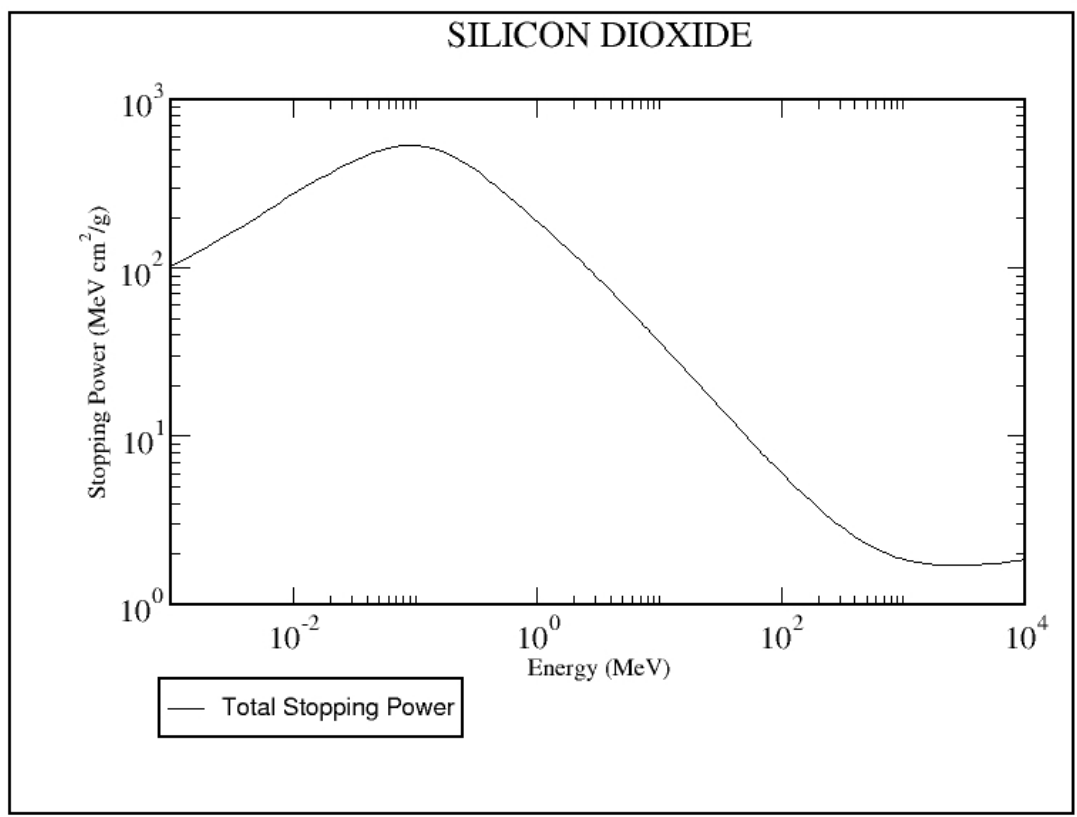}
\caption{Stopping power divided by mass density, $\rho$, of a proton
in silicon dioxide (SiO$_{2}$) as a function of proton energy. Taken
from the NIST {\it PSTAR} database (see text).
\label{si02stopping}
}
\end{figure}

\newpage
\clearpage

\begin{figure}
\figurenum{7.6}
\includegraphics[width=15cm]{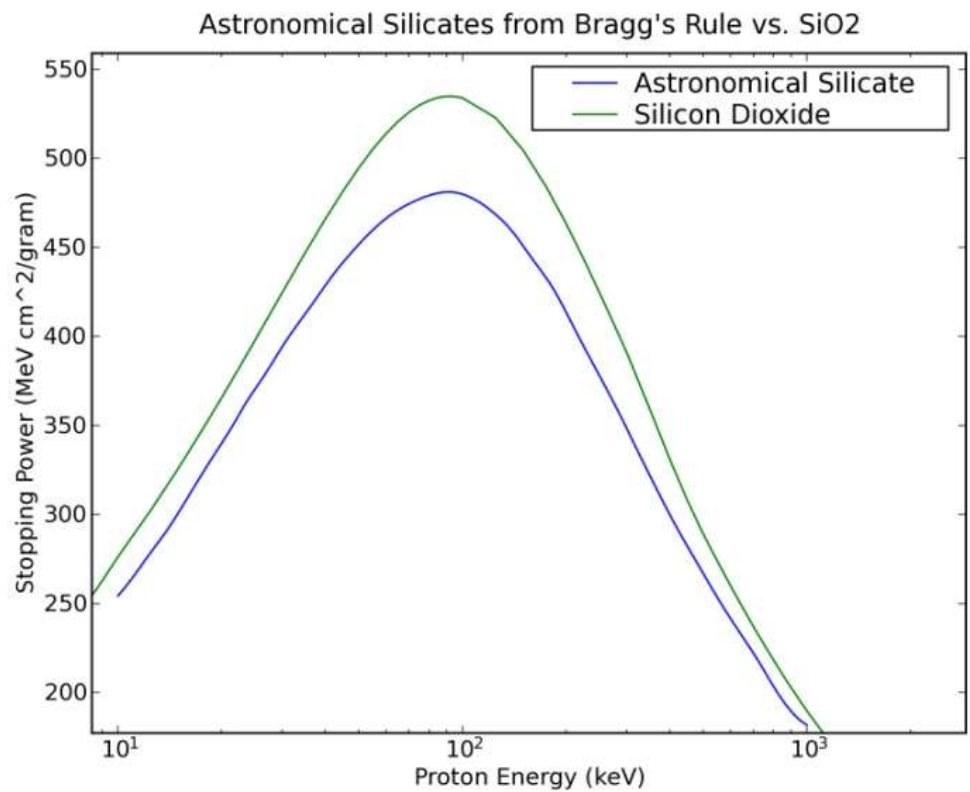}
\caption{Stopping power of astronomical silicate, as calculated from
Bragg's Rule (see text). For comparison, the stopping power of
SiO$_{2}$ obtained from the NIST {\it PSTAR} database is also shown.
\label{braggsrule}
}
\end{figure}

\newpage
\clearpage

\begin{figure}
\figurenum{7.7}
\includegraphics[width=16cm]{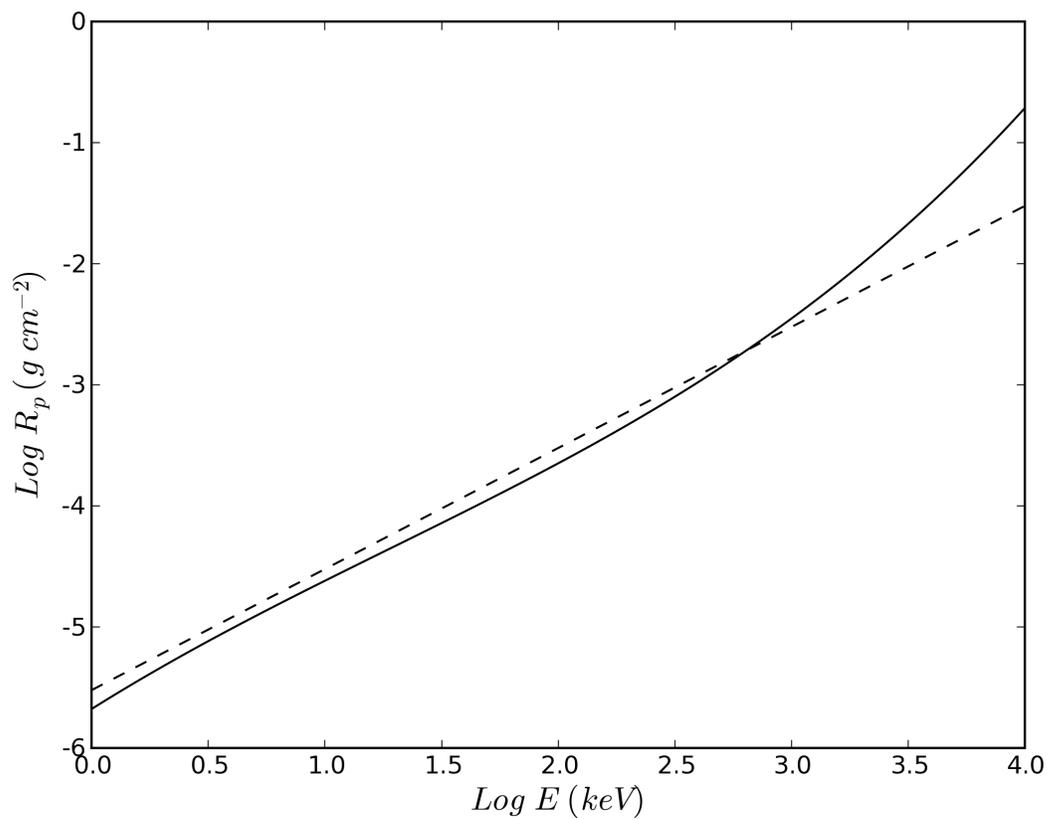}
\caption{Projected range of a proton for MgFeSiO$_{4}$ as a function
of energy. Solid curve is calculation described in text using Bragg's
rule; dashed curve is approximation of \citet{draine79} that is valid
at E $<$ 100 keV. The slight discrepancy at low energy is result of
approximations in Bragg's rule and in \citet{draine79}.
\label{proj_range}
}
\end{figure}

\newpage
\clearpage

\begin{figure}
\figurenum{7.8}
\includegraphics[width=16cm]{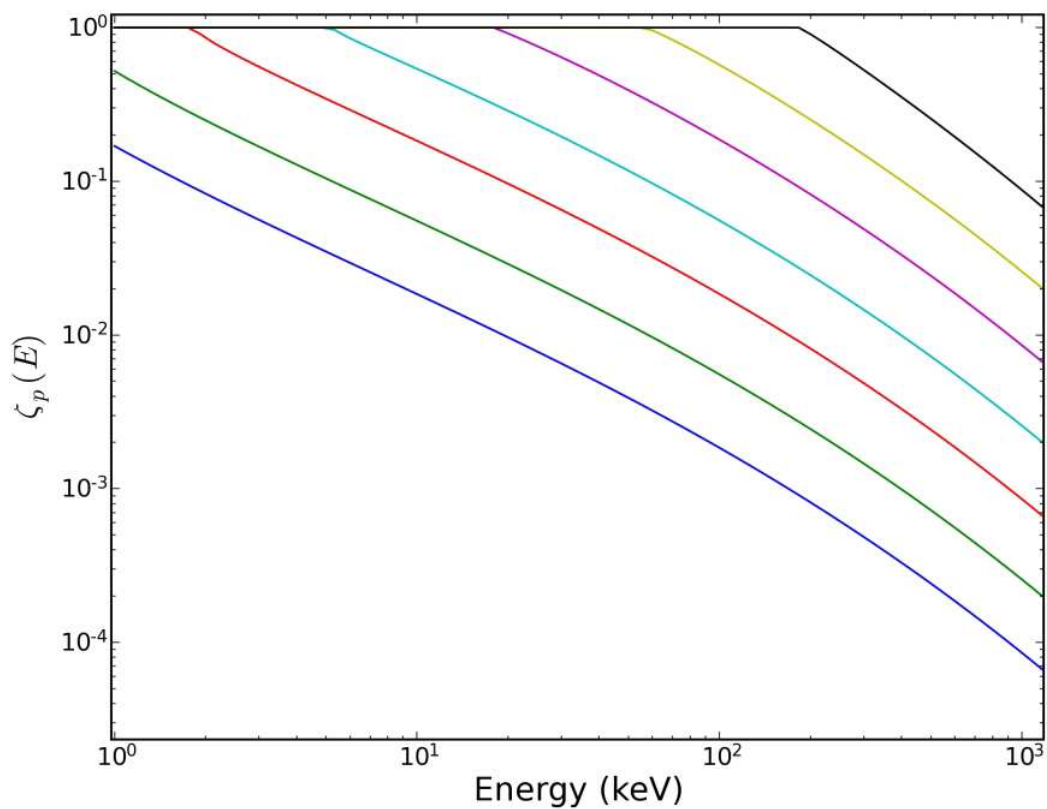}
\caption{$\zeta_{p}(a,E)$, the fractional energy deposited by a proton
into a silicate grain as a function of particle energy. From
bottom-left (blue) to top-right (black), curves represent grains of
radius 0.001, 0.003, 0.01, 0.03, 0.1, 0.3, and 1.0 $\mu$m.
\label{zeta_multiple}
}
\end{figure}

\newpage
\clearpage

\begin{figure}
\figurenum{7.9}
\includegraphics[width=16cm]{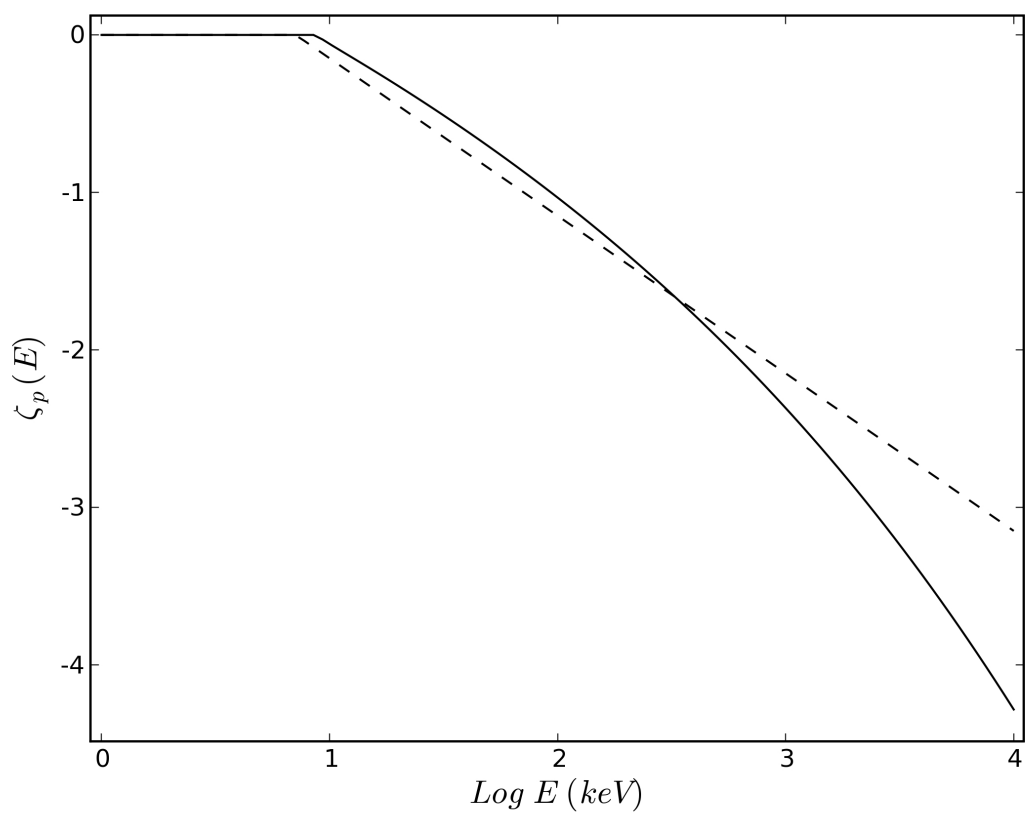}
\caption{Zeta function for a single grain of 0.05 $\mu$m radius. Solid
curve is calcuation described in text, dashed curve is approximation
of \citet{dwek81} (see text).
\label{zeta_single}
}
\end{figure}

\newpage
\clearpage

\begin{figure}
\figurenum{7.10}
\includegraphics[width=16cm]{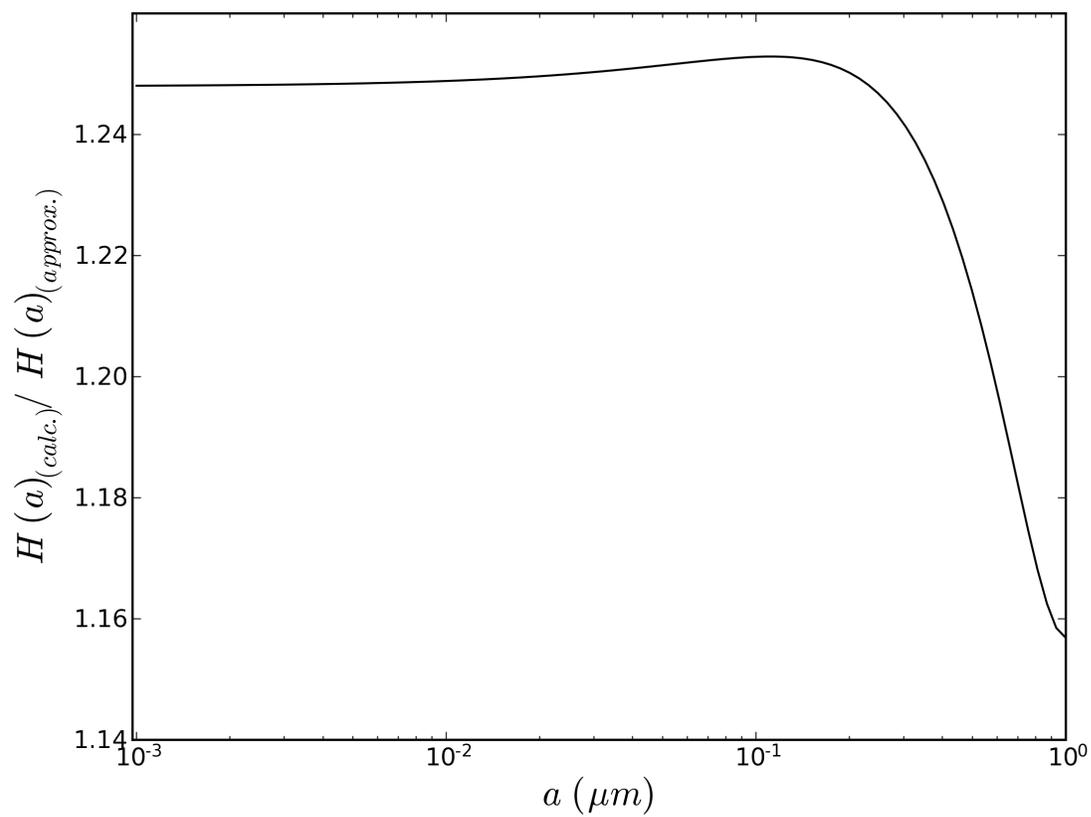}
\caption{Ratio of proton heating rate of a silicate grain from
calculations in this work to approximations from \citet{draine79} \&
\citet{dwek81} for an impinging particle energy of 100 keV, as a
function of grain radius.
\label{heatingrateratio}
}
\end{figure}

\newpage
\clearpage

\begin{figure}
\figurenum{7.11}
\includegraphics[width=16cm]{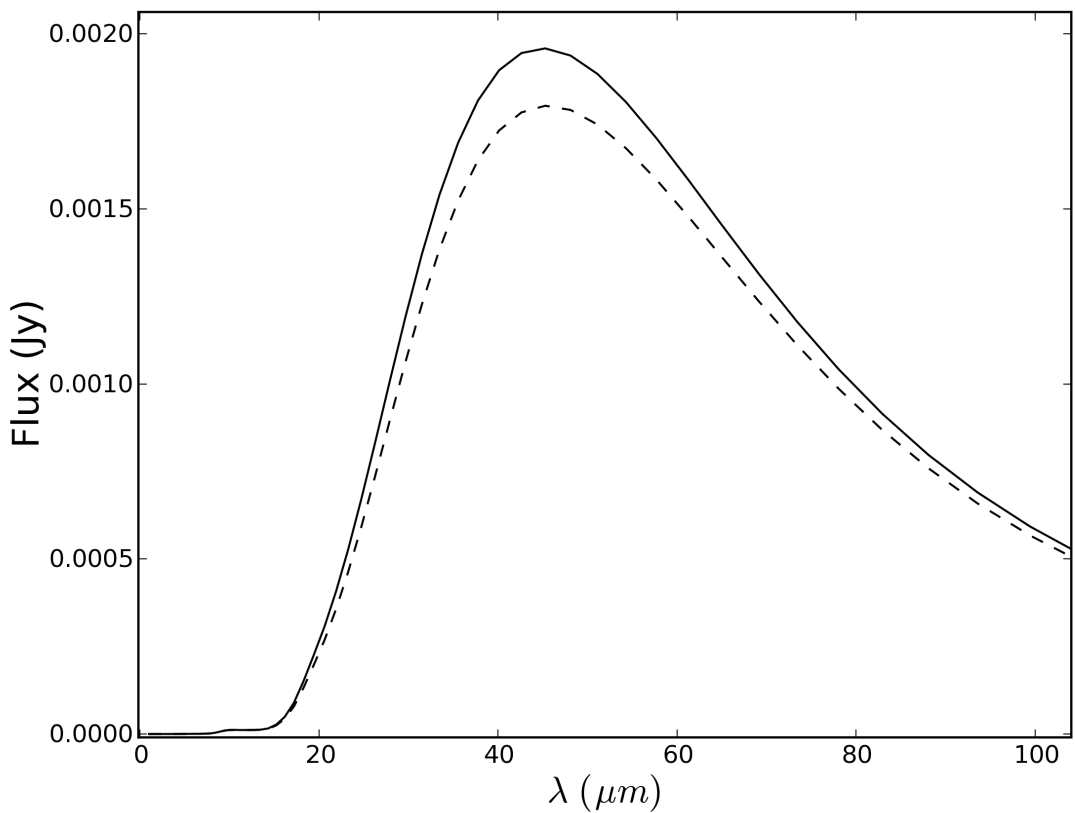}
\caption{Comparison of spectra produced in dust heating model for
protons of 100 keV. Solid line: spectrum assuming proton projected
ranges calculated in Section 7.5. Dashed line: spectrum assuming
analytical approximation to proton projected range from
\citet{draine79}. Electron heating ($T_{e}$ = 2 keV) is included in
this model.
\label{spectracompare}
}
\end{figure}

\newpage
\clearpage

\begin{figure}
\figurenum{7.12}
\includegraphics[width=16cm]{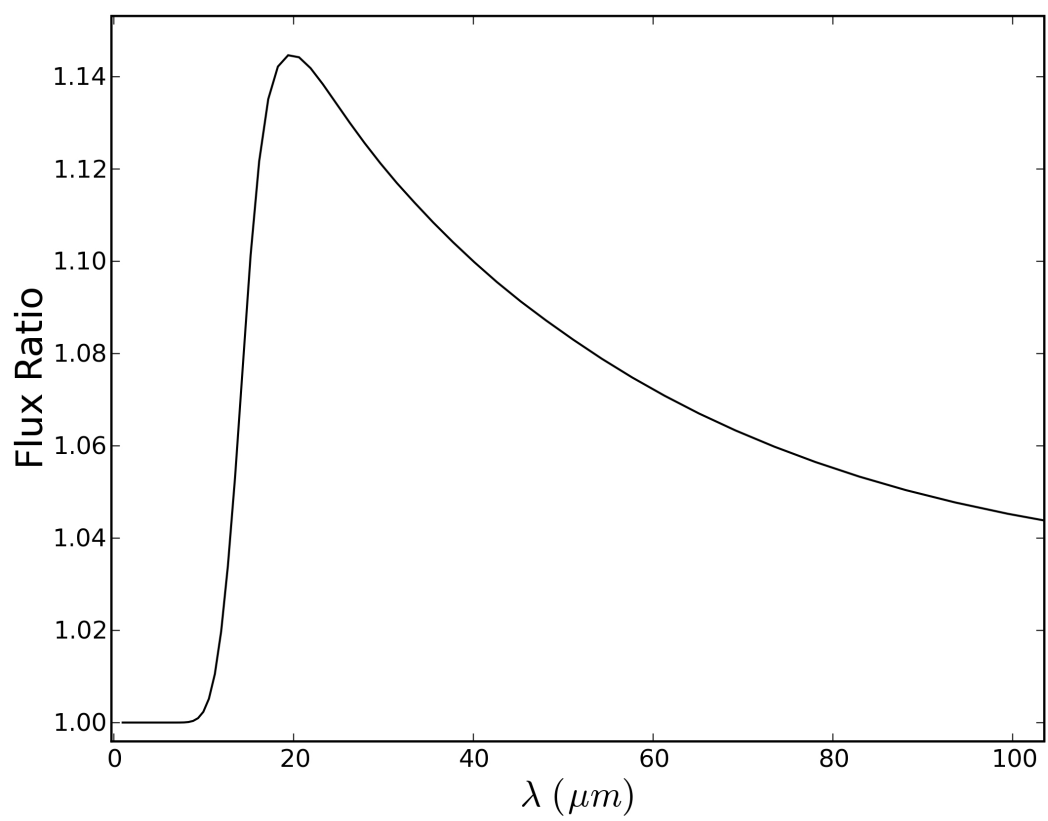}
\caption{Ratio of spectra from Figure 7.11.
\label{spectraratio}
}
\end{figure}

\newpage
\clearpage

\begin{figure}
\figurenum{7.13}
\includegraphics[width=14cm]{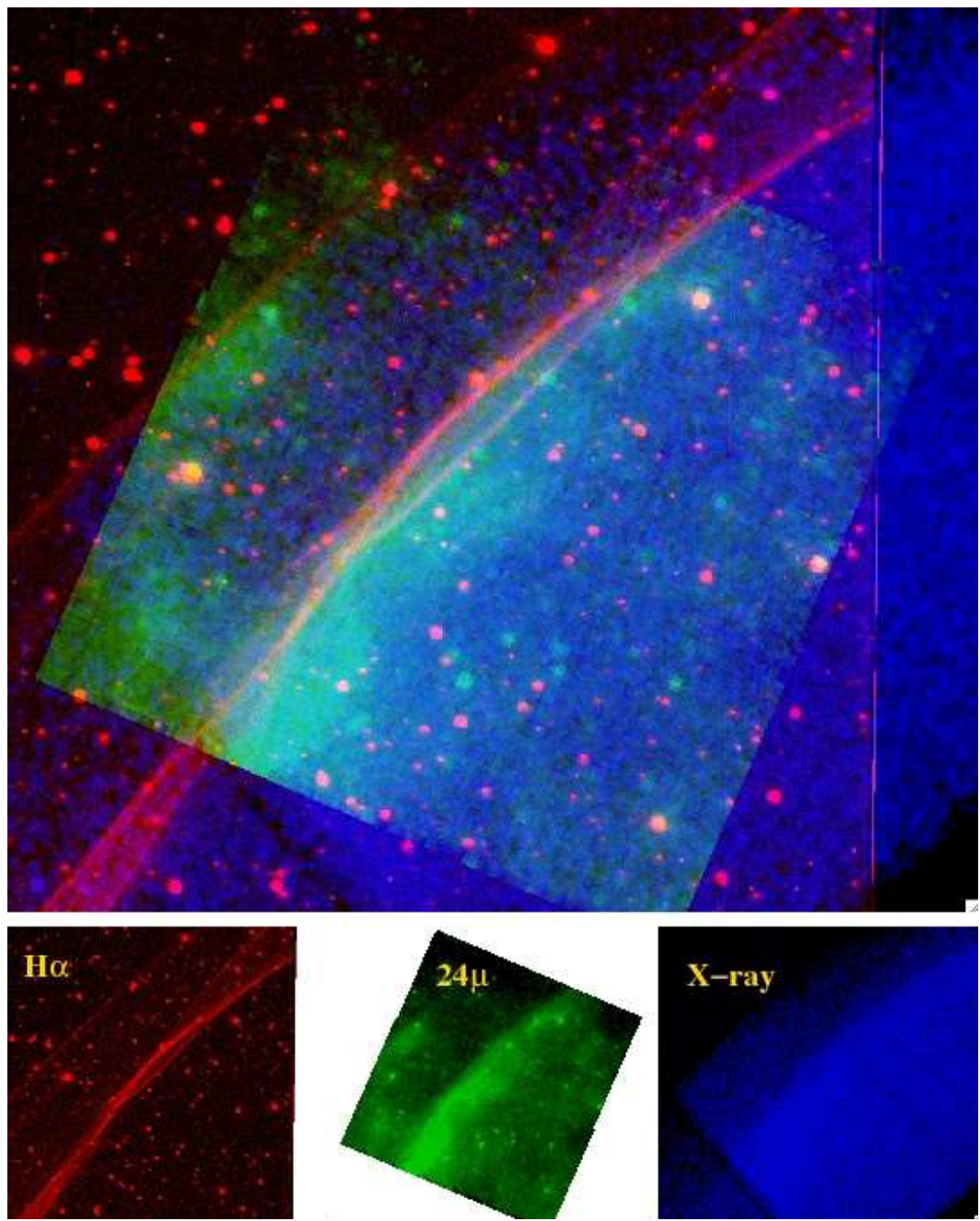}
\caption{3-color image of the P7 region of the Cygnus Loop. Red:
H$\alpha$; Green: {\it Spitzer} 24 $\mu$m; Blue: {\it Chandra} X-ray.
\label{cygloop3color}
}
\end{figure}

\newpage
\clearpage

\begin{figure}
\figurenum{7.14}
\includegraphics[width=15cm]{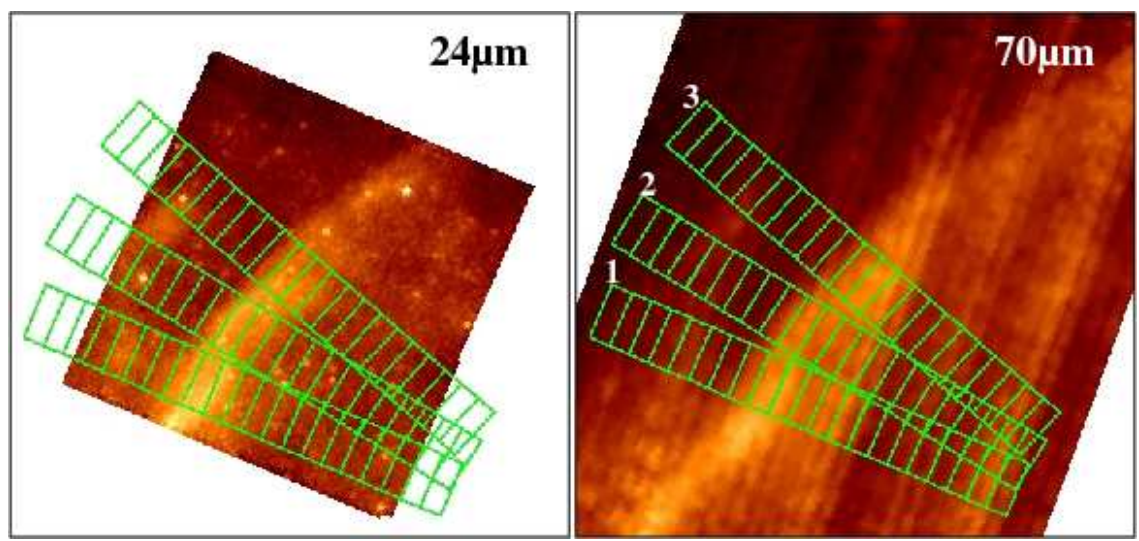}
\caption{{\it Spitzer} 24 and 70 $\mu$m images of the Cygnus Loop,
with analysis regions overlaid.
\label{cygloopspitzer}
}
\end{figure}

\newpage
\clearpage

\begin{figure}
\figurenum{7.15}
\includegraphics[width=16cm]{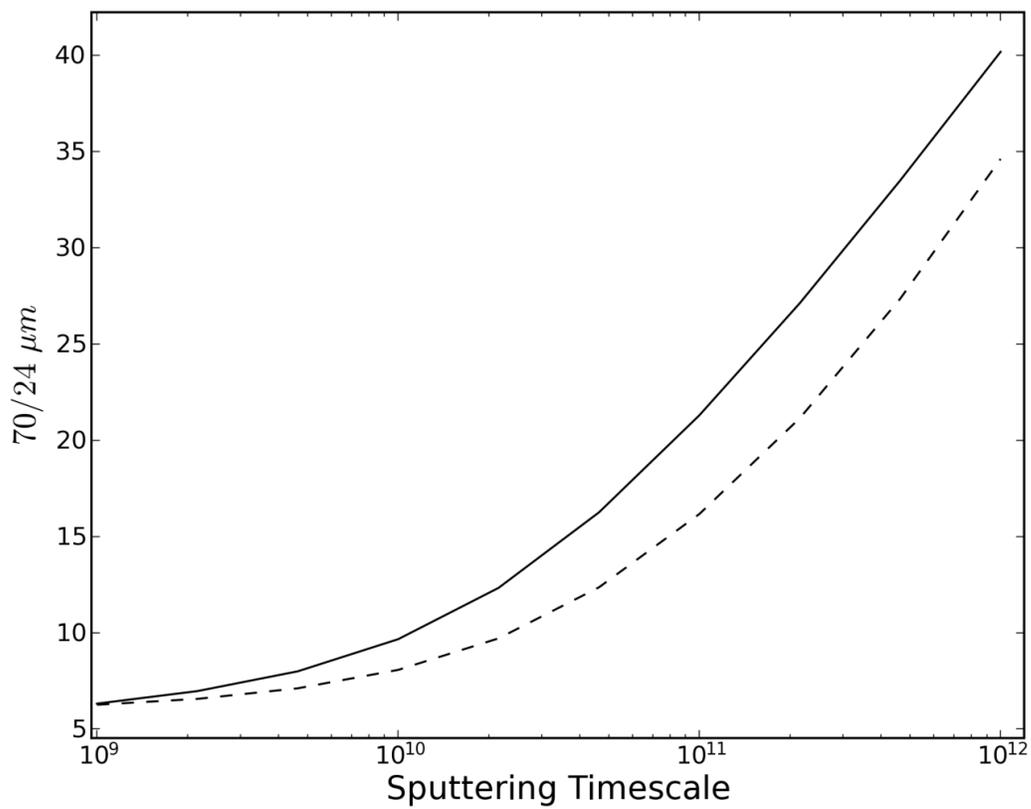}
\caption{70/24 $\mu$m flux ratio as a function of sputtering timescale
for a temperature of $T_{p}$ = $T_{e}$ = 0.3 keV, appropriate for the
Cygnus Loop. Solid black curve includes effects from both thermal and
non-thermal sputtering, dashed black line includes only thermal
sputtering.
\label{thermvsnontherm}
}
\end{figure}

\newpage
\clearpage

\begin{figure}
\figurenum{7.16}
\includegraphics[width=16cm]{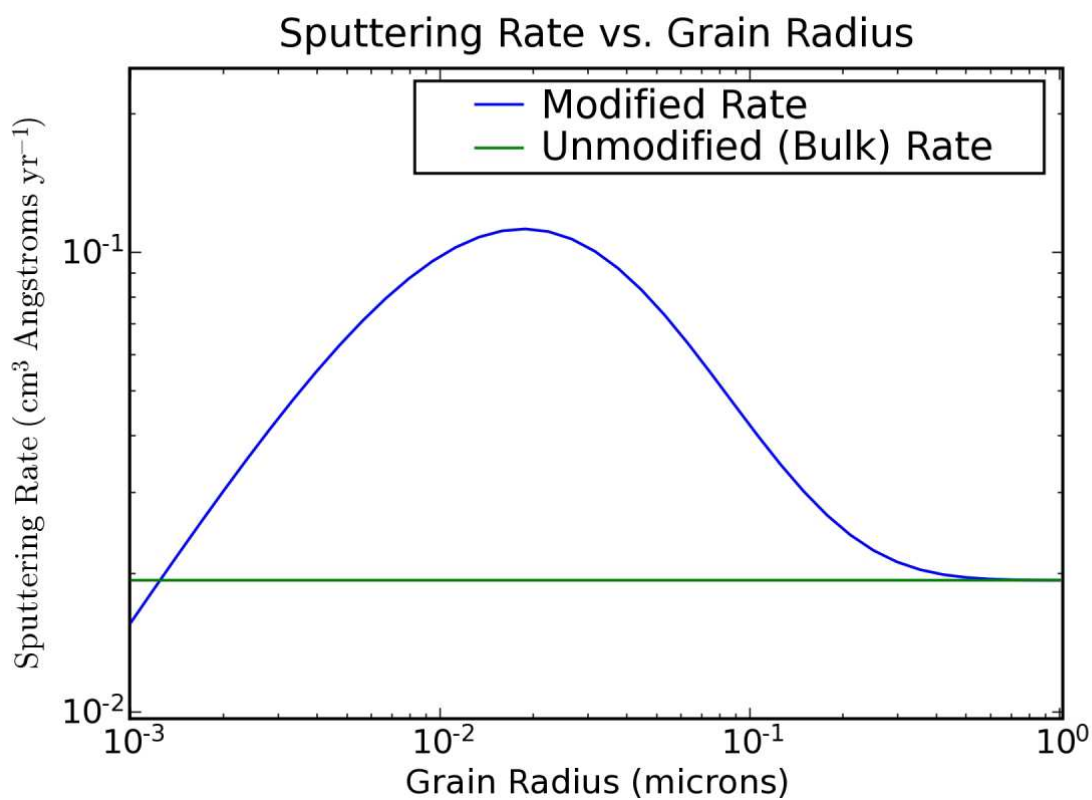}
\caption{Sputtering rate as a function of grain radius, where
sputtering is done by protons with energy 10 keV. Rates used in code
shown in blue, bulk sputtering approximation shown in green. Increase
in rate from $10^{-2} - 10^{-1}$ $\mu$m due to enhancement of
sputtering for small grains, dropoff short of $\sim 10^{-2}$ due to
decreased energy deposition rate for energetic particles. Same as
Figure 2.8.
\label{cygloopspitzer}
}
\end{figure}

\newpage
\clearpage

\begin{figure}
\figurenum{7.17}
\includegraphics[width=12cm]{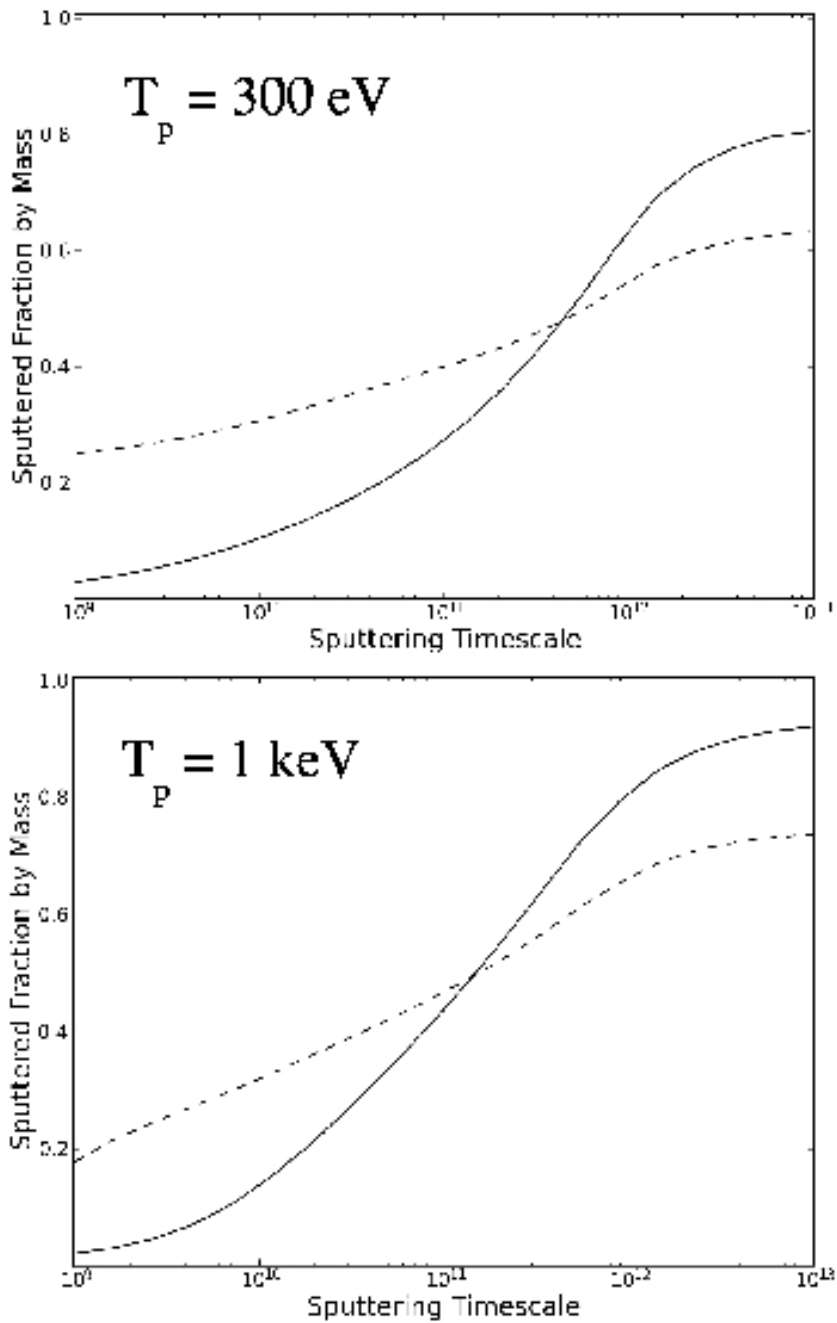}
\caption{Sputtered fraction, by mass, for grains as a function of
sputtering timescale, $\tau_p=\int_0^t n_p dt$. Solid black line:
Silicate grains; Dashed line: Graphite grains. Plots
are for dust in the ISM of the Milky Way.
\label{300-1000}
}
\end{figure}

\newpage
\clearpage

\begin{figure}
\figurenum{7.18}
\includegraphics[width=12cm]{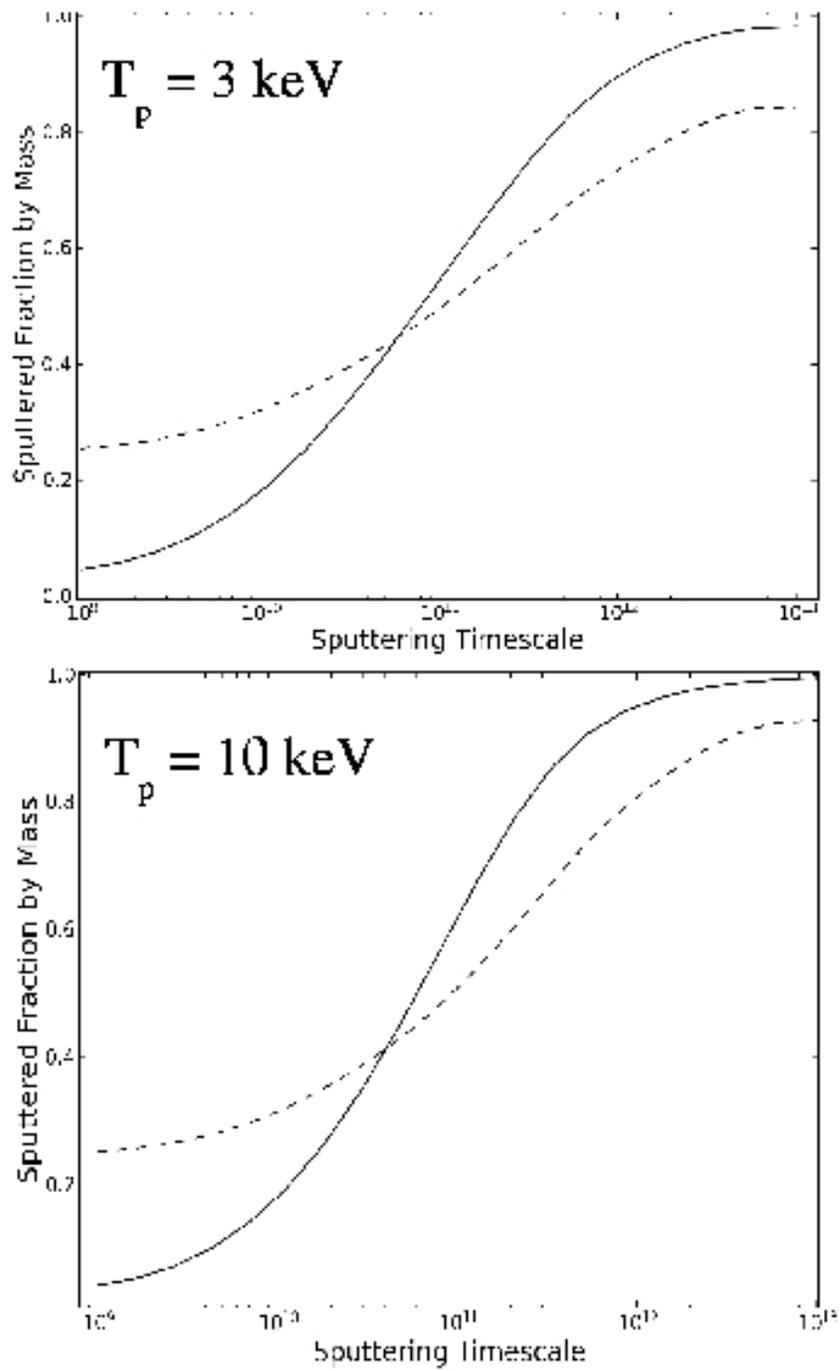}
\caption{Same as Figure 7.17, but for post-shock proton temperatures
of 3 and 10 keV.
\label{3000-10000}
}
\end{figure}

\newpage
\clearpage

\section{{\it Spitzer} IRS Observations of Two Young Type Ia Supernova
Remnants in the LMC}

This chapter is part of a manuscript in preparation for submission to
the Astrophysical Journal.

\subsection{Introduction}

Supernova remnants (SNRs) provide a laboratory to study various
aspects of interstellar medium (ISM) evolution across the whole
electromagnetic spectrum. With the advent of high spatial resolution
telescopes in both the X-ray and infrared (IR) regimes, it is possible
to probe the interaction of the rapidly moving shock wave with the
dust and gas of the surrounding ambient medium, in addition to
studying the ejecta products of the SN itself. The expanding shockwave
sweeps up and heats gas to $10^{6}-10^{7}$ K, causing it to shine
brightly in X-rays. Dust grains embedded in the hot, shocked plasma
are collisionally heated, causing them to radiate at IR
wavelengths. The same collisions with hot ions slowly destroy grains
in the process via sputtering. Because the physical processes behind
X-ray and IR emission are related, a combined approach to studying
SNRs using both energy ranges can reveal more information than either
could on its own.

To better characterize IR emission from SNRs, we conducted an imaging
survey with the {\it Spitzer Space Telescope} of $\sim 40$ known SNRs
in the Large and Small Magellanic Clouds. The Clouds were chosen
because of their known distance and relatively low Galactic IR
background. We conducted spectroscopic follow-up with the Infrared
Spectrograph (IRS) on {\it Spitzer} on several remnants during cycle 4
of observations. We report here on IRS observations of two of these
objects, SNRs B0509-67.5 (0509) and B0519-69.0 (0519). Both are
remnants of thermonuclear SNe \citep{smith91} that have fast,
non-radiative shocks (several thousand km s$^{-1}$)
\citep{tuohy82,ghavamian07}.  There is no evidence in either for
slower, radiative shocks. They are both located in the LMC, at the
known distance of $\sim 50$ kpc. In addition, both have ages
determined from light echoes \citep{rest05} to within 33\%, 0509 being
400 $\pm 120$ and 0519 being 600 $\pm 200$ years old. They are nearly
identical in size, having an angular diameter of $\sim 30$'', which
corresponds to a physical diameter of $\sim 7.3$ pc.

In \citet{borkowski06} hereafter Paper I, we used photometric
detections of both SNRs to put limits on the post-shock density and
the amount of dust destruction that has taken place behind the shock
front, as well as put a limit on the dust-to-gas mass ratio in the
ambient medium, which we found to be a factor of several times lower
than the standard value for the LMC of $\sim 0.25$\% (Weingartner \&
Draine 2001, hereinafter WD01). These limits, however, were based on
only one IR data point, the 24 $\mu$m detection. With full
spectroscopic data, we can place much more stringent constraints on
the dust destruction and dust-to-gas mass ratio in the ISM. We also
explore alternative dust models, such as porous and composite
grains. Additionally, we examine archival data from the Reflection
Grating Spectrometer (RGS) onboard the {\it XMM-Newton} X-ray
observatory. The high spectral resolution of RGS allows us to separate
lines from the ambient medium from those arising from reverse-shocked
ejecta, giving us an emission measure (EM), or an upper limit, of the
shocked ISM. The EM allows us to infer a pre-shock density, given
information derived from IR fits.

\subsection{Observations and Data Reduction}
\label{obs}

Both objects were fully mapped by the long wavelength (14-38 $\mu$m),
low-resolution ($\Delta\lambda/\lambda$ $\sim 100$) (LL) spectrometer
on {\it Spitzer's} IRS. For 0509, we stepped across the remnant in
seven LL slit pointings, stepping perpendicularly 5.1$''$ each time
(5.1$''$ is half the slit width). We then shifted the slit positions
by 56$''$ in the parallel direction and stepped across the remnant
again, ensuring redundancy for the entire object. This process was
repeated for each of the two orders of the spectrograph. Each pointing
consisted of two 120-second cycles, for a total observing time of 6720
s.

For 0519, the process was identical in terms of number of positions
and step sizes in between mappings, but each pointing consisted of
four 30-second cycles, for a total observing time of 3360 s. The
difference was due to the fact that 0519 is several times brighter in
the wavelength range of interest, and we did not want to saturate the
detectors.

The spectra were processed at the {\it Spitzer Science Center} using
version 17.1 of the IRS pipeline. We ran a clipping algorithm on the
data which removes both hot and cold pixels that are more than
3$\sigma$ away from the average of the surrounding pixels. To extract
the spectra, we used SPICE, the {\it Spitzer} Custom Extraction
tool. Once all datasets were extracted, we stacked spectra from the
same spatial location, improving the signal-to-noise ratio of the
sources. For background subtraction, we use the off-source slit
positions that come when one of the two slit orders in on the
source. In the end, we have seven (overlapping) background-subtracted
spectra for each remnant. As we show in Section~\ref{results}, this
allows us to do spatially resolved spectroscopy, despite the fact that
the remnants are only $\sim 30''$ in diameter.

We processed archival {\it XMM-Newton} RGS data from the XMM Science
Archive with version 8.0 of the Science Analysis System (SAS)
software for XMM. 0509 was observed on 4 July 2000 (Obs. ID
0111130201, PI M. Watson) for 36 ksec. 0519 was observed on 17
September 2001 (Obs. ID 0113000501, PI A. Brinkman) for 25 ksec. Since
RGS is a slitless spectrometer, spatial information is degraded for
extended sources, and the spectrum is smeared by the image of the
source. In order to model RGS spectra, the response files generated by
SAS must be convolved with a high-resolution X-ray image. For our
purposes, we used archival broadband Chandra images, along with the
FTOOL {\it rgsrmfsmooth}, to produce new response matrices.

\subsection{Results}
\label{results}

\subsubsection{0509}

\begin{figure}
\figurenum{8.1} \includegraphics[width=12cm]{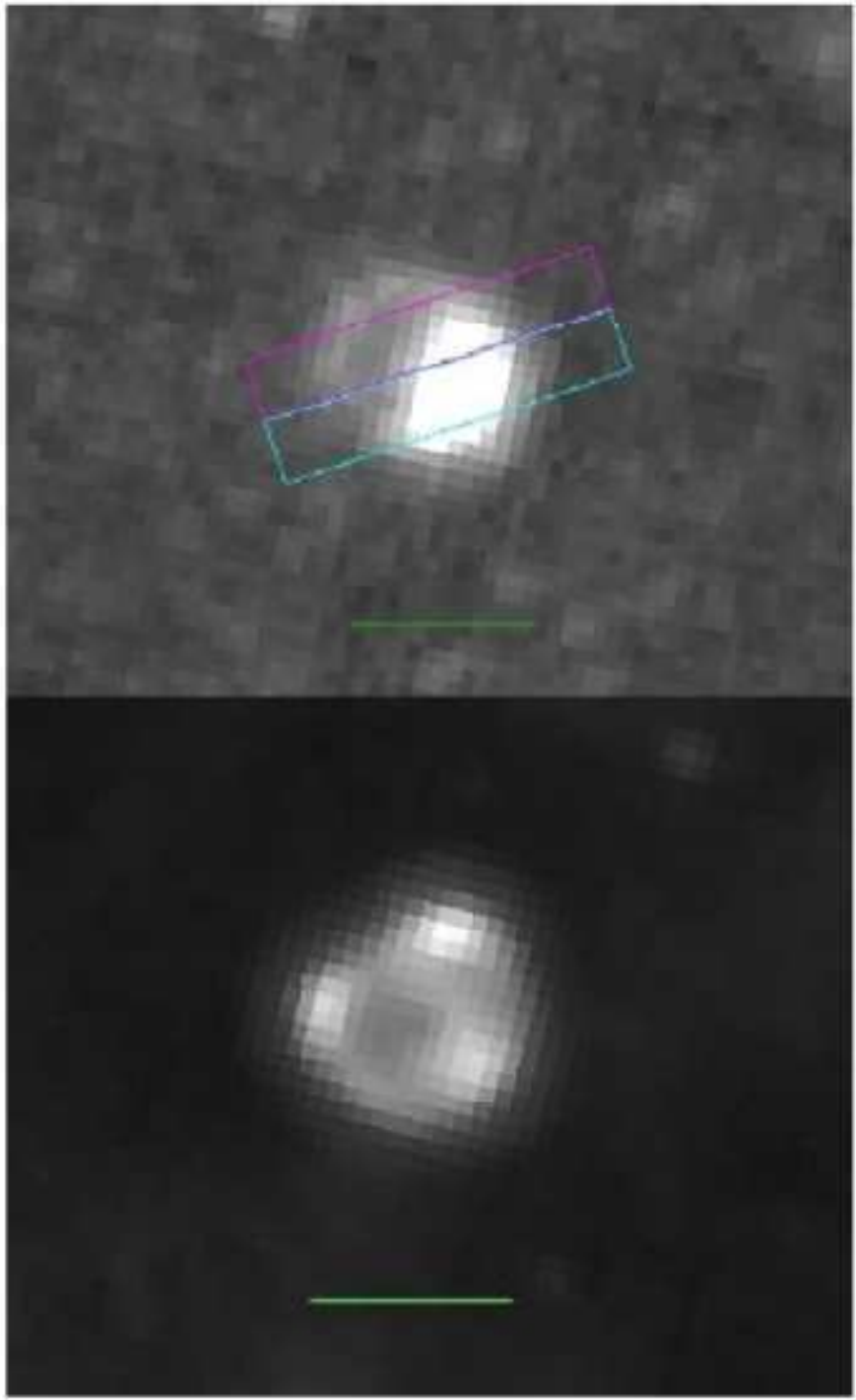}
\caption{Top: MIPS 24 $\mu$m image of SNR 0509-67.5, overlaid with
regions of spectral extraction as described in text, where magenta
region is ``faint'' region and cyan marks ``bright'' region. Bottom:
MIPS 24 $\mu$m image of SNR 0519-69.0. Green bar on both images
corresponds to 30$''$. FWHM of MIPS 24 $\mu$m PSF is approximately
7$''$.
\label{24um}
}
\end{figure}

In Figure 8.1, we show the MIPS 24 $\mu$m images of 0509 and 0519,
with overlays as described in the caption. Immediately obvious is the
large asymmetry in 0509, where the remnant brightens by a factor of 5
between the faint NE hemisphere and the brighter SW. This is a much
higher flux contrast than seen in Chandra broadband images, which show
a modest factor of 1.5 between the two hemispheres. Using IRS, we are
able to separate the remnant into 7 overlapping regions from which
spectra were extracted. For our analysis, we chose two regions (shown
on Figure~\ref{24um}) that do not overlap spatially and provide
adequate signal to noise spectra. These regions roughly correspond to
the bright and faint halves of the remnant, and the spectra are shown
in Figure~\ref{0509faint} and Figure~\ref{0509bright}. The ratio of
integrated fluxes from 14-35 $\mu$m for the two regions is $\sim
2.5$. This differs from the factor of 5 measured from the photometric
images because the slits are not exactly aligned with the bright and
faint halves of the remnant, and because of the differences in the
MIPS and IRS bandpasses.

A few obvious features immediately stand out about both
spectra. First, there is no line emission at all seen in either
spectrum. This is not unexpected, since the shocks in 0509 are some of
the fastest SNR shocks known, at $\sim 6000$ km s$^{-1}$
\citep{ghavamian07}, and one does not expect any radiative cooling in
the gas at such shock velocities. Second, though both spectra show
continuum from warm dust, there are obvious differences in the
spectra. An inflection around 18 $\mu$m can be seen in the spectrum
from the bright half, while this feature is not as clear in the faint
spectrum. We attribute this feature to the Si-O-Si bending mode in
amorphous silicate dust. In Section~\ref{disc}, we will explore
reasons for the different spectral shapes and the differences in
brightness between the two halves of the remnant.

\subsubsection{0519}

0519 does not show the large scale asymmetries that 0509 does,
although three bright knots can be seen in the 24 $\mu$m image. These
same three knots can be identified in both H$\alpha$ and X-ray images,
where {\it Chandra} broadband data show that the three knots are
prominent in the 0.3-0.7 keV band. From the MIPS 24 $\mu$m image, we
measured the total flux from the three knots added together and found
that collectively they represent only about 20\% of the
flux. Nevertheless, spectra extracted from slits that contain a bright
knot do show differences in continuum slope from those extracted where
no knots are present. We discuss our interpretation of these knots in
Section~\ref{disc}. As with 0509, there are no lines seen in the IRS
data (see Figure 8.4), as shock speeds are also quite high in this
object.

\subsection{IRS Fits}

We now turn our attention to modeling the emission seen in IRS with
numerical models of collisionally heated dust grains. We follow a
procedure identical to that followed in our previous work on these
objects; see previous chapters. A more complete description of models
can be found there; see also \citet{williams06,blair07}. Briefly, we
model warm dust heated by collisions with electrons and ions, taking
into account sputtering by ions, which destroys small grains and
sputters material off large grains. We use a plane-shock model which
superimposes regions of increasing shock age (or sputtering timescale)
$\tau_p=\int_0^t n_p dt$, where $n_{p}$ is post-shock proton density,
while keeping temperature behind the shock constant. Inputs to the
model are an assumed pre-shock grain-size distribution, grain type and
abundance, proton and electron density, ion and electron
temperature. The code calculates the heating and sputtering for grains
from 1 nm to 1 $\mu$m, producing a unique grain temperature and
spectrum for each grain size. The spectra are then added in proper
proportions according to the post-sputtering size distribution to
produce a model spectrum that can be compared with observations. We
use sputtering rates from \citet{nozawa06} with enhancements for small
grains from \citet{jurac98}. We assume that sputtering yields are
proportional to the amount of energy deposited by incoming particles,
accounting for partial transparency of grains to protons and alpha
particles at high energies \citep{serradiazcano08}, particularly
relevant for 0509 and 0519. We assume that grains are compact spheres,
but we also report results and implications if grains are porous
(i.e. contain a filling fraction of vacuum $>$ 0).

In general, our method is to fix as many of the input parameters as
possible, based on what is known from other observations. For
instance, \citet{rest05} used optical light echoes to constrain the
ages of both remnants, yielding ages of $\sim 400$ yrs. for 0509 and
$\sim 600$ yrs. for 0519. Shock velocities are known approximately for
both objects ($\sim 6000-7000$ km s$^{-1}$ for 0509 and $\sim 3800$ km
s$^{-1}$ for 0519) \citep{tuohy82,ghavamian07} from measurements of
proton temperatures. The main ``fitted'' parameter that remains is the
{\it post}-shock density of the gas, $n_{H}$, which can then be tuned
to reproduce the shape of the observed IR spectra. The overall
normalization of the model to the data (combined with the known
distance to the LMC) provides the mass in dust. Since our models
calculate the amount of sputtering that takes place in the shock, we
can then determine the amount of dust present in the pre-shock
undisturbed ISM. While we are sensitive only to ``warm'' dust, and
could in principle be underestimating the mass in dust if a large
percentage of it is too cold to radiate at IRS wavelengths, it is
difficult to imagine a scenario where any significant amount of dust
goes unheated by such hot gas behind the high-velocity shocks present
in both remnants, particularly given the upper limits at 70 $\mu$m
reported in \citet{borkowski06} (and the fact that there are no
radiative shocks observed anywhere in these remnants).

\begin{figure}
\figurenum{8.2}
\includegraphics[width=16cm]{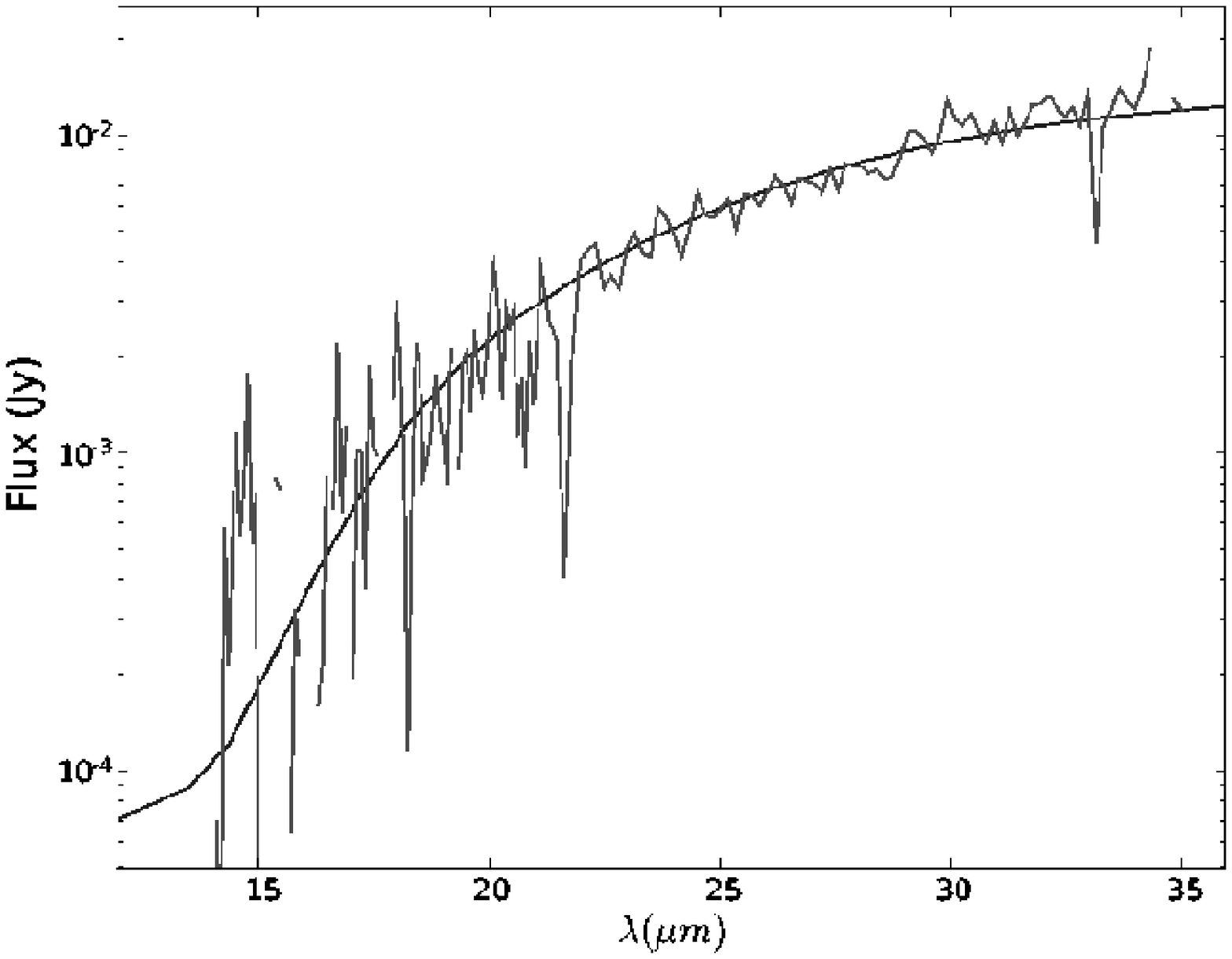}
\caption{14-35 $\mu$m IRS spectrum of the ``faint'' region of 0509, overlaid with model fit.
\label{0509faint}
}
\end{figure}

\begin{figure}
\figurenum{8.3}
\includegraphics[width=16cm]{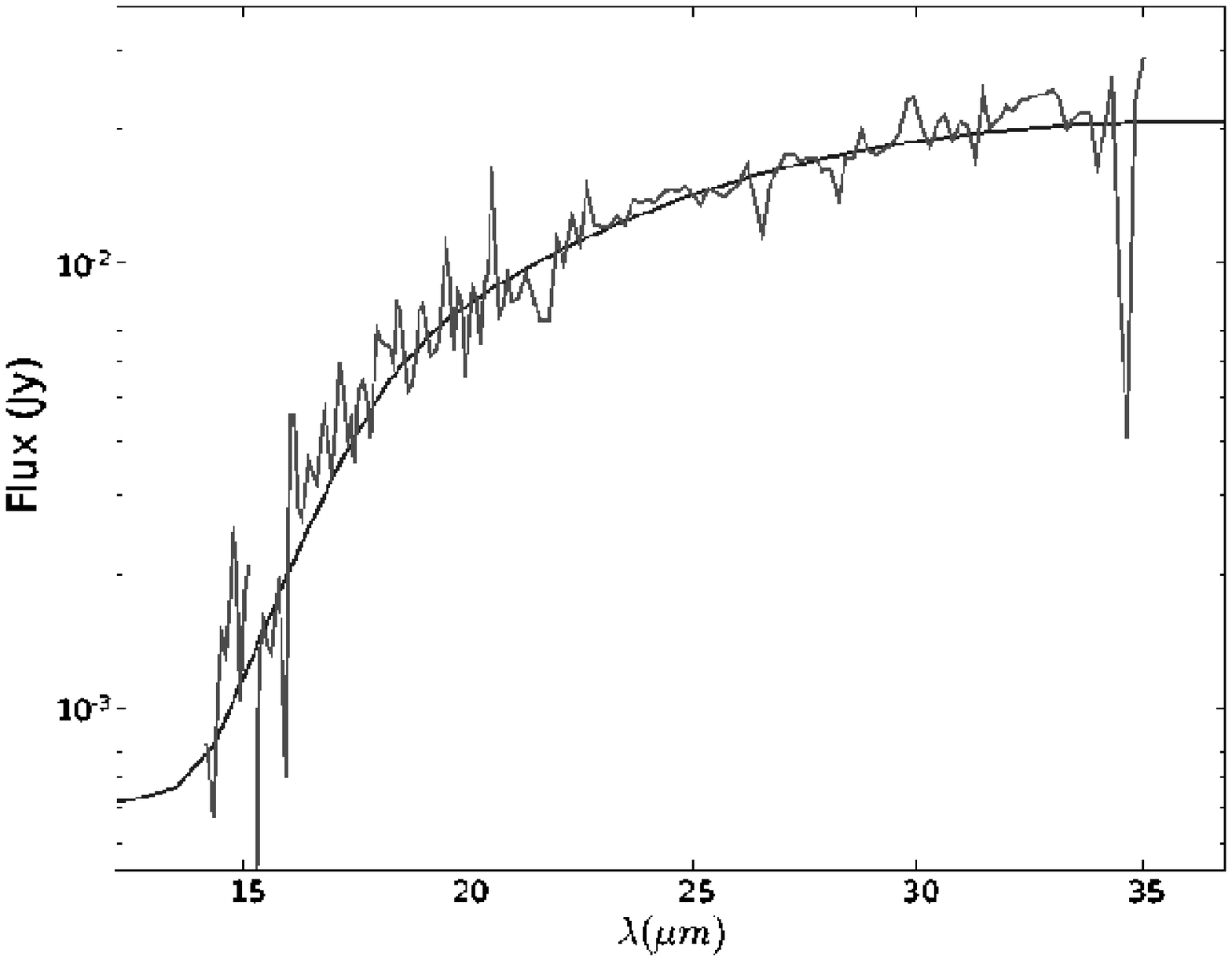}
\caption{14-35 $\mu$m IRS spectrum of the ``bright'' region of 0509, overlaid with model fit.
\label{0509bright}
}
\end{figure}

Although we have full spectral mapping of both objects, the width of
the IRS slit at 10.5$''$ severely limits the amount of spatially
resolved spectroscopy we can do on a remnant that is only 30$''$ in
diameter, as both 0509 and 0519 are. Nevertheless, we do have several
unique slit positions, and can use these to obtain spectra from
various regions to explain the various asymmetries observed in both
remnants. Because dust radiates as a modified blackbody spectrum,
emission from a small amount of warmer dust can overwhelm that from a
larger amount of colder dust, particularly at the wavelengths of
interest here. We show in Figure~\ref{0509faint} and
Figure~\ref{0509bright} extracted spectra from two regions of 0509,
which we label the ``bright'' and ``faint'' portions of the
remnant. If each of these two regions represents half of the object,
this requires a density contrast of $\sim 4$ in the post-shock gas,
with the higher density required in the ``bright'' region (higher
densities means hotter dust, hence more short wavelength
emission). Although it may be possible that a density gradient from
the NE to the SW of this order does actually exist in the ambient
medium surrounding 0509 (the angular size of 30$''$ in the LMC
corresponds to a linear diameter of about 7 pc), the implications from
this model lead to several scenarios which are unlikely and require an
appeal to special circumstances.

In comparing the MIPS image with images of the remnant at other
wavelengths, one can clearly see an enhancement in the {\it Chandra}
broadband image \citep{warren04} on the west side of the remnant. This
enhancement is mostly present in the energy range containing Fe
L-shell emission lines. The H$\alpha$ image, shown in
Figure~\ref{halpha}, shows a uniform periphery around the remnant,
with the exception of a brightness enhancement in the SW, relatively
well constrained to a small region a few arcseconds in size. If the
H$\alpha$ image is smoothed to the resolution of the MIPS 24 $\mu$m
image ($\sim 7''$), the result is morphologically similar to the 24
$\mu$m image. From these comparisons, we conclude that there is not an
overall NE-SW density gradient in the ISM, rather, the ISM is mostly
uniform except for the SW, where the remnant is running into a
localized region of higher density. To obtain the conditions for the
whole of the remnant, we fit a model to the ``faint'' region, freezing
all parameters to those reported in Table 8.1 and allowing post-shock
density to vary (which also causes the sputtering timescale to
vary). If we assume a standard LMC dust model, we get a post-shock
density of $n_{H}$ = 0.88 cm$^{-3}$.

Our assumption that the bright region of 0509 is the result of a more
localized enhancement in density implies that the spectrum extracted
from that region is then a composite of emission from the uniform
parts of the remnant and from the small, denser region. {\it Spitzer}
does not have the resolution required to isolate this dense region,
but we can make a crude isolation by assuming that the two slit
positions cover an equal surface area on the remnant, and subtracting
the faint spectrum from the bright. The residual spectrum is of less
than ideal signal-to-noise. Nevertheless, a fit to this spectrum
implies a gas density of $\sim 7$ cm$^{-3}$, or about an order of
magnitude higher than the rest of the remnant. A small region of hot
dust outshines a more massive region of cooler dust at 24 $\mu$m,
which explains the factor of 5 ratio in the flux between the two
sides. X-rays from this object are dominated by ejecta, not swept-up
ISM, and there are several factors to consider beyond density when
considering the flux of H$\alpha$ coming from a region. Higher
resolution observations at all wavelengths will shine more light on
this issue.

\begin{figure}
\figurenum{8.4}
\includegraphics[width=16cm]{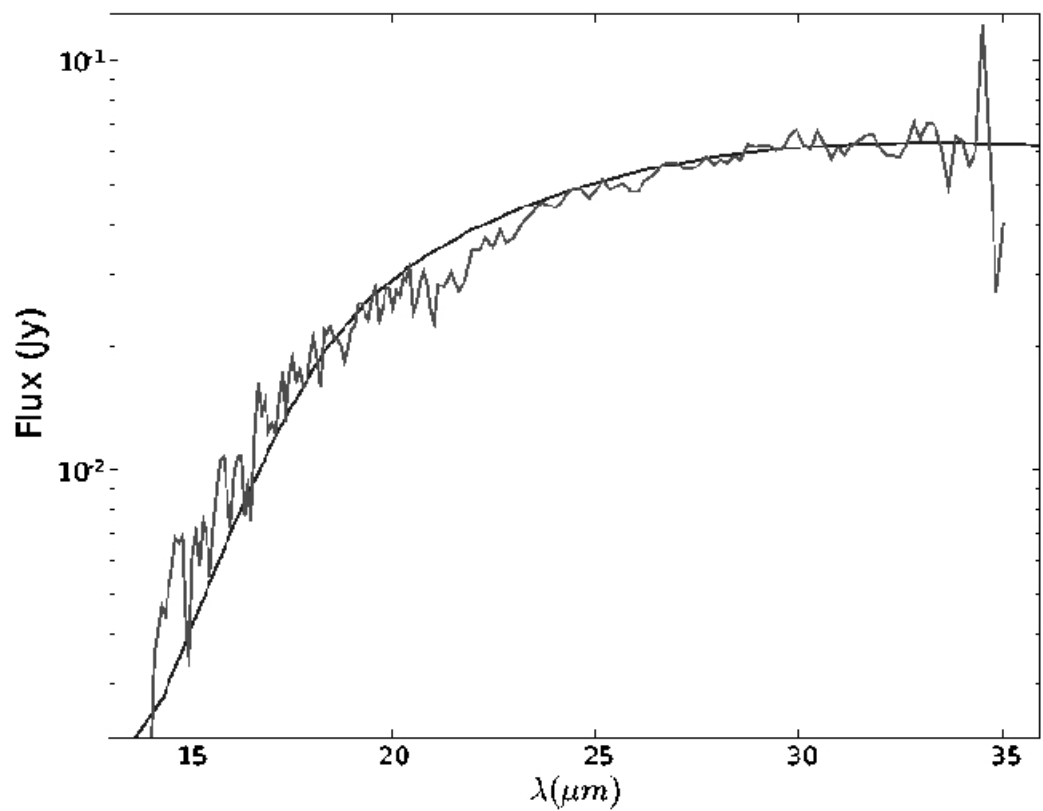}
\caption{14-35 $\mu$m IRS spectrum of SNR 0519-69.0, extracted from a
slit placed across the middle of the remnant, free of emission from
the bright knots seen in the 24 $\mu$m image; with model overlaid.
\label{0519irs}
}
\end{figure}

For 0519 we have a similar scenario, except that instead of one bright
region, we have three bright knots. As noted previously, these knots
correspond spatially with knots seen in both X-rays and H$\alpha$ (see
Figure~\ref{halpha}). We adopt an identical strategy here, isolating
spectra from slit positions that do not overlap with one of the bright
knots. As it turns out, there is only one slit position in this object
that is nearly completely free of emission from a knot, a slit
position that goes directly across the middle of the remnant. The
spectrum from this region is shown in Figure~\ref{0519irs}. We assume
that the conditions within this slit position are indicative of the
remnant as a whole. Using parameters found in Table 8.1, we obtain a
post-shock density of $n_{H}$ = 8 cm$^{-3}$. We do not have adequate
signal-to-noise to isolate one of the bright knots alone. However, we
approximate the density required by noting that a fit to the entire
slit position that contains the brightest of the three knots (the
northernmost knot) requires a density that is higher by a factor of
$\sim 2$. Given that this spectrum contains emission from both the
uniform parts of the remnant and the bright knots, it is likely that
the density in the knots themselves are perhaps a factor of $\sim 3$
higher that the more uniform parts of the remnant.

\subsection{X-ray Modeling of RGS Data}

Although dust models in the IR are a powerful diagnostic of the
post-shock gas, they are insensitive to pre-shock gas conditions,
while the total swept mass depends on this. X-ray modeling of SNR
spectra, on the other hand, can provide a handle on the pre-shock gas
through the emission measure (EM) of the gas, defined as EM =
$\int_{0}^{V} fn_{e}n_{H} dV$, where $n_{e}$ and $n_{H}$ are the
post-shock electron and hydrogen densities, respectively, $V$ is the
emitting volume of hot gas, and $f$ is the filling fraction of the
material. If the densities do not vary greatly over the emitting
volume, the EM can be rewritten as EM = $n_{e} n_{H} V$. Determining
the density directly from the EM can be done, but with the caveat that
any clumping of the gas will bias the result, owing to the factor of
$n^{2}$ in the definition.

If $n_{e}$ can be independently determined (such as through IR
observations of warm dust), then it can be divided out of the emission
measure, leaving only the product of n$_{H}$ and $V$, which,
regardless of any clumping effects, is simply the mass in gas swept up
by the forward shock. The volume of these remnants is well-known,
thanks to their resolved angular size and location in the LMC ($D$ =
50 kpc). Thus, it is a straightforward calculation to determine the
average pre-shock density that the forward shock has encountered over
its lifetime. A caveat to this approach is that the swept-up gas must
still be hot enough to emit X-rays, but for remnants at a relatively
early stage of evolution, as is the case for both 0509 and 0519, this
is not a concern.

The difficulty in this approach lies in disentangling X-ray emission
from the shocked ambient medium from that arising from reverse-shocked
ejecta. For large remnants, this could be done spatially using CCD
spectra, but both 0509 and 0519 are $\sim 30''$ in diameter with the
ejecta not well separated from the shocked ISM, rendering such an
approach impossible. For 0509, previous studies
\citep{warren04,badenes08} have used Chandra CCD spectra to model
emission from the remnant, but these studies are subject to the
inherent difficulty in using CCD spectra to disentangle ISM and ejecta
emission. An alternative approach is to take high-resolution spectra
from grating spectrometers, such as RGS on {\it XMM-Newton}. If lines
can be identified as arising only from shocked ambient medium, then
model fits to these lines would provide an EM for the ISM. This can be
most easily done with N lines, as virtually no N is produced in type
Ia explosions. \citet{kosenko08} used this approach to fit the 24.8
\AA\ N Ly-$\alpha$ line in the spectrum of 0509, but we find that this
line is very weak and unconstraining to the fitting procedure used
here. Here we fit O and Ne lines seen in RGS spectra and attribute
them to the shocked ambient medium, since neither element appears to
be produced in great abundance in type Ia SNe \citep{marion09}.

\subsubsection{0509}

The preferred explosion model for 0509, DDTa \citep{badenes08},
synthesizes 0.04 $M_\odot$ of O, the least of any of the explosion
models considered for that remnant. In Figure~\ref{rgsspectra}, we
show RGS spectra from both remnants. Fe L-shell lines dominate the
spectra, but strong lines from both H- and He-like O are clearly
visible in both cases at 18.97 and 21.7 \AA. Ne lines blend together
with Fe lines between 12 and 14 \AA; we attempt to disentangle them as
described below. Our procedure for modeling lines is similar to that
in \citet{blair07} for Kepler's SNR; see Appendix A for a more
detailed description. An additional step we took here was to first fit
a model to only the data between 14-18 \AA, which are dominated by
iron lines. We then added a component ({\it vpshock} model with fixed
normal LMC abundances with ionization timescale and normalization
free) to fit the data around Ne and O lines, fitting the data from
9-23 \AA\ with the two models. These two models can be thought of as
the ``ejecta'' model and the ``ISM'' model. We use two absorption
components, one resulting from galactic absorption column density,
$N_{H}$, equal to 5 $\times$ 10$^{20}$ cm$^{-2}$, and one from the LMC
(at fixed LMC abundances), equal to 2 $\times$ 10$^{20}$ cm$^{-2}$
\citep{warren04}. The ``ejecta'' model is important only for
disentangling Ne and Fe emission, it has no significant effect on
fitting O lines. In the ``ISM'' model, we freeze the electron
temperature to 2 keV, roughly the number one gets from Coulomb heating
models behind the shock (the dependence of the emission measure of the
gas on the electron temperature is small). We introduce an additional
line smoothing parameter to both models to allow for the fact that
lines are broadened by Doppler motions. For 0509, we find that
relative fluxes in \ion{O}{7}, \ion{O}{8}, \ion{Ne}{9}, and
\ion{Ne}{10} are well fit (reduced $\chi^{2}$ = 1.2 for 235 d.o.f.) by
a model with $\tau_{i}=\int_0^t n_e dt$ = 3.5 (3.1, 4.0) $\times$
10$^{10}$ cm$^{-3}$ s and EM = 2.0 (1.9, 2.2) $\times$ 10$^{58}$
cm$^{-3}$, where all errors are 90\% confidence limits. The oxygen
line width obtained from fitting is $v_{\sigma}$ = 5250 (4775, 5800)
km s$^{-1}$.

Given our independent measurement of electron density from IR fits, we
are now able to divide it out of the EM, leaving only the swept-up
mass in gas. RGS data do not provide the spatial information required
to separate the two halves of the remnant, so RGS derived EMs are
spatially integrated. Using $n_{e}$ of 0.9 cm$^{-3}$ yields a swept-up
gas mass of 17 $M_\odot$. When combined with the physical size of the
remnant (3.6 pc in radius at a distance of 50 kpc), the compact grain
model yields a volume averaged pre-shock density of $\rho_{0}$ $\le $
5.6 $\times$ 10$^{-24}$ g cm$^{-3}$ ($n_{0}$ $\le$ 2.4 cm$^{-3}$). It
is important to note that the pre-shock densities and swept-up gas
masses determined in this fashion should be considered upper limits,
as any ejecta contribution to the O and Ne lines would reduce the
emission measure of the ISM model. The type Ia explosion models of
\citet{badenes03} contain some unburned O and C as ejecta products.
However, recent near-IR observations of type Ia SNe \citep{marion09}
suggest that the entire progenitor is burned in the explosion and that
O and Ne are byproducts of carbon burning.

If we divide the post-shock $n_{H}$ from IR fits by the pre-shock
$n_{0}$ from X-ray fits, we can get a lower limit on the compression
ratio for the two models. Under standard ``strong-shock'' jump
conditions, the compression ratio for gas swept-up by the shock wave
is 4, but cosmic-ray modification of the shock front will raise this
ratio \citep{jones91,berezhko99}. For the numbers presented above,
$n_{H}/n_{0}$ $\ge$ 0.35. Obviously this result is unphysical
(although it is only a lower limit) as one does not expect a
``rarefaction'' wave; we present in Section 8.6 an alternative
scenario that can resolve this issue.

\subsubsection{0519}

\begin{figure}
\figurenum{8.5}
\includegraphics[width=13.5cm]{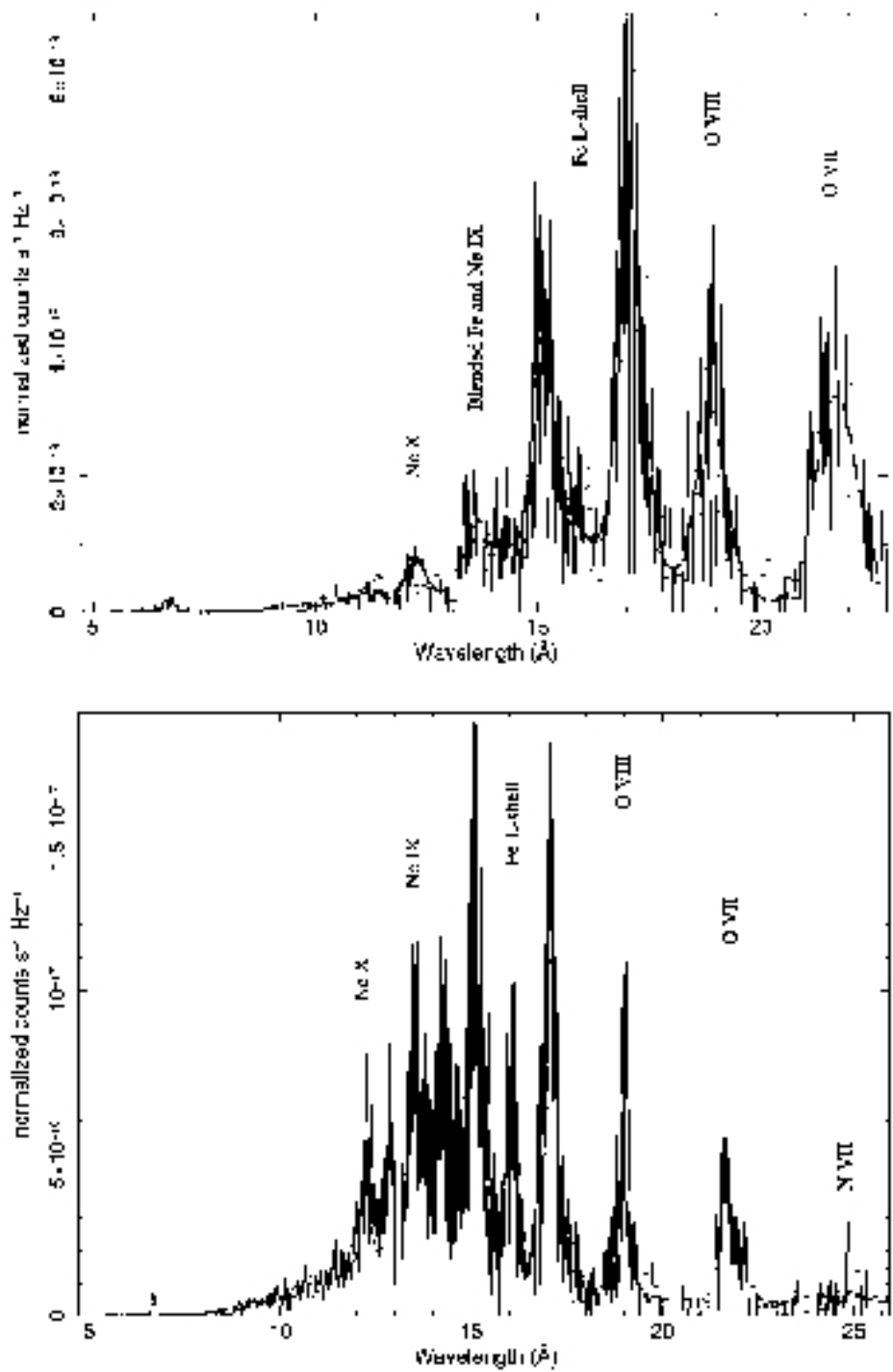}
\caption{Top: {\it XMM-Newton} RGS spectrum of 0509-67.5, from 5-23
\AA. Bottom: RGS spectrum of 0519-69.0, from 5-27 \AA; models overlaid
in both as described in text.
\label{rgsspectra}
}
\end{figure}

\begin{figure}
\figurenum{8.6}
\includegraphics[width=15cm]{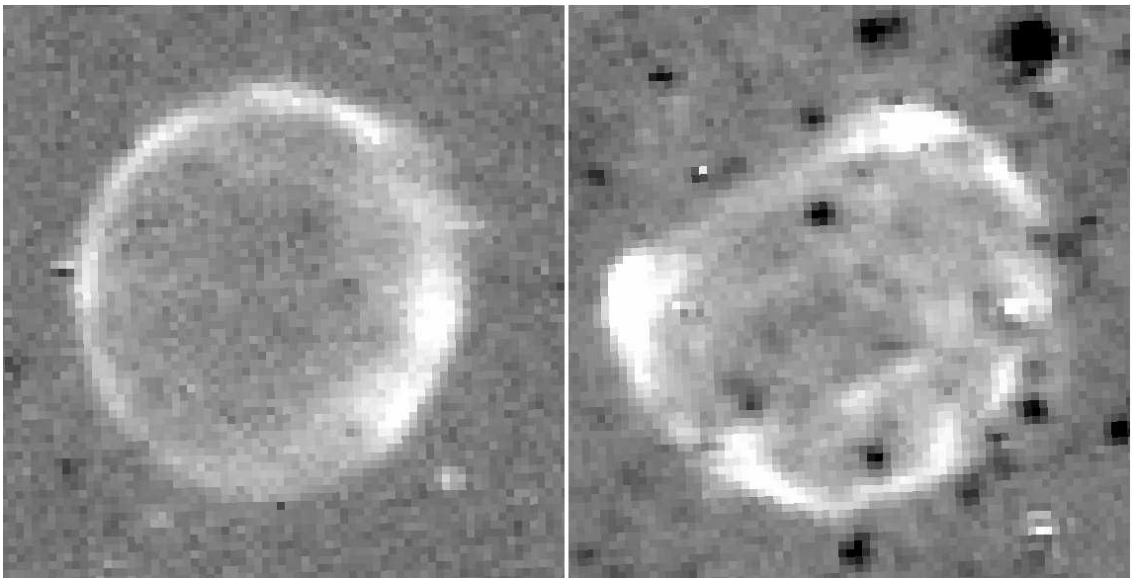}
\caption{Star-subtracted H$\alpha$ images of 0509 (left) and 0519 (right). Note
brightness enhancement in SW for 0509, and location of bright knots in
0519.
\label{halpha}
}
\end{figure}

Our technique for fitting RGS data from 0519 was identical, except
that we extended the fitting range out to 27 \AA\ to account for the
fact that we had better signal-to-noise at long wavelengths in this
case. The fit to the data can be seen in Figure~\ref{rgsspectra}. We
used two absorption components; one for the galaxy with $N_{H}$ = 6
$\times$ 10$^{20}$ cm$^{-2}$ and one for the LMC with $N_{H}$ = 1.5
$\times$ 10$^{21}$ cm$^{-2}$. We find an ionization timescale of 2.26
(2.17, 2.46) $\times$ 10$^{11}$ cm$^{-3}$ s, and an EM of 1.60 (1.47,
1.62) $\times$ 10$^{59}$ cm$^{-3}$. We obtain a line width for oxygen
lines of $v_{\sigma}$ = 1475 (1350, 1700) km s$^{-1}$. Using our
IR-derived value of $n_{e}$ = 9.6 cm$^{-3}$, we obtain a swept-up gas
mass of 13.9 $M_\odot$. The remnant has a nearly identical radius to
that of 0509 (3.6 pc) at 50 kpc, which yields a pre-shock density of
$\rho_{0}$ $\le$ 4.67 $\times$ 10$^{-24}$ g cm$^{-3}$ ($n_{H_{0}}$
$\le$ 2.0 cm$^{-3}$).  As with 0509, these estimates for pre-shock
density are upper limits, as any ejecta contribution to O lines would
lower the EM. The lower limit on the compression ratio at the shock is
then $n_{H}/n_{H_{0}}$ $\ge$ 4.

\subsection{Discussion}
\label{disc}

While the values of density for 0519 seem reasonable, there is a clear
issue in the case of 0509, where the inferred pre-shock density is
higher than the post-shock density. Of course, the pre-shock density
is only a lower limit, and will increase if there is a contribution
from the ejecta to oxygen and neon lines. While this is possible and
cannot be excluded, bringing the pre-shock density down to a
reasonable value by this method would require a large amount of oxygen
expelled in the explosion, and this scenario might be at odds with
both explosion models and recent observations of type Ia SNe, as
discussed above. Even if this were the case, the pre-shock density
would have to be lowered to a value of n$_{0}$ = 0.21, which is a
factor of 2 lower than reported from hydrodynamic simulations by
\citet{badenes08}.

We present here an alternative possibility that is based on relaxing
the assumption that dust grains are compact spheres with a filling
fraction of unity. We have updated our dust codes to include the
porosity of dust grains (where porosity, $\cal P$, is the volume
fraction of the grain that is filled with vacuum, instead of grain
material), as well as allowing for grains to be composite in nature,
i.e. a grain made up of silicates, graphite, amorphous carbon, and
vacuum. The details of this model are discussed in Chapter 7. What
remains here is to choose a grain ``recipe'' to use to fit the
data. Since there are few constraints from the ISM at this time, we
have adopted here a value of $\cal P$ = 50\%.

Our ``{\it compact}'' model, which we used to derive results given
above, assumes separate populations of non-porous silicate and
graphite grains with appropriate size distributions for the LMC
(WD01). For our ``{\it porous}'' grain model, we use a slightly
modified version of the model of \citet{clayton03}, where grains are
composed of 50\% vacuum, 33.5\% silicate, and 16.5\% amorphous carbon,
distributed from 0.0025 to 1.5 $\mu$m. We also report results in Table
8.2 from an older porous grain model, that of \citet{mathis96}, where
we assume a grain composed of 50\% vacuum, 42\% silicate, and 8\%
graphite. The proportion of silicate to graphite in this model is
roughly what one expects for the overall grain populations in the
LMC. The Mathis model contains a distribution of porous grains from
0.01-1.0 $\mu$m, which constitute 85\% of the population by mass, and
a small population of compact silicate grains below 0.01 $\mu$m
containing the remaining 15\% of the mass
\citep{mathis96}. \citet{heng09} calculated X-ray scattering
properties of various composite porous grains and found that grains
with $\cal P$ $\ge$ 0.55 are ruled out. Thus, in considering models
that are both fully compact and 50\% porous, we should be bracketing
the range of values seen in the ISM. The narrow bandwidth of the IRS
data does not allow us to determine which model fits the data better,
as we obtained equally good fits from all models (albeit with a
different value of density), and report results from all. A caveat to
using these porous grain distributions is that both of them were
optimized for the galaxy, and not the LMC. To the best of our
knowledge, however, a size distribution appropriate for porous grains
in the LMC has not yet been developed.

The results are as follows. For 0509, the ``porous'' model of
\citet{clayton03} yielded a best fit to the data with a post-shock
density of $n_{H}$ = 3.0 cm$^{-3}$. When combined with the EM from
RGS, the inferred shocked gas mass is $\le$ 4.65 $M_\odot$, which
gives a pre-shock density of $\rho_{0}$ $\le$ 1.54 $\times$ 10$^{-24}$
g cm$^{-3}$ ($n_{H_{0}}$ $\le$ 0.66 cm$^{-3}$). This gives
$n_{H}/n_{H_{0}}$ $\ge$ 4.5, where the lower limit again assumes no
contribution to O and Ne lines from ejecta. This result is roughly
consistent with a standard strong shock wave, though it does not rule
out a cosmic-ray modified shock. This value of density is also quite
close to the favored value of \citet{badenes08} of $n_{0}$ = 0.43
cm$^{-3}$.

For 0519, the porous model gives a post-shock density of $n_{H}$ = 22
cm$^{-3}$. This yields a shocked-gas mass of $\le$ 5.0 $M_\odot$ and a
pre-shock density of $\rho_{0}$ $\le$ 1.7 $\times$ 10$^{-24}$ g
cm$^{-3}$ ($n_{H_{0}}$ $\le$ 0.72 cm$^{-3}$). The value of
$n_{H}/n_{H_{0}}$ $\ge$ 30 is quite high, partially a result of using a
porosity of 50\%, which is at the upper end of the preferred values of
grain porosity in the ISM. It is possible that the compression ratio
in the shock is actually higher than 4 due to cosmic-ray modification,
but the current data are insufficient to determine this conclusively.

So which of the dust models listed above is the ``correct'' one? We
seem to be in a situation where the spectrum and dynamics of 0509 are
better fit with a porous grain model with highly porous ({\cal P}
$\sim 50$\%) grains, while those of 0519 are better fit with a compact
grain model assuming solid grains. The current data do not allow us to
definitively resolve this issue, and it is conceivable that either
model could be correct for either case, since neither is ruled out by
physics. 

\subsubsection{Dust-to-Gas Mass Ratio}

Of particular note in all results presented in Table 8.2 is that the
inferred dust-to-gas mass ratios are always lower than what is
expected in the LMC ($\sim 0.25$\%, WD01), even after accounting for
grain destruction. This is consistent with our previous studies of
both LMC and galactic SNRs \citep{borkowski06,williams06,blair07}. We
would, however, like to point out that while our derived dust masses
are model dependent, the dependency is small, and one would obtain a
similar result regardless of the model used. As a check of this, we
made fits to the data using a greatly simplified model, that of a
single grain at a single temperature, ignoring heating, sputtering,
and grain-size distribution. \citet{dwek87} provides approximate
analytic expressions for the mass of dust, dependent only on the
temperature of the grain and the observed IR luminosity. To within a
factor of 2, these results agree with results presented here, and
still fall well short of the expected dust-to-gas mass ratio expected
for the LMC. A similar result was found for SN 1987A
\citep{bouchet06}, although the shocks there probe circumstellar
material, and have not yet reached the ambient ISM (but see Dwek et
al., 2008).

Porous dust grains require a higher density relative to compact grains
to reproduce the same spectrum, but they are also more efficient
radiators of energy than compact grains. For this reason, the inferred
dust mass from the porous grain models, seen in Table 8.2, is lower by
a factor of $\sim 5-10$. However, for a fixed X-ray EM, the shocked
gas mass that these higher densities imply is also lower by a factor
of $\sim 3$, so the dust-to-gas mass ratio in the ambient medium would
be lower by a factor of $\sim 2$ in the porous model as compared to
the compact model.

One possibility that has been suggested for SNRs of both type Ia and
CC origin \citep{gomez09,dunne03} is that large amounts of dust are
indeed present, but are too cold (T$_{d}$ $\sim 16$ K) to be
detectable at {\it Spitzer} wavelengths and are only visible in the
sub-millimeter regime. If this is true, then the total dust-to-gas
mass ratio for the ISM could be quite different from what we report
here, which is based only on ``warm'' dust. This claim has been
disputed for Kepler's SNR \citep{blair07} based on a detection at 70
$\mu$m and an upper limit at 160 $\mu$m, but we lack detection at
either wavelength here, and upper limits are unconstraining. {\it
Herschel} observations would provide the key wavelength coverage
between the mid-IR and sub-mm necessary to settle this issue.

\subsection{Conclusions}

We present new mid-IR spectral observations of two young SNRs in the
LMC, 0509-67.5 and 0519-69.0, as well as analysis of archival
high-resolution X-ray spectra. By fitting dust heating and sputtering
models to IR spectra, we can determine the post-shock gas density. If
the emission from the shocked ISM can be disentangled from that from
ejecta in X-ray spectra, this post-shock density can be divided out of
the emission measure to determine the amount of shocked gas, which is
a direct measurement of the pre-shock density of the ISM. Of course,
this method requires that the size of the remnant be known, but this
is the advantage of using SNRs in the Magellanic Clouds, which have a
well-known distance. In principle, this method could be a powerful
tool in measuring the compression ratio at the shock front.

There are several sources of uncertainty in modeling emission from
warm dust grains, however. First and foremost, we are limited by both
the spatial resolution of {\it Spitzer} for such small SNRs and the
low signal-to-noise ratio of the spectra. We use a plane-shock model
with constant plasma temperature to model dust emission, which is not
ideal for a spherical blast wave. Estimates of swept gas mass assume
constant ambient density ahead of the shock. Sputtering rates for such
high temperature protons and alpha particles are probably only
accurate to within a factor of 2 (but see Chapter 7), and we use the
simple approximation that the sputtering rate varies according to the
amount of energy deposited into a grain of given size and
composition. Lastly, but possibly most importantly, we do not know the
composition of interstellar dust grains, in terms of the degree of
porosity and mixing of various grain types.

High-resolution {\it XMM-Newton} RGS spectra allow us to disentangle
ejecta emission from shocked ambient medium emission. We consider all
O and Ne lines seen in RGS to come from shocked ISM, and we derive an
emission measure for this component, which, combined with our
post-shock electron density from IR fits, allows us to make a direct
measurement of the pre-shock density. This approach is subject to the
uncertainty that it is not clear whether any part of the O lines come
from ejecta, but any ejecta contribution would only serve to lower the
EM of the ISM component, making our values for pre-shock density
determined here upper limits.

The dust-to-gas mass ratios reported here are lower by about an order
of magnitude from what is generally expected in the ISM. Since the
general properties of dust in the ISM come from optical/UV absorption
line studies averaged over long lines of sight, it is possible that
there exist large local variations on parsec scales. The ratios
presented here probe these small spatial scales. Another potential
explanation is that the sputtering rate for dust grains is
significantly understated in the literature. However, since the
deficit of dust is about an order of magnitude, sputtering rates would
also have to be increased by this amount to account for the
discrepancy. Also, since sputtering rates typically assume a compact
grain, further work in this field will be needed to determine if these
rates are appropriate for porous grains.

Studies such as this will benefit greatly from the increases in both
resolution and sensitivity of future generations of telescopes, such
as the {\it James Webb Space Telescope}. Being able to spatially
separate the dust spectra right behind the shock from that further
inside the shell will be crucial to reducing some of the uncertainties
listed above. From an X-ray point of view, studies such as this will
be greatly improved with the high spectral resolution of the {\it
International X-ray Observatory}, which will further aid in
disentangling line emission from Fe L-shell lines with those of other
elements.

\newpage
\clearpage

\begin{deluxetable}{lccccc}
\tablenum{8.1}
\tablecolumns{6}
\tablewidth{0pc}
\tabletypesize{\footnotesize}
\tablecaption{Model Input Parameters}
\tablehead{
  \colhead{Object} & $V_{s}$ (km s$^{-1}$) & $T_e$ (keV) & $T_p$ (keV) & Age (yrs.) & Ref.}

\startdata

0509-67.5 & 6000 & 2.0 & 90 & 400 & 1, 2\\
0519-69.0 & 3800 & 1.5 & 36 & 600 & 1, 2\\

\enddata

\tablecomments{References: (1) Ghavamian et al 2007, (2) Rest et al. 2005}
\label{inputtable}
\end{deluxetable}

\newpage
\clearpage

\begin{deluxetable}{lcccccc}
\vspace{-0.3truein}
\tablenum{8.2}
\tablecolumns{4}
\tablewidth{0pc}
\tabletypesize{\footnotesize}
\tablecaption{Model Results}
\tablehead{
\colhead{Object} & $n_{H}$ (cm$^{-3}$) & M$_{G}$ ($M_\odot$) & $n_{0}$ (cm$^{-3}$) & M$_{D}$ 
& M$_{D}$/M$_{G}$ & $n_{H}/n_{0}$ }

\startdata

0509-67.5 (com.)\tablenotemark{a} & 0.88 & $\le 16.7$ & $\le 2.4$ & 1.4 $\times$ 10$^{-3}$
& $\ge 8.7 \times$ 10$^{-5}$ & $\ge 0.35$ \\
0509-67.5 (Clay.)\tablenotemark{b} & 3.0 & $\le 4.65$ & $\le 0.66$ & 3.0 $\times$ 10$^{-4}$
& $\ge 6.5 \times$ 10$^{-5}$ & $\ge 4.5$ \\
0509-67.5 (Math.)\tablenotemark{c} & 2.3 & $\le 6.1$ & $\le 0.86$ & 4.1 $\times$ 10$^{-4}$
& $\ge 6.7 \times$ 10$^{-5}$ & $\ge 2.67$ \\
0519-69.0 (com.)\tablenotemark{a} & 8.0 & $\le 13.9$ & $\le 2.0$ & 1.4 $\times$ 10$^{-3}$
& $\ge 1.0 \times$ 10$^{-4}$ & $\ge 4.0$ \\
0519-69.0 (Clay.)\tablenotemark{b} & 22.0 & $\le 5.0$ & $\le 0.72$ & 3.4 $\times$ 10$^{-4}$
& $\ge 6.8 \times$ 10$^{-5}$ & $\ge 30.5$ \\
0519-69.0 (Math.)\tablenotemark{c} & 14.0 & $\le 7.9$ & $\le 1.13$ & 5.7 $\times$ 10$^{-4}$
& $\ge 7.2 \times$ 10$^{-5}$ & $\ge 12.4$ \\

\enddata

\tablenotetext{a}{Compact grain model, see text for details}
\tablenotetext{b}{Porous grain model of \citet{clayton03}, see text
for details} \tablenotetext{c}{Porous grain model of \citet{mathis96},
see text for details} \tablecomments{Column 2, {\it post-shock}
density as determined by model fit to IR data; column 3, mass in gas
swept up by forward shock, as determined from model fits to X-ray
data; column 4, {\it pre-shock} density from
M$_{G}$/(4$\pi$R$^{3}$/3), where R is radius of forward shock; column
5, mass in dust swept by forward shock, after effects from sputtering
are considered; column 6, inferred dust-to-gas mass ratio in the
ambient ISM.; column 7, inferred compression ratio of shock. All
limits are based on assumption that O lines are produced entirely by
the shocked ISM.}
\label{resultstable}
\end{deluxetable}

\newpage
\clearpage

\newpage
\clearpage

\newpage
\clearpage

\newpage
\clearpage

\newpage
\clearpage

\newpage
\clearpage

\newpage
\clearpage

\section{Summary}

Shock waves generated by supernova explosions play a crucial role in
the evolution of the ISM. By heating gas and dust, spreading new
elements throughout the cosmos, and reshaping the dust and gas-phase
abundances, SNRs are one of the dominant feedback mechanisms for
galaxies, and can mark not only the endpoint, but also the beginning
of stellar evolution. By studying SNRs, we can learn about both the
conditions and dynamics of the remnants themselves and their impact on
their surrounding medium. This work has focused on the need for a
multi-wavelength approach to studying SNRs, with emphasis on the IR
portion of the spectrum, where we see emission from warm dust grains
heated via collisions with hot protons and electrons in the post-shock
gas. IR studies of SNRs can shed light on one of the primary mysteries
of ISM physics: How much dust is present in the ISM, and where was it
formed?

\subsection{Dust-to-Gas Mass Ratio}

SNRs studied in this document are in both the Milky Way galaxy and the
LMC, of both core-collapse and type Ia origin, and span a range in
ages from young (e.g. 0509-67.5, Kepler) to middle aged (e.g. DEM L71,
N132D) to old (e.g. N49B, Cygnus Loop). The shock ages, pre-shock
densities, post-shock densities, and amount of material swept vary
over orders of magnitude for various objects. However, every remnant
studied shares one curious property: the dust-to-gas mass ratio
inferred for the ambient ISM is lower than what is inferred from
general absorption studies along lines of sight to bright UV sources
in the galaxy and Magellanic Clouds. 

Dust masses are determined from photometric and/or spectroscopic
observations of IR emission from SNRs. When a model spectrum is fit to
data, the overall normalization of the model is provided by only two
parameters: the radiating dust mass and the distance to the
object. For LMC remnants, the distance is well known to be $\sim 50$
kpc. For galactic remnants there is a bit more uncertainty in the
distance, but for Kepler and the Cygnus Loop these uncertainties are
likely less than 25\%. Since SNR shock waves destroy dust in addition
to lighting it up, I calculate the amount of sputtering that has taken
place for a given set of shock parameters and use this to infer the
initial mass in dust warmed by the forward shock. The dust mass
arrived at via this method is the amount of dust present in the {\it
ambient ISM}.

Gas masses swept up by the forward shock can be determined in several
ways. The X-ray emissivity of a hot plasma is proportional to the
product of electron density and the total mass in gas present. If the
density can be determined, then the swept gas mass can be
calculated. Alternatively, if one has a measure of the pre-shock
density in the ambient ISM from optical/UV observations, one can
multiply by the volume to get the total gas mass overtaken by the
blast wave, either for the entire remnant or for a small portion of
the shock.

Dust-to-gas ratios derived in this work are between a factor of 3 and
20 lower than what is expected for the ISM of both the Milky Way and
LMC. Obviously, the dust masses derived from model fits are model
dependent, but unless major new physics is involved, varying
parameters in current models cannot produce enough variation in
results to accommodate conventional dust-to-gas ratios. {\em Even
simple analytic expressions that relate the observed IR luminosity to
the total amount of radiating dust reproduce dust masses within 50\%
of those derived using the more sophisticated modeling techniques
described here.} In recent work on the Cygnus Loop \citep{sankrit10},
we took an alternative approach, determining the path length through
the emitting region for a small section in the NE corner of the
Loop. By comparing model fluxes with observed values, we found that a
low dust-to-gas ratio is required to derive a path length consistent
with FUSE and ROSAT data. The ratio determined in this fashion was
0.38\%, a factor of 2-3 lower than the standard galactic value.

It is possible that some dust remains cold even behind the shock, and
would escape detection in the short-wavelength MIPS and IRS bandpasses
on {\it Spitzer}. I will discuss this possibility more below as it
relates to the total amount of dust produced in SNe. Regarding the
dust-to-gas mass ratio issue, most of the derived values reported in
this work include a detection from the forward shock in the 70 $\mu$m
{\it Spitzer} band. For a significant amount of dust to be present, it
would have to be quite cold to escape detection at 70 $\mu$m. For dust
in the post-shock environment to be cold, it must be contained in
dense clumps. Such densities would imply radiative shocks, which are
easily detectable via optical emission. It is difficult to imagine a
scenario where dust could exist at such cold temperatures while
immersed in the hot plasma of a non-radiative shock.

\subsection{Ejecta Dust in SNRs}

IR and sub-millimeter observations of high redshift galaxies imply
large amounts of dust present in the early universe
\citep{nozawa03}. The source of this dust remains a mystery, and it is
believed that SNe must play a major role in synthesizing dust
grains. \citet{nozawa03} considered Population III SNe and concluded
that CC SNe in the early universe would need to convert 2-5\% of the
mass of the progenitor into dust grains to account for
observations. In their models, a 20 $\msun$ progenitor would produce
$\sim 0.6$ $\msun$ of dust, while a 30 $\msun$ progenitor would yield
$\sim 1.3$ $\msun$. For pair-instability SNe with progenitor masses $>
140 \msun$, 20-80 $\msun$ of dust may be produced. In more recent
work, \citet{cherchneff10} find that their models of dust formation in
ejecta produce about a factor of 5 less dust than in Nozawa et al.,
with a 20 $\msun$ progenitor star only producing $\sim 0.1-0.15$
$\msun$ of new dust. Since we cannot make direct observations of SNe
at that distance, we are left to examine more local SNe to determine
the amount of dust formed in the ejecta.

{\em I find no evidence for a large amount of ejecta dust in any SNR
studied thus far.} Although in some cases the possibility of a large
quantity of cold dust cannot be eliminated, we can conclusively rule
out a significant amount of warm dust in the ejecta in every
object. In Kepler's SNR, the 400-year old remnant of a type Ia
explosion, we find an upper limit of 0.1 $\msun$ based on both 70 and
160 $\mu$m {\it Spitzer} data. Further observations at longer IR
wavelengths will be needed to confirm this result. In 0540-69.3, the
young remnant of a core-collapse SN, we find only $\sim 10^{-3} \msun$
of dust in the inner ejecta, although it is possible that more could
be unshocked.

While it is certainly possible that cold dust could exist in the
ejecta and remain undetected by {\it Spitzer's} instruments if it were
unshocked, it is difficult to conceive of a scenario where dust passes
through the reverse shock and remains cold. This again would require a
radiative shock, which would be easily detectable. Examination of
Figure 4.1 shows no visual evidence of dust morphologically associated
with X-ray emitting ejecta for remnants of CC SNe. To measure this
quantitatively, I use upper limits at 70 $\mu$m for several of the
middle-aged remnants of CC SNe reported in Chapter 4. To get an upper
limit on the amount of ejecta dust, it is necessary to choose a
temperature for the dust. \citet{rho08} report observations of
0.02-0.05 $\msun$ of freshly-formed ejecta dust in Cas A. The
temperatures of the various ``cold'' dust components observed range
from 40-150 K. I adopt a temperature of 50 K for dust in the ejecta
that has been through the reverse shock, and report upper limits to
the dust mass that could be present. I consider the entire volume of
the remnant interior to the inner edge of the forward shock.

For N132D, which is by far the brightest SNR in the IR in the sample
considered, the upper limit on dust at 50 K is $\sim 0.15 \msun$. This
is comparable to the amount predicted from CCSNe by
\citet{cherchneff10}. \citet{france09} used UV data from N132D to
derive a progenitor mass of 50 $\msun$. \citet{nozawa03} predict a
dust formation mass of 2-5\% of the progenitor mass from CC SNe,
leading to a mass of 1-2.5 $\msun$ from a progenitor of 50 $\msun$. In
order for this mass of dust to be present in N132D, and be below upper
limits at 70 and 160 $\mu$m from {\it Spitzer}, it would have to be
below $\sim 35$ K. For N49B and 0453-68.5, both of which are fainter
by nearly an order of magnitude, the upper limit on dust masses are
$\sim 0.03$ and $\sim 0.015 \msun$, respectively. These are only
intended to be rough estimates, as the composition and temperature of
ejecta dust is unknown. Nonetheless, it appears unlikely that current
SNe are the prodigious dust producers that Pop. III SNe may have once
been. It is possible that large amounts of dust are initially {\it
formed} in the ejecta from supernovae, but do not survive passage
through the reverse shock, but dust created and destroyed in this
fashion would not increase the overall dust abundance in a galaxy. The
data from {\it Spitzer} do now allow us to determine this, and
observations with the longer wavelength detectors of {\it Herschel},
with its better resolution and sensitivity than {\it Spitzer} at
wavelengths beyond 100 $\mu$m, are necessary to confirm these results,
as well as examine the possibility of large amounts of cold dust
previously undetected.

\subsection{Future Work}

Typical models for ISM grains assume spherical particles with solid
material filling fractions of unity all made of one material. This is
probably unrealistic, as real grains almost certainly contain some
degree of {\bf porosity (i.e. part of their volume is taken up by
vacuum)} (Okamoto et al. 1994). It is also likely that grain formation
mechanisms allow for {\bf composite grains, i.e. grains made up of
multiple types of materials} (Shen et al. 2009). Grains that are
porous, composite, or both have properties that are different from
their ``compact'' counterparts. Their optical constants are different,
and must be approximated with either an effective medium theory (EMT),
where different optical constants are added together by various means
(see Bohren and Huffman, 1983 for a review of various EMTs), or
something like the multilayer sphere approach developed by
Voshchinnikov \& Mathis (1999), or the discrete dipole approximation
developed by Draine and Flatau (1994). In work done on this front thus
far, I have found an EMT to be the most straightforward way to
approximate the dielectric constants for arbitrary materials. I have
already developed code that is capable of calculating the optical
constants and absorption coeffecients for grains of any arbitrary
combination of astronomical silicates (Draine \& Lee 1984), graphite,
amorphous carbon, and vacuum. This code can be extended, if necessary,
to include other grain ingredients.

One also needs to modify {\bf rates of energy deposition and
sputtering} for porous and composite grains to properly calculate emission
from these grains. The field of porous and composite grains is a
relatively new and wide open field of astrophysics, and opportunities
to test models have in the past been few and far between. SNR
properties inferred from a porous grain model can be significantly
different than those from a compact model, as shown in Chapter 8. The
potential to learn about the nature of dust in the ISM {\bf and} the
nature of SNRs is great. The case of the young SNR 0509-67.5 (0509) is
one example where porous grains may show a better fit to data than do
compact grains.

For supernovae and young SNRs, dust emission is often seen at {\bf
short wavelengths}, such as in IRAC energy bands. All attempts to
obtain a self-consistent model to explain this emission in relation to
that in the mid-IR have been unsuccessful (Blair et al. 2007). The
problem is related to modeling the {\bf stochastic nature of small
grains}, which are primarily responsible for short wavelength
emission. For sufficiently small grains with small heat capacities,
grain temperatures fluctuate between subsequent collisions, and grains
emit most of their energy at the highest temperatures. These small
grains are particularly sensitive to the {\bf energy deposition rates
of impinging particles}, which regulate both the heating and
destruction of grains. These rates are typically assumed to be simple
functions of the properties of the grain and the incoming particle
(Dwek \& Werner 1981), but laboratory data have shown that such an
approach is not valid under extreme conditions, such as those seen in
young SNRs. Since small grains are also {\em by far} the most numerous
grains in the ISM, any model of dust must have more realistic models
for the physics involved. For young SNRs, like 0509, ion temperatures
are sufficiently high compared to electrons (Warren \& Hughes 2004,
Ghavamian et al. 2007) that {\bf heating of grains by ions} is
non-negligible. I have already begun exploring this, but as with the
porous grain calculations, I believe I have only scratched the surface
on this front. It will be absolutely crucial to understand
short-wavelength IR emission before {\it JWST} is launched, as it will
only observe from $\sim 1-20$ $\mu$m, and thus will only be sensitive
to emission from small grains. I will develop and implement more
physically realistic approaches to energy deposition rates,
particularly for small grains.

As grains are sputtered, their constituent elements are {\bf liberated
back into the gaseous phase}. Although dust, by mass, makes up a tiny
fraction of the ISM, the depletion of heavy elements onto dust grains
can be significant. Any attempt to model X-ray spectra as a function
of post-shock ionization timescale should take these increasing
abundances for elements like Fe, Mg, and Si into account (Vancura et
al. 1994). Our code does allow us to follow this liberation, and I
would implement this effect in X-ray spectral modeling of SNRs, while
exploring ways to make such information available to the community as
a whole.

Work remains to be done on both the theoretical and observational side
to better understand SNRs in both the IR and X-ray bands. Models need
to be further tested against archival data and results from laboratory
astrophysics, particularly in regard to heating and sputtering of
porous grains. With the cold phase of {\it Spitzer} now over, its
usefulness in this particular field is minimal, but huge datasets of
observations of the galactic plane and Small and Large Magellanic
Clouds were completed and archived before the cryogen was
exhausted. New observatories are already operating or will be in the
next few years. The diffraction-limited optics of {\it Herschel} (a
factor of 4 better than {\it Spitzer}), combined with its increased
sensitivity at long wavelengths, make it an ideal candidate to
differentiate between various grain models and find any cold dust that
is present in SNRs. At shorter wavelengths, {\it SOFIA}, while not as
sensitive as {\it Spitzer}, will provide much clearer views of bright
galactic objects, allowing tighter constraints on dust destruction
behind the shock. The launch of {\it JWST} in the middle of this
decade will truly usher in a new era of IR astronomy, with a
resolution nearly an order of magnitude better than {\it
Spitzer}. Since {\it JWST} will focus on the 1-25 $\mu$m range of the
spectrum, much work needs to be done in the modeling of emission from
small grains, including those which thermally fluctuate between
collisions. And although we are still more than a decade away from
{\it IXO}, both {\it Chandra} and {\it XMM-Newton} are still
operating, and the amount of archival data from both of these
observatories is vast. The last 6 years of {\it Spitzer} observations
have revolutionized our understanding of the IR window on the
universe, and we have only scratched the surface of what the next 10
years will bring.

\newpage
\clearpage

\appendix

\begin{center}
{\Huge{\bf Appendices}}
\end{center}
\include{appendixa}
\newpage
\include{appendixb}
\newpage
\include{appendixc}

\end{document}

%% file: appendixa.tex
\section{X-ray Emission Measure of the Shocked CSM in Kepler's SNR} 

We have used archival {\it XMM-Newton} data to estimate the emission
measure of the shocked CSM around Kepler's SNR. While Kepler's X-ray
spectrum is dominated by ejecta emitting strongly in lines of heavy
elements such as Fe, Si, and S, Ballet (2002) noted a good match
between an RGS1 image around the O Ly$\alpha$ line and the optical
H$\alpha$ images. This suggests that the low-energy X-ray emission
dominated by N and O originates in the shocked CSM. The Ly$\alpha$
lines of N and O, and the He$\alpha$ line complex of O, are well
separated from strong Fe L-shell lines in the RGS spectra (see Figs. 1
and 2 in Ballet 2002). We used these spectra to arrive at the
following N and O line fluxes: $2.5 \times 10^{-13}$ ergs cm$^{-2}$
s$^{-1}$ for \ion{N}{7} $\lambda$24.779, $7.5 \times 10^{-13}$ ergs
cm$^{-2}$ s$^{-1}$ for the \ion{O}{7} He$\alpha$ complex at $\sim
21.7$ \AA, and $2.8 \times 10^{-12}$ ergs cm$^{-2}$ s$^{-1}$ for
\ion{O}{8} $\lambda$18.967.  The measured \ion{O}{8} $\lambda$18.967
flux may include a non-negligible contribution from \ion{O}{7}
He$\beta$ $\lambda$18.627, as these two lines blend together in the
RGS spectra.

We have used a nonequilibrium-ionization (NEI) thermal plane-parallel
shock without any collisionless heating at the shock front to model N
and O line fluxes. This plane shock model is available in XSPEC as
{\tt vnpshock} model (Arnaud 1996; Borkowski et al. 2001). We assumed
that N and O lines are produced in fast (2000-2500 km s$^{-1}$;
Sollerman et al. 2003) nonradiative, Balmer-dominated shocks with a
mean post-shock temperature of $\sim 5$ keV. The ISM extinction
$E(B-V)$ toward Kepler is equal to 0.90 (Blair et al. 1991), and with
$R_V = A_V/E(B-V) = 3.1$ and $N_H = 1.79 \times 10^{21} A_V$ cm$^{-2}$
(Predehl \& Schmitt 1995), $N_H$ is equal to $5.0 \times 10^{21}$
cm$^{-2}$. We assumed solar abundances for O (from Wilms et al. 2000).
We can reproduce \ion{O}{7} He$\alpha$ and \ion{O}{8} $\lambda$18.967
(+ \ion{O}{7} He$\beta$ $\lambda$18.627) fluxes with an emission
measure $EM = n_e M_{g}$ equal to $10 M_\odot$ cm$^{-3}$ (at the
assumed 4 kpc distance) and a shock ionization age of $10^{11}$
cm$^{-3}$ s. By matching the measured \ion{N}{7} $\lambda$24.779 flux,
we arrive at an oversolar (1.6) N abundance, confirming again the
nitrogen overabundance in the CSM around Kepler.

The shock ionization age cannot be estimated reliably based on the
measured N and O line fluxes alone. These lines are produced close to
the shock front, and their strengths depend only weakly on the shock
age. Shocks with different ages can satisfactorily reproduce the
observed O and N line fluxes, with the emission measure $EM$ inversely
proportional to the shock age in the relevant shock age range from $5
\times 10^{10}$ cm$^{-3}$ s to $4 \times 10^{11}$ cm$^{-3}$ s.
Additional information is necessary to constrain the shock ionization
age, such as fluxes of Ne and Mg lines or the strength of the
continuum at high energies.  Because the X-ray spectrum of Kepler is
dominated by ejecta at higher energies, it is very difficult to
separate CSM emission from the ejecta emission in the
spatially-integrated {\it XMM-Newton} spectra. We note, however, that
shocks with ages of $2 \times 10^{11}$ cm$^{-3}$ s and longer produce
more emission than seen in Kepler.  In particular, the Mg lines and
the high energy continuum are too strong. For ionization ages as short
as $5 \times 10^{10}$ cm$^{-3}$ s, the O He$\alpha$/O Ly$\alpha$ line
ratio becomes excessive (0.36) in the model versus observations
(0.28).  Such short ionization ages may still be plausible if shock
ages are as short as $\sim 80$ yr and postshock electron densities are
$\sim 20$ cm$^{-3}$ as implied by infrared data.

We conclude that a reasonable estimate of the CSM emission measure,
equal to $10 M_\odot$ cm$^{-3}$, is provided by a plane shock with an
age of $10^{11}$ cm$^{-3}$ s. However, the $EM$ is known only within a
factor of 2 because of the poorly known shock ionization age and
uncertain (perhaps spatially varying) absorption. (A 10\%\ range in
absorption listed by Blair et al. (1991) results in 35--40\%\ error in
the derived EM.)  It is also possible, even likely, that X-ray
emission is produced in a variety of shocks with different speeds
driven into gas with different densities. In this case, low energy
X-ray emission would be predominantly produced in slow shocks, while
high energy emission would originate in fast shocks. A single shock
approximation used here might then underestimate the true CSM emission
measure. A future X-ray study based on available high
spatial-resolution {\it Chandra} data will help in resolving such
issues, and will result in better estimates of the CSM emission
measure.

%% file: appendixb.tex
\section{Photoionization Calculation for 0540-69.3}

There are two sources of ionizing photons that can pre-ionize the
material ahead of the shock; ionizing photons produced behind
radiative shocks, and those produced by relativistic electrons in the
form of synchrotron radiation. We examine each of these in
turn. Detailed photoionization calculations would require modeling
that is beyond the scope of this paper, and we present calculations
that are only intended to be rough estimates. Since we do not have a
detailed, multi-dimensional model that provides the shock dynamics
after it encounters the iron-nickel bubble, we here detail the
calculations done in the absence of the bubble, assuming the models of
C05 describe the global shock encountering the inner ejecta. We intend
only for this rough calculation to show that photoionization is a
plausible mechanism for ionizing material out to $8''$.

First, it is necessary to determine the amount of ionizing radiation
emergent from behind the shock. \citet{shull79} give emergent photon
number fluxes per incoming hydrogen atom as a function of shock
speed. Since we know both the density and the shock speed in 0540 as a
function of time for a given density profile, we are able to calculate
the number of ionizing photons emerging from the shock over the
lifetime of the remnant. Here we consider the $m=1.06$ case. We count
all photons with energies above 13.6 eV as ionizing. However, Shull \&
McKee only considered shocks up to 130 km s$^{-1}$. By calculating the
cooling time from equation (2), we see that shocks in 0540 are
radiative up to speeds of over 150 km s$^{-1}$. In order to
extrapolate the numbers given in Shull \& McKee, we use figure 13 of
\citet{pun02}, and assume a single constant factor as the relationship
between the total number of H$\alpha$ photons and the total number of
ionizing photons. We then simply integrate the total number of
ionizing photons throughout the lifetime of the nebula.  We exclude
early times when densities were high enough that recombination times
were shorter than the age of the remnant (about the first 450
years). Using the same conditions as were used above for modeling the
nebula, we find that photoionization from radiative shocks can ionize
0.53 $\msun$.

Next, we calculated the ionizing flux from the synchrotron nebula
itself. We used the optically determined synchrotron power-law of
$\alpha = -1.1$, and considered photons from the Lyman alpha limit up
to 1 keV, though the choice of the upper limit has little effect due
to the steep drop of the synchrotron spectrum. In order to integrate
the luminosity of the nebula over time, it was necessary to use the
time evolution power-law index of $l=0.325$ \citep{reynolds84}. We
considered the emission from the nebula from after the time that
recombinations were important up through the presumed age of the
remnant (450-1140 yrs.) We find enough ionizing photons to ionize 0.21
$\msun$.

We then calculated how far out the ionization front would extend, to
see if this could account for the [O III] emission at $8''$ observed
by Morse et al. Using $m=1.06$ as the power-law index for the ejecta
density profile, the relation between the mass and radius of the
ionization front to that of the shock front can be written as

\begin{equation}
({M_{if}\over M_{sh}})^{0.515} = {R_{if}\over R_{sh}},
\end{equation}

where $M_{if}$ is the mass ionized by both mechanisms, plus the mass
swept up during the early stages of the remnant when recombinations
were occuring, and $M_{sh}$ is the mass swept by the shock, given
above as 0.75 $\msun$. With $M_{if} = 1.25 \msun$ and $M_{sh} = 0.75
\msun$, we find a ratio of $R_{if}$ to $R_{sh}$ of 1.3. While this is
not quite enough to account for the observed [O III] emission at 1.8
pc, this is almost certainly an underestimate of the amount of
photoionized material.

The same calculations for the case of a flat density profile yield the
following values. UV photons from the radiative shocks can photoionize
0.83 $\msun$, while the synchrotron photons from the nebula can ionize
0.18 $\msun$ (the difference in this number is due to the fact that
the recombination timescale is slightly longer for the higher
densities involved in this case, thus fewer photons are included in
the final photon count). The shock itself sweeps up 0.95 $\msun$, and
the relation between the mass interior to the ionization front and the
mass interior to the shock front is given by 

\begin{equation}
({M_{if}\over M_{sh}})^{1/3} ={R_{if}\over R_{sh}}. 
\end{equation}

We find that the ionization front is 1.2 times
farther out than the shock front. Again, while this is not enough to
account for what is observed, it can be considered a lower limit.

As a possible resolution to this, we return to the issue of heavy
element abundances. The calculations above assume standard solar
abundances, but, one would clearly expect the shock encountering the
ejecta to be overtaking material that is higher in metallicity than
solar. If the ejecta that the shock is running into is enriched in
helium and other heavier elements, more mass can be ionized per
ionizing photon (differences in ionization potential
notwithstanding). Since, for the case of $m=1.06$, a modest factor of
${\sim 2}$ in the amount of shock ionized mass would account for the
emission seen at $8''$, this is an entirely plausible explanation.

%% file: appendixc.tex
\section{Spherical Model of 0540-69.3}

We include here the results from our spherically symmetric
model. Although these results indicate that such a model is not able to
account for line emission, it was nonetheless an important starting
point for our more complete models.

In order to model the PWN, it is necessary to determine the inner
ejecta density profile. \cite{matzner99} examined the relationship
between the progenitor of a core-collapse supernova and the resulting
density distribution of the ejecta. They find that core-collapse SNe
lead to density profiles that are best fit by two components, an inner
component that is relatively flat, and an outer component that is
extremely steeply dropping. In the case of a red supergiant (RSG), the
flat inner ejecta correspond to the mass contained in the helium core
of the progenitor star, a few solar masses of material. In
approximating these results for the cases of type Ib/c and type IIP
supernovae, C05 uses the expression $\rho_{SN} = At^{-3}(r/t)^{-m}$,
where $m=0.0$ and $1.06$ for the inner ejecta of type IIP and type
Ib/c SNe, respectively. He concludes that 0540 is the result of an
explosion of a Wolf-Rayet star, and thus should have little or no H in
the inner ejecta. However, in light of recent optical observations
that have detected H lines in the inner ejecta
\citep{serafimovich04,morse06}, it is now believed \citep{chevalier06}
that 0540 is a type IIP, the result of a red supergiant.

However, the power-law approximations of C05 do not take into account
any mixing of ejecta. Even if the progenitor star did explode as a
type IIP, any mixing of ejecta would steepen the power-law index from
a flat distribution to one that declines as a function of radius. We
therefore consider values of $m$ of both 0 and 1.06 here.

We assume the standard picture of a pulsar emitting magnetic-dipole
radiation at the spin frequency, slowing down with a constant
braking index, $n$, defined by ${\dot \Omega}
\propto -\Omega^{n}$.  Then the total pulsar energy loss ${\dot E}(t)$
is given by

\begin{equation}
{\dot E(t)} = {{\dot E}_0 \over (1 + {t\over \tau})^{(n+1)/(n-1)}}
\label{edot}
\end{equation}

where $\tau$ is a slowdown timescale related to the characteristic
time $t_{\rm ch} \equiv P/2{\dot P}$ by 

\begin{equation}
\tau = {2t_{\rm ch}\over n-1} -t.
\end{equation}

Several different values for the braking index have been reported in
recent years; we adopt the most recent measurement of $n=2.14$
\citep{livingstone05}. Assuming an age of $t= 1140$ yr, $P=50$ ms,
${\dot P}=4.8 \times 10^{-13}$ s s$^{-1}$, and characteristic time
$t_{\rm ch} = 1655$ yr, we find $\tau = 1770$ yr.  We assume a current
pulsar spindown energy input of ${\dot E}=1.5 \times 10^{38}$ ergs
s$^{-1}$. From this we can calculate ${\dot E_{0}}$ according to
Equation~\ref{edot}.

Using the X-ray determined radius of $5''$, or approximately 1.2 pc,
we apply the model of C05 for the accelerating PWN bubble driven into
the cold ejecta. We first consider a model with a perfectly flat inner
ejecta density profile, i.e. $m=0$. The model yields a shell velocity
$V_{\rm shell}$ that is currently 1170 km s$^{-1}$, with a shock
velocity $V_{\rm shock}$ (that is, the difference in the shell
velocity and the free-expansion velocity of the ejecta) of 150 km
s$^{-1}$. The current pre-shock density of the ejecta, $\rho_{0}$, is
$9.2 \times 10^{-24}$ g cm$^{-3}$, and the shock has swept up a total
mass in gas, $M_{\rm swept},$ of 0.95 $\msun$. For the $m=1.06$ case,
we find a somewhat higher shell and shock velocity, as would be
expected since the shell is encountering less dense material as it
expands, relative to $m=0$. We find $V_{\rm shell} = 1200$ km s$^{-1}$
and $V_{\rm shock} = 190$ km s$^{-1}$, with $\rho_{0} = 4.7 \times
10^{-24}$ g cm$^{-3}$ and $M_{swept} = 0.75$ $\msun$.

As we will show, the data favor the case of $m=1.06$, and in fact
argue for an even steeper density profile. A flat distribution would
overpredict certain optical lines, as discussed below. In addition, we
show in Appendix B a rough estimate of the amount of ionizing
radiation available (both thermal and synchrotron) 
to produce the [O III] halo seen out to $8''$. 
For the case of $m=0$, we need nearly 5 times more ionizing
photons to account for the material seen at $8''$. For $m=1.06$, we
only need a factor of $\sim 2$. 
While our estimates are probably only good to a factor of 2, the models
clearly prefer steeper density profiles.

Line strengths can also help distinguish between ejecta density
profiles. \citet{chevalier92} investigated the cooling time of the
post-shock gas in an SNR. For the case of $m=0.0$, the density ahead
of the shock is high enough that the cooling times for the remnant are
short compared with the age of the remnant. Enhancements in heavy
element abundances shorten the cooling times further. Because of this,
the shock quickly becomes radiative, and a fast ($\sim 150$ km
s$^{-1}$), radiative shock will significantly overpredict several
lines, including [O III] and [Fe VII]. It is possible that Fe is over
abundant, but then the observed [Fe II] IR line would have to come
from somewhere else.

As a resolution to this problem, we explore the effect of different
density profiles on the power radiated in lines behind the shock from
shocked gas in the process of cooling. Assuming \citep[as
in][]{mckee87} that the cooling curves of \cite{raymond76} can be
approximated as $\Lambda \propto T^{-1/2}$, we use the following
expression for the radiated power from the cooling layer behind the
shock:

\begin{equation}
P \propto \int_{shell} \rho_e \rho_H \Lambda(T) dV.
\end{equation}

Since the models of C05 give the density of material entering the
shock, we were able to numerically integrate the radiated power over
the thickness of the cooling layer, where we define the limits of
integration of the cooling layer as the thickness of the layer in
which the gas cools from its immediate post-shock temperature down to
10$^{4}$ K. In terms of relative power, the $m=1.06$ model radiated
about 45\% less power. We also ran a model with $m=2.0$, and found a
factor of about 3.5 less energy radiated. We do not use this model to
favor a particular value of $m$, only to demonstrate that any mixing of
the inner ejecta, which would likely lead to a value of $m$ for the
average density greater than 0, would reduce the amount of emission
radiated in lines. 

The spherically symmetric model is thus insufficient to describe the
data in two ways. Densities are not high enough to account for
observed optical and IR lines, and fast radiative shocks would
overpredict lines that are not seen, such as [O III] and [Fe VII]. Our
model discussed in the main text provides a potential solution to both
problems in the form of an iron-nickel bubble in the inner
ejecta. Because the fast shock initially propagated through the
low-density medium of the bubble, [O III] and [Fe VII] lines should
not be strong, and the passage of the shock through the high-density
bubble wall would provide the dense environment necessary for lines
that are observed.